\documentclass[twocolumn,aps,prb,showpacs,amsmath,superscriptaddress,nolongbibliography,notitlepage]{revtex4-1}
\usepackage{amssymb}
\usepackage{mathrsfs}
\usepackage{graphicx}
\usepackage{float}
\usepackage[caption=false]{subfig}
\usepackage[pdftex,colorlinks=red]{hyperref}
\usepackage{verbatim}
\usepackage{txfonts}
\usepackage{epstopdf}
\usepackage[normalem]{ulem}
\usepackage{xcolor}
\usepackage{bm}
\usepackage[utf8]{inputenc}

\usepackage{braket}
\usepackage{url}
\usepackage{amsmath}

\definecolor{brightmaroon}{rgb}{0.92, 0.03, 0.08}
\def\purp{\textcolor{purple}}

\def\scite#1{[\onlinecite{#1}]}

\bibpunct{[}{]}{,}{n}{}{}

\begin{document}

\title{Topological Non-Hermitian skin effect}
\author{Rijia Lin}
\thanks{These authors contributed equally to this work}
\affiliation{Guangdong Provincial Key Laboratory of Quantum Metrology and Sensing $\&$ School of Physics and Astronomy, Sun Yat-Sen University (Zhuhai Campus), Zhuhai 519082, China}
\author{Tommy Tai}
\email{tommytai@mit.edu}
\thanks{These authors contributed equally to this work}
\affiliation{Department of Physics, Massachusetts Institute of Technology, Cambridge, MA, USA}
\affiliation{Department of Physics, National University of Singapore, Singapore 117542}
\author{Mengjie Yang}
\affiliation{Department of Physics, National University of Singapore, Singapore 117542}
\author{Linhu Li}\email{lilh56@mail.sysu.edu.cn}
\affiliation{Guangdong Provincial Key Laboratory of Quantum Metrology and Sensing $\&$ School of Physics and Astronomy, Sun Yat-Sen University (Zhuhai Campus), Zhuhai 519082, China}
\author{Ching Hua Lee}\email{phylch@nus.edu.sg}
\affiliation{Department of Physics, National University of Singapore, Singapore 117542}

\date{\today}

\begin{abstract}
This article reviews recent developments in the non-Hermitian skin effect (NHSE), particularly on its rich interplay with topology. The review starts off with a pedagogical introduction on the modified bulk-boundary correspondence, the synergy and hybridization of NHSE and band topology in higher dimensions, as well as, the associated topology on the complex energy plane such as spectral winding topology and spectral graph topology. Following which, emerging topics are introduced such as non-Hermitian criticality, dynamical NHSE phenomena, and the manifestation of NHSE beyond the traditional linear non-interacting crystal lattices, particularly its interplay with quantum many-body interactions. Finally, we survey the recent demonstrations and experimental proposals of NHSE.\\
Note: this arxiv version is an experimental ``living'' review article that will be continually updated to keep abreast of recent developments in the field. Please cite the published version of this review at \url{https://link.springer.com/article/10.1007/s11467-023-1309-z}.
\end{abstract}

\maketitle
\tableofcontents

\section{Introduction}

The past two decades have witnessed a burgeoning interest in the intriguing properties of non-Hermitian Hamiltonians~\cite{Bender1998nonH,bender2007making}. 
Compared with Hermitian Hamiltonians with real eigenenergies, which represent unitary time evolution of isolated quantum systems,
non-Hermitian Hamiltonians provide an effective physical description of many non-conservative systems,
including open quantum systems \scite{Rotter2009non},
solid-state systems with finite lifetime induced by interactions~\cite{yoshida2018non,shen2018quantum,yamamoto2019theory},
and  acoustic/photonic systems with gain and loss~\cite{ma2016acoustic,cummer2016controlling,zangeneh2019active,feng2017non,el2018non,longhi2018parity,ozawa2019topological}.
Most early investigations of non-Hermitian Hamiltonians have been devoted to systems with parity-time (PT) symmetry~\cite{Bender1998nonH,feng2017non,el2018non,xiao2020non}.
As a specific form of pseudo-Hermiticity~\cite{mostafazadeh2002pseudo}, PT symmetry allows a non-Hermitian Hamiltonian with balanced gain and loss to have a real spectrum, enabling a stable unitary time-evolution for eigenstates of the system.
On the other hand, a PT symmetric Hamiltonian does not guarantee that the symmetry is also possessed by each of the eigenstates.
In the so-called PT-broken phase,  a PT symmetric Hamiltonian can have pairs of complex-conjugate eigenenergies, whose eigenstates can be transformed into each other through a PT symmetry operation.
The transition between PT-symmetric and PT-broken phases is accompanied by the emergence of exceptional points (EP) for the system, 
a type of spectral degeneracies of non-Hermitian systems 
where two or more eigenstates coalesce into one and the Hamiltonian matrix becomes rank-deficient~\cite{gunther2007projective,Rotter2009non,regensburger2012parity,miri2019exceptional,meng2024exceptional,torres2019perspective,sayyad2022realizing}.
Besides providing a signal of PT transition, EPs are physically interesting in their own good,  
as they can lead to various exotic features of non-Hermitian systems, such as
unidirectional invisibility~\cite{lin2011unidirectional,feng2013experimental} and enhanced sensitivity~\cite{jan2014enhancing,liu2016metrology,hodaei2017enhanced,chen2017exceptional} in photonics, 
and unusual topological~\cite{dembowski2001experimental,gao2015observation,Mailybaev2005geometric,Lee2016nonH,leykam2017edge,Yin2018nonHermitian,shen2018topological,li2019geometric} and dynamical properties~\cite{Hu2017EP,Hassan2017EP} for non-Hermitian Hamiltonians encircling an EP.

Among the many fascinating aspects of non-Hermitian Hamiltonians, one noteworthy route that triggers an enormous number of extensive studies over the years 
is how non-Hermiticity affects the topological properties of lattice systems.
In the study of topological phases of matter, a most fundamental principle is the bulk-boundary correspondence (BBC), i.e. the number of edge states under the open boundary conditions (OBCs) has an one-on-one correspondence to a topological invariant defined for the bulk states under the periodic boundary conditions (PBCs)~\cite{hasan2010colloquium,qi2011topological,bernevig2013topological}.
In early studies, it was commonly believed that the BBC still holds for non-Hermitian systems, despite that non-Hermitian topological phases may behave differently from their Hermitian counterparts~\cite{rudner2009topological,hu2011absence,esaki2011edge,diehl2011topology,zhu2014pt,malzard2015topologically,harter2016mathcal}.
However, in 2016, T. E. Lee noticed that non-Hermiticity can break conventional BBC for topological phases, evidenced by an inconsistency between the spectra of a system under the periodic and open boundary conditions (PBCs and OBCs)~\cite{Lee2016nonH}. 
This phenomenon was firstly understood as a consequence of a half-integer winding number for non-Hermitian Hamiltonians encircling an
EP in momentum space~\cite{Lee2016nonH,leykam2017edge,Yin2018nonHermitian}, but this winding number, whose definition involves only PBC Hamiltonians, does not predict the number of topological edge states under the OBCs accurately when the system is closed to a topological phase transition.
Another interpretation is to consider the evolution of the system from PBCs to OBCs, which changes the topological structure of the spectrum by passing through a series of EPs~\cite{xiong2018does}, 
but it does not provides a solution to restore the BBC either. 
This problem is eventually solved in 2018 by the sensational formulation of the non-Hermitian skin effect (NHSE),
under which all eigenstates are spatially localized at the boundary of a system under the OBCs~\cite{alvarez2018non,yao2018edge}.
Independently discovered by two research groups~\cite{alvarez2018non,yao2018edge}, the NHSE has motivated the so-called non-Bloch band theory,
recovering a generalized BBC for non-Hermitian Hamiltonians~\cite{yao2018edge,yokomizo2019non,lee2020unraveling,fu2022degeneracy}. 
A parallel formulation of the non-Bloch Hamiltonian is to consider the PBC-OBC spectral evolution with a tunable hopping strength across the boundary, effectively giving rise to an imaginary flux to the system~\cite{xiong2018does,lee2019anatomy}. Ref.~\cite{lee2019anatomy}, which pre-dated the detailed study of the generalized Brillouin zone (GBZ) of 1D models with arbitrarily far hoppings, also developed a criterion for the presence of topological zero modes in generic 1D 2-band PH-symmetric models with arbitrarily far hoppings, requiring only information on the PBC dispersion poles/zeros, and not the GBZ.
It is also worth noticing that the eigenstates of non-Hermitian Hamiltonians are not necessarily mutually orthogonal, and the biorthogonal condition needs to be taken into account to obtain a proper topological invariant from bulk states~\cite{kunst2018biorthogonal}. A recent work demonstrates a numerically efficient relationship between EPs and biorthogonality~\cite{zou2022measuring}.

Ever since its discovery, tremendous progress has been made in various aspects of NHSE that differ drastically from Hermitian systems, which we will review at length in this article. 
To mention a few examples, essentially being a one-dimensional (1D) directional phenomenon, NHSE is responsible for several unidirectional physical effects~\cite{song2019non, 
Wanjura2020,wanjura2021correspondence,xue2021simple, 
longhi2022self,
xue2022non
};
the massive accumulation of eigenstates at the boundaries hints an extreme sensitivity to weak boundary couplings~\cite{budich2020sensor,guo2021exact,li2021impurity} and its associated abnormal critical phenomenon~\cite{li2020critical,yokomizo2021scaling,liu2020helical};
in two-dimension (2D) or higher, richer geometric structures of defects and boundaries also lead to different varieties of NHSE~\cite{sun2021geometric,bhargava2021non,schindler2021dislocation,zhang2022universal,panigrahi2022non,manna2022inner}.
Other than these theoretical advances, NHSE has also been realized and examined in various experimental platforms, including RLC circuit lattices~\cite{helbig2020generalized,hofmann2020reciprocal,liu2021non,zou2021observation,zhang2021observation,shang2022experimental}, acoustic~\cite{zhang2021observation,zhang2021acoustic,gao2022non} and photonic lattices~\cite{weidemann2020topological,song2020two}, single-photon quantum walks~\cite{xiao2020non,xiao2021observation,wang2021detecting}, mechanical metamaterials~\cite{Brandenbourger2019,ghatak2020observation}, and cold atoms where NHSE is manifested in momentum space~\cite{gou2020tunable,liang2022observation}.
NHSE model design has also been aided by other seemingly unrelated fields, such as machine learning used to classify topological phases~\cite{scheurer2020unsupervised,yu2021unsupervised} and optimize higher-order NHSE circuit measurements~\cite{shang2022experimental}, and an interesting analogy of the NHSE problem with classical electrostatics that allows for reverse engineering NHSE Hamiltonians of any desired spectra and state localization profile~\cite{yang2022designing}. 

In this review, after introducing the generalized BBC of the non-Bloch band theory for systems with NHSE,
we will further discuss how NHSE affects localization of topological edge states.
That is, beyond modifying the BBC of topological phases, NHSE can also change the  direction of localization of topological edge states as it provides another localization mechanism for eigenstates~\cite{oztas2019schrieffer,zhu2021delocalization,cheng2022competition,okuma2022non}.
This phenomenon is also associated with the single (defective) edge state reported in Ref.~\cite{Lee2016nonH}, which can be viewed as two exceptionally degenerate edge states in the thermodynamic limit~\cite{Yin2018nonHermitian,wang2020defective}. 
A more intriguing development of the interplay between NHSE and conventional topological properties is their simultaneous effect of NHSE  in 2D or higher dimensions, which leads to a new class of hybrid skin-topological higher-order boundary modes, living in boundaries with a co-dimension of two or higher~\cite{Lee2019hybrid,li2020topological,li2022gain,zhu2022hybrid}. It is found later that this hybrid skin-topological effect can be regarded as a special kind of a new class of higher-order NHSE, where conventional topological protection is not a necessary condition for these higher-order boundary modes~\cite{kawabata2020higher,okugawa2020second,fu2021nonH}. 
Besides affecting conventional topological properties, NHSE itself has also been found to be a signature of a spectral winding topology unique in non-Hermitian systems with complex spectrum~\cite{zhang2020correspondence,borgnia2020nonH,okuma2020topological}, which is the second main theme of this review.
A quantized response of the spectral winding is later found in the steady-state response of such systems, without involving the linear response theory based on many-body ground states~\cite{Li2021,liang2022anomalous}.
On the other hand, nontrivial spectral winding is also found during the investigation of non-Hermitian quasicrystal, 
the spectral winding topology of which is found to not necessarily correspond to NHSE in real space, but in reciprocal space for systems without translational symmetry~
{\cite{longhi2019topological,jiang2019interplay,longhi2019metal,zeng2020winding,zeng2020topological,liu2020nonH,zhai2020many,cai2021boundary,liu2021exact1,liu2021localization,claes2021skin,longhi2022non4}}. 
In \cite{zhang2021tidal}, interesting mathematical relations are further found between the NHSE, knot theory and spectral winding topology in the context of 3D exceptional metals. In the later half of this article, we also review how the NHSE can non-perturbatively modify state dynamics as well as signatures of criticality, in both single-body and interacting many-body systems.
We also note that while this review will mainly focus on theoretical aspects, we will devote the last section to reviewing experimental demonstrations and proposals of NHSE-related phenomena.

\section{Modified band topology from the NHSE}
Topological BBC plays a central role in describing topological phases of matter, as it associates topological boundary modes with bulk topological invariants, analogous to the order parameters for conventional quantum phases and phase transitions.
Without a doubt, the breakdown of BBC in non-Hermitian systems has attracted great attention since its discovery, and many efforts have been made in recovering it with different methods.
For examples, in 1D non-Hermitian systems,
BBC have been restored between topological winding numbers and different properties of the systems,
including zero-energy edge states in an semi-infinite system~\cite{Yin2018nonHermitian},
a singular spectrum obtained from a singular-value decomposition~\cite{herviou2019defining}, 
and spatial growth of the bulk Green function~\cite{zirnstein2021bulk}, to mention a few.
In this section, we will introduce the method of non-Bloch band theory and NHSE, 
which recovers a modified BBC between boundary states under OBC and topological invariants defined in a so-called generalized Brillouin zone (GBZ) - a complex continuation of the conventional Brillouin Zone.

\subsection{NHSE and non-Bloch band topology: minimal non-Hermitian SSH model}

\begin{figure}
    \includegraphics[width=1\linewidth]{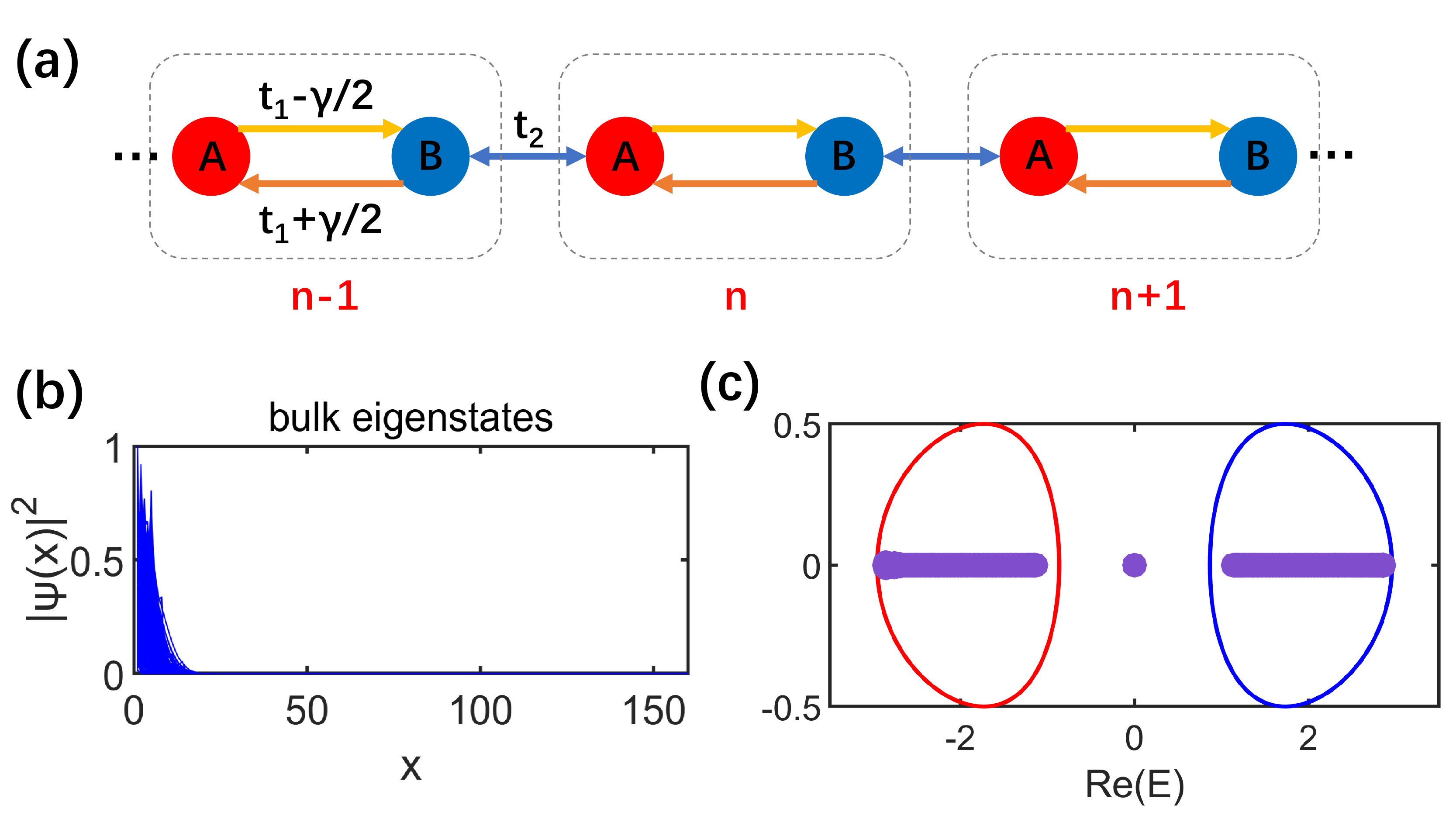}
    \caption{(a) Schematic of the non-Hermitian SSH model.
    (b) Squared wavefunction amplitude $|\psi(x)|^2$ plots, showing that the bulk modes are localized near the left boundary.
    (c) Energy spectra of the non-Hermitian SSH model with open(purple) and periodic(red and blue solid lines denote different bands) boundary conditions.
    Parameters: $t_2=2,\gamma=1,t_{1}=1;L=160$. Figures reproduced from the model in Ref.~\cite{yao2018edge}.} 
    \label{fig:nonH-SSH}
\end{figure}

As a starting point, we shall first give a brief review of NHSE in a non-Hermitian Su-Schrieffer-Heeger (SSH) model~\cite{su1979} [see Fig.\ref{fig:nonH-SSH}(a)], which is a minimal model for demonstrating the generalized bulk-boundary correspondence and non-Bloch band theory for non-Hermitian band topology, by reproducing the results in Ref.~\cite{yao2018edge}.
A comprehensive study of non-Hermitian extensions to another representative 1D topological model, namely the Creutz model~\cite{Creutz1999}, can be found in Ref.~\cite{liang2022topological}.
The real-space and Bloch Hamiltonians of the non-Hermitian SSH model are respectively given by
\begin{eqnarray}
    H_{\rm nonH-SSH}&=&\sum_n\left[(t_1+\gamma/2)\hat{a}^\dagger_n\hat{b}_n+(t_1-\gamma/2)\hat{b}^\dagger_n\hat{a}_n\right]\nonumber\\
    &&+\sum_n \left[t_2\hat{b}^\dagger_n\hat{a}_{n+1}+h.c.\right]
\end{eqnarray}
with $\hat{\alpha}^\dagger_n$ the creation operator acting on $\alpha$ sublattice of the $n$th unit cell, and
\begin{eqnarray}
    h_{\rm nonH-SSH}(k)&=&h_x\sigma_{x}+\left(h_{y}+i\frac{\gamma}{2}\right)\sigma_{y},\label{eq:H_SSH}
\end{eqnarray}
where $h_{x}=t_{1}+t_{2}\cos k,h_{y}=t_{2}\sin k$,and $\sigma_{x,y}$ are the Pauli matrices.
Non-Hermiticity is introduced through nonzero $\gamma$, which is responsible for non-reciprocity of this system.
Under the OBC, the bulk eigenstates of this non-Hermitian SSH model are localized near the left boundary, known as the "non-Hermitian skin effect" (NHSE).
[see Fig.\ref{fig:nonH-SSH}(b)].
These eigenstates are different from Bloch waves which are eigenstates under PBC and extended in lattice.
OBC and PBC spectra are also qualitatively different, forming lines and loops on the complex energy planes respectively, as shown in Fig.\ref{fig:nonH-SSH}(c).
Thus the gap-closing point of $H_{\rm nonH-SSH}(k)$ (touching of the two complex energy bands) in parameter space may not correspond to that of the open-boundary hamiltonian $H_{\rm OBC}$, indicating the breakdown of bulk-boundary correspondence (BBC).

The difference of this model between OBC and PBC can be understood with a similarity transformation of the Hamiltonian, which maps the model under OBC into a Hermitian one while keeping its eigenvalues unchanged.
To see this, we first write down the eigenequation as 
\begin{eqnarray}
H_{\rm nonH-SSH}\ket{\psi}=E\ket{\psi}, \label{eq:eigenequation}
\end{eqnarray}
with real-space eigenvectors taking the form of $\ket{\psi}=(\psi_{1,A},\psi_{1,B},\psi_{2,A},\psi_{2,B},\cdots,\psi_{L,A},\psi_{L,B})^{T}$.
Under OBC, we can apply a similarity transformation 
\begin{eqnarray}
\bar{H}_{\rm OBC}=S^{-1}H_{\rm nonH-SSH,OBC}S
\end{eqnarray}
with $S={\rm diag}\{ 1,r,r,r^2,r^2,\cdots,r^{L-1},r^{L-1},r^L \}$, $r=\sqrt{ \left| (t_1+\gamma/2) / (t_1-\gamma/2) \right|}$ and $L$ the total number of unit cells of the 1D chain, so that $\bar{H}_{\rm OBC}$ becomes the Hermitian SSH model with intracell and intercell hoppings
given by $\bar{t}_{1}=\sqrt{(t_1-\gamma/2)(t_1+\gamma/2)},\bar{t}_{2}=t_{2}$ respectively.
Therefore, $\bar{H}_{\rm OBC}$ shall have the same bulk spectrum as that of this Hermitian SSH model under PBC ($H_{\rm PBC}$), and possesses extended bulk eigenstates $|\bar{\psi}\rangle$.
The gap-closing point of this Hermitian SSH model is $\bar{t}_{1}=\bar{t}_{2}$, which is also the gap-closing point of $H_{\rm OBC}$.
Moreover, it's now straightforward to see that eigenstates $\ket{\psi}=S\ket{\bar{\psi}}$ of the non-Hermitian SSH model are localized near the boundary, as the similarity transformation of $S$ has an exponentially increasing modifier for unit cells from $n=1$ to $n=L$.

In the non-Hermitian SSH model, a generalized BBC is restored between Hermitian Hamiltonians $H_{\rm OBC}$ and $H_{\rm PBC}$ through the similarity transformation.
However, in general non-Hermitian Hamiltonians have complex OBC spectrum and cannot be transformed into Hermitian ones with a similarity transformation, 
thus we need a more universal approach to establish a generalized BBC of non-Hermitian Hamiltonians.
We first explicitly write down the bulk eigenequation Eq.\eqref{eq:eigenequation} under OBC,
\begin{eqnarray}
\begin{aligned}
t_2\psi_{n-1,B}+\left(t_1+\dfrac{\gamma}{2}\right)\psi_{n,B}=E\psi_{n,A},    \\
\left(t_1-\dfrac{\gamma}{2}\right)\psi_{n,A}+t_{2}\psi_{n+1,A}=E\psi_{n,B},  \label{eq:recursive}
\end{aligned}
\end{eqnarray}
where $E$ is eigenenergy and the index of unit cell $n=1,2,\cdots,L-1$,
with boundary conditions
\begin{eqnarray}
\begin{aligned}
    \left( t_{1}+\frac{\gamma}{2} \right) \psi_{1,B}-E\psi_{1,A}=0,   \\
    \left( t_{1}-\frac{\gamma}{2} \right)\psi_{L,A}-E\psi_{L,B}=0.
\end{aligned}\label{eq:boundary_con_SSH}
\end{eqnarray}
Due to the translation symmetry of the bulk equation, we can take an ansatz of the eigenvectors, $\ket{\psi}=\sum_{j}\ket{\phi^{(j)}}$ 
with $(\phi^{(j)}_{n,A},\phi^{(j)}_{n,B})=z^{n}(\phi^{(j)}_{A},\phi^{(j)}_{B})$,
with $j$ indexes different solutions of $\phi$.
Therefore Eq. \eqref{eq:recursive} becomes
\begin{eqnarray}
\begin{aligned}
    \left[ \left( t_{1}-\dfrac{\gamma}{2} \right) +t_{2}z^{-1} \right] \phi_{B}^{(j)} = E\phi_{A}^{(j)},    \\
    \left[ \left( t_{1}+\dfrac{\gamma}{2} \right) +t_{2}z \right] \phi_{A}^{(j)} = E\phi_{B}^{(j)},
    \end{aligned}\label{eq:phi}
\end{eqnarray}
or
\begin{eqnarray}
    \phi^{(j)}_{A}=\frac{E}{t_1-\gamma/2+t_{2}z_{j}}\phi_{B}^{(j)}, ~
    \phi^{(j)}_{B}=\frac{E}{t_{1}+\gamma/2+t_{2}z_{j}^{-1}}\phi_{A}^{(j)}.
\end{eqnarray}
On the other hand, 
Eq. \eqref{eq:phi} also leads to
\begin{eqnarray}\label{characteristic equation}
    E^2= \left[ \left( t_{1}-\dfrac{\gamma}{2} \right) +t_{2}z^{-1} \right] \left[ \left( t_{1}+\dfrac{\gamma}{2} \right) +t_{2}z \right],
\end{eqnarray}
which has two solutions $z_{1,2}$ satisfying
\begin{eqnarray} 
z_{1}z_{2}=\dfrac{t_{1}+\frac{\gamma}{2}}{t_{1}-\frac{\gamma}{2}}.
\end{eqnarray}
Since $z_{1,2}$ correspond to the same eigenenergy $E$,
a linear combination of $(\phi^{(j)}_{A} \quad \phi^{(j)}_{B})^{T}$ with $(j=1,2)$ is also an eigenvector of the Hamiltonian, given by
\begin{eqnarray}\label{wavefunction}
\begin{pmatrix}
\psi_{x,A}  \\
\psi_{x,B}        
\end{pmatrix}
=
z_{1}^{x}
\begin{pmatrix}
    \phi_{A}^{(1)}  \\
    \phi_{B}^{(1)}
\end{pmatrix}
+
z^{x}_{2}
\begin{pmatrix}
    \phi_{A}^{(2)}  \\
    \phi_{B}^{(2)}
\end{pmatrix}.
\end{eqnarray}
Combining with the boundary conditions of Eq.~\eqref{eq:boundary_con_SSH},
we shall arrive at
\begin{eqnarray}\label{beta}
z_{1}^{L+1}(t_{1}-\gamma/2+t_{2}z_{2})=z_{2}^{L+1}(t_{1}-\gamma/2+t_{2}z_{1}).
\end{eqnarray}
The solution of this equation is nontrival only when $|z_{1}|\neq|z_{2}|$.
For example, if $|z_{1}|>|z_{2}|$, this condition becomes
\begin{eqnarray}
z_{1}^{L+1}(t_{1}-\gamma/2+t_{2}z_{2})=0
\end{eqnarray} 
in the thermodynamic limit ($L\to\infty$).
It is equivalent to $z_{1}=0$ or $t_{1}-\gamma/2+t_{2}z_{2}=0$,
and in this situation there is only one pair of $(z_{1},z_{2})$ in Eq. (\ref{wavefunction}),
which is independent of the system size $L$ and gives only a single solution of eigenenergy $E$.
But obviously the number of the solutions of Eq. (\ref{beta}) should be proportional to the system size $L$,
otherwise the system under OBC would not have continuum bands.  
Therefore we obtain the condition of continuum bands,
\begin{eqnarray}\label{non-Bloch band theory}
    |z_{1}|=|z_{2}|=r \equiv \sqrt{ \left| \frac{t_{1}-\gamma/2}{t_{1}+\gamma/2} \right|}.\label{eq:z1z2}
\end{eqnarray} 

\begin{figure}
    \includegraphics[width=1\linewidth]{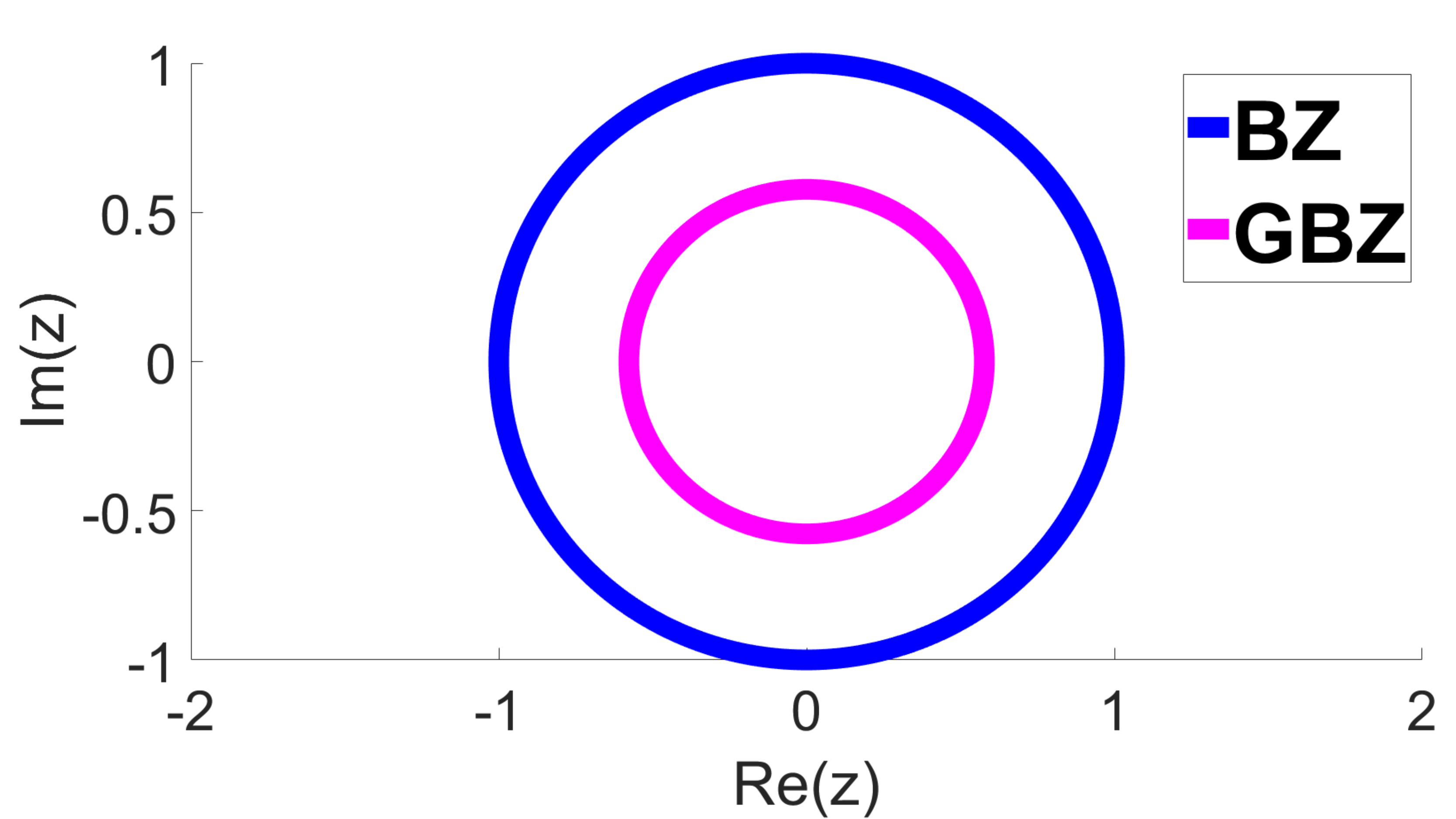}
    \caption{The Brillouin zone (blue loop, a unit circle) and generalized Brillouin zone(pink loop) of the non-Hermitian SSH model, which lies within the Brillouin zone. In general, the generalized Brillouin zone is usually not circular for more general settings.
    Parameters are $t_2=2,\gamma=1,t_{1}=1$. Figure reproduced from the non-Bloch band theory introduced in Refs.~\cite{yao2018edge,yokomizo2019non}.} 
    \label{fig:nonH-SSH_GBZ}
\end{figure}

Eq.~\eqref{eq:z1z2} is also the kernel of non-Bloch band theory~\cite{yokomizo2019non}.
That is, since it only restricts the absolute values of $z_{1,2}$, 
the continuum bands can be given by Eq.~\eqref{characteristic equation} with
\begin{eqnarray}
z=z_{\rm GBZ}=e^{ik}/r,
\end{eqnarray}
with $k$ varying from $0$ to $2\pi$. In other words, 
the phase factor $e^{ik}$ in usual Bloch waves of Hermitian systems is now replaced by $z_{\rm GBZ}$.
Here GBZ stands for the ``generalized Brillouin zone", a concept analogous to the Brillouin zone (BZ) with $z_{\rm BZ}=e^{ik}$, but describing OBC spectrum of non-Hermitian systems (see Fig.\ref{fig:nonH-SSH_GBZ} for a comparison of BZ and GBZ). In the literature, $z_{\rm GBZ}$ is sometimes denoted as $\beta$. It can also be viewed  as a complex deformation of the momentum, $k\rightarrow k+i \kappa$,
with $\kappa=\ln r$ describing the inverse localization length of skin modes~\cite{yokomizo2019non,zhang2020correspondence,lee2019anatomy}. For this model, $\kappa$ is constant, but generically $\kappa$ would be a function of $k$, which we will discuss in the next subsection.

{Recently, the topological phases of the non-Hermitian SSH model and the non-Bloch band theory has been studied in terms of the electronic polarization~\cite{PhysRevResearch.2.043046,masuda2022relationship,masuda2022electronic}.}
Beyond OBC systems, the GBZ has also be used to describe systems with other boundary conditions such as a domain wall~\cite{deng2019non}, on-site disorder~\cite{liu2022modified} or a strong local impurity~\cite{li2021impurity}.

\subsection{Non-Bloch band theory for general cases}
For more general cases, the generalized Brillouin zone may not be a unit circle in the complex plane, i.e. the imaginary flux $\kappa$ may be $k$-dependent, meaning that different PBC eigenstates correspond to OBC skin states with different decay rates.
In such cases, it is impossible to find a similarity transformation to remove the NHSE of the whole system, as each $k$-eigenstate corresponds to a different localization length.
Nevertheless, the GBZ $z_{\rm GBZ}$ can still be determined through an analyzing the characteristic polynomial of the Hamiltonian.
Here we shall briefly summarize the method
for a general 1D single-band non-Hermitian system with non-reciprocal hoppings, described by the Hamiltonian
\begin{eqnarray}
H=\sum_{x=1}^L\sum_{j=-m}^n t_j\hat{c}^\dagger_x\hat{c}_{x+j},\label{eq:single-band}
\end{eqnarray}
with $\hat{c}_x$ the annihilation operator acting on site $x$, and $N$ the total number of lattice sites.
By solving the eigenvalue equation 
$H\Psi_s(E)=E\Psi_s(E)$ with an ansatz for an eigenstate with eigenenergy $E$ as
\begin{eqnarray}
\Psi_s(E)=(z_s,z_s^2,...,z_s^L)^T,
\end{eqnarray}
we obtain a characteristic polynomial equation for $H$ with a given $E$,
\begin{eqnarray}
f(z_s,E):=\sum_{j=-m}^n t_jz_s^j-E=H(z_s)-E=0,
\end{eqnarray}
where $H(z_s)$ is the same as the Hamiltonian in momentum space for $z_s=e^{ik}$, and there shall be $m+n$ solutions of $z_s$ marked by its $s$ index.
Note that so far we only assume $x$ starts from $1$, but have not considered boundary conditions of the system.
As a matter of fact, these solutions of $z_s$ correspond to eigenstates under the semi-infinite boundary conditions with $x$ ranging from $1$ to $L$ and $L$ tends to infinity, provided $|z_s|<1$~\cite{okuma2020topological}.

To obtain the GBZ of the system, one needs to take into account OBC for the system, namely setting $t_j=0$ in Eq. \eqref{eq:single-band} for $x+j>L$ or $x+j<1$.
An eigensolution of the OBC Hamiltonian can be constructed from a linear combination of $\Psi_s(E)$, $\Psi(E)=\sum_s c_s\Psi_s(E)$, which also needs to satisfy the OBC. Since $L$ can be arbitrarily large, in order for there to always be at least two surviving eigensolutions, we arrive at the conclusion that a nontrivial solution of $\{c_s\}$ exists only when $|z_m(E)|=|z_{m+1}(E)|$ for some $m$, which limits the allowed values of eigenenergies $E$~\cite{yokomizo2019non,lee2019anatomy,zhang2020correspondence}. These allowed values are precisely those that form the OBC spectrum.

Although the above conclusion does not directly give us an analytical relation between the imaginary flux $\kappa(k)$ and the momentum $k$, but it already provides an efficient scheme for obtaining the GBZ numerically.
That is, for any complex value $E$, the solutions $z_m$, $z_{m+1}$ of the characteristic polynomial equation $f(z_s,E)=0$ form a pair of GBZ solutions if they have the same absolute value.
OBC spectrum of the system is then reproduced by collecting all values of $E$ satisfying this condition.
Alternatively, an analytical solution of the GBZ can be obtained by introducing a concept of auxiliary GBZ and solving the corresponding algebraic equation~\cite{yang2020non}. The Hamiltonian $H(k+i\kappa(k))$, which is defined on the GBZ and which thus no longer experiences the NHSE~\cite{lee2020unraveling,tai2022zoology}, is known as the surrogate Hamiltonian. We also mention in passing that the difficult problem for solving for the GBZ and surrogate Hamiltonian may be circumvented by mapping the NHSE problem onto an equivalent electrostatics problem, which we will disucss in Sect.~\ref{sec:electro}.
{A modified GBZ theory has also been introduced to describe NHSE and bulk-boundary correspondence in disordered systems~\cite{ZHANG2023}.}

\subsection{Transition of topological localization direction and half quantized winding numbers}
In addition to modifying the BBC, NHSE provides another mechanism to induce exponential localization of eigenmodes, and hence can alter the localization properties of topological edge modes~\cite{chen2019finite,oztas2019schrieffer,zhu2021delocalization,rafi2022unconventional,cheng2022competition}, as already experimentally demonstrated in a mechanical setting~\cite{wang2022non1,wang2022extended}.
To see this, let us revisit the non-Hermitian SSH model in Eq. \eqref{fig:nonH-SSH}, which is topologically nontrivial when $|t_1|<\sqrt{(t_{2})^2+(\gamma/2)^2}$. Within this parameter regime, a pair of zero-energy edge modes appear and localize at the two ends of the 1D chain [see Fig.\ref{fig:topo_skin_local}(a)], analogous to the Hermitian counterpart.
When increasing $t_{1}$, non-Hermiticity becomes increasingly dominant, and one of the topological edge modes becomes 
extended at $t_1=1.5$ [Fig.\ref{fig:topo_skin_local}(b)], and 
localized at the same edge as the other for $t_1>1.5$ [Fig.\ref{fig:topo_skin_local}(b)].

This transition of localization direction can be understood from a competition between NHSE localization and topological localization~\cite{zhu2021delocalization,cheng2022competition}. 
This is because here NHSE induces a unidirectional exponential localization of eigenmodes, and topological localization induces a bidirectional one.
Therefore one topological edge mode will be pumped to the other end of the 1D chain when the NHSE dominates.
A transition point thus emerges where the two localization mechanisms are balanced, and this topological edge mode becomes delocalized~\cite{longhi2018non}.
Generally speaking, this transition occurs if the eigenenergy of this edge mode coincides with the PBC spectrum of the same system~\cite{zhu2021delocalization,tang2022symmetry}.
In the non-Hermitian SSH model, these topological edge states are pinned at zero energy as required by sublattice symmetry,
\begin{eqnarray}
\sigma_zh_{\rm nonH-SSH}(k)\sigma_z=-h_{\rm nonH-SSH}(k).
\end{eqnarray}
As a result, the localization transition accompanies an EP of the two bands, which also represents a transition between separable and inseparable bands, as shown in Fig.\ref{fig:topo_skin_local}(c) to (e).

From the topological aspect, this transition can also be related to a half-quantized winding number~\cite{Yin2018nonHermitian,zhu2021delocalization}.
In 1D Hermitian systems, a sublattice symmetry ensures a $\mathbb{Z}$-class topology characterized by an integer winding number $\nu$, 
which counts the number of times that the Hamiltonian vector $\mathbf{h}(k)$ in the momentum space encircling the origin of $\mathbf{h}(k)=0$.
Explicitly, this winding number is defined as
\begin{eqnarray}
\nu=\frac{1}{2\pi}\oint _{\rm BZ} \partial_k \phi dk,
\end{eqnarray}
with $\tan \phi=h_y/h_x$ for the Hermitian SSH model [i.e. Eq. \eqref{eq:H_SSH} with $\gamma=0$].
Here the origin of $\mathbf{h}(k)=0$ represents a degenerate point of the two energy bands, which splits into two EPs in the pseudospin space when $\gamma\neq 0$ in the non-Hermitian SSH model [pink star and red triangle in Fig.\ref{fig:topo_skin_local}(g) to (i)].
Therefore two winding numbers can be defined regarding each of the two EPs,
\begin{eqnarray}
\nu_{\pm}=\frac{1}{2\pi}\oint _{\rm BZ} \partial_k \phi_{\pm} dk,
\end{eqnarray}
with
\begin{eqnarray}
\tan{\phi_{\pm}}=\frac{h_{yr}\pm h_{xi}}{h_{xr}\mp h_{yi}},
\end{eqnarray}
$h_x=h_{xr}+ih_{xi}$ and $h_y=h_{yr}+ih_{yi}$.
The total winding number is shown to be given by 
\begin{eqnarray}
\nu=(\nu_++\nu_-)/2.
\end{eqnarray}
hence the winding number $\nu$ takes a half-quantized value when the trajectory of the real part of $\mathbf{h}(k)$, denoted as $\mathbf{h}_r(k)$, encircles only one of the two EPs.~\cite{Yin2018nonHermitian}.
At the transition point between integer and half-quantized winding numbers, $\mathbf{h}_r(k)$ passes one EP in the pseudospin space, resulting in an EP between the two energy bands. Thus the half-quantized winding number also characterizes the localization transition and band-structure transition in the non-Hermitian SSH model.

In generic two-band models in the absence of symmetries, $\mathbf{h}_r(k)$ of generic two-band models traces a curve in a 3D pseudospin space, where the two 0D EPs evolve into a 1D singularity ring~\cite{li2019geometric}]. The winding of $\mathbf{h}_r(k)$ around an EP now becomes a linkage between two 1D trajectories.

\purp{ 
\begin{figure}
    \includegraphics[width=1\linewidth]{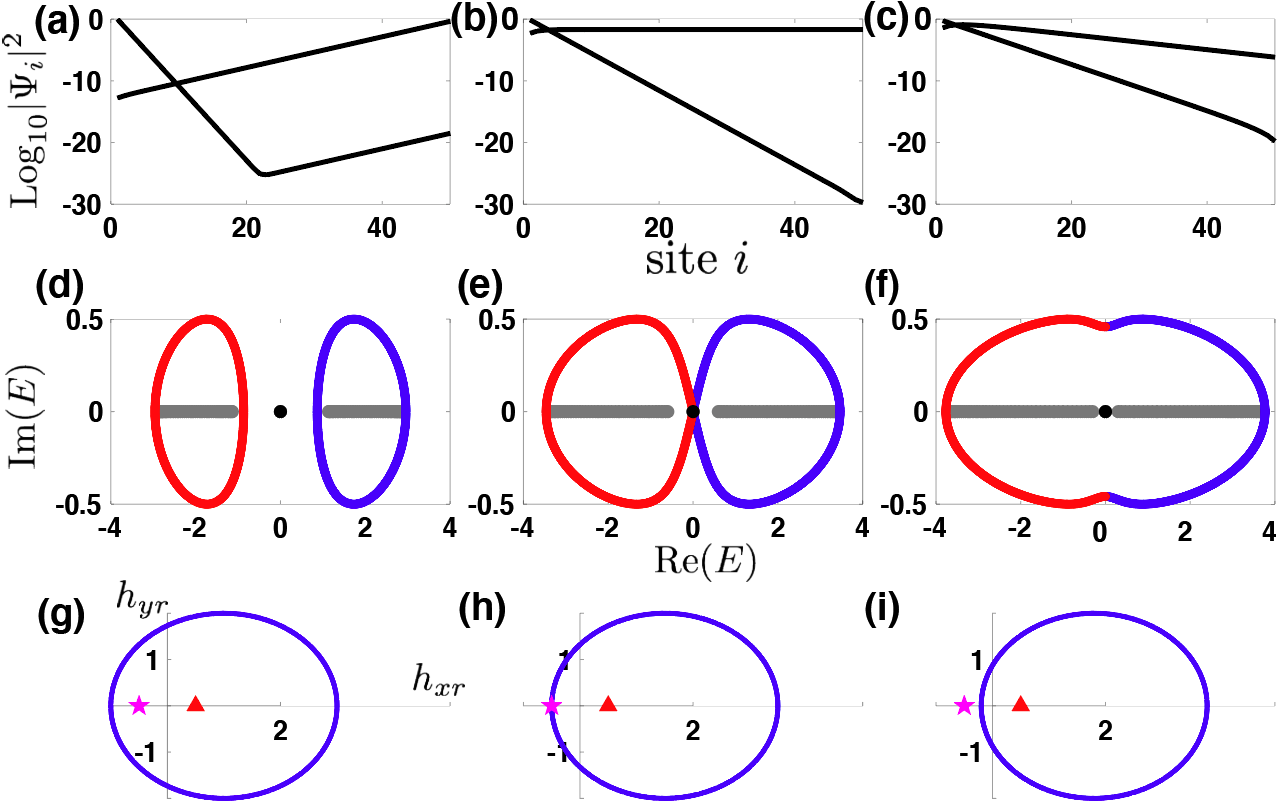}
    \caption{Transition of topological localization direction and half quantized winding numbers in the non-Hermitian SSH model in Eq. \eqref{fig:nonH-SSH}. (a)-(c) The squared wavefunction amplitude $|\psi(x)|^2$ of zero-energy modes. (d)-(f) The PBC (red,blue) and OBC (gray) spectra of the non-Hermitian SSH model with PBC spectra are seen to deform from two loops into one, while OBC ones remain gapped with a pair of zero-energy eigenmodes. (g)-(i) Winding trajectories of $\mathbf{h}_r(k)=(h_{xr},h_{yr})$. Pink star and red triangle represent EPs given by $\mathbf{h}_r(k)=(\pm h_{yi},\mp h_{xi})=(\pm \gamma/2,0)$, for the model with $h_{x}=t_{1}+t_{2}\cos(k)$  and $h_{y}=t_{2}\sin(k)+i\gamma/2$.
    Parameters are $t_2=2$, $\gamma=1$, $L=160$ (number of unit cells), and $t_{1}=1$, $1.5$, and $1.65$ from left to right respectively. Figures reproduced from the methods in Refs.~\cite{Yin2018nonHermitian,zhu2021delocalization} for the model of Eq. \eqref{fig:nonH-SSH}.}
    \label{fig:topo_skin_local}
\end{figure}
}

\subsection{Topological characterization without the GBZ}
\label{sec:anatomy}

Even though one can in principle always (numerically) evaluate the GBZ and compute the correct topological numbers on the GBZ, for generic models this is in practice often fraught with subtleties due to GBZ singularities, as well as the amplification of floating-point errors due to the NHSE. Fortunately, for particle-hole symmetric systems of the momentum-space form $H^{\text{PH}}[\{r_{a/b}\}; \{p_{a/b}\}](z)=$ 
\begin{equation}
\left(\begin{matrix}
0 & a(z) \\
b(z) & 0 \\
\end{matrix}\right)
=\left(\begin{matrix}
0 & z^{r_a}\prod_i^{p_a} \frac{(z-a_i)}{z\sqrt{a_i}} \\
z^{r_b}\prod_i^{p_b} \frac{(z-b_i)}{z\sqrt{a_i}}  & 0 \\
\end{matrix}\right)
\label{Ham}
\end{equation}
where $z=e^{ik}$, it is possible to determine whether topological zero modes exists just from the polynomial structure of $a(z)$ and $b(z)$ since only off-diagonal elements of the 2-by-2 matrix Hamiltonian exist, \emph{without} computing the GBZ at all. This was developed in Ref.~\cite{lee2019anatomy}, which pre-dated the formal development of the GBZ. Here coefficients $a_i,b_i$ 
are the complex roots of Laurent polynomials $a(z),b(z)$, both of which can be rescaled without changing the topology. 

Ref.~\cite{lee2019anatomy} provided three equivalent formulations of the criterion for having topological zero modes. 

\begin{enumerate}

\item The first formulation is a combinatorial condition on the coefficients $a_i$ and $b_i$ of the polynomials $a(z)$ and $b(z)$, which can take generic forms:
\\
\\
{\it \indent An isolated topological zero mode exists if and only if the $r_a+r_b$ largest coefficients \emph{do not} contain $r_a$ members from $\{a_1,...,a_{p_a}\}$ and $r_b$ members from $\{b_1,...,b_{p_b}\}$.} \\ \\
Even though this formulation is very explicit in terms of the roots $a_i,b_i$, as well as the pole orders $r_a,r_b$, its geometric and topological interpretation is not explicit.

\item The second formulation below recasts the above formulation in terms of topological windings. An isolated topological zero mode exists if and only if
\begin{equation}
\exists\, R\in(0,\infty) \ \ \ \text{such that}\ \   W_a(R)W_b(R)<0,
\label{winding}
\end{equation}
where the windings $W_a(R)$ and $W_b(R)$ are given by
\begin{align}
W_{g}(R)&=\oint_{|z|=R} \frac{d(\log g(z))}{2\pi i}=\#Z_g(R)-\#P_g
\label{windings_ab}
\end{align}
with $g=a,b$. Hence $W_g$ counts the number of zeros $\#Z_g(R)$ minus the number of poles $\#P_g$ encircled by a circle $|z|=R$ of radius $R\in\mathbb{R}$. 
Due to the skin localization in NHSE, we do not perform the winding integral on the typical $|z|=1$ contour, i.e. like in the usual BZ. The criterion adroitly does away the need to evaluate the GBZ directly, which can have a radius that depends on $\text{arg}(z)$ in an arbitrarily complicated manner.

\item The previous formulation can also be expressed in terms of the energy surface vorticity and eigenmode winding $V(R),W(R)=(W_a(R)\pm W_b(R))/2$~\cite{shen2018topological,yao2018edge}, 
such that the winding condition becomes
\begin{equation}
\exists\, R\in(0,\infty) \ \ \ \text{such that}\ \  |V(R)|<|W(R)|.
\label{winding2}
\end{equation}
\end{enumerate}
From the above topological criterion, which is applicable to PH-symmetric models with arbitrarily complicated hoppings (the coefficients of $a(z), b(z)$), we see that while evaluating the full PBC spectrum requires knowledge of the GBZ, finding out whether they are topological only requires the above winding conditions.

\section{NHSE in higher dimensions}

\subsection{Chern lattices with NHSE}

Lattice systems with Chern topology underlie the highly sought-after quantum anomalous Hall effect~\cite{haldane1988model,qi2008,chang2013experimental,lee2014lattice,neupert2015fractional}, and have thus been thoroughly studied and reviewed. As such, we will just briefly mention some recent advances.

As in other lattice NHSE systems, the effective band structure and its properties have to be evaluated in the GBZ, not the usual BZ. In the GBZ, the NHSE is effectively ``gauged away'', and the Chern number of a non-Hermitian Chern insulator\cite{Yao2018nonH2D,kawabata2018anomalous,lee2020unraveling,xiao2022topology} can be computed as usual via the formula 
\begin{equation}
\mathcal{C}=\frac1{2\pi i}\int_\text{GBZ}\left[\langle \partial_{k_y}\psi|\partial_{k_x}\psi\rangle-\partial_{k_x}\psi|\partial_{k_y}\psi\rangle\right]d^2k
\end{equation}
where the ket $|\psi\rangle$ ranges over the right occupied states and $\langle \psi|$ is its corresponding biorthogonal bra state. What can be different from Hermitian cases is that the occupied band can have its own spectral singularities and topological configurations. As such, the Berry curvature and band metric can be discontinuous in the GBZ, leading to semiclassical response kinks, as demonstrated in a model where the Chern bands exhibit singularities with 3-fold rotational symmetry in the complex spectral plane~\cite{lee2020unraveling}.

Even though non-Hermitian Chern bands can be mathematically described in the same way in their GBZ as Hermitian Chern bands in their BZ, their phenomenology exhibits important differences. In particular, because the biorthogonal bras and kets are not really Hermitian conjugates of the ``same'' states, under NHSE, a quantized topological invariant may not give rise to quantized transport:~\cite{liu2021supermetal,philip2018loss} gives a non-quantized contact effect from the NHSE. That said, it is still possible to formulate chiral anomalies rigorously for non-Hermitian systems~\cite{sayyad2022non}, although a distinction must be made between directed state amplification by the NHSE, and chiral pumping by the chiral anomaly. The interplay between these two effects lead to the non-Hermitian chiral magnetic skin effect predicted by Ref.~\cite{bessho2021nielsen}, which also extends the known Nielsen-Ninomiya theorem for nonchiral gapless fermions protected by symmetry. Furthermore, since the chiral boundary states of a topological insulator can possess their own GBZ, it is possible to design the chiral edge/hinge states of a non-Hermitian topological insulator passive without gain/loss, and hence immune to the NHSE~\cite{wang2022hermitian}.

Closely related to non-Hermitian Chern lattices are non-Hermitian lattices with magnetic flux. The key mathematical difference is only that of the size of the magnetic unit cell. NHSE lattices with flux exhibit exotic properties illustrating the interplay of time-reversal breaking and the NHSE. At the intuitive semi-classical level, the cyclotron trajectories of the wave packets in a 2D lattice always form closed orbits in four-dimensional (4D) phase space~\cite{lee2018electromagnetic,petrides2018six}, and the semi-classical quantization rules remain valid despite the nonreciprocity, with the propensity to preserve real Landau levels~\cite{shao2022Cyclotron}. See~\cite{deng2022non} for a generic phenomenological approach to the interplay of flux and the NHSE, and~\cite{denner2022magnetic} for a comprehensive treatment of the flux response under nonhermiticity via localized flux intersections on a single plaquette. Interestingly, skin states localized at the boundary can be pushed back into the bulk by an applied magnetic field, thereby leading to flux-suppressed NHSE~\cite{lu2021magnetic}.

Since wavepackets at the edge of a Chern lattice are steered by chiral topological pumping, through the Chern polarization, the extent of gain/loss experienced by an edge wavepacket can be controlled. Additionally, by weakly coupling two Chern topological layers with oppositely directed gain, it is possible to devise a topological guiding mechanism where there is asymptotic gain when a topological island becomes sufficiently wide. This has been shown to give rise to a percolation-induced PT transition in a disordered setting~\cite{yang2023percolation}.

\subsection{Non-Hermitian Weyl and exceptional metals}

\begin{figure*}
    \includegraphics[width=\linewidth]{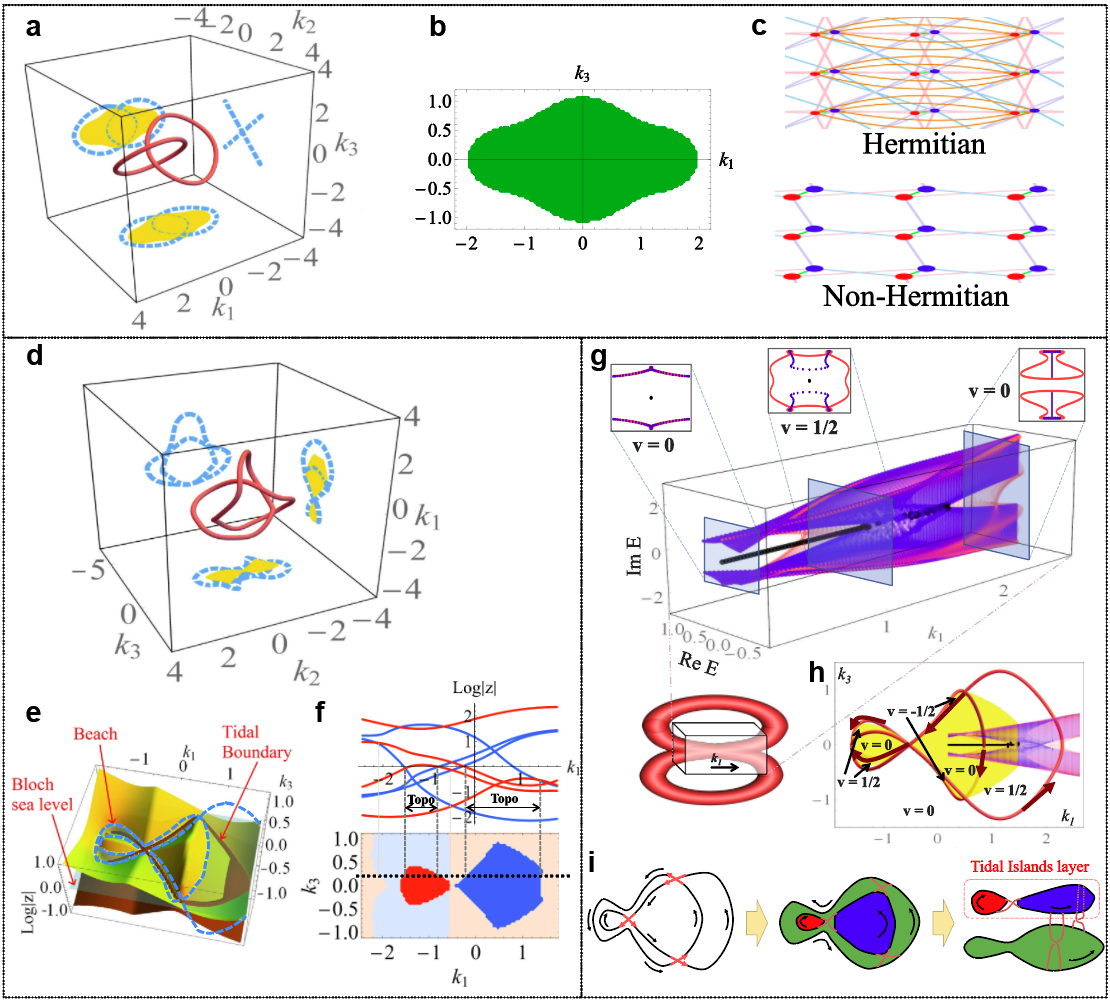}
    \centering
    \caption{Topological surface states of non-Hermitian nodal knot  metals. (a) The disagreement between the surface state projections with (yellow) and without non-Hermiticity (blue) for the Hopf link. (b) Analytically solved surface state of the Hopf link with $\bold k_\perp = k_2$. (c) Schematic comparing the tight-binding implementations of the Hermitian and non-Hermitian nodal knots. (d) Similar to (a) but for the trefoil knot. (e) Non-reciprocal similarity transforms can rescale $z=e^{ik_\perp}$, leading to fluctuations of $\log|z|=0$ level, identified as the `sea level'. The skin gap closures correspond to the surface states and manifest as `tidal boundary', whereas the band intersections with the sea level recover the Hermitian drumhead surface states, with a close analogy to `shore'. (f) More detailed construction of the tidal phase boundary of the trefoil knot. (g) Evolution of the PBC and OBC spectra with $k_1$, revealing intricate relations between the complex energy bands, vorticity and the `tidal surface states'. (h) The `tidal region' is demarcated by the half-integer vorticity boundary. (i) A direct relationship between the tidal islands and the Seifert surface of the dual nodal knot metal. Figures reproduced from~\cite{zhang2021tidal}.  }
    \label{tidal}
\end{figure*}

In a higher dimensional system, the gap between non-Hermitian bands can close along trajectories known as exceptional rings~\cite{zhen2015spawning,xu2017weyl,liu2022experimental,cerjan2019experimental,tang2022realization}, in analogy to the nodal rings of Hermitian systems, some with non-abelian structure~\cite{bzduvsek2016nodal,huh2016long,li2016topo,li2017chiral,li20172pi,yan2017nodal,li2017type,li2017engineering,zhou2018hopf,pezzini2018unconventional,zhang2018nodal,lee2018floquet,unal2020topological,bouhon2020non,lee2020enhanced,ang2020universal,yang2020observation,lee2020imaging,tai2021anisotropic,lenggenhager2021triple,wang2022experimentala,slager2022floquet,peng2022phonons,bouhon2022multi,park2022topological}. Likewise, topological surface or edge states can appear at surface/edge terminations, since for a Hamiltonian $H(\bold k)=H(\bold k_\parallel,k_\perp)$, gap closures at certain $\bold k_\parallel$ can behave like topological phase boundaries in the parameter space of $\bold k_\perp$.

When the NHSE is present under OBCs in the $k_\perp$ direction, the topological invariant in $k_\perp$ must be computed in the GBZ, and not the ordinary BZ. As such, the projection (shadow) of the exceptional trajectory onto the boundary may not necessarily correspond to the topological phase boundary on the surface, demarcated by the bulk nodal lines~\cite{zhang2021tidal,yang2022non}. This is schematically illustrated in Fig.~\ref{tidal}a,d for the non-Hermitian Hopf link and trefoil knots.

Nodal or exceptional structures with intricate momentum space structures possess interesting topological linkages or knots. However, their models necessarily contain higher harmonics in momentum space, which correspond to long-ranged real space couplings. A common approach for realistic implementation of such Hamiltonians is through Floquet engineering, where quenching between different simple nodal/exceptional configurations can give rise to much more intricate Dirac/nodal/exceptional structures~\cite{li2018realistic,yan2017floquet,carlstrom2018EL,chen2018floquet,carlstrom2019EL,kim2019floquet,staalhammar2019hyperbolic,salerno2020floquet,zhang2021tidal,meng2022terahertz,qin2022light}. Floquet engineering can also generate gap closures on demand, such as non-Hermitian Weyl semimetals~\cite{wu2022non}. Providentially, the inclusion of non-Hermiticity favourably enlarges the parameter space for nodal knot and facilitates their implementation via with Hamiltonians $H(\bold k)=\bold h(\bold k)\cdot\sigma$ with local hoppings. This is schematically illustrated in Fig.~\ref{tidal}c. The advantage conferred by non-Hermicity can be understood by considering the criteria for gap closures, which is more easily satisfied via not one but two conditions $|\mathrm{Re}~\bold h|=|\mathrm{Im}~\bold h|$ and $\mathrm{Re}~\bold h\cdot \mathrm{Im}~\bold h=0$.

Now, the aforementioned disagreement (if present) between drumhead surface states and their non-Hermitian analog fundamentally stems from the NHSE. In higher-dimensional settings, the NHSE manifests as a  `inward' compression of the drumhead surface states into a smaller area contained within the drumhead region. This smaller region is coined the `tidal state' region as it can be derived from a marine analogy from Ref.~\cite{zhang2021tidal}, as described below. Crucially, the boundary of these novel `tidal states' corresponds to the gap closures of the skin states.

To understand the mathematical and geometric foundation of this tidal analogy, we examine its derivation from Ref.~\cite{zhang2021tidal}. There, a formal criterion to determine the `tidal state' boundaries was established for particle-hole symmetric Hamiltonians (of the form given by Eqn.~\ref{Ham}), which encompasses a large class of systems. Since $\bold k_\parallel$ coordinates are just spectators when taking the OBCs along the $k_\perp$ direction, we may directly invoke the formulation (Eqn.~\ref{winding}) from Ref.~\cite{lee2019anatomy} (described in Sect.~\ref{sec:anatomy}) to find topological zero modes~\cite{lee2019anatomy}. We restate this below:
\begin{equation}
\exists\, R\in(0,\infty) \ \ \ \text{such that}\ \   W_a(\bold k_\parallel,R)W_b(\bold k_\parallel,R)<0,
\label{winding2}
\end{equation}
To illustrate this procedure, we graphically work out the topological phase boundary for the simplest non-Hermitian nodal knot metal - the Hopf link, given by the Hamiltonian $\mathbf{h}=2(\sin k_3+i(\cos k_1+\cos k_2+\cos k_3-m),\sin k_1+i\sin k_2,0)$ with $\bold k_\parallel$ being $(k_1,k_3)$. It follows that the Hamiltonian can be recasted into the non-reciprocal SSH form, with $a(z;\bold k_\parallel)=2i(1/z - t_+)$, $b(z;\bold k_\parallel)=2i(z - t_-)$ and $t_\pm=m\pm\sin k_1-\cos k_1-e^{-ik_3}$. The winding numbers involved in the criterion for topological modes are
\begin{equation}
    W_a(\bold k_\parallel,R)=-\Theta(t_+^{-1}-R)\label{windinghopf1}
\end{equation}
\begin{equation}
    W_b(\bold k_\parallel,R)=\Theta(R-t_-)\label{windinghopf2}
\end{equation}
where we have the `tidal boundary' demarcated by the inequality $|t_+t_-|=\prod_\pm[(m\pm\sin k_1-\cos k_1-\cos k_3)^2+\sin^2k_3]<1$. This region is plotted in Fig.~\ref{tidal}b and is clearly distinct from the corresponding Hermitian drumhead surface state (Fig.~\ref{tidal}a).

To illuminate the nomenclature of this tidal construction, we consider the trefoil knot. By graphically plotting the inverse skin depth $\kappa=\log|z|$ (Fig.~\ref{tidal}e-f), we see that these surfaces intersect precisely along the `tidal boundaries', i.e. the skin mode solutions experience gap closure, which facilitates topological phase transitions in momentum space. To reconcile with our PBC-OBC interpolation, we identify the `sea level' in our marine landscape analogy as the Bloch states characterized by $\log|z|=0$, i.e. purely real momenta. This `sea level' is non-unique due to the rescaling given by the non-reciprocal similarity transformation, i.e. fluctuations of the sea level analogous to tides. The well-known `drumhead states' in the corresponding Hermitian system are what we call the `beaches' and this coincides with the intersection trenches between the energy bands and the sea level, i.e. `tidal' boundaries (Fig.~\ref{tidal}e). This completes the marine landscape analogy from~\cite{zhang2021tidal}.

To gain a broader perspective of nodal knot metals, we highlight an elegant relation between vorticity, the complex energy bands, and the Seifert surfaces in non-Hermitian nodal knots. Also uncovered in Ref.~\cite{zhang2021tidal}, this stems from the bulk-boundary correspondence between the OBC `tidal' region shape and the bulk point-gap topology (measured by the vorticity). As illustrated earlier, the former is constrained within the interior of PBC bulk nodal loop projections. Along this surface-projected nodal line, we can construct a director $\bold u(\bold k)=\nabla_{\bold k}a(\bold k)\times\nabla_{\bold k}b(\bold k)$ and count the number of times any point is encircled anticlockwise by $\bold u$. This is the half-integer vorticity $v(\bold k_\parallel)=(W_a(\bold k_\parallel,1)+W_b(\bold k_\parallel,1))/4\pi$  and is thus a hallmark of skin gap closure. However, we remark that non-trivial vorticity is insufficient to guarantee skin gap closure, and one has to inspect the complex energy band crossings.

By considering any closed path in the trefoil knot's surface BZ, we uncover the PBC loci to be a cobordism of one or more conjoined tube/s along this chosen path, flanked by an interior skeleton of skin states (Fig.~\ref{tidal}g). These tubes will join at their ends to form a Riemann surface indicative of the vorticity structure (Fig.~\ref{tidal}h). Crucially, tidal regions are topologically constrained to contain islands of vanishing vorticity.

Lastly, to further highlight the topological significance of the `tidal states', we can appropriately reverse the directors $\bold u$ at each crossing in the knot diagram, so as to construct a `dual' knot that bounds a Seifert surface. The dual Seifert surface has an intimate relation to the knot topology. Remarkably, this resultant layer structure of the Seifert surface resembles our celebrated `tidal islands'. All in all, non-Hermitian nodal knot metals are indeed rich in various forms of topology, and have been simulated on various experimental platforms~\cite{PhysRevLett.127.026404, zhang2021tidal}.

\subsection{Higher-order NHSE and hybrid skin-topological systems}

Higher dimensions also offer a fertile setting for topological and NHSE states to hybridize.
In purely Hermitian settings, higher-order topological insulators (HOTIs)~\cite{benalcazar2017HOTI} result from the nontrivial interplay of topological localization in 2 or more directions, arranged such that the first-order (1D) topological polarization cancels. In non-Hermitian settings, the NHSE provides another distinct avenue for localization~\cite{PhysRevB.99.081302}. Hybrid skin-topological systems, first proposed in Ref.~\cite{Lee2019hybrid}, result from the non-trivial interplay of topological and NHSE localization in different directions. Finally, one can also define higher-order NHSE systems~\cite{fu2021nonH} characterized by NHSE localizations in more than one direction. Such higher-order NHSE have been proposed in suitably designed non-Hermitian superconductors~\cite{ghosh2022non}, and also lead to new avenues for disorder induced phase transitions in 2D~\cite{kim2021disorder}.

\subsubsection{Hybrid skin-topological modes - Original proposal in a non-reciprocal lattice}\label{sec:HSTE_origin}

In two or higher dimensions, the non-Hermitian skin effect and Hermitian topological boundary localization can be treated on equal footing, leading to the hybrid skin-topological effect in net reciprocal lattices~\cite{Lee2019hybrid}. Simply put, such systems support opposite non-reciprocity for different sublattices, which cancel each other in each unit cell, resulting in extended bulk states free from the NHSE. On the other hand, topological boundary modes are usually sublattice-polarized, hence they experience a spontaneous breaking of reciprocity~\footnote{Note that asymmetric couplings are not always need to break reciprocity~\cite{zhu2022hybrid}.} and are further pushed to lower dimensional boundaries (e.g. 1D edge states are pushed to 0D corners for a 2D system). At a more sophisticated level, the interplay of topological and skin localizations can be exploited for the engineering of ``PT-activated'' states~\cite{lei2024activating}.

\begin{figure*}
    \includegraphics[width=1\linewidth]{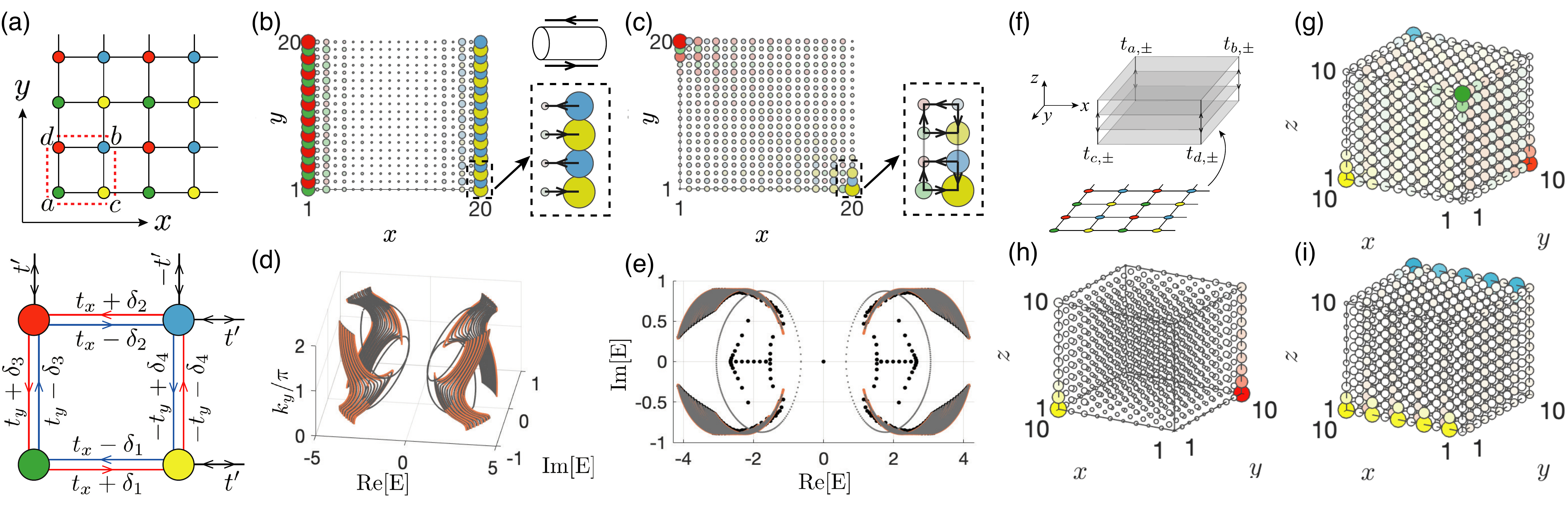}
    \caption{Hybrid skin-topological effect in a non-Hermitian Benalcazar-Bernevig-Hughes model and its 3D extension. (a) The model contains four sublattices in each unit cell, with non-reciprocal real hopping parameters $\delta_{1,2,3,4}$ along $x$ and $y$ directions. (b) Emergence of 1D boundary modes under $x$-OBC/$y$-PBC (cylinder geometry). (c) Boundary modes accumulate towards the corners under double OBC due to the hybrid skin-topological effect.
    (d) and (e) Spectra of the system with PBC (brown), $x$-OBC/$y$-PBC (gray), and double OBC (black).
    (f) A 3D lattice from stacks of the 2D model in (a), which are coupled through extra non-reciprocal hoppings along $z$ direction.
    (g) STT corner modes, (h) SST corner modes, and (i) ST0 hinge modes in the 3D model, originated from different hybridizations of NHSE and topological localization along different directions.
    The figures are reproduced from Ref.~\cite{Lee2019hybrid}.}
    \label{fig:skintopo_1}
\end{figure*}

This higher-order (lower-dimensional) boundary localization was first unveiled in a non-Hermitian extension of the 2D Benalcazar-Bernevig-Hughes model~\cite{benalcazar2017HOTI}, which can also be viewed as a mesh of two different nonreciprocal Su-Schrieffer-Heeger model along each of $x$ and $y$ directions, as shown in Fig.~\ref{fig:skintopo_1}(a)~\cite{Lee2019hybrid}. 
In this model, non-Hermiticity is introduced by two non-reciprocal real hopping parameters $\delta_{1}$ and $\delta_2$ ($\delta_{3}$ and $\delta_4$) along $x$ ($y$) direction, which balance each other and lead to a net reciprocity when they are chosen to be the same.
For properly chosen parameters, topological boundary modes appear and distribute evenly along the two 1D boundaries of a cylinder geometry (with PBC taken along $y$ direction) of the lattice [Fig.~\ref{fig:skintopo_1}(b)], which further evolve into 0D corner modes when OBC is taken along both directions (i.e. a double OBC) [Fig.~\ref{fig:skintopo_1}(c)], representing a NHSE acting on these boundary modes.
On the other hand, bulk modes in this system remain extended and are free of NHSE.
This can be seen from the spectral properties in~\ref{fig:skintopo_1}(d) and (e), where bulk bands are almost identical for the system under different boundary conditions.
Similar to 1D systems, it indicates that no imaginary flux shall be introduced when changing the boundary conditions, a sign of the absence of NHSE for the bulk modes. 
In contrast, as seen in~\ref{fig:skintopo_1}(e), eigenenergies of boundary modes show a typical feature of the emergence of NHSE, i.e. they form a loop-like spectrum for $x$-OBC/$y$-PBC (cylinder geometry), which shrinks into some open lines under double OBC.

Inspired by this explicit 2D construction, hybrid skin-topological phases can be straightforwardly generalized to higher dimensions, which support much richer classes of skin-topological boundary modes, as each dimension can contribute skin (S) or topological (T) boundary modes, or neither (0).
For instance, as shown in Fig.~\ref{fig:skintopo_1}(f), a 3D system realizing the hybrid skin-topological effect can be obtained by stacking a series of the 2D model in 
Fig.~\ref{fig:skintopo_1}(a). In different parameter regimes, this model supports all the three possible classes of hybrid skin-topological modes in 3D, namely the STT corner modes, SST corner modes, and ST0 hinge modes, as demonstrated in Fig.~\ref{fig:skintopo_1}(g) to (i).

Soon after its discovery, experiments of hybrid skin-topological effect have been proposed using cold atoms loaded in optical lattices~\cite{li2020topological}, and physically realized in circuit lattices~\cite{zou2021observation}.
In the non-Hermitian BBH model, non-reciprocity and its resultant hybrid skin-topological effect are induced by asymmetric hopping amplitudes, which are difficult to realize in quantum platforms such as cold atoms.
Alternatively, in cold atoms, non-Hermiticity can be experimentally implemented through an extra resonant optical beam transferring the atoms to an excited state, representing an atom loss for the concerned system~\cite{Jiaming2019gainloss}. 
In particular, Ref.~\cite{li2020topological} considers 
a two-orbital two-sublattice optical lattice
with an orbital-dependent atom loss, which interplay with other inter- and intra- orbital hoppings and induces an effective non-reciprocity along $x$ direction. By considering a specific two-tone shaking of the lattice, the non-reciprocity is induced toward opposite directions for the two sublattices, leading to a cancellation of non-reciprocity in the bulk.
Thus hybrid skin-topological effect emerges in this system when topological boundary modes appear, which manifests as a topological switch to control NHSE along the system's boundary.
On the other hand, in circuit lattices, non-reciprocal hoppings can be induced and precisely tuned by negative impedance converter through current inversion, allowing for a direct realization of the non-Hermitian BBH model and its 3D extension supporting different types of hybrid skin-topological modes~\cite{zou2021observation}. We note in passing that 2D higher-order lattice can also acquire their requisite coupling values through Floquet engineering~\cite{wu2021floquet,zhou2022q}, based on an approach first developed in 1D~\cite{cao2021non}.

\subsubsection{Hybrid skin-topological effect in a honeycomb lattice without asymmetric couplings}
\begin{figure}
    \includegraphics[width=1\linewidth]{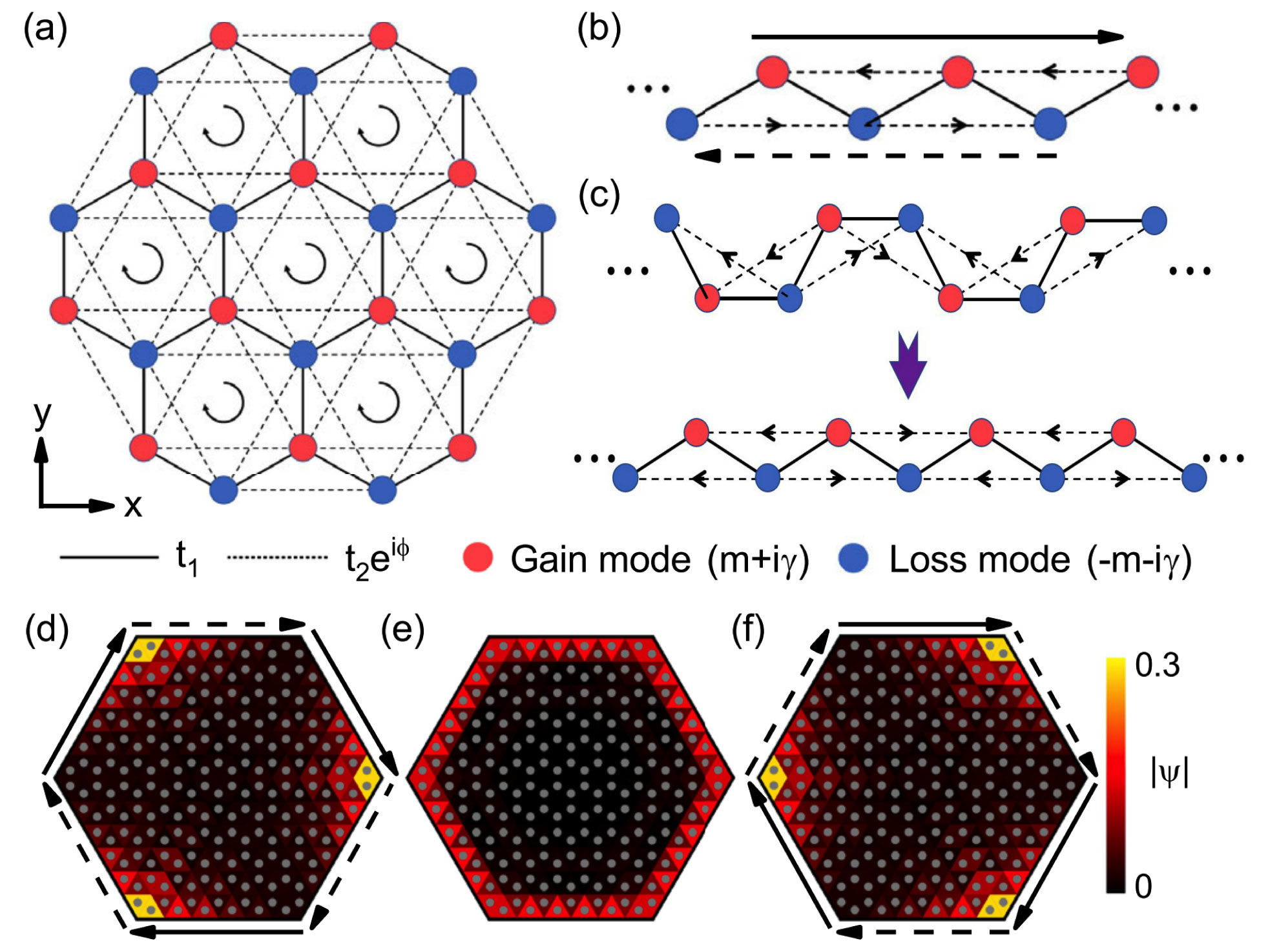}
    \caption{ Hybrid skin-topological effect in a non-Hermitian Haldane model. (a) The model contains real nearest neighbor hopping $t_1$, and complex next-nearest neighbor hopping $t_2e^{\pm \phi}$, with arrows indicating the direction of hopping with a positive phase. Non-Hermiticity is introduced by the complex on-site potential $m\pm i\gamma$ on the two sublattices respectively.
    (b) Zigzag boundary of the system. Long solid and dashed arrows [also in (d) to (f)] represent the chiral edge current for the two sublattices, which are toward and opposite to the localization direction of NHSE respectively.
    (c) Armchair boundary of the system. Arrows connecting different sites in (b) and (c) represent the same property as that in (a).
    (d) to (f) Distribution of edge modes in the model with negative, zero, positive $\gamma$ respectively.
    The figures are reproduced from Ref.~\cite{li2022gain}.}
    \label{fig:skintopo_2}
\end{figure}
Recently, a class of hybrid skin-topological modes has been discovered in a non-Hermitian Haldane model~\cite{haldane1988model} with gain and loss, without the need for asymmetric couplings~\cite{li2022gain,zhu2022hybrid}.

As shown in Fig.~\ref{fig:skintopo_2}(a), the model is given by a honeycomb lattice with real nearest neighbor hopping ($t_1$) and complex next-nearest-neighbor hopping with an amplitude $t_2$ and a phase $\phi$. Complex on-site mass terms $m\pm i\gamma$ are added to the two sublattices, which interplay with the chiral edge currents induced by the flux of $\phi$ and lead to a non-reciprocal pumping along the system's 1D boundary. An interesting property of this model is that the hybrid skin-topological skin effect emerges only along the zigzag boundary, but not the armchair boundary. 
These two types of boundaries are demonstrated in Fig.~\ref{fig:skintopo_2}(b) and (c) respectively. 
and the only difference between them is the direction of flux. For the armchair boundary, the next-nearest-neighbor couplings have phases with alternate signs,
thus the fluxes of neighbor triangular plaquettes cancel each other, and the NHSE is absent.
In contrast, the zigzag boundary supports a non-vanishing flux, which induces the hybrid skin-topological modes in the presence of sublattice-dependent gain and loss, as shown in Fig.~\ref{fig:skintopo_2}(d) to (f).
Such a gain-loss-induced hybrid skin-topological effect is discovered independently in Refs.~\cite{li2022gain,zhu2022hybrid}.
The former focuses mainly on different types of boundary and a PT phase transition of skin-topological modes,
and the latter extends their results to a Floquet realization of the hybrid skin-topological effect induced by gain and loss, and proposes an auxiliary Hermitian Hamiltonian of higher-order topological phases to understand this effect from a different angle.

\subsubsection{Higher-order NHSE}

    \begin{figure}
        \includegraphics[width=1\linewidth,height=7.8cm]{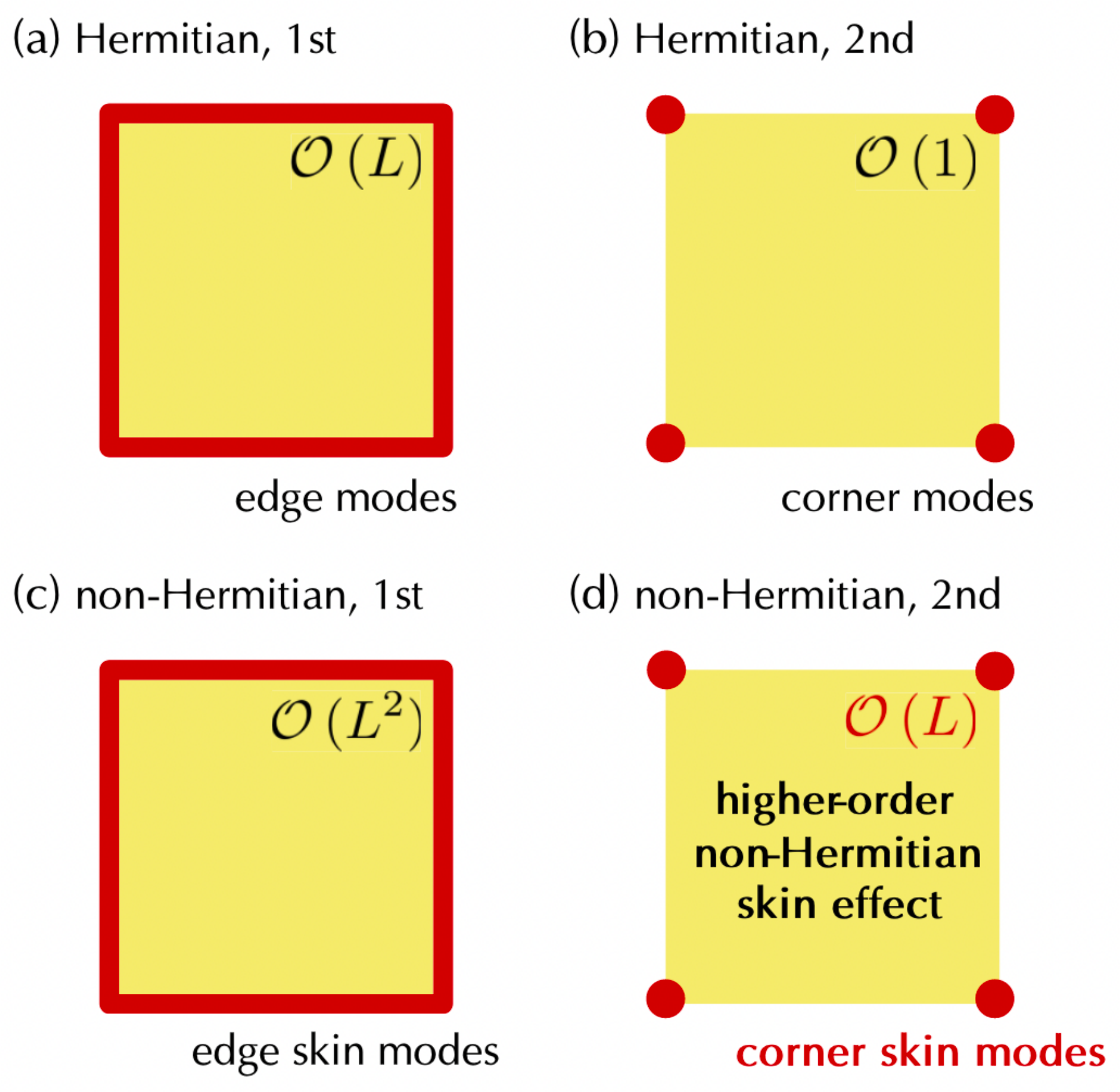}
        \caption{Schematics contrasting higher-order NHSE and higher-order Hermitian topological phases for 2D systems with $L\times L$ sites.
        (a) Hermitian first-order topological insulator. $\mathcal{O}(L)$ chiral or helical
        modes appear at the edges.
        (b) Hermitian second-order topological insulator. $\mathcal{O}(1)$ zero modes appear at the corners.
        (c) First-order non-Hermitian skin effect. $\mathcal{O}(L^2)$ skin modes appear at the edges.
        (d) Second-order non-Hermitian skin effect. $\mathcal{O}(L)$ skin modes appear at the corners.
        Figures are adopted from Ref.~\cite{kawabata2020higher}.
} 
        \label{fig:higher-order NHSE}
    \end{figure}
When the NHSE localization occurs in more than one direction (i.e. SS or SST modes), we obtain a higher-order NHSE system~\cite{kawabata2020higher,okugawa2020second,fu2021nonH,kim2021disorder}, as experimentally reported in Ref.~\cite{PhysRevB.99.201411,PhysRevB.99.121411,zhang2021observation,shang2022experimental}. 
    In conventional (Hermitian) higher-order topological phases, e.g. $d$-dimensional $n$th-order topological insulators, 
    topologically protected boundary modes appear at the $(d-n)$-dimensional boundaries, with their number scaling as $\mathcal{O}(L^{d-n})$, as sketched in Fig.\ref{fig:higher-order NHSE}(a)(b).
    In contrast, $n$th-order NHSE $d$ dimensions induces $\mathcal{O}(L^{d-n+1})$ number of skin modes localized at the $(d-n)$-dimensional boundaries, while 
    bulk modes remains delocalized in the system. In this sense, the hybrid skin-topological effect discussed above also represents a type of higher-order NHSE, where higher-order skin modes are also protected by conventional topological properties.
    Note that the terminology ``Higher-order NHSE" has also been used to describe different phenomena in different literatures. For example, in Ref.~\cite{Lee2019hybrid}, it refers to corner NHSE where all bulk modes localize at a corner of a 2D lattice.

    Conventionally, first-order NHSE (i.e. bulk NHSE) originates from from intrinsic non-Hermitian spectral winding topology of the Bloch Hamiltonian $H(k)$~\cite{zhang2020correspondence,borgnia2020nonH,okuma2020topological,Li2021}. Alternatively, this non-Hermitian topology can also be mapped to the band topology of a Hermitian Hamiltonian in extended space:
    \begin{eqnarray}
    \tilde{H}(k,E_r) := 
    \begin{pmatrix}
        0   &   H(k)-E_r  \\
        H^{\dagger}(k)-E_r^{*}    &   0   \\
    \end{pmatrix}.
    \end{eqnarray}
    This extended Hermitian Hamiltonian $\tilde{H}(k,E)$ respects chiral symmetry with a Pauli matrix $\sigma_z$, $\sigma_z \tilde{H}(k,E)\sigma_z=-\tilde{H}(k,E)$.
    If the spectral winding of $H(k)$ for a given reference energy $E_r$ is topologically nontrivial, the skin effect occurs for the non-Hermitian system,
    and the extended Hermitian Hamiltonian $\tilde{H}(k,E)$ also possesses nontrivial (Hermitian) topology which supports zero-energy 
    edge modes under the open boundary conditions.
    It associates the zeros modes of the chiral-symmetric Hermitian Hamiltonian with a nonzero spectral winding number of $H(k)$~\cite{okuma2020topological}.
    With this mapping between Hermitian and non-Hermitian topology, 
    a higher-order NHSE can be mapped
    to a chiral-symmetric Hermitian Hamiltonian with higher-order Hermitian topology. As higher-order topological phases are usually protected by spatial symmetries (e.g., inversion, mirror, and rotation symmetries)~\cite{okuma2020topological,song2017HOTI,khalaf2018HOTI,trifunovic2021higher},
    higher-order NHSE originated from intrinsic non-Hermitian topology can also be associated with these symmetries.

    \begin{figure}
        \includegraphics[width=1\linewidth]{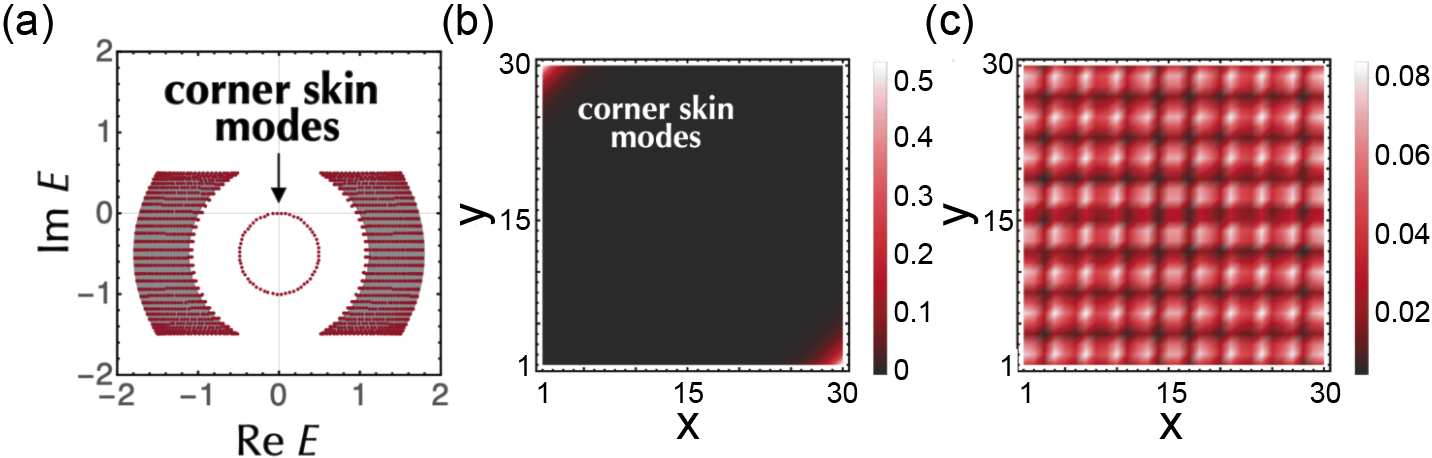}
        \caption{Second order non-Hermitian skin effect for the model of Eq. (\ref{second-order NHSE model}).
        The parameters are $\lambda=1,\gamma=0.5$. Open boundary conditions are
        imposed along both of the directions.
        (a) The complex spectrum, (b) corner skin modes ($E=-0.027-0.0008i$), and (c)
        delocalized bulk modes ($E=-1.64-0.94i$) of the model are
        demonstrated for $30\times30$ sites. Figures are adopted from Ref.~\cite{kawabata2020higher}.
        } 
        \label{fig:higher-order NHSE model}
    \end{figure}
  Below, we briefly introduce another model proposed in Ref.~\cite{kawabata2020higher}, which exhibits the second-order NHSE. Its Bloch Hamiltonian reads
    \begin{eqnarray}\label{second-order NHSE model}
        H(\mathbf{k}) = &-i& (\gamma+\lambda\cos k_{x}) + \lambda(\sin k_{x})\sigma_{z} \nonumber \\
                        &+& (\gamma+\lambda\cos k_{y})\sigma_{y} + \lambda(\sin k_{y})\sigma_{x},
    \end{eqnarray}
    where $\gamma$ and $\lambda$ are real parameters, and $\sigma_{i}$'s($i=x,y,z$) are Pauil matrices.
    Its extended Hermitian Hamiltonian is given by
    \begin{eqnarray}\label{BBH model}
        \tilde{H}_{\rm BBH}(\mathbf{k}) &=& 
                                    \begin{pmatrix}
                                        0 & H(\mathbf{k})   \\
                                        H^{\dagger}(\mathbf{k}) &   0   \\
                                    \end{pmatrix}   \nonumber\\
                                    &=& (\gamma+\lambda\cos k_{x})\tau_{y} + \lambda(\sin k_{x})\sigma_{z}\tau_{x} \nonumber \\
                                    & &    +(\gamma+\lambda\cos k_{y}) \sigma_{y}\tau_{x} + \lambda(\sin k_{y}) \sigma_{x}\tau_{x},
    \end{eqnarray} 
    where $\tau_{i}$'s($i=x,y,z$) are Pauli matrices that describe an additional pseudospin-1/2 degree of freedom.
    This Hamiltonian is known as the 2D Benalcazar-Bernevig-Hughes model and describes a second-order topological insulator~\cite{benalcazar2017HOTI}).
    The connection between the second-order NHSE in Eq.(\ref{second-order NHSE model}) and second-order Hermitian topology in 2D BBH model can help us 
    understand the origin of the second-order non-Hermitian skin effect.
    For example, in 2D BBH model, under the open boundary conditions along both directions, zero-energy modes appear at the corners for $|\gamma/\lambda|<1$.
    Correspondingly, in the model of Eq. (\ref{second-order NHSE model}),
    eigenmodes isolated from bulk bands emerge when $|\gamma/\lambda| < 1$ [Fig.\ref{fig:higher-order NHSE model}(a)], which localize at two corners of the 2D lattice [Fig.\ref{fig:higher-order NHSE model}(b)], manifesting the second-order NHSE. Meanwhile, eigenmodes in the bulk bands remain delocalized in the bulk, as shown in Fig.\ref{fig:higher-order NHSE model}(c). According to Ref.~\cite{kawabata2020higher}, the second-order NHSE in this model is protected by a rotation-type symmetry, represented by 
    $-i\sigma_yH^\dagger(k_x,k_y)=H(-k_y,k_x).$
    {Based on this model, Ref. \cite{li2022enahncement} shows that a perpendicular magnetic field can enhance this second-order NHSE, or even induce it in the otherwise trivial parameter regime.}

\subsection{Other NHSE phenomena unique to higher dimensions}

Here, we review some other interesting phenomena resulting from the interplay of the NHSE with the enlarged spatial degrees of freedom in higher dimensions.

\subsubsection{Translation-invariant bulk}

We first discuss scenarios where the skin modes accumulate at the boundaries of a clean $D$-dimensional lattice. Implementing the $D$ different OBCs sequentially, one can generalize the known result that the NHSE occurs if and only if the PBC spectrum covers nonzero area (i.e. possess nontrivial spectral winding) in the complex energy plane~\cite{zhang2020correspondence,borgnia2020nonH,okuma2020topological,Li2021}.

However, the generalization in Ref.~\cite{zhang2022universal} may fail to apply when the operation of taking OBCs in the various directions do not commute. This can happen, for instance, when the lattice cannot be "disentangled" or broken down into the sum of 1D NHSE lattices~\cite{jiang2022dimensional}. In such scenarios, the spectrum obtained by first taking the GBZ in, say, the $x$-direction, and then followed by the iterated GBZ in the $y$-direction, is different from that obtained by taking the $y$- and then $x$-direction GBZ. This indicates a breakdown in the conventional paradigm of taking 1D GBZs, forcing us to take a different approach that considers a fundamentally multi-dimensional GBZ, as introduced by Ref.~\cite{jiang2022dimensional}. In Ref.~\cite{jiang2022dimensional}, it was further shown that a multi-dimensional nontrivially "entangled" lattice can in general possess a "transmutated" or dimensionally-reduced GBZ which consists of a union of low-dimensional subspaces of the GBZ under Hermitian settings.

Further violations of the nonvanishing spectral area criterion~\cite{zhang2022universal} occur when there are GBZ singularities, such that certain pairs of biorthonormal eigenvectors become orthogonal. This can allow an open boundary to change the spectrum from real to complex, even though it is usually the case that an open boundary makes a complex spectrum real due to arrested NHSE pumping. As a specific example, the 2D model with ``rank-2'' chirality~\cite{zhu2022higher} exhibits a real spectrum when open boundaries are opened in one direction, but the spectrum becomes complex upon opening boundaries in both lattice directions. A qualitatively similar observation was also observed at sufficiently high dimensions in Ref.~\cite{song2021non}.

\subsubsection{Non translation-invariant bulk}

We now discuss the NHSE arising right in the "bulk" due to translation breaking. In general, disorder such as dislocations can act like localized defect boundaries, such that the NHSE causes skin mode accumulation against them~\cite{schindler2021dislocation,bhargava2021non}, albeit not necessarily in the same way as open boundaries, which possess translation invariance in the directions within the boundaries~\cite{panigrahi2022non}.

Suitably designed disorder can even admit skin modes of all decay lengths, such that they are described by a large range of $\text{Im}(k)$. This could allow for the disordered spectrum to "fill up" the interior of the PBC spectrum loop~\cite{jiang2022filling}, recapitulating the scenario of semi-infinite boundaries which also admit skin modes of all decay lengths~\cite{kawabata2019symmetry}.

Lattices with fractal structures poses the interesting scenario where the dimensionality may not even be well-defined~\cite{theiler1990estimating,qi2013exact,lee2016exact,gu2016holographic,manna2022inner}. in Ref.~\cite{manna2022inner}, 2D fractal lattices are shown to host  ``inner skin effects'', where details at different scales lead to different levels of skin mode accumulation and thus different levels of skin localization. Interestingly, mirror crystalline symmetry interplays with NHSE to result in sensitivity of the energy spectrum on the boundary condition only along mirror invariant lines - the mirror skin effect~\cite{PhysRevResearch.2.022062}.

We will revisit the interplay of NHSE with breaking of translation invariance in Sec. \ref{sec:dis}.

\section{Topology on the complex energy plane}\label{Sec.eng_topo}

\subsection{Point-gap and line-gap topology}
    \begin{figure}
        \includegraphics[width=1\linewidth]{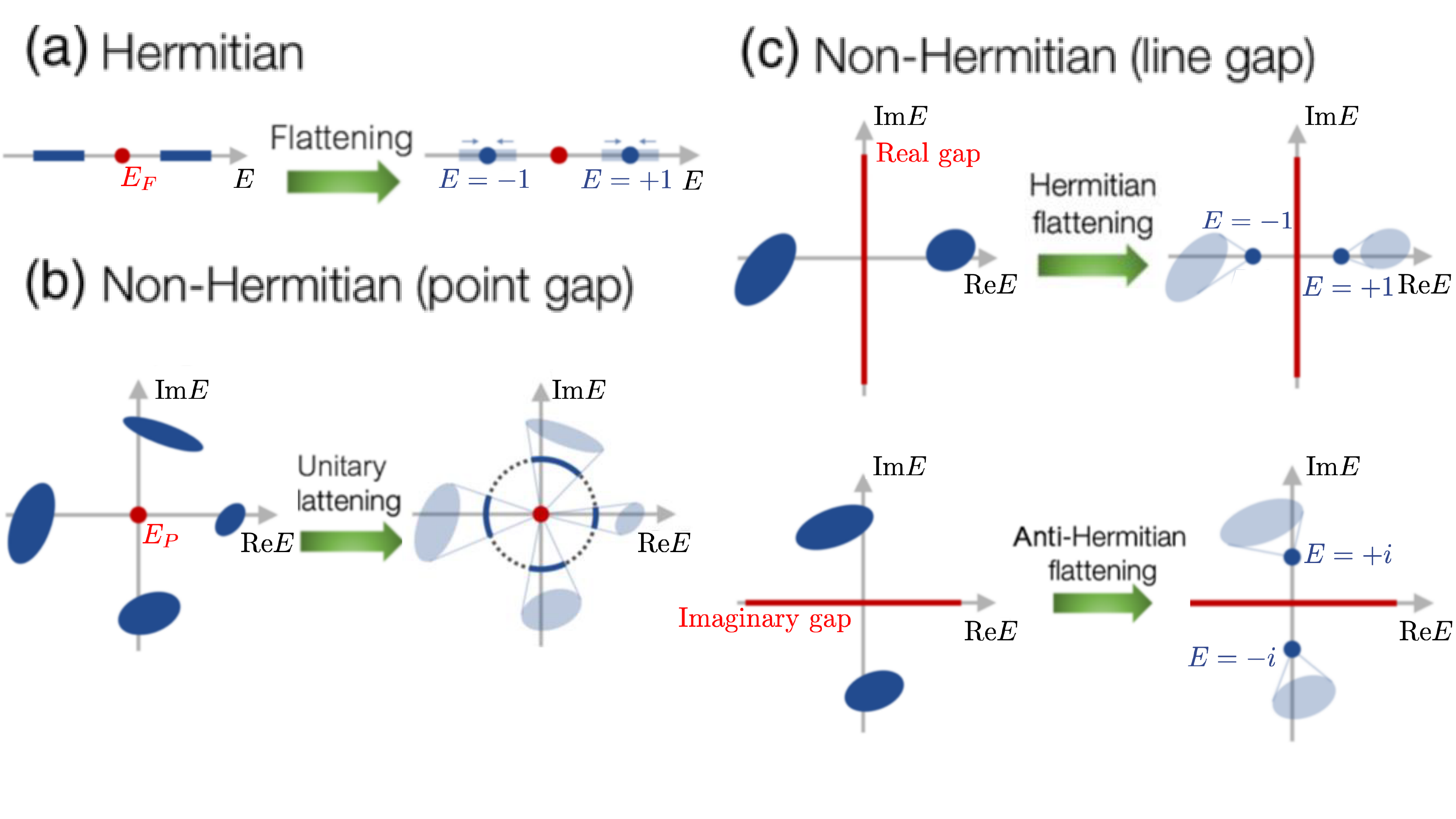}
        \caption{Qualitatively different types of energy gaps in Hermitian and non-Hermitian systems. (a) Two gapped Hermitian bands can be flattened to two points at $E=\pm 1$ along the energy axis, with a flattened Hamiltonian $H^2=1$.
        (b) Complex energy bands with a point gap can be flattened to a unit circle in the complex energy plane, with the system's Hamiltonian becoming a unitary one, $H^\dagger H=1$.
        (c) Hermitian and anti-Hermitian flattening of complex energy bands with a line gap. A non-Hermitian Hamiltonian with a real (an imaginary) line gap can be flattened to a Hermitian (an anti-Hermitian) Hamiltonian with $H^2=+1$ ($H^2=-1$).
        Figures adopted from Ref.~\cite{kawabata2019symmetry}.} 
        \label{fig:gap-flatten}
    \end{figure}
    Since gap closing is typically associated with a topological phase transition,
    it is important to provide a precise definition of energy gaps in different systems. Here we shall briefly review the definition of point gaps and line gaps for non-Hermitian Hamiltonians in Ref.~\cite{kawabata2019symmetry}.
    It is clear that for a Hermitian Hamiltonian,
    its real spectrum is defined to have an energy gap if and only if its energy bands do not cross 
    a zero-dimensional point $E=E_{F}$, which is called the Fermi energy [see Fig.~\ref{fig:gap-flatten}(a)].
    In contrast, 
    the spectrum of a non-Hermitian Hamiltonian spans the 2D complex energy plane,
    hence a complex-energy gap can be either a zero-dimensional point or a one-dimensional line. Similarly, a non-Hermitian Hamiltonian can be defined to have a zero-dimensional point gap
     (one-dimensional line gap) if and only if its complex-energy bands do not cross a reference point $E=E_P$ (reference line) 
    in the complex plane, as shown in Fig.~\ref{fig:gap-flatten}(b) and (c).
More precisely, taking into account the restrictions on eigenenergies from different symmetry (e.g., eigenenergies 
    come in $(E,E^{*})$ pairs due to the time-reversal symmetry), it is convenient to choose $E_{P}=0$, and a non-Hermitian Hamiltonian $H(\mathbf{k})$ is defined to have a point gap if and only if it is invertible (i.e.,$\forall \mathbf{k}$ $\det H(\mathbf{k})\neq 0$) and all the eigenenergies 
    are nonzero (i.e.,$\forall \mathbf{k}$  $E(\mathbf{k})\neq 0$ ).
Thus, a non-Hermitian Hamiltonian $H(\mathbf{k})$ is defined to have
    a line gap in the real (imaginary) part of its complex spectrum, denoted as a real (imaginary) gap, if and only if it is
    invertible (i.e.,$\forall \mathbf{k}$ $\det H(\mathbf{k}) \neq 0$) and the real (imaginary) part of all the 
    eigenenergies is nonzero [i.e.,$\forall \mathbf{k}$ ${\rm Im}[E(\mathbf{k})] \neq 0$~(${\rm Im}[E(\mathbf{k})]\neq 0$)].

These enriched definitions of energy gaps have led to several unusual topological properties for non-Hermitian systems.
    Intuitively, if a non-Hermitian Hamiltonian has a point (line) gap, it can be continuously deformed into
    a unitary (Hermitian/anti-Hermitian) matrix while keeping the point (line) gap and its symmetries, which means that
    the two Hamiltonians before and after the deformation are topological equivalent [see Fig.\ref{fig:gap-flatten}(b)(c)]. 
    These properties have been proven in Ref.~\cite{kawabata2019symmetry}, which play a crucial role in {the complete topological classification of 38 symmetry classes for point gaps, and 54 symmetry classes for non-Hermitian line gaps~\cite{liu2019topological} and Floquet systems~\cite{liu2022symmetry}.}
    The studies of point-gap topology in non-Hermitian systems have also be generalized to systems with crystal symmetries such as an inversion symmetry, where topological invariants can be determined from high-symmetric momenta~\cite{okugawa2021non,PhysRevB.103.L201114}.

        As another example, an immediate topological description of a point gap is the spectral winding number defined as in Eq. \eqref{eq:winding}, which vanishes together with the point gap when Anderson localization is induced to the system by spatial disorder~\cite{gong2018topological}. 
    A nontrival spectrum winding number associated with a point gap under
    PBCs also result in NHSE under OBC~\cite{zhang2020correspondence,okuma2020topological}, as discussed in Sec.~\ref{Sec:winding_NHSE}.
    
\begin{figure}
        \includegraphics[width=1\linewidth]{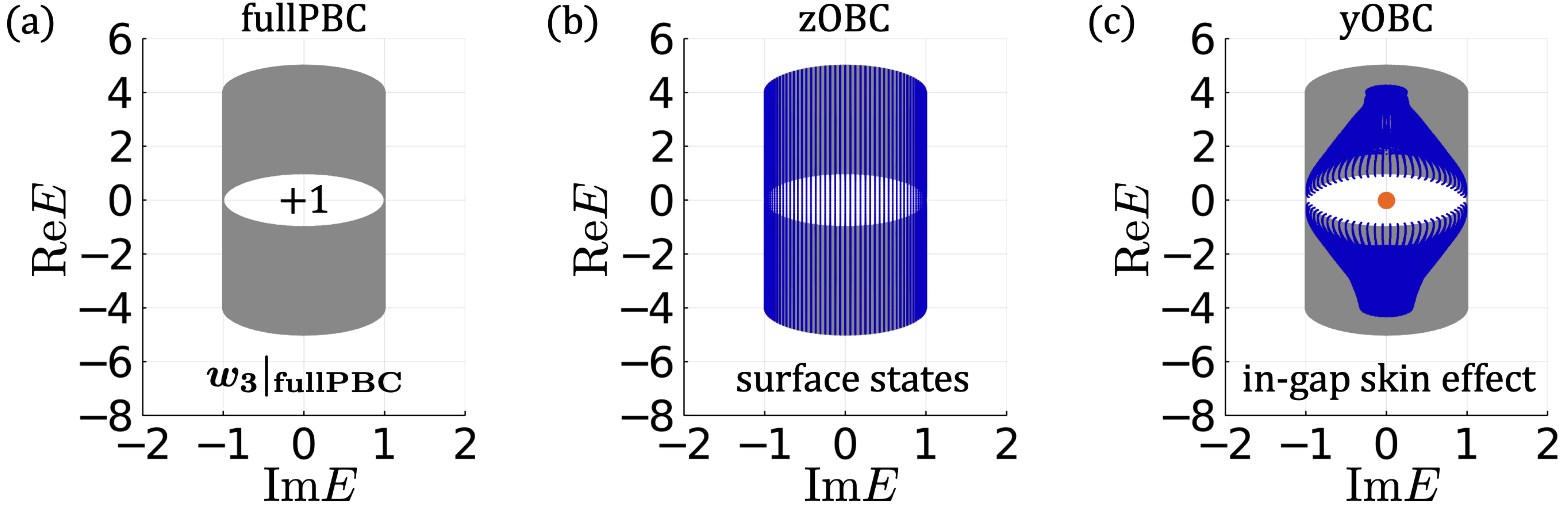}
        \caption{A 3D system with different bulk and surface behaviors under different boundary conditions.
        Gray color indicate the PBC spectrum, and blue and orange colors correspond to the spectra with boundary conditions specified in the figures.
        (a) The PBC spectrum has a point gap with a nontrivial 3D winding number $+1$. (b) When OBC is taken along $z$ direction, surface states cover the point-gapped region with a nontrivial 3D winding number.
        (c) When OBC is taken along $y$ direction, a sharp change of the spectrum indicates the occurrence of NHSE, and in-gap skin modes (orange) appear.
        Figures are adopted from Ref.~\cite{nakamura2022bulk}.} 
        \label{fig:3D_point_gap}
    \end{figure}

In 3D topological phases, it has also been suggested that topological surface states originates from a 3D winding
    number under PBC, which is the point-gap topological number for general 3D systems~\cite{denner2021exceptional,PhysRevB.103.L201114}.
    On the other hand, a more recent study unveils that a system with a nontrivial 3D winding number may have distinguished behaviors for boundary states with OBC along different directions, as shown in Fig.~\ref{fig:3D_point_gap}~\cite{nakamura2022bulk}.
    Following the idea of real space topological invariants for non-Hermitian topological systems~\cite{song2019realspace,Tang2020,PhysRevB.104.L161116,PhysRevB.104.L161117},
    Ref.~\cite{nakamura2022bulk} defines real space winding numbers for different OBCs, which establishes the bulk-boundary correspondence in the point-gap topology (with certain symmetry protection) of 3D non-Hermitian systems.

\subsection{Topological origin of NHSE}\label{Sec:winding_NHSE}

We next describe an important connection between nontrivial spectral winding, which leads to spectral loops of nonzero area, with the presence of the NHSE in 1D non-Hermitian Hamiltonians. We define a spectral winding number
\begin{eqnarray}
w(E_r)=\int_0^{2\pi} \frac{d k}{2\pi i}\frac{d}{d k}\log {\rm det}[H(k)-E_r],\label{eq:winding}
\end{eqnarray}
with $H(k)$ a non-Hermitian Hamiltonian, and $E_r$ a chosen complex value as a reference energy.
In contrast to the winding number of $\mathbf{h}_r(k)$ in the pseudospin space discussed in the last subsection,
a spectral winding number describes the number of times that the complex spectrum of $H(k)$ encircles $E_r$ anti-clockwisely throughout the Brillouin zone.
Mathematically, $w(E_r)$ can be related to the total number of zeros and poles of $H(z)-E_r$ with $z=e^{ik}$, 
enclosed by the path of $z$ on the complex plane with $k$ varying from $0$ to $2\pi$, i.e. the BZ.
Explicitly, 
\begin{eqnarray}
w(E_r)=N_{\rm zeros}-N_{\rm pole},
\end{eqnarray}
where $N_{\rm zeros}$ and $N_{\rm pole}$ are the counting of zeros and poles weighted by their respective orders.
Replacing $H(k)$ with $H(z_{\rm GBZ})=H(k+i\kappa(k))$ in Eq. \eqref{eq:winding}, a winding number can be defined for the GBZ, which has been proven to always enclose $m$ zeros and a pole of order $m$, resulting in a zero winding number for the GBZ spectrum (equivalent to the OBC one exact for topological edge states)~\cite{zhang2020correspondence}.
This means that if a system supports nonzero spectral winding under the PBC, its GBZ must have $\kappa(k)=0$, 
so that OBC eigenstates have nonzero decaying rates and exhibit NHSE.
Alternatively, this topological bulk-boundary correspondence between OBC NHSE and PBC spectral winding number has also been unveiled with a doubled Green’s function approach~\cite{borgnia2020nonH}, PBC to OBC interpolations~\cite{edvardsson2022stability} and detailed studies of Toeplitz matrices~\cite{okuma2020topological}. 
Such spectral properties of systems with NHSE have also been studied by mathematicians decades ago~\cite{bottcher2005spectral}.

Further insight into the NHSE in terms of the detailed structure of the eigenstate trajectories, not just the eigenvalue trajectories, can be obtained through the use of Majorana stars~\cite{bartlett2021illuminating,teo2020topological}. A $N$-component state vector can be mapped onto $N-1$ Bloch spin vectors (stars), such that its complex degrees of freedom can be visualized in terms of the real angles and real correlations between these $N-1$ vector directions. These visualizations provide additional geometric and topological pictures into generic multicomponent non-Hermitian models.

\subsection{Quantized response from spectral winding topology}

As discussed in the previous Sec.~\ref{Sec:winding_NHSE}, a nontrivial spectral winding is responsible for the emergence of NHSE under the OBCs, representing a new type of bulk-boundary correspondence of spectral winding topology.
A one-on-one correspondence has been established between the spectral winding and number edge states under the semi-infinite boundary conditions~\cite{okuma2020topological}, which naturally arises for the Fock space mapped onto a real-space lattice~\cite{pan2021point}.
Experimental observation of arbitrary spectral winding has also been carried out by visualizing the frequency band structure of optical frequency modes~\cite{wang2021generating}.

\begin{figure*}
    \includegraphics[width=0.8\linewidth]{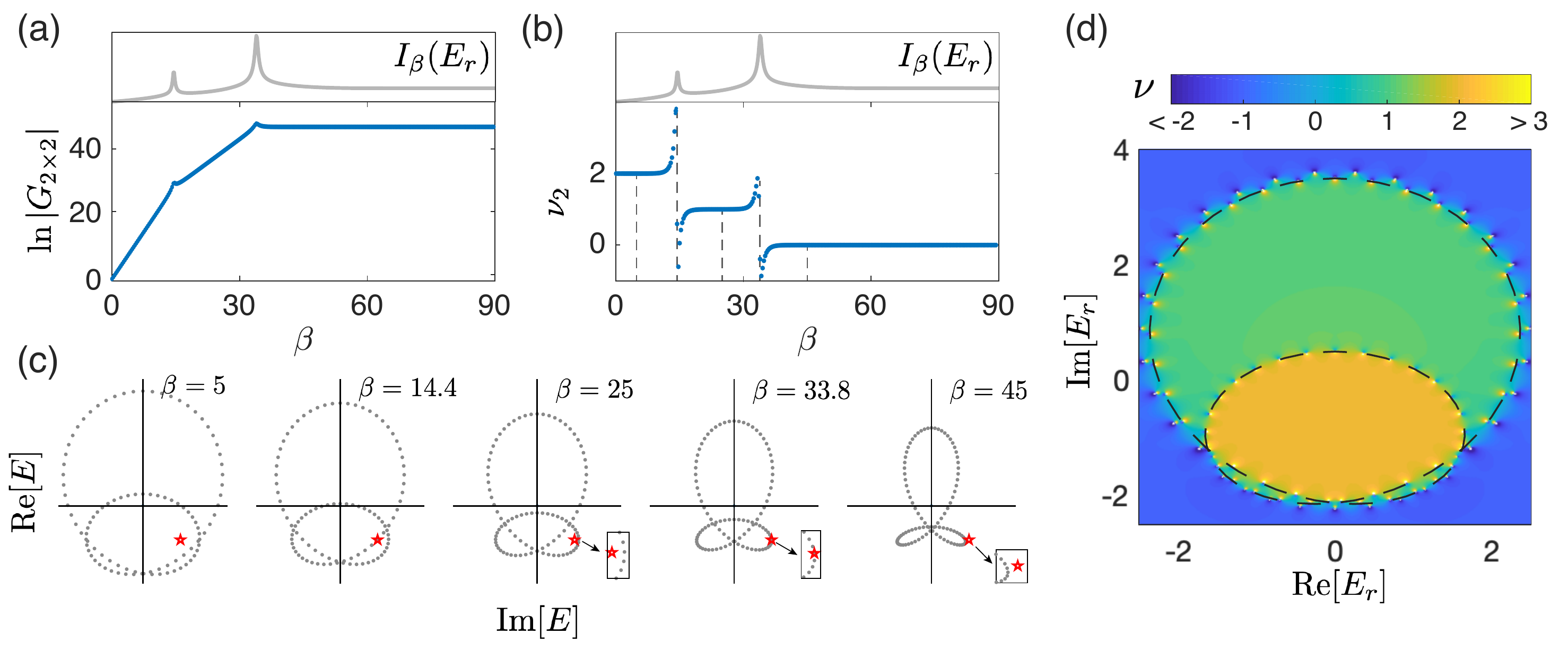}
    \caption{Quantized response of non-Hermitian spectral winding topology. 
    Results are obtained from the model of Eq. \eqref{eq:single-band}, with $t_1=1$, $t_{-1}=0.5$, $t_2=2$, and $t_j=0$ for other values of $j$. 
    (a) and (b) the two quantities extracted from the Green's function as functions of $\beta$, the parameters determining the boundary conditions. (c) Spectra at different values of $\beta$, corresponding to the five dashed lines in (b) respectively. Red star indicates the chosen reference energy $E_r=-0.96+i$ for (a) and (b).
    (d) The quantized response quantity $\nu={\rm Max}[\nu_1,\nu_2]$, as spectral winding number ranges from $0$ to $2$ for the chosen system. Figures reproduced from~\cite{Li2021}.
    }
    \label{fig:quantized_response}
\end{figure*}

In the framework of conventional band topology, Hermitian topological systems exhibit quantized physical phenomena that hinges on their respective topological invariants. Celebrated examples include the quantized charge transport in Thouless pumps for 1D systems, and the quantum Hall effect for 2D systems. It is thus natural to ask if non-Hermitian systems exhibit quantized response?
Recently, a topological quantized response with a one-on-one correspondence to non-Hermitian spectral winding topology has been proposed by considering the Green's function of the Hamiltonian with PBC-OBC interpolations,
\begin{eqnarray}
G(\beta)=\frac{1}{E_r-H(\beta)},
\end{eqnarray}
with $E_r$ a chosen reference energy, $H(\beta)$ the Hamiltonian, and $\beta$ a parameter controlling the boundary conditions~\cite{Li2021}. Namely, for a general single-band Hamiltonian of Eq. \eqref{eq:single-band},
the hopping parameters are set to $t_j\rightarrow t_j e^{-\beta}$ for $x+j>L$ or $x+j<1$.
Therefore the system is under the PBC when $\beta=0$, and OBC when $\beta$ tends to infinity.
Note that a finite-size non-Hermitian system behaves as under the OBC when $\beta$ exceeds a finite value associated with the non-Hermitian parameters~\cite{budich2020sensor,li2021impurity}.
Finally, for a spectral winding number $w(E_r)=m$, the quantized response is defined as 
\begin{eqnarray}
\nu_m=d \ln |G_{m\times m}|/d \beta,
\end{eqnarray}
with $G_{m\times m}$ the $m\times m$ block of the top-right (bottom-left) corner of $G(\beta)$ for positive (negative) $m$.
A typical example for a model with $w(E_r)=0,1,2$ for different values of $E_r$ is shown in Fig.\ref{fig:quantized_response}.
The quantities of $\ln|G_{2\times 2}|$ and $\nu_2$ for a reference energy $E_r$ with $w(E_r)=2$ 
as a function of $\beta$ are displayed in Fig.~\ref{fig:quantized_response}(a) and (b), where clear quantized plateaus from $2$ to $0$ are seen in the latter.
The correspondence betwen $\nu_2$ and  the spectral winding regarding the chosen reference energy (red star) is further verified by comparing with the complex spectra in Fig.\ref{fig:quantized_response}(c).
Another way to demonstrate the topological quantized response is to consider $\nu_m$ at $\beta\gtrsim 0$, which reflects the spectral winding topology of the PBC system, as shown in Fig.\ref{fig:quantized_response}(d).

Physically, the quantized response quantity $\nu_m$ can be associated with the directional signal amplification for non-Hermitian systems~\cite{Wanjura2020,xue2021simple,wanjura2021correspondence}. 
Namely, 
for a local steady-state driving field (input signal) $\epsilon_x(\omega)$ with a frequency $\omega$ at a given location $x$, a response field (output signal) $\psi_{x'}(\omega)$ can be obtained as $\psi_{x'}(\omega)=G_{xx'}\epsilon_x(\omega)$~\cite{xue2021simple}, with $G_{xx'}$ an element of the Green's function.
In other words, $G_{xx'}$ describes the amplification ratio between sites $x$ and $x'$, thus the quantized response quantity $\nu_m$ describes the changing rate of this ratio between the first and last $m$ sites of a 1D chain, during a PBC-OBC transition of the system.
Alternatively, $\nu_m$ may also be detected by measuring the two-point impedance $Z_{xx'}$ between the two sites in circuit lattices, which is related to the Green’s function via $Z_{xx'}=G_{xx'}+G_{x'x}-G_{xx}-G_{x'x'}$~\cite{Li2021,lee2018topolectrical}.

We end the section with a few remarks.
First, instead of turning off the boundary coupling, a local on-site potential can also act as a boundary of the system, and leads to the same quantized response~\cite{liu2021exact2}.

Secondly, in a system with two weakly coupled non-Hermitian chains, it is found that spectral winding topology of one chain can be detected by the quantized response solely of the other chain, reflecting an anomalous hybridization of spectral winding topology of the two chains~\cite{liang2022anomalous}.

Thirdly, nontrivial spectral winding has also been found to emerge for boundary states in 2D lattices, which can also be captured by the quantized response defined for the boundary of the systems~\cite{ou2023nonH.107.L161404}. Physically, the boundary spectral winding originates from the interplay between non-Hermitian non-reciprocal pumping and conventional topological localization, analogous to the mechanism behind hybrid skin-topological effect introduced in Sec.~\ref{sec:HSTE_origin}.

Fourthly, while the Green's functions defined above are designed to capture signatures of the NHSE, bulk Green's functions in general i.e.~\cite{okuma2022boundary} do not necessarily capture the onset of NHSE pumping~\cite{zirnstein2021bulk}, at least for a sufficient large system~\cite{mao2021boundary}.

\subsection{Complex band evolution as braiding processes}

Another topological feature that can arise from the complex eigenenergies of non-Hermitian systems is the braiding between different energy bands i.e. braiding between the trajectories of complex eigenenergies as a parameter~\cite{hu2021knots,PhysRevB.106.195425}. 
Braids can be closed to form knots~\cite{kauffman1991knots}, which exhibit extremely rich topology since there is an infinite number of knot configurations that cannot be deformed into each other~\cite{kauffman1991knots,li2018realistic,staalhammar2019hyperbolic,hu2022knot,li2022non_knots}.

Unlike spectral winding or point-gap topology, nontrivial braiding emerges only in non-Hermitian multiband systems with separable bands, i.e. $E_i(k)\neq E_j(k)$ for all band indices $i\neq j$ and momentum $k$.
Conceptually, as a 1D momentum $k$ varies from $0$ to $2\pi$, eigenenergy trajectories of different bands may wind around each other and form a ``braid".
The simplest example is that the trajectories of two bands exchange once and connect to each other, giving rise to a energy vorticity unique in non-Hermitian systems~\cite{shen2018topological}.

\begin{figure}
    \includegraphics[width=1\linewidth]{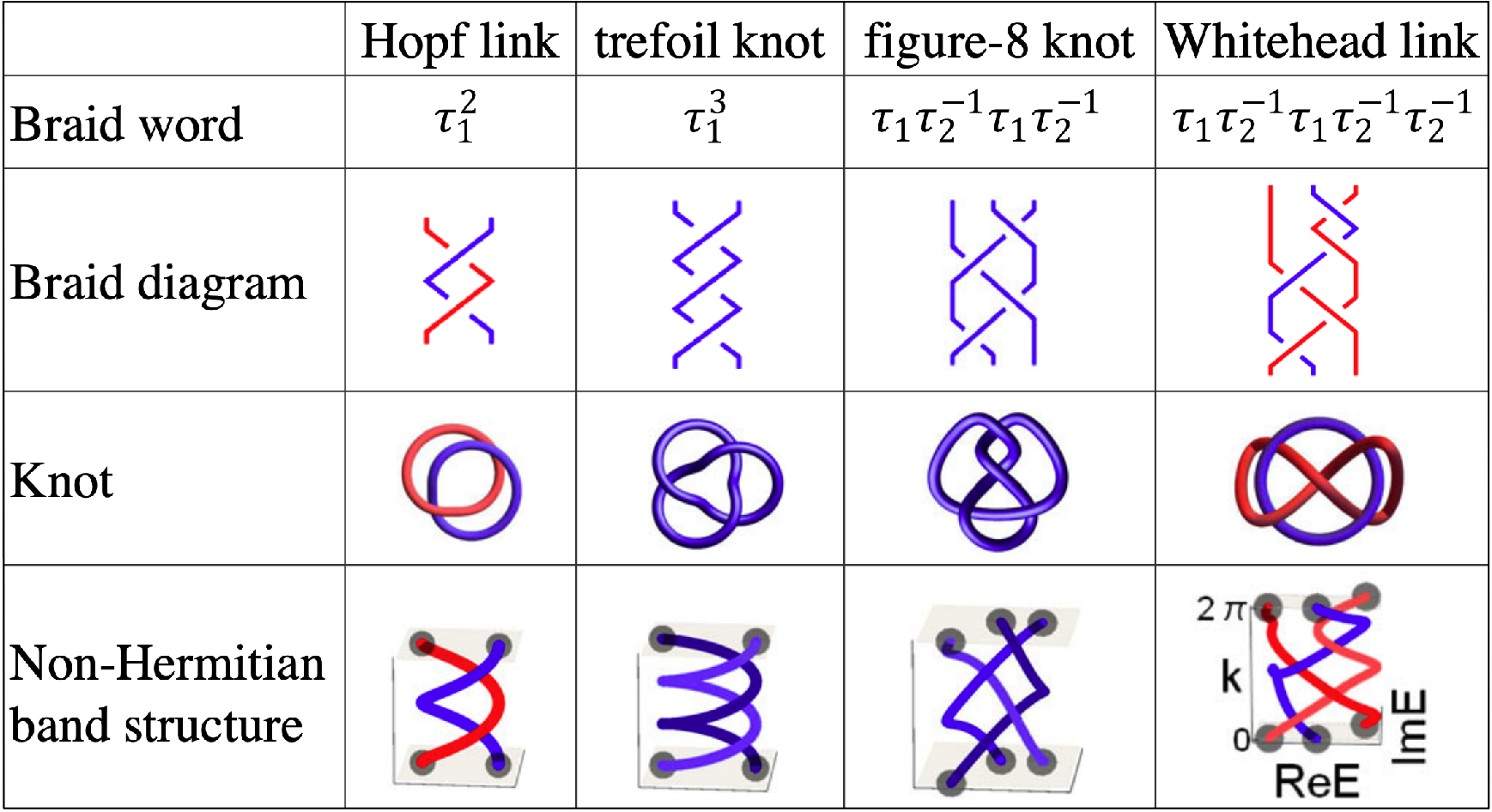}
    \caption{Several examples of topologically distinct braids, their knot closures and their realizations in 1D non-Hermitian Bloch bands. $\tau_i$ and $\tau_i^{-1}$ denote braid operators of the $i$th string crossing over and under the $(i+1)$th string from the left, respectively. Braid diagrams in the second row are mapped to the knots in the third row when connected top to bottom. 
    The fourth row demonstrate typical structures of energy bands for these braids and knots in the 3D space of momentum and complex energy.
		The figure is adopted from Ref.~\cite{hu2021knots}.
    }
    \label{fig:knots}
\end{figure}

Different braids of these trajectories cannot be continuously deformed into each other without touching between different bands (usually some EPs), representing a topological feature unique in non-Hermitian multiband systems.
Due to the periodicity of Brillouin zone, eigenenergy trajectories of separable bands are closed loops, hence their braids can be mapped to knots in the 3D energy-momentum space. 
Several examples of nontrivial energy braids and knots are demonstrated in Fig.~\ref{fig:knots} adopted from Ref.~\cite{hu2021knots}. 
In addition, Ref.~\cite{hu2021knots} also develops an algorithm to construct tight-binding Hamiltonian for any desired knot, and propose a scheme to probe the knot structure via quantum quench.
Such nontrivial braiding of non-Hermitian Bloch bands has been experimentally observed in coupled ring resonators with phase and amplitude modulation,
by 
extracting complex band structure from measured transmission signals measured from the resonators,
and reproducing the complex spectrum of a non-Hermitian lattice system in a frequency synthetic dimension~\cite{wang2021topological}. Braiding of exceptional arcs have also been demonstrated~\cite{tang2022experimental,tang2020exceptional}.

In a more recent study, it has been shown that nontrivial braiding also exists for non-Bloch bands describing non-Hermitian systems under OBC~\cite{PhysRevB.106.195425}. 
The braiding of non-Bloch bands can be topologically different from that of Bloch bands of the same system, since the presence of the NHSE leads to different band structures under PBC and OBC.
It also further reveals that the spectral winding and braiding of separable bands are essentially different types of topology, as the former is always trivial for non-Bloch bands, despite that they both originate from the paths of eigenenergies changing with momentum $k$ in the complex energy plane.

\subsection{Emergent spectral graph topology}

\begin{figure*}
\includegraphics[width=\linewidth]{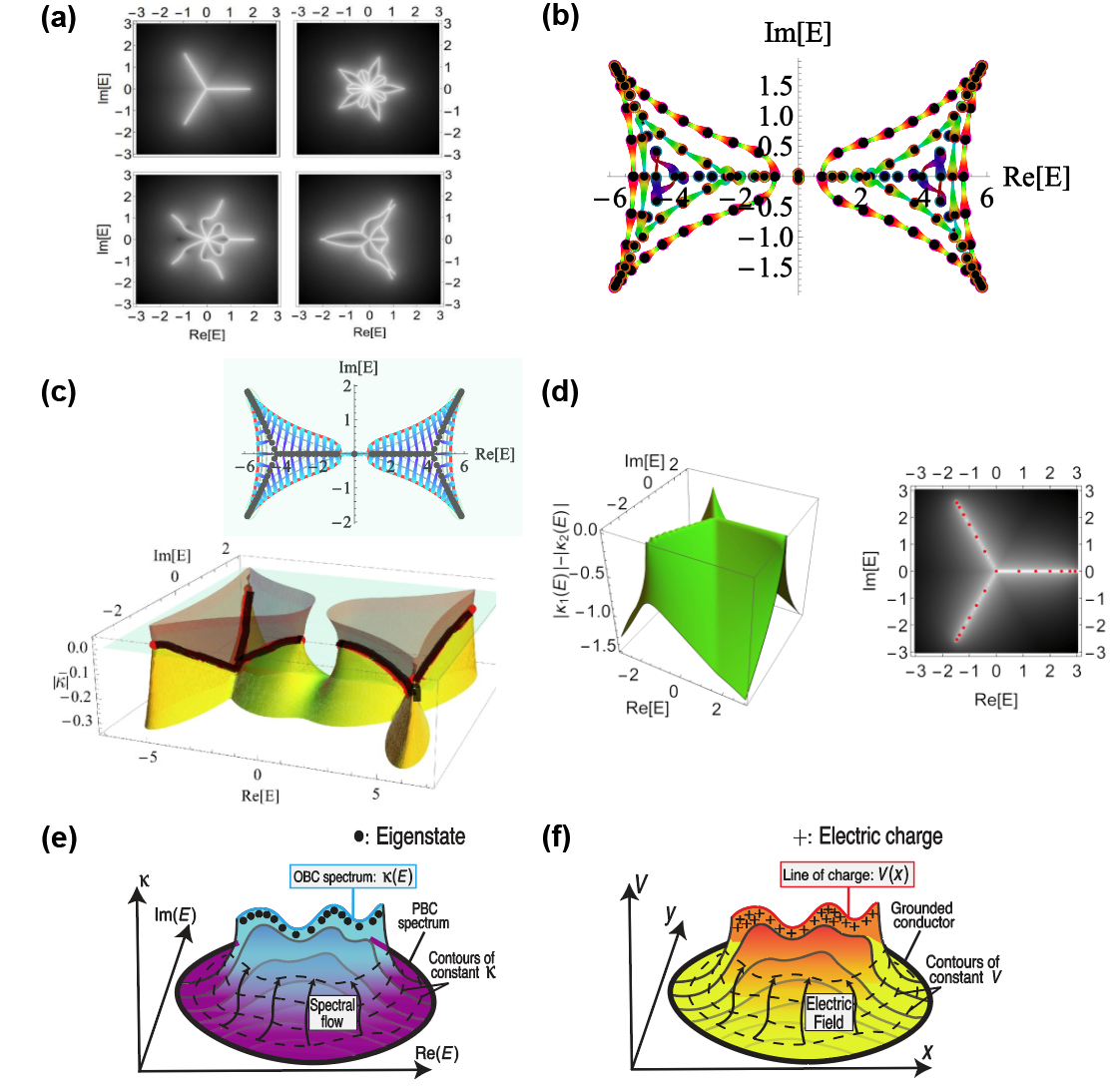}
    \caption{Emergent spectral graph topology. (a) An assortment of OBC spectra displaying a rich variety of spectral graph topology.
    (b) Eigenvalues of Eq.~\ref{Hspectralflow} flow into each other as a real boundary flux $\text{Re}(\phi)$ is threaded over a period, for the model given in Eq.~\ref{Hspectralflow}~\cite{lee2019anatomy}. The outermost loop describes the PBC loop; successively smaller inner loops represent the spectrum as the boundary coupling is decreased through increasing attenuation factors $e^{-\text{Im}(\phi)}$. (c) Illustration of PBC-OBC spectral flow of Eq.~\ref{Hspectralflow}, with PBC bulk eigenvalues (red) flow along the blue magenta curves upon threading $\kappa\propto\mathrm{Im}[\phi]$, which eventually converge to the OBC eigenvalues (black)~\cite{lee2019anatomy}. (d) An equivalent representation would be plotting the intersections of the inverse skin depth solution surfaces $|\kappa_i|$~\cite{tai2022zoology}. (e) 
    The PBC-OBC spectral flow can be generalized to all non-Hermitian lattices, where the intimate relationship between the complex energies $E$ and the inverse spatial decay lengths $\kappa$ draw a parallel with the electrostatic potential landscape, with PBC and OBC spectral loci corresponding to grounded conductors ($V=0$) and lines of induced charges respectively~\cite{yang2022designing}. Figures (a,d), (b,c) and (e,f) are taken from Refs.~\cite{tai2022zoology}, \cite{lee2019anatomy} and \cite{yang2022designing} respectively.} 
    \label{fig:spectral_graph}
\end{figure*}

While the complex spectrum under PBCs is characterized by winding numbers as the momentum $k$ is traversed over a period from $0$ to $2\pi$, it generically collapses into lines, curves, closed loops and branches as OBCs are introduced, as shown in Fig.~\ref{fig:spectral_graph}a. As such, OBC NHSE spectra are generically characterized by spectral \emph{graph} topology, which remain invariant under conformal transformations in the energy spectrum as first systematically studied in Ref.~\cite{lee2020unraveling}, and further characterized and classified in Ref.~\cite{yang2020non,tai2022zoology,fu2022anatomy}. This spectral graph topology on the complex energy plane is an emergent feature of NHSE with no Hermitian analog.

OBC NHSE spectra generically assume graph-like structures because they are the ``shrunken'' versions of PBC spectral loops. This can be understood through complex flux pumping arguments, as initially suggested by Ref.~\cite{xiong2018does}, and subsequently expanded on in Ref.~\cite{lee2019anatomy}. As explained in Ref.~\cite{lee2019anatomy}, let us first interpolate between PBCs and OBCs by multiplying the boundary couplings by $e^{i\phi}$, $\phi\in\mathbb{C}$. For real $\phi$, this is equivalent to threading a real flux through the lattice, as can be seen via a gauge transform; if we let $\text{Im}(\phi)\rightarrow \infty$, we obtain the OBC limit. Intuitively, the effect of threading $\text{Re}(\phi)\rightarrow\text{Re}(\phi) + 2\pi$ diminishes as $\text{Im}(\phi)$ increases, since the boundary coupling would be exponentially suppressed in magnitude and its phase should therefore exhibit diminishing influence on the whole system. This is illustrated in Fig.~\ref{fig:spectral_graph}(b) for the illustrative model with next-nearest-neighbor (NNN) hopping (Eqn.~\ref{Hspectralflow}), which hosts a richer graph structure compared to the prototypical models with only one non-reciprocal length scale like the non-Hermitian SSH model. 
\begin{equation}
H_{NNN}(z)=\frac{9}{4}\sigma_x-3z\sigma_-+3\left(1-\frac{1}{z}-\frac{1}{z^2}\right)\sigma_+,
\label{Hspectralflow}
\end{equation}
where $\sigma_\pm=(\sigma_x\pm i\sigma_y)/2$ and $z=e^{ik}$ as usual. When there is no suppression to the boundary coupling i.e. $\text{Im}(\phi)=0$, we have the PBC loop, and threading the real flux over one cycle maps one eigenvalue to the next. This cyclic permutation of the eigenvalues still occurs when $\text{Im}(\phi)>0$, but because the boundary couplings are now much weaker, the eigenvalue flow should also be correspondingly diminished: as seen in Fig.~\ref{fig:spectral_graph}(b), this is indeed achieved by having smaller, ``shrunken'' loops in the interior of the original PBC loops. Extrapolating, we expect that in the OBC limit, the effect of pumping real flux $\text{Re}(\phi)$ should entirely vanish - and this can only be possible if the spectral loop have somehow ``shrunken'' until they become degenerate i.e. enclosing zero area, such that for every eigenvalue, there exists another one infinitesimally close in the complex plane. This spectral flow can also be represented as a spatial flow along the eigenspectra surfaces (of generalized boundary conditions) as we interpolate from PBC to OBC, as shown in Fig.~\ref{fig:spectral_graph}c. Crucially, at the OBC limit, the spectra is the intersection of two $|z|$ surfaces in the complex plane (Fig.~\ref{fig:spectral_graph}d) since in order to simultaneously satisfy the OBCs at both ends where the wavefunctions vanish, the OBC eigenstate should be a superposition of two or more degenerate generalized Bloch solutions that decay equally fast.

\subsubsection{The characteristic polynomial and corresponding allowed spectral graphs}
Spectral graph topology was first briefly studied in Ref.~\cite{lee2020unraveling}, where the branching pattern of simple non-Hermitian models is related to the number of coexisting non-reciprocal length scales. By starting from the simplest cases containing one or two non-reciprocal length scales, the OBC spectra are analytically worked out via a complex momentum deformation $k\rightarrow k+i\kappa(k)$. In this representation, the spatial eigenmode accumulation due to NHSE is nullified via a spatial basis rescaling from the original Hamiltonian $H(k)$ to the surrogate Hamiltonian $\overline{H}(k)=H(k+i\kappa(k))$. By doing this, the PBC spectra of $\overline{H}(k)$ will recover the desired OBC spectrum of the original Hamiltonian $H(k)$, at the expense of inducing an emergent non-locality in real space as a trade-off. Concretely, for the simplest non-trivial case with two non-reciprocal length scales, we have the bivariate characteristic polynomial to be
\begin{equation}
    P(E,z)=F(E)-z^2-\frac{b}{z}\label{twononrec}
\end{equation}

\begin{figure*}
    \includegraphics[width=\linewidth]{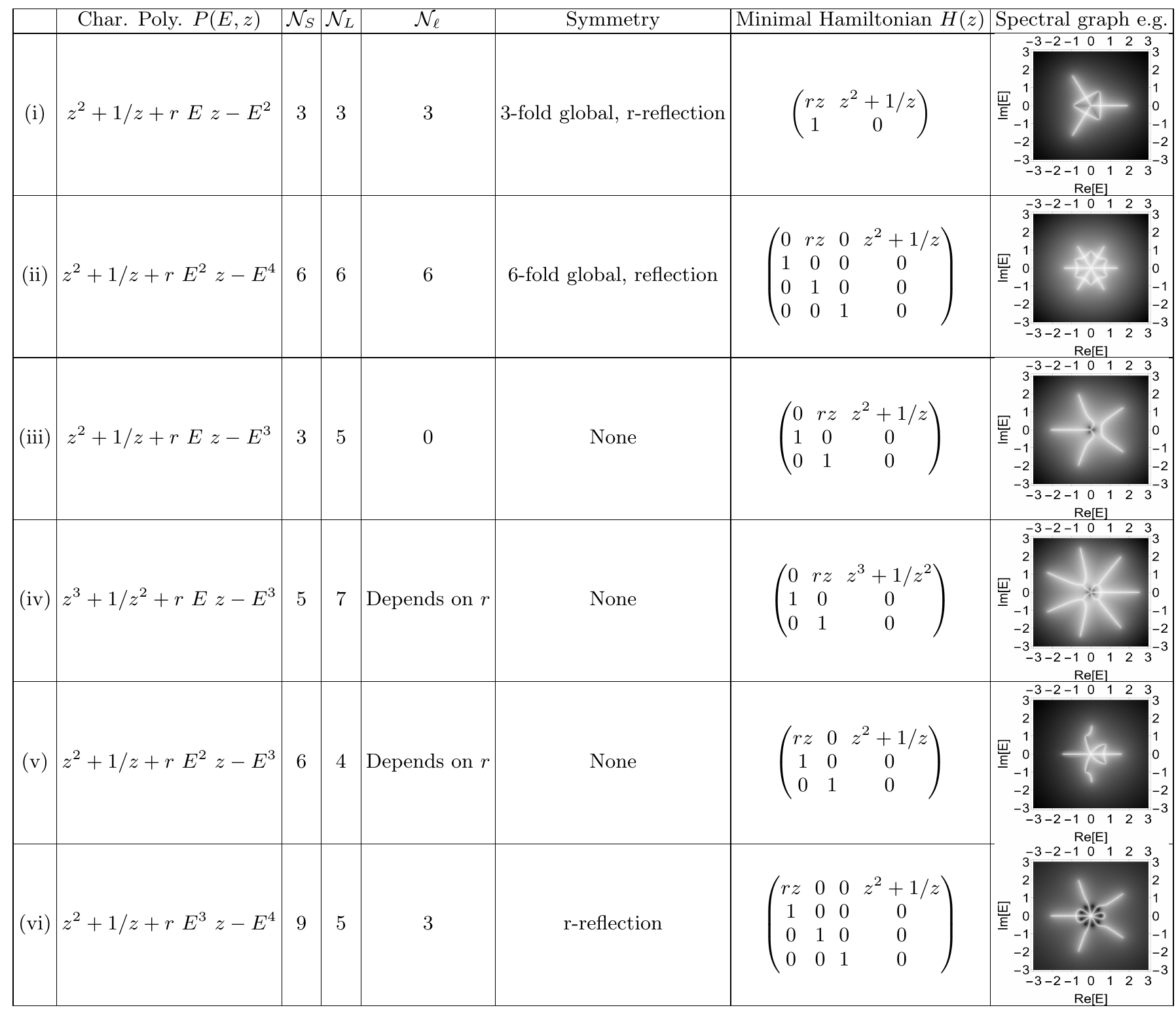}
    \caption{Classification table to illustrate the complexity of spectral graph topology, as adopted from~\cite{tai2022zoology}. For each form of the canonical dispersion $P(E,z)$, one can 1. associate a non-unique minimal Hamiltonian $H(z)$, 2. identify emergent global symmetries of the spectral graph $\bar E$ not necessarily present in $H(z)$, and 3. characterize its spectral graph topology with its number of branches $\mathcal{N}_S, \mathcal{N}_L$ and loops $\mathcal{N}_\ell$, as well as its adjacency matrix.}
    \label{zoologytable}
\end{figure*}
which we can solve for the eigenspectra $\overline{\epsilon}(k)$ of the surrogate Hamiltonian $\overline{H}$ to satisfy
$$F(E)\propto(b/2)^{2/3}\omega_j$$
where the graph topology adopts the shape of three straight lines radiating from the origin of the complex $F(E)$ plane aligned along the cube roots of unity $\omega_j$. With this approach, the smallest complex deformation $|\kappa(k)|$ needed to recover the surrogate Hamiltonian $\overline{H}(k)$ may also be worked out:
$$\kappa(k)=-\frac{1}{3}\log\bigg|\frac{b}{2\cos(k-2\pi j/3)}\bigg|$$
where $j$ is chosen to give the branch the smallest complex deformation $|\kappa(k)|$. The key takeaway is that the OBC loci $\overline{\epsilon}(k)$ is only dependent on the structure of the couplings (e.g. number of non-reciprocal length scales, coupling parameters) and independent of the form of $F(E)$ (which tells us e.g. number of bands). By extending this to generic non-reciprocal couplings of the characteristic polynomial form
$$E^N=z^p+\frac{b}{z^q}$$
with $p,q>0$, the OBC spectrum $\overline{\epsilon}(k)$ can be shown numerically to take the shape of a $N(p+q)$-pointed star, i.e. the Brillouin Zone is folded $p+q$ number of times. A systematic analysis was subsequently done by Ref.~\cite{tai2022zoology}, where a very rich graph topology of generic bounded non-Hermitian spectra is uncovered, distinct from the topology of conventional band invariants and spectral winding familiar in both Hermitian and non-Hermitian settings. Here, the goal was to uncover the deep mathematical relationships between spectral graph topology of non-Hermitian systems and the algebro-geometric properties of the energy-momentum dispersion, i.e. the bivariate Laurent polynomial $P(E,z)$, which also control the localization of Wannier functions~\cite{kohn1959analytic,he2001exponential,brouder2007exponential,lee2016band,monaco2018optimal}. A more generic and sophisticated energy dispersion of the form
\begin{equation}
P(E,z)=Q(z)+r\,G(E)J(z)-F(E)\label{genericbivariate}
\end{equation}
was considered, containing a term involving the product of an energy-band related term $E$ and the momenta $z=e^{ik}$. Eqn.~\ref{genericbivariate} is sufficiently generic and encompasses a wide class of non-Hermitian Hamiltonians since a vast group of previously unrelated Hamiltonians are now tied together via a conformal transformation in the complex energy $E\rightarrow f(E)$ for some analytical function $f$. Here, a kaleidoscope of interestingly shaped spectral graphs resembling stars, flowers or insects was uncovered. The simplest examples are tabulated in Fig.~\ref{zoologytable}, together with the minimal non-unique Hamiltonian, its adjacency matrix representation, as well as, the emergent global symmetries of the eigenspectra.

Similar to how conventional eigenstate topology manifests as linear response quantization, the topological transition between different spectral graphs physically manifests as linear response kinks~\cite{lee2020unraveling,qin2023kinked}, with the different parts of the eigenstates mixing abruptly, leading to enigmatic gapped marginal transitions with no Hermitian analog, giving rise to emergent Berry curvature discontinuities~\cite{lee2020unraveling,qin2023kinked} with physically measurable response signatures, as elaborated in~\cite{qin2023kinked} for an ultracold atomic system.

\subsection{Electrostatics approach to solving the NHSE problem}
\label{sec:electro}

Ref.~\cite{yang2022designing} established a correspondence between the NHSE problem and the age-old problem of the electrostatic field of charge configurations. By mapping a NHSE problem onto an electrostatics problem, one can circumvent direct numerical evaluation of the NHSE spectrum, which for sophisticated spectral graphs may quickly require too many real-space sites, leading to rapid accumulation of floating point numerical errors. This also circumvents the difficulty associated with complicated and perhaps unsolvable algebraic equations to determine the GBZ by completely doing away with them, and instead only requires solving a simpler boundary-valued Poisson equation (Fig.~\ref{fig:spectral_graph}e-f).

Specifically, Ref.~\cite{yang2022designing} demonstrated the correspondence between (i) PBC spectral loops with equipotential conductors, (ii) OBC spectral eigenvalues with electric charges, (iii) PBC-OBC spectral flow as electric field lines and (iv) density of states in the complex $E$ plane with charge density. These relations follow from identifying the inverse skin depth $\kappa$ with the electrical potential $V$,as illustrated in Fig.~\ref{fig:spectral_graph}.

The duality between charge density and spectral density can be understood by considering the DOS along an arbitrary curve $\epsilon$ in the complex $E$ plane of a lattice with $L$ sites, obtained via the Cauchy-Riemann relations:
\begin{equation}
    \rho_\epsilon=\frac{L}{2\pi}|\hat{\epsilon}\times\nabla_E\kappa(k)|\label{DoS}
\end{equation}
This bears mathematical resemblance to the induced charge density on a plane with discontinuous field strength
\begin{equation}    \sigma_\epsilon=\mp2\varepsilon_0|\hat{\epsilon}\times\nabla V|
\end{equation}
This elegant analogy allows us to tackle the difficult inverse problem - engineer a non-Hermitian Hamiltonian $H(k)$ with desired OBC spectral properties and desired spatial profile $\kappa(k)$. A toolbox of familiar methods can be employed here - superposition of point charges and the method of images, further enriching the utility of this approach by including scenarios with non-Bloch band collapse~\cite{PhysRevLett.124.066602}, etc.

The electrostatic analogy also provides intuition to the phenomenon of non-Hermitian pseudo-gaps and pseudo-bands~\cite{li2022non}: Under the NHSE, the PBC and OBC bands may not be in one-to-one correspondence. To understand how, consider the spectral flow as PBCs are deformed into OBCs. In the electrostatics picture, the spectral flow corresponds to electric field lines, and for conductor geometries that are "too sharp", it in conceiveable that neighboring field lines may diverge and group themselves into different pseudo sub-bands. Physically in finite-size OBC systems, this may result in the appearance of topological edge states and bands that ostensibly do not correspond to the topological indices~\cite{li2022non}.



\section{Emergent criticality from the NHSE}
Beyond significantly modifying the band structure and topology of a system, the NHSE also introduces a new length scale, the skin decay length $\kappa^{-1}$. This extra degree of freedom nontrivially affects the behavior of critical systems, as we will review below.

\subsection{Critical non-Hermitian skin effect}
Intriguingly, when two non-Hermitian systems with different inverse skin lengths (or more generally GBZs) are coupled together, a novel critical behaviour is observed - the critical non-Hermitian skin effect (CNHSE). First introduced in Ref.~\cite{li2020critical}, the notion of CNHSE arose after the celebrated GBZ formalism (which we previously established in Section IIB to restore the conventional bulk-boundary correspondence) which only holds in the thermodynamic limit, was rigorously challenged. Fundamentally, the non-Hermiticity effects contribute their share of long-ranged influences which is crucial in critical phenomena. Consequently, critical skin states can even exhibit scale-free behavior while decaying exponentially in space, contrary to conventional critical states which are almost synonymous with power-law spatial decay. They also possess unusual size-dependent entanglement entropy behavior, which challenges the usual approaches for characterizing critical states through their entanglement entropy scaling~\cite{li2020critical}.

\begin{figure*}
    \includegraphics[width=0.9\linewidth]{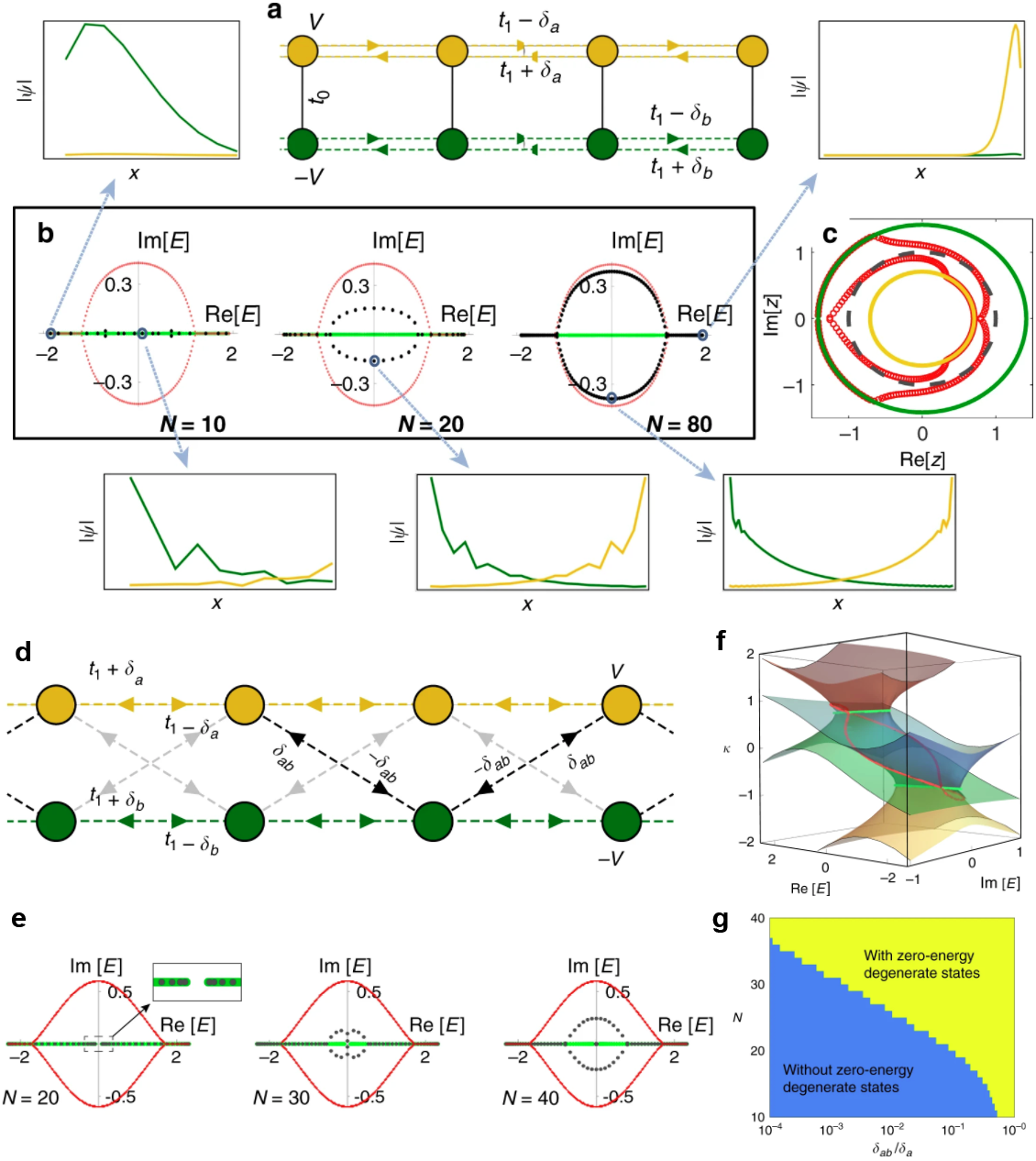}
    \caption{Emergent criticality in finite non-Hermitian lattices, as adopted from~\cite{li2020critical}. (a-c) Critical behaviour in asymmetrically coupled systems involving two non-Hermitian systems with different inverse skin lengths. As the size of the lattice increases, the open boundary spectrum transitions from the decoupled thermodynamic limit (green spectra) to the coupled thermodynamic limit (red spectra). (d-g) Similarly, a critical topological phase transition can be obtained with cross inter-chain non-reciprocal couplings as the size of the lattice increases. This is robust with exponentially weak, but non-zero inter-chain coupling.}
    \label{criticalskin}
\end{figure*}

Although the GBZ formalism holds in the thermodynamic limit, the spectra agreement with that of finite systems only holds far away from the critical point. At these critical points, the characteristic polynomial $f(E,z)$ cannot be reduced to two systems because the two very different subsystems are coupled~\cite{li2020critical}. This is concretely illustrated with a minimal model of two coupled non-Hermitian Hatano-Nelson chains with only non-reciprocal nearest-neighbor hoppings, described by the Hamiltonian
\begin{equation}
    H_{\mathrm{2-chain}}(z)=\begin{pmatrix}g_a(z)&t_0\\t_0&g_b(z)\\\end{pmatrix}\label{criticaleqn1}
\end{equation}
with $g_a(z)=t_a^+z+t_a^-/z+V$, $g_b(z)=t_b^+z+t_b^-/z-V$, and $t_{a/b}^{\pm}=t_1\pm\delta_{a/b}$, as illustrated in the schematic Fig~\ref{criticalskin}a. When the chains are decoupled, i.e. $t_0=0$, the characteristic polynomial $f(z,E)=(g_a(z)-E)(g_b(z)-E)$ is reducible such that each factor of $f(z,E)$ determines the skin eigensolutions of its respective chain. Yet, when the coupling is switched on, $t_0\neq 0$, $f(z,E)$ is no longer irreducible. For the simplest case where $t_a^+=t_b^-=1$ and $t_a^-=t_b^+=0$, we have the irreducible bivariant characteristic polynomial
\begin{equation}
    f(z,E)=E^2-E(z+z^{-1})+(z+V)(z^{-1}-V)-t_0^2\label{criticaleqn2}
\end{equation}
The resulting eigenenergy roots $E=\cos k\pm\sqrt{t_0^2+(V+i\sin k)^2}$ are no longer Laurent polynomials in $z=e^{ik}$ that can be separately interpreted as de facto subsystems with local hoppings. To obtain the OBC spectrum (in the thermodynamic limit), we set $|z_a|=|z_b|$:
\begin{equation}
    E_\infty^2=\frac{1-\eta^2}{1+\eta^2}+V^2+t_0^2\pm2\sqrt{t_0^2-\eta^2+\eta^2t_0^2}/(1+\eta^2),\quad\eta\in\mathbb{R}\label{criticaleqn3}
\end{equation}
which does not reduce to the above-mentioned OBC spectrum of the two decoupled chains in the limit $t_0\rightarrow 0$ (Fig.~\ref{criticalskin}b). Likewise, the $t_0\rightarrow 0$ limit of the coupled GBZ loci are qualitatively different with the collapsed GBZs of the decoupled case (Fig.~\ref{criticalskin}c). The corresponding OBC $E_\infty$ spectrum and the GBZ for $t_0\neq 0$ are qualitatively different.

In critical systems such as the above example, the eigenstates are formed from the superpositions of eigenstates from dissimilar subsystems. This can be understood from a more intuitive perspective. In the GBZ picture, the physical local hoppings are replaced with effectively non-local ones so as to `unravel' the real-space eigenstate accumulation due to NHSE~\cite{lee2020unraveling}. In other words, the NHSE "renormalizes" the hopping strengths such that they increase dramatically with system size, such that the same bare physical couplings can be tuned into the strong or weak coupling regimes just by changing the system size. 
While mathematically, the CNHSE arises when the energy eigenequation $f(E,z)$ exhibits an algebraic singularity involving dissimilar auxiliary GBZ across the transition, physically, this manifests as a discontinuity in eigenenergies and eigenstates in the thermodynamic limit. In physical finite systems, this discontinuity would have to be manifested as some type of finite-size scaling behavior.

As a result, the spectrum exhibits a strong finite-size scaling and the simplistic GBZ picture no longer holds. 
The scaling rule was analytically worked out in Ref.~\cite{yokomizo2021scaling} with a minimal model and its universality was demonstrated for multiband models~\cite{yokomizo2021scaling}. Exact solutions exhibiting boundary scaling behaviour were also worked out in Ref.~\cite{guo2021exact}. Here, they transcended the difficulties and ambiguities presented in conventional numerical methods and via their analytical results, they uncovered the origin of size-dependent NHSE and quantitatively demonstrated the interplay effect of boundary hopping terms and lattice size~\cite{guo2021exact}. Recently, the scaling rule was shown to apply much more generically~\cite{qin2022universal}, with the GBZ shown to explicitly depend on the system size $N$ according to $|\beta|\sim |\beta|_{N\rightarrow \infty}+b^{1/(N+1)}-1$, $b$ a function of the model parameters.
{Finite-size spectral properties have also been noticed earlier in Ref.~\cite{chen2019finite}, 
where the energy gap of a non-Hermitian SSH chain exhibits an oscillating exponential decay or a real-imaginary transition as the system's size grows, depending on how Hermiticity is introduced to the system}

The paradigmatic example of two coupled dissimilar non-Hermitian chains is further studied in detail in Ref.~\cite{rafi2022critical} by considering the interplay with inter-chain coupling and different types of skin mode localization. Moreover, topologically-protected zero modes arise, even when the individual chains do not harbor such zero modes, and exhibit critical phenomena as well. This was also analytically studied in detail, with a proposal to realizing it in topolectrical circuit lattices~\cite{rafi2022system}. A proposal to realize the CNHSE in open quantum systems was given in Ref.~\cite{liu2020helical}, by explicitly considering the Lindblad master equation. This work unraveled the $\mathbb{Z}_2$ skin effect from the CNHSE, showcasing how both the dynamical CSE and the anomalous CSE arise from the modified GBZ equation.

CNHSE is most saliently revealed in size-dependent topological phase crossovers, where the system only exhibits topological modes at certain system sizes. To study that, one can build upon the prototypical example - the two-chain model (Eqn.~\ref{criticaleqn1}), but instead of non-reciprocal intra-chain couplings, we have non-reciprocal inter-chain couplings between adjacent unit cells, which is illustrated schematically in Fig.~\ref{criticalskin}d, and described by the Hamiltonian:
\begin{equation}
    H_{\mathrm{CNHSE-SSH}}(z)=\begin{pmatrix}h_0(z)+h_z(z)&-ih_y(z)\\ih_y(z)&h_0(z)-h_z(z)\\\end{pmatrix}\label{criticaleqn4}
\end{equation}
where $h_y(z)=i\delta_{ab}(z+1/z)$, $h_z(z)=V+\delta_-(z-1/z)$, $h_0(z)=t_1(z+1/z)+\delta_+(z-1/z)$, with $\delta_{\pm}=0.5(\delta_a\pm\delta_b)$. Similar to before, the OBC spectra transitions discontinuously from the decoupled spectra to the coupled spectra as $N$ increases (Fig.~\ref{criticalskin}e,f). This manifests as a gap closure before the emergence of a point gap with two zero-energy degenerate modes lying in its centre - a paradigmatic example of a topological phase transition, but with an intriguing size-induced effect. Notably, this phenomenon is robust for exponentially weak inter-chain coupling for sufficiently large $N$, as illustrated in the phase diagram in Fig.~\ref{criticalskin}g.




\subsection{Exotic non-Hermitian critical behavior}

In general, criticality occurs whenever the bands become gapless. In non-Hermitian systems, gapless points can be more interesting either because of the richer variety of gaps (i.e. point and line gaps), or due to the defectiveness of exceptional gapless points.

The critical properties of non-Hermitian gapless points have been extensively studied, for example, their fidelity susceptibility~\cite{sun2022biorthogonal}, disorder effects~\cite{bao2021topological} and their thermodynamic scaling~\cite{arouca2020unconventional}.  The criticality of two paradigmatic models, the extended Non-Hermitian SSH~\cite{aquino2022critical} and Kitaev models~\cite{rahul2022topological}, were studied concretely by tracking the evolution of the gapless zero energy edge states. In turn, this unravels the relation between EPs and criticality as gap closing points are associated with the appearance of EPs, which in interacting contexts are deeply related to non-unitary conformal field theories (Yang-Lee singularities)~\cite{jian2021yang,shen2023proposal}. In Ref.~\cite{li2022non}, it was also noticed that certain non-Hermitian gaps may appear to host topological in-gap modes even when the topological index is trivial, due to the phenomenon of non-Hermitian pseudo-gaps.

Other than explicitly working out the critical boundaries of the phase diagrams of various paradigmatic models, a natural approach to describe criticality would be to use renormalization group (RG). Yet, the presence of non-Hermiticity poses considerable challenge to applying conventional RG theory directly since conventional RG flow may drive a critical state towards a non-critical state. To transcend this difficulty, the work~\cite{PhysRevB.102.205116} proposes a novel real-space block decimation RG scheme which is much more natural given that the critical hypersurfaces of non-Hermitian systems are obtained in real space under open boundary conditions.
With their distinctive properties compared to conventional criticality, these rich non-Hermitian critical behaviors also inspire further investigations, such as driven dynamics associated with the Kibble-Zurek mechanism~\cite{yin2017KZ}.

\subsubsection{non-Hermitian quantum entanglement}

Remarkably, the effect of non-Hermiticity on quantum criticality and entanglement phase transition is profound, as first comprehensively studied in Ref.~\cite{okuma2021quantum}. The subsequent work~\cite{kawabata2022entanglement} studied the impact of NHSE on the entanglement dynamics and non-equilibrium phase transitions in open quantum systems. Firstly, they showed the NHSE suppresses the entanglement propagation, leading to a non-equilibrium steady state characterized by the area law of entanglement entropy, in contrast with the volume law for thermal equilibrium states. Secondly, they revealed a new type of entanglement phase transition induced by the NHSE, arising from the competition between coherent coupling and nonreciprocal dissipation; the non-equilibrium steady state exhibits the volume law for small dissipation but the area law for large dissipation, between which the entanglement entropy grows subextensively (i.e., logarithmically with respect to the subsystem size). Anomalously, this non-equilibrium quantum criticality is characterized by a nonunitary conformal field theory whose effective central charge is extremely sensitive to boundary conditions. This originates from an EP in the non-Hermitian Hamiltonian, as also previously put forth in Ref.~\cite{lee2022exceptional}, and the concomitant scale invariance of the skin modes localized according to a power law instead of exponential localization. Moreover, the NHSE leads to the purification and the reduction of von Neumann entropy even in Markovian open quantum systems described by the Lindblad master equation.

The effect of non-Hermiticity on entanglement entropy can be completely understood at the single-particle level for free fermion systems. Through Peschel's formula~\cite{peschel2009reduced}, one expresses the entanglement entropy as $S=-\text{Tr}[\bar P\log\bar P+(\mathbb{I}-\bar P)\log(\mathbb{I}-\bar P)]$, where $ P=\sum_{\mu\in\text{occ.}}|\psi_\mu^R\rangle\langle \psi^L_\mu|$ is the \emph{single-particle} projector onto the occupied bands and $\bar P$ is the truncation of $P$ onto a region demarcated by real-space entanglement cuts. In the Hermitian context, this formula has enabled the explicit identification of topological spectral flow with the flow of $\bar P$ eigenvalues~\cite{qi2011generic,hughes2011inversion,alexandradinata2011trace,lee2015free}. In the recent years, similar results have been extended to non-Hermitian models~\cite{herviou2019entanglement,chen2021entanglement}, with an enigmatic discovery of negative entanglement entropy at a phase transition within the non-Hermitian SSH model that can be linked to the bc-ghost non-unitary conformal field theory~\cite{chang2020entanglement}. In this work, the entanglement spectra are concretely studied with the PT-symmetry SSH model and the non-Hermitian Chern insulator model, and some attempt was made in linking the effective $c=-2$ central charge from entanglement entropy scaling with a bc-ghost non-unitary conformal field theory.

\subsubsection{Exceptional bound states}

It was subsequently found that negative entanglement entropy is a generic feature of lattice models containing EPs~\cite{lee2022exceptional}, because the defectiveness at an EP leads to singularities of the two-point function such that $\bar P$ exhibits special eigenvectors known as ``exceptional bound states''. The effective central charge $c$ in $S\sim\frac{c}{3}\log N$ depends linearly on the order of the EP. Interestingly, these exceptional bound states corresponding to $\bar P$ eigenvalues that typically lie way outside the interval $[0,1]$, and thus contribute strongly to negative entanglement entropy. Due to this spectral gap, they are very robustly protected by the existence of the exceptional band crossing, and constitute a new class of robust bound states distinct from topological and NH skin states. 

That said, exceptional bound states exist solely within topological bands, such that they are topologically-protected by construction~\cite{xue2024topologically}. While Hermitian topologically bands are typically orthogonal to each other even when they cross energetically, being localized at opposite boundaries, they can still experience considerable overlap under the non-Hermitian skin effect. Indeed, in Ref~\cite{xue2024topologically}, they overlap so well such that they become geometrically defective, and even exhibits super-volume law negative entanglement scaling in the presence of the macroscopic degeneracy of topological flat bands.

While exceptional bound states may seem rather physically elusive, being initially defined as the eigenstates of non-Hermitian Fermi gas propagators, they can also exist as the eigenstates of any physical Hamiltonian or graph Laplacian that are mathematically equivalent to a $\bar P$ operator of another parent system. Ref.~\cite{zou2023experimental} reports the first experimental detection of such exceptional bound states in a classical electrical circuit setup.

\section{NHSE state dynamics}
The NHSE non-trivially influences the dynamical properties of the system, giving rise to novel phenomena such as wave self-healing~\cite{longhi2022self}, non-Hermitian edge burst~\cite{xue2022non}, chiral tunneling~\cite{PhysRevLett.124.066602}, the dynamic skin effect~\cite{li2022wave}, wave self-acceleration~\cite{li2022wave,PhysRevB.105.245143}, non-Bloch quench dynamics~\cite{PhysRevResearch3023022}, anharmonic Rabi oscillations~\cite{lee2020ultrafast}, direction reversal of NHSE via coherent coupling~\cite{li2022direction}, as well as, manipulating directional amplification and funneling via electric fields~\cite{peng2022manipulating}.

\begin{figure*}
    \centering
    \includegraphics[width=\linewidth]{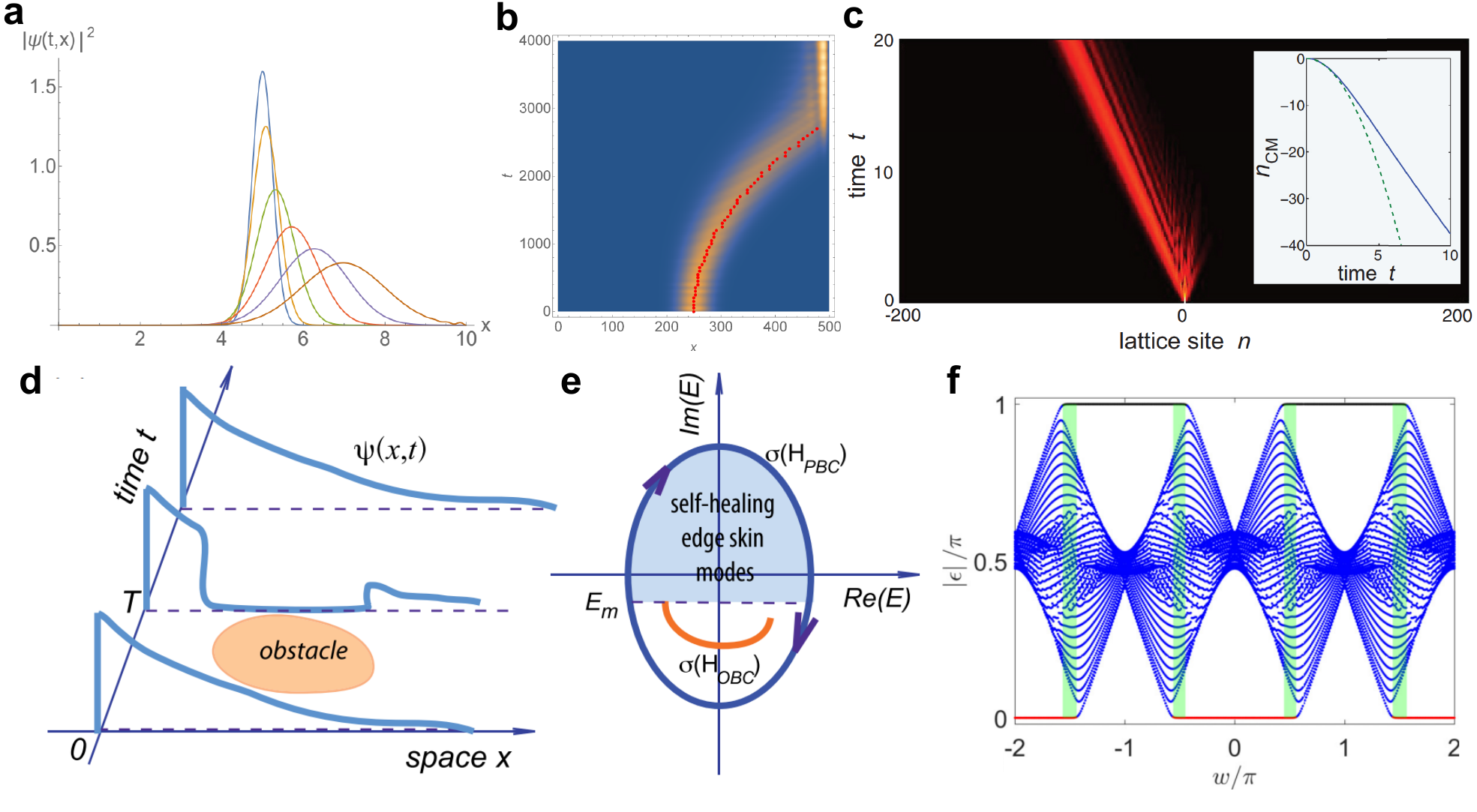}
    \caption{Dynamical effects associated with the NHSE. (a-b) Wavepacket spreading and acceleration of an initially stationary localized wavepacket in a non-Hermitian lattice~\cite{li2022wave}. Upon traversing to the boundary, the wavepacket does not get reflected, suggesting an inelastic scattering. (c) More generally, the self-acceleration plotted is proportional to the area enclosed by the PBC spectrum~\cite{PhysRevB.105.245143}. (d) Self-healing of a wavepacket upon striking an arbitrary space-time potential. (e) Topological criterion to self-healing edge skin modes~\cite{longhi2022self}. (f) Example of Floquet edge states in a periodically driven non-Hermitian lattice~\cite{zhang2020non}. Figures (a,b), (c), (d,e) and (f) are reproduced from Refs.~\cite{li2022wave}, \cite{PhysRevB.105.245143}, \cite{longhi2022self} and \cite{zhang2020non} respectively.}
    \label{fig:dynamical}
\end{figure*}

In the dynamical skin effect, the wave packet acceleration
and inelastic scattering are explained by the interplay
of the NHSE and the Hermitian wave packet spreading (Fig.~\ref{fig:dynamical}a)~\cite{li2022wave}. Fundamentally, this acceleration is transiently induced by the non-reciprocal hoppings in the lattice as the wavepacket traverses through the non-Hermitian lattice. Intriguingly, a localized stationary wavepacket can be accelerated by the inherent non-Hermiticity of the lattice and reach the boundary without being reflected (Fig.~\ref{fig:dynamical}b). The self-acceleration of the wavepacket, in the early time dynamics of a system that exhibits NHSE, is further studied by Ref.~\cite{PhysRevB.105.245143} and was shown to be proportional to the area enclosed by the energy spectrum of the Bloch Hamiltonian under periodic boundary conditions (Fig.~\ref{fig:dynamical}c). Intriguingly, non-Hermitian skin modes in semi-infinite lattices can self-reconstruct their shape after being scattered off by a space-time potential, via a phenomenon dubbed as `self-healing'~\cite{longhi2022self}, as illustrated by the schematic Fig.~\ref{fig:dynamical}d. The work further proves that in a non-Hermitian semi-infinite lattice with a left boundary, any topological edge skin mode at energy $E$ with winding $W(E)<0$ and $\mathrm{Im}(E)$ larger than the largest imaginary part of the OBC energies, is a self-healing wavefunction (as illustrated in Fig.~\ref{fig:dynamical}e).

Another novel dynamic phenomenon associated with the NHSE is the non-Hermitian edge burst, which arises from the interplay between NHSE and imaginary (dissipative) gap closure. This is first demonstrated on a lossy lattice with a Bloch Hamiltonian
\begin{equation}
H(k)=(t_1+t_2\cos k)\sigma_x+\bigg(t_2\sin k+i\frac{\gamma}{2}\bigg)\sigma_z-i\frac{\gamma}{2}\mathbb{I}\label{edgeburst1}
\end{equation}
with loss occurring only on one sublattice. Unlike typical quantum-walk models, this features the NHSE. This is true since Eqn.~\ref{edgeburst1} is simply the non-Hermitian SSH model with left-right asymmetric hopping. The model can also effectively arise from an open system governed by the quantum master equation:
\begin{equation}
    \frac{d\rho}{dt}=-i[H_{\mathrm{eff}},\rho]+\sum_x(L_x\rho L_x^\dag -\frac{1}{2}\{L_x^\dag L_x\rho\})\label{edgeburst2}
\end{equation}
where the effective non-Hermitian Hamiltonian is $H_{\mathrm{eff}}=\sum_{i,j}c_i^\dag h_{ij}c_j-\sum_x\frac{1}{2}L_x^\dag L_x$, where $h$ is the Hermitian part of Eqn.~\ref{edgeburst1}, while $L_x=\sqrt{2\gamma}c_x^B$ is the dissipator. The edge burst always occur whenever the imaginary (dissipative) gap closes, i.e. $|t_1|\leq|t_2|$, except at $t_1=0$. This correspondence has been previously worked out in Refs.~\cite{kawabata2019symmetry,gong2018topological,shen2018topological}: the existence of NHSE is related to the non-zero area enclosed by the complex energy spectrum. All in all, the edge burst manifests as a substantial loss on the boundary. More fundamentally, its origin can be identified as a universal bulk-edge scaling relation derived via Green's functions~\cite{xue2022non}, with algebraic decay stemming from the closure of the imaginary gap, and the small decay exponent at the boundary stemming from the NHSE.

The above example highlights the remarkable fact that the intimate relationship between the bulk and the edge continues to hold in the state dynamics. Concretely, it was shown that the Lyapunov exponent in the long-time behavior of bulk wave dynamics (far from the edges) can generally reveal non-Bloch symmetry-breaking phase transitions and the existence of the non-Hermitian skin effect~\cite{longhi2019probing}.

Apart from topological toy models, there are surprises associated with NHSE found in open quantum systems. In~\cite{PhysRevResearch.4.023160}, the Lindblad master equation was exactly solved for a dissipative topological SSH chains of fermions. The sensitivity on the boundary conditions is reflected in the rapidities governing the time evolution of the density matrix giving rise to a Liouvillian skin effect, which leads to several intriguing phenomena including boundary sensitive damping behavior, steady state currents in finite periodic systems, and diverging relaxation times in the limit of large systems. In a system with quantum jumps and stochasticity, both the short- and long-time relaxation dynamics provide a hidden signature of the skin effect found in the semiclassical limit~\cite{PhysRevB.102.201103}. Even more remarkably, the directed funneling of light at an interface was shown to be possible purely from stochastic fluctuations, even though the hoppings are reciprocal on average~\cite{longhi2020stochastic}. A

A comprehensive understanding of the effects of NHSE on state dynamics will allow us to better design sensors and devices. For example, it was shown that trapping light at a topological interface with NHSE depends significantly on the initial state ~\cite{longhi2021bulk}. Moreover, the implications of non-Bloch band theory to particle Bloch dynamics are profound and lead to new physics. For instance, at the collapse of non-Bloch bands, electrons irreversibly tunnel between Bloch bands in a chiral fashion, contrary to Hermitian systems where Zener tunnelling is oscillatory~\cite{PhysRevLett.124.066602}. Tangentially, the Hartman effect --  the independence of the phase tunneling time on the barrier width, can exist without any PT symmetry requirements whenever the barrier itself exhibits NHSE~\cite{longhi2022hartman}. Finally, the NHSE can be selectively turned on and off under the presence of static or time-dependent electric fields; this is a consequence of the interplay between Stark localization and dynamic localization, and the NHSE~\cite{peng2022manipulating}.




The presence of the NHSE renders the study of quench dynamics of these non-Hermitian topological models challenging. NHSE dictates the collective localization of states on the boundary under OBCs, therefore coupling the dynamics of different momentum states. Furthermore, the eigenenergy spectra of both the initial and final Hamiltonian (related by the quantum quench) will have generically complex energy, particularly loops in the complex plane. Ref.~\cite{PhysRevResearch3023022} circumvents this issue by projecting the quench dynamics onto the generalized momentum sectors of the GBZ, in turn revealing the dynamic skyrmions in the generalized momentum-time domain, which are intimately related to the non-Bloch topological invariants of the pre- and post-quench Hamiltonians. This formalism would facilitate the direct detection of non-Bloch topological invariants in experiments.

\subsection{Real spectra and asymptotic non-divergent states from the NHSE}

The directed amplification from unbalanced couplings causes a generic initial state to evolve and spread out such that it is amplified more in one direction than the other. As such, under OBCs, it would eventually encounter a boundary and be unable to propagate further. Since directed propagation and amplification are tied to each other in such a NHSE lattice, the state's amplification would also be significantly suppressed by the boundary or even completedly stalled. In the latter case, this would correspond to an energy spectrum that is entirely real due to the NHSE. The reality of the eigenspectra thus has significant implications on the dynamical behaviour, i.e. the eigenstates do not blow up after long time-evolution (but also see~\cite{xue2022non}).

Non-Hermitian models with real eigenenergies are highly sought-after for their stability. There are many proposals and methods to engineer such systems. The most common way to guarantee real spectra is to enforce parity-time (PT) symmetry on the Hamiltonian, such that the gains and losses conspire to lead to eigenstates with conserved total amplitude~\cite{Bender1998nonH,el2018non,feng2017non,stegmaier2021topological}. Yet, having a PT-symmetric Hamiltonian is neither a necessary nor a sufficient condition because the PT symmetry itself has to remain unbroken~\cite{mostafazadeh2002pseudo}. Alternatively, one could obtain real spectra from pseudo-Hermitian systems, which exist only in the a priori unknown engineered real spectrum system~~\cite{mostafazadeh2002pseudo}. There are also specific attempts at engineering real spectra by tuning model parameters, to induce real-complex transitions~\cite{PhysRevB.103.054203,suthar2022non}, which will not be the focus of this review. We will like to highlight more general approaches to engineering real spectra via the NHSE itself, without necessarily invoking symmetries of the lattice couplings or their corresponding momentum-space Hamiltonians.

In Ref.~\cite{zhang2022real}, it was shown that real OBC eigenenergies correspond to intersections of the inverse skin depth $\kappa$ curves at purely real energies (called $\kappa$ crossings), which can exist even if the PBC eigenenergies are already complex, i.e. if the $\kappa$ curves cross $\kappa=0$ at $\mathrm{Im}[E]\neq 0$. Hence, as long as the symmetry of the $\kappa$ curves themselves are preserved, the OBC spectra will be real. Through a systematic investigation, the work~\cite{zhang2022real} gives examples of simple ansatz models with purely real OBC spectrum, that are also as local as possible. More generally, via an electrostatics approach~\cite{yang2022designing} (as outlined in Section IV), one can reverse engineer a parent Hamiltonians for any desired real OBC spectrum and skin localization. All in all, these works open the door to a plethora of models with stable eigenenergies, beyond the use of conventional symmetries.

\subsection{NHSE in Floquet systems}
Floquet topological phases are extensively studied in time-periodic Hermitian Hamiltonians~\cite{rechtsman2013photonic,flaschner2016experimental}. The richness of Floquet topological phases in non-Hermitian systems was first explored in the work~\cite{Longwen2018nonHFloquet}. Here, non-Hermiticity-induced Floquet topological phases with unlimited winding numbers with arbitrarily many real zero- and $\pi$-quasienergy edge states are engineered. This is achieved by subjecting a one-dimensional ladder-geometry lattice with a piecewise time-periodic quench. Effectively, periodic driving can induce long-range hoppings, thereby giving rise to emergent physics. In one period of the proposed driving, the quench alternates between a non-Hermitian Hamiltonian $H_1$ (with asymmetric couplings between sublattices in the same unit cell) and a Hermitian Hamiltonian $H_2$, given by
\begin{equation}
    H_1=\sum_n[ir_y(|n+1\rangle\langle n|-\mathrm{h.c.})+2i\gamma|n\rangle\langle n|]\otimes\sigma_y\label{floquet1}
\end{equation}
\begin{equation}
    H_2=\sum_n[ir_x(|n\rangle\langle n+1|+\mathrm{h.c.})+2\mu|n\rangle\langle n|]\otimes\sigma_x\label{floquet2}
\end{equation}

A systematic study was done later on the interplay of NHSE and periodic driving~\cite{zhang2020non}.
Here, a new phenomenon dubbed the Floquet non-Hermitian skin effect (FNHSE) was discovered. The non-Hermiticity not only splits each spectral degenerate point of the parent Hermitian Floquet system into two EPs, but also induce many other EPs. Moreover, the quenched Hamiltonians comprising the periodic quench protocol do not need to exhibit NHSE in order for the periodically quenched system to exhibit FNHSE. The FNHSE does break the BBC but under certain parameter regimes where low-order truncation of the
characteristic polynomial can be done with negligible error. Finally, the existence of two different types of Floquet edge modes can still be predicted exactly by introducing the generalized Brillouin zone (GBZ) in two time-symmetric frames. As a remark, the unsupervised identification of Floquet topological phase boundaries has been successful for Hermitian models, and holds the promise to be extended to non-Hermitian models~\cite{PhysRevResearch.4.013234}.

The study of non-Hermitian Floquet systems was extended to disordered systems possessing non-Hermitian Floquet topological Anderson insulator phases~\cite{PhysRevB.102.041119}, as well as, with spatial modulated on-site potential~\cite{zhang2021floquet}. The NHSE, in the presence of the NHSE, also results in exponentially enhanced Rabi frequencies due to the exponentially large amplified states\cite{lee2020ultrafast}.

Typically, the characterization of topological phases of the time-periodic open quantum systems is done via the use of a dynamical winding number~\cite{PhysRevA.100.053608,PhysRevB.100.184314} or the frequency-space Floquet Hamiltonian. For the latter, the eigen-energy forms the celebrated Wannier-Stark ladder, and at each frequency space lattice site, repeated Floquet bands are observed~\cite{rudner2013anomalous}. Floquet topological phase transitions are thus identified by the collapsing of repeated Floquet band gap. Yet, this is not observed in non-Hermitian systems. In the work~\cite{wang2022nonf}, the authors proposed the non-Floquet theory, which features a temporally non-unitary transformation on the Floquet state, to restore the Wannier-Stark ladders in the Floquet spectrum. Another work proposed a dual topology characterization scheme, so as to circumvent the need to construct the GBZ~\cite{PhysRevB.103.L041404}.

Floquet effects in systems with NHSE can be engineered for applications. In a lattice of coupled ring resonators~\cite{gao2022non}, for example, by unconventionally fixing the on-site gain or loss in each ring and by identifying the lattice modes as Floquet eigenstates, the anomalous Floquet NHSE can be realized. Here, skin modes exist at every Floquet quasienergy, allowing for broadband asymmetric transmission, akin to anomalous Floquet insulators~\cite{zhang2021superior}. In another work~\cite{wu2022non}, it was shown that the presence of NHSE can transform Weyl semimetals into Weyl-exceptional-ring semimetals, as well as a zoo of exotic non-Hermitian topological phases. Finally, it was also proposed that Floquet driving a non-Hermitian topological superconductor can yield multiple Majorana edge modes, useful for realizing environmentally robust Floquet topological quantum computations~\cite{PhysRevB.101.014306}.

Periodic driving can also generate nontrivial "mixed" higher-order topology~\cite{PhysRevB.102.094305}. In Ref.~\cite{liu2022mixed}, Floquet driving in anomalous Floquet topological insulators are show to generate intrinsic topologically non-trivial non-Hermitian boundary states which are furthermore scale-invariant. The entire system is thus characterized by an unprecedented `mixed' higher-order topology, where a bulk system with Floquet topology induces a non-Hermitian topology on the boundary.

\section{NHSE beyond linear non-interacting crystal lattices}
\begin{figure*}
    \includegraphics[width=0.8\linewidth]{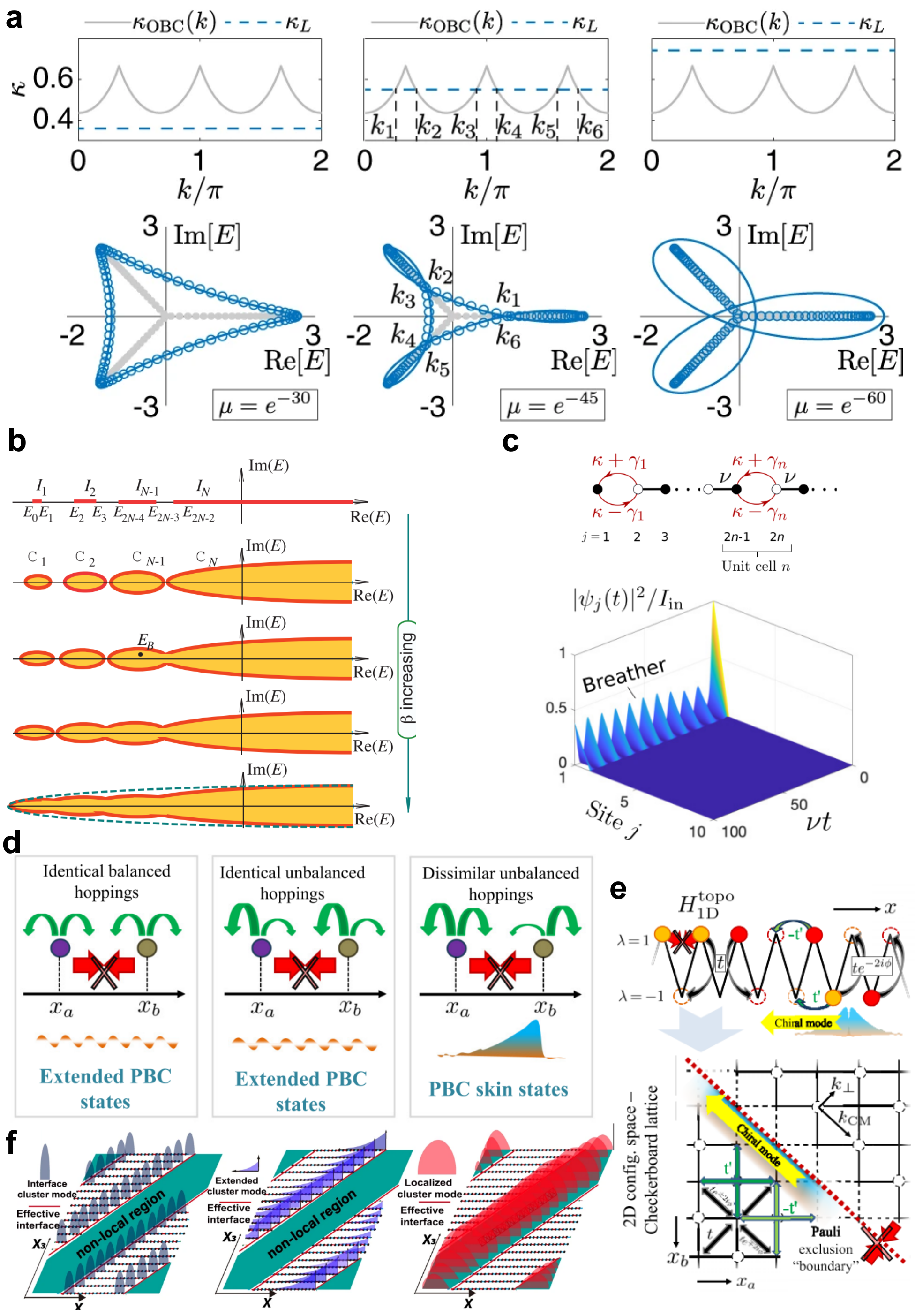}
    \centering
    \caption{Unconventional manifestations of the NHSE. (a) Lattices with non-reciprocal impurities can result in the coexistence of scale-free accumulation and NHSE~\cite{li2021impurity}. (b) Introducing non-Hermiticity, the imaginary vector potential can result in the NHSE even under PBC~\cite{longhi2021non}. (c) With both nonlinearity and nonreciprocal non-Hermiticity, a novel oscillatory soliton, a topological end breather, is formed and is strongly localized to a self-induced topological domain near the end of the lattice~\cite{lang2021non}. (d) A combination of particle statistics and suitably engineered many-body interactions can result in boundaries (sites of disallowed occupation) in the many-body configuration space. Topological and skin states can thus form without physical boundaries~\cite{lee2021many}. (e) An example of a 1D chain with well-designed two-body interactions that result in a chiral propagating state along the diagonal boundary in the two-body configuration space~\cite{lee2021many}. (f) This principle can be generalized to the strongly interacting limit, which results in the formation of localized non-Hermitian skin clusters~\cite{shen2022non}, shaped by the connectivity structure of the many-body Hilbert space instead of the real-space lattice. Figures (a), (b), (c), (d,e) and (f) are reproduced from Refs.~\cite{li2021impurity}, \cite{longhi2021non}, \cite{lang2021non}, \cite{lee2021many} and \cite{shen2022non} respectively.}
    \label{fig:unconventionalNHSE}
\end{figure*}

Typically, the NHSE is formulated on a translation-invariant lattice described by a linear non-interacting tight-binding model. But in fact, the NHSE requires just non-reciprocal breaking of Hermiticity. More interesting incarnations abound when we relax these conditions of translation invariance, linearity, single-particle physics, etc, as we will review below.


\subsection{Breaking of translation invariance - NHSE interplaying with disorder or impurities}\label{sec:dis}
The prototypical way to break translational invariance is to introduce disorder, which acts as partial ``boundaries'' that acquire non-local and non-perturbative influences due to the NHSE~\cite{bao2021topological}. Surprisingly, the overall decay scaling is independent of the system's size, in spite of the NHSE which exponentially localizes steady states~\cite{li2021impurity,liu2021exact1}.

To understand this, consider a local impurity represented as a modified coupling between the first and last sites, i.e.
\begin{equation}
    H=\sum_{x=0}^{L-1}(e^\alpha c_x^\dag c_{x+1}+e^{-\alpha}c_x^\dag c_{x-1})+\mu_+c_Lc_0+\mu_-c_0^\dag c_L\label{SFA}
\end{equation}
where $\mu_\pm=\mu e^{\pm\alpha}$, with $\mu$ controlling the local impurity, $\alpha>0$, and $x=0,1,\dots,L$ being the lattice site index. This goes beyond the typical interpolation between PBCs and OBCs, with $\mu\in[0,\infty)$ instead of just $\mu\in [0,1]$. When one analytically solves the eigenstates for Eqn.~\ref{SFA}, we yield two isolated strongly localizing eigenstates at both ends, with the other eigenstates exponentially decaying with a common decay constant
\begin{equation}
    \kappa_L=\frac{\ln\mu-2\alpha}{L-1}\label{SFA2}
\end{equation}
When $\mu=e^{2\alpha}$, i.e. quasi-PBC, it is possible to `gauge' away to recover the original Hamiltonian under PBC. For $\mu>e^{2\alpha}$, the leftwards hopping from $x=L$ to $x=0$ is further enhanced whereas the opposite is further suppressed. This accumulation is $|\psi_{L,n}|/\psi_{1,n}|=e^{2\alpha/\mu}$, i.e. scale-free decay profile. The reverse occurs for $0<\mu<e^{2\alpha}$.

Going beyond hopping with one length scale, we can consider the following
\begin{equation}
    H_{\mathrm{NNN}}=\sum_{x=0}^{L-1}e^\alpha c_x^\dag c_{x+1}+\mu e^\alpha c_L^\dag c_0+\sum_{x=0}^L e^{-\alpha}c_x^\dag c_{x-2}\label{SFA3}
\end{equation}
which yields a three-fold symmetric spectrum for all values of $\mu$. The impurity-free version of such models have been studied in Refs.~\cite{lee2020unraveling,tai2022zoology,zhang2022real,fu2022anatomy}. While the inverse decay lengths $\kappa_L(\mu)$ of SFA states, as induced by the impurity, are insensitive to the exact configuration of non-reciprocal hoppings in the bulk, the OBC skin modes have $k$-dependent inverse decay lengths. The two phenomena compete and may coexist for some critical $k$-values, as shown in Fig.~\ref{fig:unconventionalNHSE}a, leading to a range of qualitatively different edge-localized regimes beyond usual NHSE states, and also dualities between strong and weak boundary couplings.

Beyond this simple impurity toy model, the seminal phenomenon of Anderson localization is modified in the presence of non-Hermiticity, giving rise to the coexistence of localized and extended states even in one- and two-dimensional lattices~\cite{cai2021boundary,sarkar2022interplay,yuce2022coexistence,roccati2021non}. More profoundly, chiral hinge states of higher-order non-Hermitian topological insulators remain robust once disorder is switched on, and transitions to surface states as the disorder strength is increased~\cite{wang2022chiral}. Moreover, disorder can drive the system into a HOTI phase~\cite{kim2021disorder}. One could also construct a real-space topological invariant for strongly-disordered non-Hermitian systems and in turn, predicting the non-Hermitian Anderson skin effect where the skin effect solely arises from the presence of disorder~\cite{claes2021skin}. The evolution of the mobility edge (the energy band boundary between localized states and extended states) and the competition between NHSE and localization effects is exemplified concretely using the Hatano-Nelson model with unidirectional hopping under on-site potential uncorrelated disorder~\cite{longhi2021spectral}. To fully encapsulate the effects of (particularly strong) on-site disorder in an analytical fashion, a ``modified GBZ theory'' has been suggested - essentially involving the search of the minimum of the polynomial $F(E,\beta)=|{\rm det}[E-H_{\mathrm{PBC}}(\beta)] - {\rm det}[E-H_{\mathrm{OBC}}]|$~\cite{liu2022modified}. This yields an interval instead of a single point, in turn desirably restoring the bulk-boundary correspondence for disordered samples. The additional restrictions presented by the ``modified GBZ theory'' also correctly describes the interplay of NHSE and the magnetic field, where the latter similarly breaks translational invariance~\cite{liu2022modified}. Other examples of disordered systems with NHSE that have been recently studied include the many-body coupled Hatano-Nelson chains in the presence of a random disorder potential~\cite{suthar2022non}, and a quasiperiodic lattice (the disorder is emulated by the incommensurate quasi-periodic on-site potential)~\cite{PhysRevB.105.L121115} with Rashba Spin-Orbit interaction~\cite{chakrabarty2022skin}.

Disorder may also be used to adroitly realize an effective semi-infinite 1D lattice system, with a complex eigenspectrum that completely fills up the interior of a PBC loop. This is done by concatenating Hatano-Nelson chain segments with random couplings. The result is equivalent to an ensemble of Hatano-Nelson chains with different inverse skin lengths $\kappa$~\cite{jiang2022filling}.

The interplay of the point gap topology (responsible for giving closed PBC loops) and topological defects give rise to more interesting NHSE phenomena. In Ref.~\cite{bhargava2021non}, two dislocations were introduced into a two-dimensional weak Hatano-Nelson lattice. Concurrently, a skin and anti-skin effect is realized on each defect - a macroscopic localization of states towards one dislocation and a concomitant depletion of states away from the other. A topological invariant for dislocation vector, with state accumulation from the dislocation. A topological invariant is identified, which takes the form of a $\mathbb{Z}_2$ Hopf index that depends on the Burgers vector characterizing the dislocations. Crucially, the anti-skin effect uncovers an additional knob for tailoring the positions of eigenstates in non-Hermitian systems, which is pertinent for applications.

In higher dimensions, mismatch between the macroscopic symmetry and the lattice symmetry itself could also be a means to realize NHSE, even on a reciprocal system~\cite{wang2023experimental}. This skin effect is solely dependent on the geometry of the system, hence facilitating new routes for wave structuring.

Random impurities can also manifest as fluctuating hopping amplitudes in a spatially-ordered lattice. The NHSE can still be realized in a stochastic system, dubbed the stochastic NHSE, even if the couplings are symmetric on the average~\cite{longhi2020stochastic}. The stochastic skin effect stems from the point-gap topology of the Lyapunov exponents under PBC.

\subsection{NHSE without a real-space lattice}

Although almost always formulated as arising from asymmetric lattice couplings, the NHSE is just a result of the interplay between non-reciprocity and non-Hermitian gain/loss, and does not require a lattice. Indeed, Ref.~\cite{longhi2021non} suggested that the appearance of the NHSE by an imaginary gauge field $\beta$ in a finite crystal with OBC is a very general feature, that holds beyond the usual tight-binding models. As discussed, $\beta$ has the interpretation of an inverse decay length and is responsible for the localization of wavefunctions under OBCs. Staying with PBC, one can introduce non-Hermiticity via an imaginary vector potential $\beta$ in the Schr\"{o}dinger's equation. This is equivalent to complexifying the Bloch wave number, resulting in the energy spectrum to transition from purely real intervals to arbitrarily closed loops in the complex plane. As $\beta$ increases, the individual closed curves will increase in area and eventually merge with adjacent curves, leading to an open curve in the complex energy plane, approaching the free-particle dispersion curve in the large $\beta$ limit. This is illustrated in Fig.~\ref{fig:unconventionalNHSE}b.

The imaginary vector potential is crucial in formulating an intriguing duality between non-Hermiticity and curved spacetime~\cite{lv2022curving,wang2022duality,lv2022emergent}. Specifically, by mapping the continuum limit of non-Hermitian lattice models to the Schr\"{o}dinger equation on a Poincar\'{e} half-plane, the inverse localization length $\kappa$ manifests as an imaginary vector potential which curves the space. The significance of this result is profound. Theorists can study curved spaces on easily accessible experimental non-Hermitian systems such as electrical circuits. Correspondingly, experimentalists can employ readily accessible curved spaces, such as hyperbolic surfaces~\cite{gao2015topological,kollar2019hyperbolic,lenggenhager2022simulating}, to realize experimentally challenging non-Hermitian models with more easily implementable non-Hermitian building blocks.

\subsection{Nonlinear NHSE systems}
The interplay with the NHSE and the classical non-linearity leads to intriguing new phenomena, such as trapping effects~\cite{ezawa2022dynamical,yuce2021nonlinear}, breathers, and solitons~\cite{lang2021non}.

{We describe the topological end breather - a novel oscillatory soliton in a nonlinear, non-reciprocal, non-Hermitian lattice that exhibits the NHSE~\cite{lang2021non}. The end breather is strongly localized to a self-induced topological domain near the end of the lattice, in sharp contrast to the extended topological solitons in linear lattices~\cite{tuloup2020nonlinearity}. Fundamentally, this is aided by the NHSE which suppresses topologically trivial bulk states, leading to a domain wall near the lattice edge. To understand this, consider the non-reciprocal SSH-like model with Hamiltonian
\begin{equation}
    H=\sum_{n}[(\kappa+\gamma_n)a_{2n-1}^\dag a_{2n}+(\kappa-\gamma_n)a_{2n}^\dag a_{2n-1}+\nu(a_{2n}^\dag a_{2n+1}+a_{2n+1}^\dag a_{2n})]\label{nonlinear1}
\end{equation}
where the intercell hoppings are reciprocal, while the intracell hoppings are generally non-linear, of the form
\begin{equation}
    \gamma_n=\gamma_s-\frac{\gamma_s-\gamma_0}{1+(|\psi_{2n-1}(t)|^2+|\psi_{2n}(t)|^2)/I_s}\label{nonlinear2}
\end{equation}
where $I_s$ is a saturation intensity state. An example of a topological end breather (Fig.~\ref{fig:unconventionalNHSE}c) is demonstrated with parameters $\gamma_0=0$ and $\gamma_s=\sqrt{7}\nu/2$.
Such non-linear models, as opposed to interacting quantum models, can be realized in classical platforms such as electrical circuits equipped with non-linear elements~\cite{hadad2018self,wang2019topologically,kotwal2021active,hohmann2022observation} i.e. diodes or non-linear capacitors~\cite{hadad2018self,dobrykh2018nonlinear,fu2022gate,ezawa2022nonlinear}, with linear non-reciprocity provided by operational amplifiers~\cite{hofmann2019chiral}.


\subsection{NHSE in interacting many-body systems}
A plethora of condensed matter phenomena hinges on the many-body nature of quantum interactions giving rise to novel emergent phenomena with no single-particle analog. The NHSE is no exception, where new phenomena can manifest, such as an emergent real-space Fermi surface~\cite{mu2020emergent}, clustering of the eigenspectrum~\cite{zhang2022symmetry}, multifractality of many-body skin effect~\cite{hamanaka2024multifractality} and many-body distributions caused by NHSE interplaying with Fermionic repulsion~\cite{alsallom2022fate,mu2020emergent,suthar2022non}, including~\cite{shen2023observation} which was observed on a digital quantum processor. Fundamentally, the competition between repulsion and boundary state localization reshapes (Fig.~\ref{fig:unconventionalNHSE}d from Ref.~\cite{lee2021many}) how the NHSE manifests in non-Hermitian systems, as manifested in the full-counting statistics~\cite{dora2022full}, non-Hermitian Laughlin states~\cite{yoshida2019non},Kitaev-Hubbard bosons~\cite{wang2022non2}, the Lieb-Liniger Bose gas with imaginary vector potential~\cite{mao2022non} or spin chain excitations~\cite{chen2022topological}. Interestingly, the NHSE itself can also be suppressed by certain correlation/interaction effects~\cite{yoshida2022reduction,dora2022full}. 

A special class of interacting few-body quantum systems can be exactly mapped onto one-body problems in a higher-dimensional lattice. Specifically, Refs.~\cite{lee2021many,shen2022non} proposed a framework for mapping an $N$-body system in $d$ dimensions to a single-body problem in $Nd$ dimensions. In particular, it is possible to achieve robust cluster states where particles are localized next to each other, in the absence of boundaries, solely using particle statistics and appropriately engineered interactions (Fig.~\ref{fig:unconventionalNHSE}e from Ref.~\cite{lee2021many}). By identifying the original interacting chain with the single-particle Hilbert space of a higher-dimensional configuration lattice, one can consequently observe skin states aggregating at the effective ``boundaries'' in the many-body configuration space~\cite{lee2021many,shen2022non,zhang2022observation}. 
Fock space skin localization can even be engineered to enhance quantum ``scarring''~\cite{shen2024enhanced}, where the ergodicity breaking is strengthened by the accumulation of states in the direction of decreasing Hamming distance from the initial Neel state. This Fock skin effect, however, is complicated by the macroscopically large Hilbert space of states at the same Hamming distance, and do not possess a simple correspondence with the asymmetry in the physical hoppings. Beyond studying the state dynamics and phase diagrams of particular archetypal models, Ref.~\cite{kawabata2022many} generalized the topological invariants for many-body non-Hermitian systems. 

The NSHE can also emerge in the effective descriptions of various interacting models, even though the interactions themselves are not asymmetric hoppings. For instance, real-space dynamical mean-field theory reveals the NHSE in the pseudospectrum of some strongly-correlated systems~\cite{yoshida2021real}. NHSE can also occur in the synthetic field moments space of zero-dimensional bosonic quantum dimers~\cite{arkhipov2023emergent}. An new, interacting form of the NHSE also occurs in the presence of interacting impurities, as manifested by so-called squeezed polarons which are impurity-localized dipole-like density profiles that are impervious to the lattice boundaries~\cite{qin2022non}. More interesting multi-polar NHSE signatures have been suggested in ~\cite{gliozzi2024many}, and it remains to be seen if such geometric anisotropy may be generalized to give rise to quantum Hall-like states~\cite{lee2015geometric} with NHSE interplay.

Interestingly, quasi-particle excitations of a closed many-body Hermitian system may exhibit effective dissipation due to their scattering off other degrees of freedom within the considered system, leading to experimentally measurable responses for solid-state systems~\cite{micallo2023correlation}. Finally, coupling an interacting Kitaev honeycomb spin model with the environment leads to an emergent non-equilibrium phase called the exceptional spin liquid~\cite{PhysRevLett.126.077201}.

Beyond many-body Hamiltonians, the NHSE -- extensive exponential localization of "skin" modes -- also arises in the dynamics of open systems. For instance,~\cite{vznidarivc2022solvable} found an unexpected sudden transition in the purity relaxation rate in the many-body unitary dynamics of qudits, a behavior which can be traced to the asymmetric matrix elements of the Toeplitz matrix underlying the purity dynamics. The NHSE was also revealed in the anomalous behavior of quantum emitters in non-Hermitian heat baths~\cite{gong2022anomalous,gong2022bound,roccati2022exotic}, and interpreted as a Maxwell pressure demon in the many-body context~\cite{cao2021physics}. Interesting, under a space-time duality mapping, the sensitivity to initial conditions in a quantum chaotic system can reinterpreted as the sensitivity to boundary conditions in a NHSE system~\cite{zhou2022space}.\\

\section{Physical realizations of the NHSE}

\subsection{Recent experimental demonstrations}


Compared to merely realizing gain/loss, experimentally realizing the NHSE is more challenging, requiring the simultaneous presence of non-reciprocity and non-Hermitian gain/loss. It is not only until 2020 that the NHSE was first demonstrated in electrical~\cite{helbig2020generalized}, quantum optics~\cite{xiao2020non} and mechanical platforms~\cite{ghatak2020observation}. ``Topolectrical'' circuit realizations implement the non-reciprocity primarily through operational amplifiers~\cite{hofmann2019chiral,helbig2020generalized,liu2021non,xu2021coexistence,deng2022nth,zhang2023electrical,stegmaier2024realizing}, and their versatility~\cite{li2019emergence,lenggenhager2022simulating,shang2024observation} has enabled the experimental simulation of higher-order states~\cite{zou2021observation,stegmaier2021topological,shang2022experimental} in two or more dimensions, potentially aided by machine learning techniques~\cite{shang2022experimental}. Photonic setups~\cite{weidemann2020topological} and single photon quantum walks have also fruitfully demonstrated the transition between the NHSE and other condensed matter phenomena of interest, such as non-Hermitian quasicrystals~\cite{lin2022simulating} and topological Anderson insulators~\cite{lin2022observation}. Mechanical~\cite{ghatak2020observation,wang2022non1} and acoustic setups~\cite{zhang2021acoustic,zhang2021observation} rely on the intrinsic non-reciprocity of the medium, and have also been highly successful in demonstrating the 1D and higher-order NHSE. 
{In Ref.~\cite{palacios2021guided}, second-order NHSE has been observed in active matter systems of Janus particles,
manifesting as spontaneous particle edge guidance and corner accumulation of self-propelled particles.}
Although NHSE is fundamentally a single-particle phenomenon, experimental demonstrations in solid state NV-center platforms~\cite{yu2021experimental} and ultracold atomic lattices~\cite{liang2022observation} constitute significant steps towards physical investigations of the interplay of the NHSE with many-body effects.

\subsection{Experimental proposals and simulations for the NHSE}

While the NHSE has aleady been observed in a select set of experiments, there are many other proposals for future experimental demonstration in a variety of physical platforms. Below, we briefly review some existing proposals, each suited for realizing different NHSE-related phenomena. We would omit an explicit discussion of proposals based on classical circuits, since they are already well discussed in the literature~\cite{bergholtz2021exceptional,zhang2022review}.

\subsubsection{Photonics/Optics proposals}

The NHSE can be implemented in mature photonic systems such as lasers. in Ref.~\cite{zhu2022anomalous}, it was proposed that the interplay between nonlinear gain saturation and the non-Hermitian skin effect gives a  laser with the opposite behavior from usual: multimode lasing occurs at low output powers, but pumping beyond a certain value produces a single lasing mode, with all other candidate modes experiencing negative effective gain, thereby giving rise to the NHSE.

Lasers can also indirectly induce the NHSE, such as in exciton-polaritons condensates excited by circularly polarized laser driving in 1D~\cite{mandal2022topological}, 2D corners~\cite{xu2022non}, with implications on multistability~\cite{yu2021non}. The strong non-linearity in lasers 

The NHSE can also exist in waveguides or photonic crystals with the requisite loss and reciprocity breaking~\cite{yang2022concentrated,song2023observation,roccati2024hermitian}. In Ref.~\cite{yang2022concentrated}, it was shown that the NHSE can be induced and controlled by varying the positions of the atoms in a waveguide, with the atoms experiencing long-ranged effective couplings and non-Hermitian loss. Further, the NHSE can also be steerable via the gauge field in a coupled optical ring resonator array~\cite{lin2021steering}. The NHSE can also be induced in photonic crystals~\cite{zhong2021nontrivial,price2022roadmap} i.e. in chiral photonic crystals with anomalous PT symmetry~\cite{yan2021non}. The anomalous Floquet NHSE~\cite{gao2022non} as well as antihelical topological edge states~\cite{xie2023antihelical} have also been proposed for a photonic ring resonator lattice.

Finally, single-photon circuits can also simulate the NHSE. Beyond the pioneering NHSE experiment based on photonic quantum walks~\cite{xiao2020non}, there has also been other proposals in 2D~\cite{lin2022manipulating,li2021two}, which are potentially useful for also studying the interplay of photon-photonic interactions in a NSHE background, which boasts of a variety of interesting physics~\cite{pyrialakos2022thermalization}.

\subsubsection{Quantum circuit proposals}

Moving beyond few-photon quantum walks, universal quantum simulators for simulating a wide range of many-body phenomena are a rapidly developing technology. Here, we focus only on works relevant for physically demonstrating the interplay of many-body effects with the NHSE.

Implementing the NHSE requires a mechanism for loss, which is not naturally existing in unitary quantum circuits. Hence any such proposal for the NHSE must involve non-unitary evolution implemented through measurements or post-selection. in Ref.~\cite{fleckenstein2022non}, monitored quantum circuits emulate the non-Hermitian SSH model. It consists of rapidly alternating unitary evolution and measurement steups. The unitary stage provides the unitary evolution of effectively spinless electrons due to the Hermitian part of the Hamiltonian. The periodic measurements that are stochastically invoked correspond to the non-Hermitian $\sigma_y$ term. Under the Floquet-Magnus expansion, effective non-unitary evolution under the non-Hermitian SSH model can be obtained.

In popular state-of-the-art quantum simulators such as the IBM quantum computer, monitored measurements corresponding to particular Kraus operators may not be straightforwardly implemented. An alternative approach is to implement non-unitary evolution by embedding the non-unitary operator within a larger unitary operator, which is an important technique for realizing imaginary time evolution~\cite{lin2021real,liu2021probabilistic,kamakari2022digital}. It can be shown that this can be done with just one additional qubit, known as the ancilla qubit~\cite{lin2021real}. To demonstrate the NHSE on a lattice, there will be a need to for implement a tight-binding model on the chain of spin qubits; this has been implemented for the Hermitian Heisenberg model~\cite{smith2019simulating}, various 1D topological lattices~\cite{koh2022stabilizing} and Floquet time crystals~\cite{frey2022realization,chen2023robust}, the 2D Chern lattice~\cite{koh2022simulation} and even higher-order topological lattices in up to 4 dimensions~\cite{koh2023observation}. More recently, non-unitary evolution on a lattice has been implemented by this post-selection approach on a lattice~\cite{chen2022high}; note that for each qubit experiencing loss, an ancilla qubit is required. Future quantum computer implementations of the NHSE would likely involve lossy qubits as well as a flux ladder. 

Going beyond previous experiments that are restricted to preparing ground states via non-unitary imaginary-time evolution~\cite{lin2021real,liu2021probabilistic,kamakari2022digital}, the NHSE and its many-fermion ``Fermi skin'' profile are realized on current noisy intermediate-scale quantum processors~\cite{shen2023observation}, where the NHSE is observed at the many-body level for the first time.

\subsubsection{Mechanical/Acoustic proposals}

Mechanical systems allow for intuitive observation of non-Hermitian effects, particularly NHSE dynamics, as well as, the relationship between biorthogonality and the system's physical response.~\cite{schomerus2020nonreciprocal}, The NHSE may be induced through piezoelectric sensors and actuators with Floquet feedback~\cite{braghini2021non} and through flexural phonon modes~\cite{jin2022non}. Mechanical demonstrations of the NHSE may also result in useful applications, such as optomechanically induced transparency~\cite{wen2022optomechanically}.

\subsubsection{Ultracold atomic proposals}

Compared with most other proposals, ultracold atoms provide a promising platform for demonstrating and investigating intriguing quantum many-body physics.
In such systems, non-Hermiticity is usually introduced through atom loss, 
which may be induced by a resonant beam that couples atoms to an excited state~\cite{Jiaming2019gainloss,ren2022chiral}.
Realization of NHSE has been proposed in ultracold atoms for both continuous models~\cite{guo2022theoretical} and optical lattices~\cite{li2020topological,zhou2022engineering}, from the interplay of (pseudo)spin-dependent atom loss and a synthetic flux induced by periodic driving~\cite{li2020topological} or spin-orbit couplings~\cite{guo2022theoretical,zhou2022engineering,shen2023proposal}.
It has also been proposed and reported that NHSE can be realized in the momentum space of a two-component Bose-Einstein condensate of ultracold atoms~\cite{li2022BEC,liang2022observation}.

\subsubsection{NHSE in complex networks and active media}

NHSE is a phenomenon on generic directed graphs, not just crystal lattices. As such, it can also be manifested in graph networks representing real-world processes, for instance in the non-linear dynamics of a rock-paper-scissors game~\cite{yoshida2022non}. Such models are based on Lokta-Volterra population evolution models, which have been shown to give rise to interesting unexpected non-reciprocal and topological signatures~\cite{dobrinevski2012extinction,knebel2013coexistence,knebel2020topological,yoshida2021chiral,umer2022topologically}.

Since the NHSE ultimately stems from non-reciprocity and not the lattice structure per se, it will also manifest in systems devoid of any crystal structure, i.e. continuous media. In~\cite{scheibner2020non} and~\cite{fruchart2021non}, non-reciprocal effects are theoretically and experimentally investigated in active continuous media where the non-reciprocity arises from static deformations, and conserves linear momentum. In~\cite{fruchart2021non}, a general framework was presented encompassing three archetypal classes of self-organization out of equilibrium: synchronization, flocking and pattern formation. These systems exhibit collective phenomena not  lying at a configuration energy minimum, such as active time-(quasi)crystals, exceptional-point-enforced pattern formation and hysteresis. 
{Realization of NHSE has been proposed with magnetic materials, induced by chiral coupling between dipolar-coupledmagnets~\cite{yu2022giant,zeng2023radiation}.}

\subsection{Further discussion on physical NHSE signatures.}

Finally, we note that pairs of oppositely localized NHSE state can be observed in \emph{reciprocal} systems that are mathematically equivalent to appropriately coupled equal and opposite NHSE chains~\cite{hofmann2020reciprocal,franca2022non}, or~\cite{zhu2022hybrid} in the case of 2D hybrid ST system. Note that the breaking of bulk-boundary correspondence in the spectrum does not necessarily imply the same in the impedance of a circuit - for an RLC circuit with very different PBC vs. OBC Laplacian spectra, the two-point impedance between most pairs of points can still be approximately the same, whether under PBCs or OBCs~\cite{zhang2022observationb}.

Recently, an implementation of NHSE and its corresponding spectral winding topology has been proposed in electronic mesoscopic systems, with asymmetric coupling between electrons of the concerned system and a reservoir~\cite{geng2023nonreciprocal}. NHSE engineered in this way can be either charge- or spin-resolved in different setups, which can be probed by different transport measurements. On more general platforms, the various transport signatures of NHSE is thoroughly studied in Ref.~\cite{PhysRevA.104.023515}.

Fundamental constraints on the observability of non-Hermitian effects  in passive systems, including the NHSE, are derived in Ref.~\cite{schomerus2022fundamental}, which also discussed about the prospects for observing symmetry-protected edge states and EP signatures. Some observable signatures are embedded in the density of states, particularly the signatures of drastic mode nonorthogonality, which can be effectively exploited and detected by active elements in devices.



\begin{thebibliography}{444}%
    \makeatletter
    \providecommand \@ifxundefined [1]{%
     \@ifx{#1\undefined}
    }%
    \providecommand \@ifnum [1]{%
     \ifnum #1\expandafter \@firstoftwo
     \else \expandafter \@secondoftwo
     \fi
    }%
    \providecommand \@ifx [1]{%
     \ifx #1\expandafter \@firstoftwo
     \else \expandafter \@secondoftwo
     \fi
    }%
    \providecommand \natexlab [1]{#1}%
    \providecommand \enquote  [1]{``#1''}%
    \providecommand \bibnamefont  [1]{#1}%
    \providecommand \bibfnamefont [1]{#1}%
    \providecommand \citenamefont [1]{#1}%
    \providecommand \href@noop [0]{\@secondoftwo}%
    \providecommand \href [0]{\begingroup \@sanitize@url \@href}%
    \providecommand \@href[1]{\@@startlink{#1}\@@href}%
    \providecommand \@@href[1]{\endgroup#1\@@endlink}%
    \providecommand \@sanitize@url [0]{\catcode `\\12\catcode `\$12\catcode `\&12\catcode `\#12\catcode `\^12\catcode `\_12\catcode `\%12\relax}%
    \providecommand \@@startlink[1]{}%
    \providecommand \@@endlink[0]{}%
    \providecommand \url  [0]{\begingroup\@sanitize@url \@url }%
    \providecommand \@url [1]{\endgroup\@href {#1}{\urlprefix }}%
    \providecommand \urlprefix  [0]{URL }%
    \providecommand \Eprint [0]{\href }%
    \providecommand \doibase [0]{http://dx.doi.org/}%
    \providecommand \selectlanguage [0]{\@gobble}%
    \providecommand \bibinfo  [0]{\@secondoftwo}%
    \providecommand \bibfield  [0]{\@secondoftwo}%
    \providecommand \translation [1]{[#1]}%
    \providecommand \BibitemOpen [0]{}%
    \providecommand \bibitemStop [0]{}%
    \providecommand \bibitemNoStop [0]{.\EOS\space}%
    \providecommand \EOS [0]{\spacefactor3000\relax}%
    \providecommand \BibitemShut  [1]{\csname bibitem#1\endcsname}%
    \let\auto@bib@innerbib\@empty
    \bibitem [{\citenamefont {Bender}\ and\ \citenamefont {Boettcher}(1998)}]{Bender1998nonH}%
      \BibitemOpen
      \bibfield  {author} {\bibinfo {author} {\bibfnamefont {C.~M.}\ \bibnamefont {Bender}}\ and\ \bibinfo {author} {\bibfnamefont {S.}~\bibnamefont {Boettcher}},\ }\href {\doibase 10.1103/PhysRevLett.80.5243} {\bibfield  {journal} {\bibinfo  {journal} {Phys. Rev. Lett.}\ }\textbf {\bibinfo {volume} {80}},\ \bibinfo {pages} {5243} (\bibinfo {year} {1998})}\BibitemShut {NoStop}%
    \bibitem [{\citenamefont {Bender}(2007)}]{bender2007making}%
      \BibitemOpen
      \bibfield  {author} {\bibinfo {author} {\bibfnamefont {C.~M.}\ \bibnamefont {Bender}},\ }\href@noop {} {\bibfield  {journal} {\bibinfo  {journal} {Reports on Progress in Physics}\ }\textbf {\bibinfo {volume} {70}},\ \bibinfo {pages} {947} (\bibinfo {year} {2007})}\BibitemShut {NoStop}%
    \bibitem [{\citenamefont {Rotter}(2009)}]{Rotter2009non}%
      \BibitemOpen
      \bibfield  {author} {\bibinfo {author} {\bibfnamefont {I.}~\bibnamefont {Rotter}},\ }\href@noop {} {\bibfield  {journal} {\bibinfo  {journal} {Journal of Physics A: Mathematical and Theoretical}\ }\textbf {\bibinfo {volume} {42}},\ \bibinfo {pages} {153001} (\bibinfo {year} {2009})}\BibitemShut {NoStop}%
    \bibitem [{\citenamefont {Yoshida}\ \emph {et~al.}(2018)\citenamefont {Yoshida}, \citenamefont {Peters},\ and\ \citenamefont {Kawakami}}]{yoshida2018non}%
      \BibitemOpen
      \bibfield  {author} {\bibinfo {author} {\bibfnamefont {T.}~\bibnamefont {Yoshida}}, \bibinfo {author} {\bibfnamefont {R.}~\bibnamefont {Peters}}, \ and\ \bibinfo {author} {\bibfnamefont {N.}~\bibnamefont {Kawakami}},\ }\href@noop {} {\bibfield  {journal} {\bibinfo  {journal} {Physical Review B}\ }\textbf {\bibinfo {volume} {98}},\ \bibinfo {pages} {035141} (\bibinfo {year} {2018})}\BibitemShut {NoStop}%
    \bibitem [{\citenamefont {Shen}\ and\ \citenamefont {Fu}(2018)}]{shen2018quantum}%
      \BibitemOpen
      \bibfield  {author} {\bibinfo {author} {\bibfnamefont {H.}~\bibnamefont {Shen}}\ and\ \bibinfo {author} {\bibfnamefont {L.}~\bibnamefont {Fu}},\ }\href {\doibase 10.1103/PhysRevLett.121.026403} {\bibfield  {journal} {\bibinfo  {journal} {Phys. Rev. Lett.}\ }\textbf {\bibinfo {volume} {121}},\ \bibinfo {pages} {026403} (\bibinfo {year} {2018})}\BibitemShut {NoStop}%
    \bibitem [{\citenamefont {Yamamoto}\ \emph {et~al.}(2019)\citenamefont {Yamamoto}, \citenamefont {Nakagawa}, \citenamefont {Adachi}, \citenamefont {Takasan}, \citenamefont {Ueda},\ and\ \citenamefont {Kawakami}}]{yamamoto2019theory}%
      \BibitemOpen
      \bibfield  {author} {\bibinfo {author} {\bibfnamefont {K.}~\bibnamefont {Yamamoto}}, \bibinfo {author} {\bibfnamefont {M.}~\bibnamefont {Nakagawa}}, \bibinfo {author} {\bibfnamefont {K.}~\bibnamefont {Adachi}}, \bibinfo {author} {\bibfnamefont {K.}~\bibnamefont {Takasan}}, \bibinfo {author} {\bibfnamefont {M.}~\bibnamefont {Ueda}}, \ and\ \bibinfo {author} {\bibfnamefont {N.}~\bibnamefont {Kawakami}},\ }\href {\doibase 10.1103/PhysRevLett.123.123601} {\bibfield  {journal} {\bibinfo  {journal} {Phys. Rev. Lett.}\ }\textbf {\bibinfo {volume} {123}},\ \bibinfo {pages} {123601} (\bibinfo {year} {2019})}\BibitemShut {NoStop}%
    \bibitem [{\citenamefont {Ma}\ and\ \citenamefont {Sheng}(2016)}]{ma2016acoustic}%
      \BibitemOpen
      \bibfield  {author} {\bibinfo {author} {\bibfnamefont {G.}~\bibnamefont {Ma}}\ and\ \bibinfo {author} {\bibfnamefont {P.}~\bibnamefont {Sheng}},\ }\href@noop {} {\bibfield  {journal} {\bibinfo  {journal} {Science advances}\ }\textbf {\bibinfo {volume} {2}},\ \bibinfo {pages} {e1501595} (\bibinfo {year} {2016})}\BibitemShut {NoStop}%
    \bibitem [{\citenamefont {Cummer}\ \emph {et~al.}(2016)\citenamefont {Cummer}, \citenamefont {Christensen},\ and\ \citenamefont {Al{\`u}}}]{cummer2016controlling}%
      \BibitemOpen
      \bibfield  {author} {\bibinfo {author} {\bibfnamefont {S.~A.}\ \bibnamefont {Cummer}}, \bibinfo {author} {\bibfnamefont {J.}~\bibnamefont {Christensen}}, \ and\ \bibinfo {author} {\bibfnamefont {A.}~\bibnamefont {Al{\`u}}},\ }\href@noop {} {\bibfield  {journal} {\bibinfo  {journal} {Nature Reviews Materials}\ }\textbf {\bibinfo {volume} {1}},\ \bibinfo {pages} {16001} (\bibinfo {year} {2016})}\BibitemShut {NoStop}%
    \bibitem [{\citenamefont {Zangeneh-Nejad}\ and\ \citenamefont {Fleury}(2019)}]{zangeneh2019active}%
      \BibitemOpen
      \bibfield  {author} {\bibinfo {author} {\bibfnamefont {F.}~\bibnamefont {Zangeneh-Nejad}}\ and\ \bibinfo {author} {\bibfnamefont {R.}~\bibnamefont {Fleury}},\ }\href@noop {} {\bibfield  {journal} {\bibinfo  {journal} {Reviews in Physics}\ }\textbf {\bibinfo {volume} {4}},\ \bibinfo {pages} {100031} (\bibinfo {year} {2019})}\BibitemShut {NoStop}%
    \bibitem [{\citenamefont {Feng}\ \emph {et~al.}(2017)\citenamefont {Feng}, \citenamefont {El-Ganainy},\ and\ \citenamefont {Ge}}]{feng2017non}%
      \BibitemOpen
      \bibfield  {author} {\bibinfo {author} {\bibfnamefont {L.}~\bibnamefont {Feng}}, \bibinfo {author} {\bibfnamefont {R.}~\bibnamefont {El-Ganainy}}, \ and\ \bibinfo {author} {\bibfnamefont {L.}~\bibnamefont {Ge}},\ }\href@noop {} {\bibfield  {journal} {\bibinfo  {journal} {Nature Photonics}\ }\textbf {\bibinfo {volume} {11}},\ \bibinfo {pages} {752} (\bibinfo {year} {2017})}\BibitemShut {NoStop}%
    \bibitem [{\citenamefont {El-Ganainy}\ \emph {et~al.}(2018)\citenamefont {El-Ganainy}, \citenamefont {Makris}, \citenamefont {Khajavikhan}, \citenamefont {Musslimani}, \citenamefont {Rotter},\ and\ \citenamefont {Christodoulides}}]{el2018non}%
      \BibitemOpen
      \bibfield  {author} {\bibinfo {author} {\bibfnamefont {R.}~\bibnamefont {El-Ganainy}}, \bibinfo {author} {\bibfnamefont {K.~G.}\ \bibnamefont {Makris}}, \bibinfo {author} {\bibfnamefont {M.}~\bibnamefont {Khajavikhan}}, \bibinfo {author} {\bibfnamefont {Z.~H.}\ \bibnamefont {Musslimani}}, \bibinfo {author} {\bibfnamefont {S.}~\bibnamefont {Rotter}}, \ and\ \bibinfo {author} {\bibfnamefont {D.~N.}\ \bibnamefont {Christodoulides}},\ }\href@noop {} {\bibfield  {journal} {\bibinfo  {journal} {Nature Physics}\ }\textbf {\bibinfo {volume} {14}},\ \bibinfo {pages} {11} (\bibinfo {year} {2018})}\BibitemShut {NoStop}%
    \bibitem [{\citenamefont {Longhi}(2018{\natexlab{a}})}]{longhi2018parity}%
      \BibitemOpen
      \bibfield  {author} {\bibinfo {author} {\bibfnamefont {S.}~\bibnamefont {Longhi}},\ }\href@noop {} {\bibfield  {journal} {\bibinfo  {journal} {EPL (Europhysics Letters)}\ }\textbf {\bibinfo {volume} {120}},\ \bibinfo {pages} {64001} (\bibinfo {year} {2018}{\natexlab{a}})}\BibitemShut {NoStop}%
    \bibitem [{\citenamefont {Ozawa}\ \emph {et~al.}(2019)\citenamefont {Ozawa}, \citenamefont {Price}, \citenamefont {Amo}, \citenamefont {Goldman}, \citenamefont {Hafezi}, \citenamefont {Lu}, \citenamefont {Rechtsman}, \citenamefont {Schuster}, \citenamefont {Simon}, \citenamefont {Zilberberg} \emph {et~al.}}]{ozawa2019topological}%
      \BibitemOpen
      \bibfield  {author} {\bibinfo {author} {\bibfnamefont {T.}~\bibnamefont {Ozawa}}, \bibinfo {author} {\bibfnamefont {H.~M.}\ \bibnamefont {Price}}, \bibinfo {author} {\bibfnamefont {A.}~\bibnamefont {Amo}}, \bibinfo {author} {\bibfnamefont {N.}~\bibnamefont {Goldman}}, \bibinfo {author} {\bibfnamefont {M.}~\bibnamefont {Hafezi}}, \bibinfo {author} {\bibfnamefont {L.}~\bibnamefont {Lu}}, \bibinfo {author} {\bibfnamefont {M.~C.}\ \bibnamefont {Rechtsman}}, \bibinfo {author} {\bibfnamefont {D.}~\bibnamefont {Schuster}}, \bibinfo {author} {\bibfnamefont {J.}~\bibnamefont {Simon}}, \bibinfo {author} {\bibfnamefont {O.}~\bibnamefont {Zilberberg}},  \emph {et~al.},\ }\href@noop {} {\bibfield  {journal} {\bibinfo  {journal} {Reviews of Modern Physics}\ }\textbf {\bibinfo {volume} {91}},\ \bibinfo {pages} {015006} (\bibinfo {year} {2019})}\BibitemShut {NoStop}%
    \bibitem [{\citenamefont {Xiao}\ \emph {et~al.}(2020)\citenamefont {Xiao}, \citenamefont {Deng}, \citenamefont {Wang}, \citenamefont {Zhu}, \citenamefont {Wang}, \citenamefont {Yi},\ and\ \citenamefont {Xue}}]{xiao2020non}%
      \BibitemOpen
      \bibfield  {author} {\bibinfo {author} {\bibfnamefont {L.}~\bibnamefont {Xiao}}, \bibinfo {author} {\bibfnamefont {T.}~\bibnamefont {Deng}}, \bibinfo {author} {\bibfnamefont {K.}~\bibnamefont {Wang}}, \bibinfo {author} {\bibfnamefont {G.}~\bibnamefont {Zhu}}, \bibinfo {author} {\bibfnamefont {Z.}~\bibnamefont {Wang}}, \bibinfo {author} {\bibfnamefont {W.}~\bibnamefont {Yi}}, \ and\ \bibinfo {author} {\bibfnamefont {P.}~\bibnamefont {Xue}},\ }\href@noop {} {\bibfield  {journal} {\bibinfo  {journal} {Nature Physics}\ }\textbf {\bibinfo {volume} {16}},\ \bibinfo {pages} {761} (\bibinfo {year} {2020})}\BibitemShut {NoStop}%
    \bibitem [{\citenamefont {Mostafazadeh}(2002)}]{mostafazadeh2002pseudo}%
      \BibitemOpen
      \bibfield  {author} {\bibinfo {author} {\bibfnamefont {A.}~\bibnamefont {Mostafazadeh}},\ }\href@noop {} {\bibfield  {journal} {\bibinfo  {journal} {Journal of Mathematical Physics}\ }\textbf {\bibinfo {volume} {43}},\ \bibinfo {pages} {205} (\bibinfo {year} {2002})}\BibitemShut {NoStop}%
    \bibitem [{\citenamefont {G{\"u}nther}\ \emph {et~al.}(2007)\citenamefont {G{\"u}nther}, \citenamefont {Rotter},\ and\ \citenamefont {Samsonov}}]{gunther2007projective}%
      \BibitemOpen
      \bibfield  {author} {\bibinfo {author} {\bibfnamefont {U.}~\bibnamefont {G{\"u}nther}}, \bibinfo {author} {\bibfnamefont {I.}~\bibnamefont {Rotter}}, \ and\ \bibinfo {author} {\bibfnamefont {B.~F.}\ \bibnamefont {Samsonov}},\ }\href@noop {} {\bibfield  {journal} {\bibinfo  {journal} {Journal of Physics A: Mathematical and Theoretical}\ }\textbf {\bibinfo {volume} {40}},\ \bibinfo {pages} {8815} (\bibinfo {year} {2007})}\BibitemShut {NoStop}%
    \bibitem [{\citenamefont {Regensburger}\ \emph {et~al.}(2012)\citenamefont {Regensburger}, \citenamefont {Bersch}, \citenamefont {Miri}, \citenamefont {Onishchukov}, \citenamefont {Christodoulides},\ and\ \citenamefont {Peschel}}]{regensburger2012parity}%
      \BibitemOpen
      \bibfield  {author} {\bibinfo {author} {\bibfnamefont {A.}~\bibnamefont {Regensburger}}, \bibinfo {author} {\bibfnamefont {C.}~\bibnamefont {Bersch}}, \bibinfo {author} {\bibfnamefont {M.-A.}\ \bibnamefont {Miri}}, \bibinfo {author} {\bibfnamefont {G.}~\bibnamefont {Onishchukov}}, \bibinfo {author} {\bibfnamefont {D.~N.}\ \bibnamefont {Christodoulides}}, \ and\ \bibinfo {author} {\bibfnamefont {U.}~\bibnamefont {Peschel}},\ }\href@noop {} {\bibfield  {journal} {\bibinfo  {journal} {Nature}\ }\textbf {\bibinfo {volume} {488}},\ \bibinfo {pages} {167} (\bibinfo {year} {2012})}\BibitemShut {NoStop}%
    \bibitem [{\citenamefont {Miri}\ and\ \citenamefont {Alu}(2019)}]{miri2019exceptional}%
      \BibitemOpen
      \bibfield  {author} {\bibinfo {author} {\bibfnamefont {M.-A.}\ \bibnamefont {Miri}}\ and\ \bibinfo {author} {\bibfnamefont {A.}~\bibnamefont {Alu}},\ }\href@noop {} {\bibfield  {journal} {\bibinfo  {journal} {Science}\ }\textbf {\bibinfo {volume} {363}},\ \bibinfo {pages} {eaar7709} (\bibinfo {year} {2019})}\BibitemShut {NoStop}%
    \bibitem [{\citenamefont {Meng}\ \emph {et~al.}(2024)\citenamefont {Meng}, \citenamefont {Ang},\ and\ \citenamefont {Lee}}]{meng2024exceptional}%
      \BibitemOpen
      \bibfield  {author} {\bibinfo {author} {\bibfnamefont {H.}~\bibnamefont {Meng}}, \bibinfo {author} {\bibfnamefont {Y.~S.}\ \bibnamefont {Ang}}, \ and\ \bibinfo {author} {\bibfnamefont {C.~H.}\ \bibnamefont {Lee}},\ }\href@noop {} {\bibfield  {journal} {\bibinfo  {journal} {Applied Physics Letters}\ }\textbf {\bibinfo {volume} {124}} (\bibinfo {year} {2024})}\BibitemShut {NoStop}%
    \bibitem [{\citenamefont {Torres}(2019)}]{torres2019perspective}%
      \BibitemOpen
      \bibfield  {author} {\bibinfo {author} {\bibfnamefont {L.~E.~F.}\ \bibnamefont {Torres}},\ }\href@noop {} {\bibfield  {journal} {\bibinfo  {journal} {Journal of Physics: Materials}\ }\textbf {\bibinfo {volume} {3}},\ \bibinfo {pages} {014002} (\bibinfo {year} {2019})}\BibitemShut {NoStop}%
    \bibitem [{\citenamefont {Sayyad}\ and\ \citenamefont {Kunst}(2022)}]{sayyad2022realizing}%
      \BibitemOpen
      \bibfield  {author} {\bibinfo {author} {\bibfnamefont {S.}~\bibnamefont {Sayyad}}\ and\ \bibinfo {author} {\bibfnamefont {F.~K.}\ \bibnamefont {Kunst}},\ }\href@noop {} {\bibfield  {journal} {\bibinfo  {journal} {Physical Review Research}\ }\textbf {\bibinfo {volume} {4}},\ \bibinfo {pages} {023130} (\bibinfo {year} {2022})}\BibitemShut {NoStop}%
    \bibitem [{\citenamefont {Lin}\ \emph {et~al.}(2011)\citenamefont {Lin}, \citenamefont {Ramezani}, \citenamefont {Eichelkraut}, \citenamefont {Kottos}, \citenamefont {Cao},\ and\ \citenamefont {Christodoulides}}]{lin2011unidirectional}%
      \BibitemOpen
      \bibfield  {author} {\bibinfo {author} {\bibfnamefont {Z.}~\bibnamefont {Lin}}, \bibinfo {author} {\bibfnamefont {H.}~\bibnamefont {Ramezani}}, \bibinfo {author} {\bibfnamefont {T.}~\bibnamefont {Eichelkraut}}, \bibinfo {author} {\bibfnamefont {T.}~\bibnamefont {Kottos}}, \bibinfo {author} {\bibfnamefont {H.}~\bibnamefont {Cao}}, \ and\ \bibinfo {author} {\bibfnamefont {D.~N.}\ \bibnamefont {Christodoulides}},\ }\href@noop {} {\bibfield  {journal} {\bibinfo  {journal} {Physical Review Letters}\ }\textbf {\bibinfo {volume} {106}},\ \bibinfo {pages} {213901} (\bibinfo {year} {2011})}\BibitemShut {NoStop}%
    \bibitem [{\citenamefont {Feng}\ \emph {et~al.}(2013)\citenamefont {Feng}, \citenamefont {Xu}, \citenamefont {Fegadolli}, \citenamefont {Lu}, \citenamefont {Oliveira}, \citenamefont {Almeida}, \citenamefont {Chen},\ and\ \citenamefont {Scherer}}]{feng2013experimental}%
      \BibitemOpen
      \bibfield  {author} {\bibinfo {author} {\bibfnamefont {L.}~\bibnamefont {Feng}}, \bibinfo {author} {\bibfnamefont {Y.-L.}\ \bibnamefont {Xu}}, \bibinfo {author} {\bibfnamefont {W.~S.}\ \bibnamefont {Fegadolli}}, \bibinfo {author} {\bibfnamefont {M.-H.}\ \bibnamefont {Lu}}, \bibinfo {author} {\bibfnamefont {J.~E.}\ \bibnamefont {Oliveira}}, \bibinfo {author} {\bibfnamefont {V.~R.}\ \bibnamefont {Almeida}}, \bibinfo {author} {\bibfnamefont {Y.-F.}\ \bibnamefont {Chen}}, \ and\ \bibinfo {author} {\bibfnamefont {A.}~\bibnamefont {Scherer}},\ }\href@noop {} {\bibfield  {journal} {\bibinfo  {journal} {Nature materials}\ }\textbf {\bibinfo {volume} {12}},\ \bibinfo {pages} {108} (\bibinfo {year} {2013})}\BibitemShut {NoStop}%
    \bibitem [{\citenamefont {Wiersig}(2014)}]{jan2014enhancing}%
      \BibitemOpen
      \bibfield  {author} {\bibinfo {author} {\bibfnamefont {J.}~\bibnamefont {Wiersig}},\ }\href {\doibase 10.1103/PhysRevLett.112.203901} {\bibfield  {journal} {\bibinfo  {journal} {Phys. Rev. Lett.}\ }\textbf {\bibinfo {volume} {112}},\ \bibinfo {pages} {203901} (\bibinfo {year} {2014})}\BibitemShut {NoStop}%
    \bibitem [{\citenamefont {Liu}\ \emph {et~al.}(2016)\citenamefont {Liu}, \citenamefont {Zhang}, \citenamefont {{\"O}zdemir}, \citenamefont {Peng}, \citenamefont {Jing}, \citenamefont {L{\"u}}, \citenamefont {Li}, \citenamefont {Yang}, \citenamefont {Nori},\ and\ \citenamefont {Liu}}]{liu2016metrology}%
      \BibitemOpen
      \bibfield  {author} {\bibinfo {author} {\bibfnamefont {Z.-P.}\ \bibnamefont {Liu}}, \bibinfo {author} {\bibfnamefont {J.}~\bibnamefont {Zhang}}, \bibinfo {author} {\bibfnamefont {{\c{S}}.~K.}\ \bibnamefont {{\"O}zdemir}}, \bibinfo {author} {\bibfnamefont {B.}~\bibnamefont {Peng}}, \bibinfo {author} {\bibfnamefont {H.}~\bibnamefont {Jing}}, \bibinfo {author} {\bibfnamefont {X.-Y.}\ \bibnamefont {L{\"u}}}, \bibinfo {author} {\bibfnamefont {C.-W.}\ \bibnamefont {Li}}, \bibinfo {author} {\bibfnamefont {L.}~\bibnamefont {Yang}}, \bibinfo {author} {\bibfnamefont {F.}~\bibnamefont {Nori}}, \ and\ \bibinfo {author} {\bibfnamefont {Y.-x.}\ \bibnamefont {Liu}},\ }\href@noop {} {\bibfield  {journal} {\bibinfo  {journal} {Physical review letters}\ }\textbf {\bibinfo {volume} {117}},\ \bibinfo {pages} {110802} (\bibinfo {year} {2016})}\BibitemShut {NoStop}%
    \bibitem [{\citenamefont {Hodaei}\ \emph {et~al.}(2017)\citenamefont {Hodaei}, \citenamefont {Hassan}, \citenamefont {Wittek}, \citenamefont {Garcia-Gracia}, \citenamefont {El-Ganainy}, \citenamefont {Christodoulides},\ and\ \citenamefont {Khajavikhan}}]{hodaei2017enhanced}%
      \BibitemOpen
      \bibfield  {author} {\bibinfo {author} {\bibfnamefont {H.}~\bibnamefont {Hodaei}}, \bibinfo {author} {\bibfnamefont {A.~U.}\ \bibnamefont {Hassan}}, \bibinfo {author} {\bibfnamefont {S.}~\bibnamefont {Wittek}}, \bibinfo {author} {\bibfnamefont {H.}~\bibnamefont {Garcia-Gracia}}, \bibinfo {author} {\bibfnamefont {R.}~\bibnamefont {El-Ganainy}}, \bibinfo {author} {\bibfnamefont {D.~N.}\ \bibnamefont {Christodoulides}}, \ and\ \bibinfo {author} {\bibfnamefont {M.}~\bibnamefont {Khajavikhan}},\ }\href@noop {} {\bibfield  {journal} {\bibinfo  {journal} {Nature}\ }\textbf {\bibinfo {volume} {548}},\ \bibinfo {pages} {187} (\bibinfo {year} {2017})}\BibitemShut {NoStop}%
    \bibitem [{\citenamefont {Chen}\ \emph {et~al.}(2017)\citenamefont {Chen}, \citenamefont {{\"O}zdemir}, \citenamefont {Zhao}, \citenamefont {Wiersig},\ and\ \citenamefont {Yang}}]{chen2017exceptional}%
      \BibitemOpen
      \bibfield  {author} {\bibinfo {author} {\bibfnamefont {W.}~\bibnamefont {Chen}}, \bibinfo {author} {\bibfnamefont {{\c{S}}.~K.}\ \bibnamefont {{\"O}zdemir}}, \bibinfo {author} {\bibfnamefont {G.}~\bibnamefont {Zhao}}, \bibinfo {author} {\bibfnamefont {J.}~\bibnamefont {Wiersig}}, \ and\ \bibinfo {author} {\bibfnamefont {L.}~\bibnamefont {Yang}},\ }\href@noop {} {\bibfield  {journal} {\bibinfo  {journal} {Nature}\ }\textbf {\bibinfo {volume} {548}},\ \bibinfo {pages} {192} (\bibinfo {year} {2017})}\BibitemShut {NoStop}%
    \bibitem [{\citenamefont {Dembowski}\ \emph {et~al.}(2001)\citenamefont {Dembowski}, \citenamefont {Gr{\"a}f}, \citenamefont {Harney}, \citenamefont {Heine}, \citenamefont {Heiss}, \citenamefont {Rehfeld},\ and\ \citenamefont {Richter}}]{dembowski2001experimental}%
      \BibitemOpen
      \bibfield  {author} {\bibinfo {author} {\bibfnamefont {C.}~\bibnamefont {Dembowski}}, \bibinfo {author} {\bibfnamefont {H.-D.}\ \bibnamefont {Gr{\"a}f}}, \bibinfo {author} {\bibfnamefont {H.}~\bibnamefont {Harney}}, \bibinfo {author} {\bibfnamefont {A.}~\bibnamefont {Heine}}, \bibinfo {author} {\bibfnamefont {W.}~\bibnamefont {Heiss}}, \bibinfo {author} {\bibfnamefont {H.}~\bibnamefont {Rehfeld}}, \ and\ \bibinfo {author} {\bibfnamefont {A.}~\bibnamefont {Richter}},\ }\href@noop {} {\bibfield  {journal} {\bibinfo  {journal} {Physical review letters}\ }\textbf {\bibinfo {volume} {86}},\ \bibinfo {pages} {787} (\bibinfo {year} {2001})}\BibitemShut {NoStop}%
    \bibitem [{\citenamefont {Gao}\ \emph {et~al.}(2015{\natexlab{a}})\citenamefont {Gao}, \citenamefont {Estrecho}, \citenamefont {Bliokh}, \citenamefont {Liew}, \citenamefont {Fraser}, \citenamefont {Brodbeck}, \citenamefont {Kamp}, \citenamefont {Schneider}, \citenamefont {H{\"o}fling}, \citenamefont {Yamamoto} \emph {et~al.}}]{gao2015observation}%
      \BibitemOpen
      \bibfield  {author} {\bibinfo {author} {\bibfnamefont {T.}~\bibnamefont {Gao}}, \bibinfo {author} {\bibfnamefont {E.}~\bibnamefont {Estrecho}}, \bibinfo {author} {\bibfnamefont {K.}~\bibnamefont {Bliokh}}, \bibinfo {author} {\bibfnamefont {T.}~\bibnamefont {Liew}}, \bibinfo {author} {\bibfnamefont {M.}~\bibnamefont {Fraser}}, \bibinfo {author} {\bibfnamefont {S.}~\bibnamefont {Brodbeck}}, \bibinfo {author} {\bibfnamefont {M.}~\bibnamefont {Kamp}}, \bibinfo {author} {\bibfnamefont {C.}~\bibnamefont {Schneider}}, \bibinfo {author} {\bibfnamefont {S.}~\bibnamefont {H{\"o}fling}}, \bibinfo {author} {\bibfnamefont {Y.}~\bibnamefont {Yamamoto}},  \emph {et~al.},\ }\href@noop {} {\bibfield  {journal} {\bibinfo  {journal} {Nature}\ }\textbf {\bibinfo {volume} {526}},\ \bibinfo {pages} {554} (\bibinfo {year} {2015}{\natexlab{a}})}\BibitemShut {NoStop}%
    \bibitem [{\citenamefont {Mailybaev}\ \emph {et~al.}(2005)\citenamefont {Mailybaev}, \citenamefont {Kirillov},\ and\ \citenamefont {Seyranian}}]{Mailybaev2005geometric}%
      \BibitemOpen
      \bibfield  {author} {\bibinfo {author} {\bibfnamefont {A.~A.}\ \bibnamefont {Mailybaev}}, \bibinfo {author} {\bibfnamefont {O.~N.}\ \bibnamefont {Kirillov}}, \ and\ \bibinfo {author} {\bibfnamefont {A.~P.}\ \bibnamefont {Seyranian}},\ }\href {\doibase 10.1103/PhysRevA.72.014104} {\bibfield  {journal} {\bibinfo  {journal} {Phys. Rev. A}\ }\textbf {\bibinfo {volume} {72}},\ \bibinfo {pages} {014104} (\bibinfo {year} {2005})}\BibitemShut {NoStop}%
    \bibitem [{\citenamefont {Lee}(2016)}]{Lee2016nonH}%
      \BibitemOpen
      \bibfield  {author} {\bibinfo {author} {\bibfnamefont {T.~E.}\ \bibnamefont {Lee}},\ }\href {\doibase 10.1103/PhysRevLett.116.133903} {\bibfield  {journal} {\bibinfo  {journal} {Phys. Rev. Lett.}\ }\textbf {\bibinfo {volume} {116}},\ \bibinfo {pages} {133903} (\bibinfo {year} {2016})}\BibitemShut {NoStop}%
    \bibitem [{\citenamefont {Leykam}\ \emph {et~al.}(2017)\citenamefont {Leykam}, \citenamefont {Bliokh}, \citenamefont {Huang}, \citenamefont {Chong},\ and\ \citenamefont {Nori}}]{leykam2017edge}%
      \BibitemOpen
      \bibfield  {author} {\bibinfo {author} {\bibfnamefont {D.}~\bibnamefont {Leykam}}, \bibinfo {author} {\bibfnamefont {K.~Y.}\ \bibnamefont {Bliokh}}, \bibinfo {author} {\bibfnamefont {C.}~\bibnamefont {Huang}}, \bibinfo {author} {\bibfnamefont {Y.~D.}\ \bibnamefont {Chong}}, \ and\ \bibinfo {author} {\bibfnamefont {F.}~\bibnamefont {Nori}},\ }\href@noop {} {\bibfield  {journal} {\bibinfo  {journal} {Physical review letters}\ }\textbf {\bibinfo {volume} {118}},\ \bibinfo {pages} {040401} (\bibinfo {year} {2017})}\BibitemShut {NoStop}%
    \bibitem [{\citenamefont {Yin}\ \emph {et~al.}(2018)\citenamefont {Yin}, \citenamefont {Jiang}, \citenamefont {Li}, \citenamefont {L\"u},\ and\ \citenamefont {Chen}}]{Yin2018nonHermitian}%
      \BibitemOpen
      \bibfield  {author} {\bibinfo {author} {\bibfnamefont {C.}~\bibnamefont {Yin}}, \bibinfo {author} {\bibfnamefont {H.}~\bibnamefont {Jiang}}, \bibinfo {author} {\bibfnamefont {L.}~\bibnamefont {Li}}, \bibinfo {author} {\bibfnamefont {R.}~\bibnamefont {L\"u}}, \ and\ \bibinfo {author} {\bibfnamefont {S.}~\bibnamefont {Chen}},\ }\href {\doibase 10.1103/PhysRevA.97.052115} {\bibfield  {journal} {\bibinfo  {journal} {Phys. Rev. A}\ }\textbf {\bibinfo {volume} {97}},\ \bibinfo {pages} {052115} (\bibinfo {year} {2018})}\BibitemShut {NoStop}%
    \bibitem [{\citenamefont {Shen}\ \emph {et~al.}(2018)\citenamefont {Shen}, \citenamefont {Zhen},\ and\ \citenamefont {Fu}}]{shen2018topological}%
      \BibitemOpen
      \bibfield  {author} {\bibinfo {author} {\bibfnamefont {H.}~\bibnamefont {Shen}}, \bibinfo {author} {\bibfnamefont {B.}~\bibnamefont {Zhen}}, \ and\ \bibinfo {author} {\bibfnamefont {L.}~\bibnamefont {Fu}},\ }\href@noop {} {\bibfield  {journal} {\bibinfo  {journal} {Phys. Rev. Lett.}\ }\textbf {\bibinfo {volume} {120}},\ \bibinfo {pages} {146402} (\bibinfo {year} {2018})}\BibitemShut {NoStop}%
    \bibitem [{\citenamefont {Li}\ \emph {et~al.}(2019{\natexlab{a}})\citenamefont {Li}, \citenamefont {Lee},\ and\ \citenamefont {Gong}}]{li2019geometric}%
      \BibitemOpen
      \bibfield  {author} {\bibinfo {author} {\bibfnamefont {L.}~\bibnamefont {Li}}, \bibinfo {author} {\bibfnamefont {C.~H.}\ \bibnamefont {Lee}}, \ and\ \bibinfo {author} {\bibfnamefont {J.}~\bibnamefont {Gong}},\ }\href@noop {} {\bibfield  {journal} {\bibinfo  {journal} {Physical Review B}\ }\textbf {\bibinfo {volume} {100}},\ \bibinfo {pages} {075403} (\bibinfo {year} {2019}{\natexlab{a}})}\BibitemShut {NoStop}%
    \bibitem [{\citenamefont {Hu}\ \emph {et~al.}(2017)\citenamefont {Hu}, \citenamefont {Wang}, \citenamefont {Shum},\ and\ \citenamefont {Chong}}]{Hu2017EP}%
      \BibitemOpen
      \bibfield  {author} {\bibinfo {author} {\bibfnamefont {W.}~\bibnamefont {Hu}}, \bibinfo {author} {\bibfnamefont {H.}~\bibnamefont {Wang}}, \bibinfo {author} {\bibfnamefont {P.~P.}\ \bibnamefont {Shum}}, \ and\ \bibinfo {author} {\bibfnamefont {Y.~D.}\ \bibnamefont {Chong}},\ }\href {\doibase 10.1103/PhysRevB.95.184306} {\bibfield  {journal} {\bibinfo  {journal} {Phys. Rev. B}\ }\textbf {\bibinfo {volume} {95}},\ \bibinfo {pages} {184306} (\bibinfo {year} {2017})}\BibitemShut {NoStop}%
    \bibitem [{\citenamefont {Hassan}\ \emph {et~al.}(2017)\citenamefont {Hassan}, \citenamefont {Zhen}, \citenamefont {Solja\ifmmode \check{c}\else \v{c}\fi{}i\ifmmode~\acute{c}\else \'{c}\fi{}}, \citenamefont {Khajavikhan},\ and\ \citenamefont {Christodoulides}}]{Hassan2017EP}%
      \BibitemOpen
      \bibfield  {author} {\bibinfo {author} {\bibfnamefont {A.~U.}\ \bibnamefont {Hassan}}, \bibinfo {author} {\bibfnamefont {B.}~\bibnamefont {Zhen}}, \bibinfo {author} {\bibfnamefont {M.}~\bibnamefont {Solja\ifmmode \check{c}\else \v{c}\fi{}i\ifmmode~\acute{c}\else \'{c}\fi{}}}, \bibinfo {author} {\bibfnamefont {M.}~\bibnamefont {Khajavikhan}}, \ and\ \bibinfo {author} {\bibfnamefont {D.~N.}\ \bibnamefont {Christodoulides}},\ }\href {\doibase 10.1103/PhysRevLett.118.093002} {\bibfield  {journal} {\bibinfo  {journal} {Phys. Rev. Lett.}\ }\textbf {\bibinfo {volume} {118}},\ \bibinfo {pages} {093002} (\bibinfo {year} {2017})}\BibitemShut {NoStop}%
    \bibitem [{\citenamefont {Hasan}\ and\ \citenamefont {Kane}(2010)}]{hasan2010colloquium}%
      \BibitemOpen
      \bibfield  {author} {\bibinfo {author} {\bibfnamefont {M.~Z.}\ \bibnamefont {Hasan}}\ and\ \bibinfo {author} {\bibfnamefont {C.~L.}\ \bibnamefont {Kane}},\ }\href@noop {} {\bibfield  {journal} {\bibinfo  {journal} {Rev. Mod. Phys.}\ }\textbf {\bibinfo {volume} {82}},\ \bibinfo {pages} {3045} (\bibinfo {year} {2010})}\BibitemShut {NoStop}%
    \bibitem [{\citenamefont {Qi}\ and\ \citenamefont {Zhang}(2011)}]{qi2011topological}%
      \BibitemOpen
      \bibfield  {author} {\bibinfo {author} {\bibfnamefont {X.-L.}\ \bibnamefont {Qi}}\ and\ \bibinfo {author} {\bibfnamefont {S.-C.}\ \bibnamefont {Zhang}},\ }\href@noop {} {\bibfield  {journal} {\bibinfo  {journal} {Rev. Mod. Phys.}\ }\textbf {\bibinfo {volume} {83}},\ \bibinfo {pages} {1057} (\bibinfo {year} {2011})}\BibitemShut {NoStop}%
    \bibitem [{\citenamefont {Bernevig}\ and\ \citenamefont {Hughes}(2013)}]{bernevig2013topological}%
      \BibitemOpen
      \bibfield  {author} {\bibinfo {author} {\bibfnamefont {B.~A.}\ \bibnamefont {Bernevig}}\ and\ \bibinfo {author} {\bibfnamefont {T.~L.}\ \bibnamefont {Hughes}},\ }\href@noop {} {\emph {\bibinfo {title} {Topological insulators and topological superconductors}}}\ (\bibinfo  {publisher} {Princeton university press},\ \bibinfo {year} {2013})\BibitemShut {NoStop}%
    \bibitem [{\citenamefont {Rudner}\ and\ \citenamefont {Levitov}(2009)}]{rudner2009topological}%
      \BibitemOpen
      \bibfield  {author} {\bibinfo {author} {\bibfnamefont {M.~S.}\ \bibnamefont {Rudner}}\ and\ \bibinfo {author} {\bibfnamefont {L.~S.}\ \bibnamefont {Levitov}},\ }\href {\doibase 10.1103/PhysRevLett.102.065703} {\bibfield  {journal} {\bibinfo  {journal} {Phys. Rev. Lett.}\ }\textbf {\bibinfo {volume} {102}},\ \bibinfo {pages} {065703} (\bibinfo {year} {2009})}\BibitemShut {NoStop}%
    \bibitem [{\citenamefont {Hu}\ and\ \citenamefont {Hughes}(2011)}]{hu2011absence}%
      \BibitemOpen
      \bibfield  {author} {\bibinfo {author} {\bibfnamefont {Y.~C.}\ \bibnamefont {Hu}}\ and\ \bibinfo {author} {\bibfnamefont {T.~L.}\ \bibnamefont {Hughes}},\ }\href {\doibase 10.1103/PhysRevB.84.153101} {\bibfield  {journal} {\bibinfo  {journal} {Phys. Rev. B}\ }\textbf {\bibinfo {volume} {84}},\ \bibinfo {pages} {153101} (\bibinfo {year} {2011})}\BibitemShut {NoStop}%
    \bibitem [{\citenamefont {Esaki}\ \emph {et~al.}(2011)\citenamefont {Esaki}, \citenamefont {Sato}, \citenamefont {Hasebe},\ and\ \citenamefont {Kohmoto}}]{esaki2011edge}%
      \BibitemOpen
      \bibfield  {author} {\bibinfo {author} {\bibfnamefont {K.}~\bibnamefont {Esaki}}, \bibinfo {author} {\bibfnamefont {M.}~\bibnamefont {Sato}}, \bibinfo {author} {\bibfnamefont {K.}~\bibnamefont {Hasebe}}, \ and\ \bibinfo {author} {\bibfnamefont {M.}~\bibnamefont {Kohmoto}},\ }\href {\doibase 10.1103/PhysRevB.84.205128} {\bibfield  {journal} {\bibinfo  {journal} {Phys. Rev. B}\ }\textbf {\bibinfo {volume} {84}},\ \bibinfo {pages} {205128} (\bibinfo {year} {2011})}\BibitemShut {NoStop}%
    \bibitem [{\citenamefont {Diehl}\ \emph {et~al.}(2011)\citenamefont {Diehl}, \citenamefont {Rico}, \citenamefont {Baranov},\ and\ \citenamefont {Zoller}}]{diehl2011topology}%
      \BibitemOpen
      \bibfield  {author} {\bibinfo {author} {\bibfnamefont {S.}~\bibnamefont {Diehl}}, \bibinfo {author} {\bibfnamefont {E.}~\bibnamefont {Rico}}, \bibinfo {author} {\bibfnamefont {M.~A.}\ \bibnamefont {Baranov}}, \ and\ \bibinfo {author} {\bibfnamefont {P.}~\bibnamefont {Zoller}},\ }\href@noop {} {\bibfield  {journal} {\bibinfo  {journal} {Nature Physics}\ }\textbf {\bibinfo {volume} {7}},\ \bibinfo {pages} {971} (\bibinfo {year} {2011})}\BibitemShut {NoStop}%
    \bibitem [{\citenamefont {Zhu}\ \emph {et~al.}(2014)\citenamefont {Zhu}, \citenamefont {L{\"u}},\ and\ \citenamefont {Chen}}]{zhu2014pt}%
      \BibitemOpen
      \bibfield  {author} {\bibinfo {author} {\bibfnamefont {B.}~\bibnamefont {Zhu}}, \bibinfo {author} {\bibfnamefont {R.}~\bibnamefont {L{\"u}}}, \ and\ \bibinfo {author} {\bibfnamefont {S.}~\bibnamefont {Chen}},\ }\href@noop {} {\bibfield  {journal} {\bibinfo  {journal} {Physical Review A}\ }\textbf {\bibinfo {volume} {89}},\ \bibinfo {pages} {062102} (\bibinfo {year} {2014})}\BibitemShut {NoStop}%
    \bibitem [{\citenamefont {Malzard}\ \emph {et~al.}(2015)\citenamefont {Malzard}, \citenamefont {Poli},\ and\ \citenamefont {Schomerus}}]{malzard2015topologically}%
      \BibitemOpen
      \bibfield  {author} {\bibinfo {author} {\bibfnamefont {S.}~\bibnamefont {Malzard}}, \bibinfo {author} {\bibfnamefont {C.}~\bibnamefont {Poli}}, \ and\ \bibinfo {author} {\bibfnamefont {H.}~\bibnamefont {Schomerus}},\ }\href {\doibase 10.1103/PhysRevLett.115.200402} {\bibfield  {journal} {\bibinfo  {journal} {Phys. Rev. Lett.}\ }\textbf {\bibinfo {volume} {115}},\ \bibinfo {pages} {200402} (\bibinfo {year} {2015})}\BibitemShut {NoStop}%
    \bibitem [{\citenamefont {Harter}\ \emph {et~al.}(2016)\citenamefont {Harter}, \citenamefont {Lee},\ and\ \citenamefont {Joglekar}}]{harter2016mathcal}%
      \BibitemOpen
      \bibfield  {author} {\bibinfo {author} {\bibfnamefont {A.~K.}\ \bibnamefont {Harter}}, \bibinfo {author} {\bibfnamefont {T.~E.}\ \bibnamefont {Lee}}, \ and\ \bibinfo {author} {\bibfnamefont {Y.~N.}\ \bibnamefont {Joglekar}},\ }\href {\doibase 10.1103/PhysRevA.93.062101} {\bibfield  {journal} {\bibinfo  {journal} {Phys. Rev. A}\ }\textbf {\bibinfo {volume} {93}},\ \bibinfo {pages} {062101} (\bibinfo {year} {2016})}\BibitemShut {NoStop}%
    \bibitem [{\citenamefont {Xiong}(2018)}]{xiong2018does}%
      \BibitemOpen
      \bibfield  {author} {\bibinfo {author} {\bibfnamefont {Y.}~\bibnamefont {Xiong}},\ }\href@noop {} {\bibfield  {journal} {\bibinfo  {journal} {Journal of Physics Communications}\ }\textbf {\bibinfo {volume} {2}},\ \bibinfo {pages} {035043} (\bibinfo {year} {2018})}\BibitemShut {NoStop}%
    \bibitem [{\citenamefont {Alvarez}\ \emph {et~al.}(2018)\citenamefont {Alvarez}, \citenamefont {Vargas},\ and\ \citenamefont {Torres}}]{alvarez2018non}%
      \BibitemOpen
      \bibfield  {author} {\bibinfo {author} {\bibfnamefont {V.~M.}\ \bibnamefont {Alvarez}}, \bibinfo {author} {\bibfnamefont {J.~B.}\ \bibnamefont {Vargas}}, \ and\ \bibinfo {author} {\bibfnamefont {L.~F.}\ \bibnamefont {Torres}},\ }\href@noop {} {\bibfield  {journal} {\bibinfo  {journal} {Phys. Rev. B}\ }\textbf {\bibinfo {volume} {97}},\ \bibinfo {pages} {121401} (\bibinfo {year} {2018})}\BibitemShut {NoStop}%
    \bibitem [{\citenamefont {Yao}\ and\ \citenamefont {Wang}(2018)}]{yao2018edge}%
      \BibitemOpen
      \bibfield  {author} {\bibinfo {author} {\bibfnamefont {S.}~\bibnamefont {Yao}}\ and\ \bibinfo {author} {\bibfnamefont {Z.}~\bibnamefont {Wang}},\ }\href {\doibase 10.1103/PhysRevLett.121.086803} {\bibfield  {journal} {\bibinfo  {journal} {Phys. Rev. Lett.}\ }\textbf {\bibinfo {volume} {121}},\ \bibinfo {pages} {086803} (\bibinfo {year} {2018})}\BibitemShut {NoStop}%
    \bibitem [{\citenamefont {Yokomizo}\ and\ \citenamefont {Murakami}(2019)}]{yokomizo2019non}%
      \BibitemOpen
      \bibfield  {author} {\bibinfo {author} {\bibfnamefont {K.}~\bibnamefont {Yokomizo}}\ and\ \bibinfo {author} {\bibfnamefont {S.}~\bibnamefont {Murakami}},\ }\href@noop {} {\bibfield  {journal} {\bibinfo  {journal} {Physical review letters}\ }\textbf {\bibinfo {volume} {123}},\ \bibinfo {pages} {066404} (\bibinfo {year} {2019})}\BibitemShut {NoStop}%
    \bibitem [{\citenamefont {Lee}\ \emph {et~al.}(2020{\natexlab{a}})\citenamefont {Lee}, \citenamefont {Li}, \citenamefont {Thomale},\ and\ \citenamefont {Gong}}]{lee2020unraveling}%
      \BibitemOpen
      \bibfield  {author} {\bibinfo {author} {\bibfnamefont {C.~H.}\ \bibnamefont {Lee}}, \bibinfo {author} {\bibfnamefont {L.}~\bibnamefont {Li}}, \bibinfo {author} {\bibfnamefont {R.}~\bibnamefont {Thomale}}, \ and\ \bibinfo {author} {\bibfnamefont {J.}~\bibnamefont {Gong}},\ }\href {\doibase 10.1103/PhysRevB.102.085151} {\bibfield  {journal} {\bibinfo  {journal} {Phys. Rev. B}\ }\textbf {\bibinfo {volume} {102}},\ \bibinfo {pages} {085151} (\bibinfo {year} {2020}{\natexlab{a}})}\BibitemShut {NoStop}%
    \bibitem [{\citenamefont {Fu}\ and\ \citenamefont {Wan}(2022)}]{fu2022degeneracy}%
      \BibitemOpen
      \bibfield  {author} {\bibinfo {author} {\bibfnamefont {Y.}~\bibnamefont {Fu}}\ and\ \bibinfo {author} {\bibfnamefont {S.}~\bibnamefont {Wan}},\ }\href@noop {} {\bibfield  {journal} {\bibinfo  {journal} {Physical Review B}\ }\textbf {\bibinfo {volume} {105}},\ \bibinfo {pages} {075420} (\bibinfo {year} {2022})}\BibitemShut {NoStop}%
    \bibitem [{\citenamefont {Lee}\ and\ \citenamefont {Thomale}(2019)}]{lee2019anatomy}%
      \BibitemOpen
      \bibfield  {author} {\bibinfo {author} {\bibfnamefont {C.~H.}\ \bibnamefont {Lee}}\ and\ \bibinfo {author} {\bibfnamefont {R.}~\bibnamefont {Thomale}},\ }\href {\doibase 10.1103/PhysRevB.99.201103} {\bibfield  {journal} {\bibinfo  {journal} {Phys. Rev. B}\ }\textbf {\bibinfo {volume} {99}},\ \bibinfo {pages} {201103} (\bibinfo {year} {2019})}\BibitemShut {NoStop}%
    \bibitem [{\citenamefont {Kunst}\ \emph {et~al.}(2018)\citenamefont {Kunst}, \citenamefont {Edvardsson}, \citenamefont {Budich},\ and\ \citenamefont {Bergholtz}}]{kunst2018biorthogonal}%
      \BibitemOpen
      \bibfield  {author} {\bibinfo {author} {\bibfnamefont {F.~K.}\ \bibnamefont {Kunst}}, \bibinfo {author} {\bibfnamefont {E.}~\bibnamefont {Edvardsson}}, \bibinfo {author} {\bibfnamefont {J.~C.}\ \bibnamefont {Budich}}, \ and\ \bibinfo {author} {\bibfnamefont {E.~J.}\ \bibnamefont {Bergholtz}},\ }\href {\doibase 10.1103/PhysRevLett.121.026808} {\bibfield  {journal} {\bibinfo  {journal} {Phys. Rev. Lett.}\ }\textbf {\bibinfo {volume} {121}},\ \bibinfo {pages} {026808} (\bibinfo {year} {2018})}\BibitemShut {NoStop}%
    \bibitem [{\citenamefont {Zou}\ \emph {et~al.}(2022)\citenamefont {Zou}, \citenamefont {Zhou}, \citenamefont {Chen},\ and\ \citenamefont {Ye}}]{zou2022measuring}%
      \BibitemOpen
      \bibfield  {author} {\bibinfo {author} {\bibfnamefont {Y.-Y.}\ \bibnamefont {Zou}}, \bibinfo {author} {\bibfnamefont {Y.}~\bibnamefont {Zhou}}, \bibinfo {author} {\bibfnamefont {L.-M.}\ \bibnamefont {Chen}}, \ and\ \bibinfo {author} {\bibfnamefont {P.}~\bibnamefont {Ye}},\ }\href@noop {} {\bibfield  {journal} {\bibinfo  {journal} {arXiv preprint arXiv:2208.14944}\ } (\bibinfo {year} {2022})}\BibitemShut {NoStop}%
    \bibitem [{\citenamefont {Song}\ \emph {et~al.}(2019{\natexlab{a}})\citenamefont {Song}, \citenamefont {Yao},\ and\ \citenamefont {Wang}}]{song2019non}%
      \BibitemOpen
      \bibfield  {author} {\bibinfo {author} {\bibfnamefont {F.}~\bibnamefont {Song}}, \bibinfo {author} {\bibfnamefont {S.}~\bibnamefont {Yao}}, \ and\ \bibinfo {author} {\bibfnamefont {Z.}~\bibnamefont {Wang}},\ }\href@noop {} {\bibfield  {journal} {\bibinfo  {journal} {Physical review letters}\ }\textbf {\bibinfo {volume} {123}},\ \bibinfo {pages} {170401} (\bibinfo {year} {2019}{\natexlab{a}})}\BibitemShut {NoStop}%
    \bibitem [{\citenamefont {Wanjura}\ \emph {et~al.}(2020)\citenamefont {Wanjura}, \citenamefont {Brunelli},\ and\ \citenamefont {Nunnenkamp}}]{Wanjura2020}%
      \BibitemOpen
      \bibfield  {author} {\bibinfo {author} {\bibfnamefont {C.~C.}\ \bibnamefont {Wanjura}}, \bibinfo {author} {\bibfnamefont {M.}~\bibnamefont {Brunelli}}, \ and\ \bibinfo {author} {\bibfnamefont {A.}~\bibnamefont {Nunnenkamp}},\ }\href@noop {} {\bibfield  {journal} {\bibinfo  {journal} {Nature communications}\ }\textbf {\bibinfo {volume} {11}},\ \bibinfo {pages} {3149} (\bibinfo {year} {2020})}\BibitemShut {NoStop}%
    \bibitem [{\citenamefont {Wanjura}\ \emph {et~al.}(2021)\citenamefont {Wanjura}, \citenamefont {Brunelli},\ and\ \citenamefont {Nunnenkamp}}]{wanjura2021correspondence}%
      \BibitemOpen
      \bibfield  {author} {\bibinfo {author} {\bibfnamefont {C.~C.}\ \bibnamefont {Wanjura}}, \bibinfo {author} {\bibfnamefont {M.}~\bibnamefont {Brunelli}}, \ and\ \bibinfo {author} {\bibfnamefont {A.}~\bibnamefont {Nunnenkamp}},\ }\href {\doibase 10.1103/PhysRevLett.127.213601} {\bibfield  {journal} {\bibinfo  {journal} {Phys. Rev. Lett.}\ }\textbf {\bibinfo {volume} {127}},\ \bibinfo {pages} {213601} (\bibinfo {year} {2021})}\BibitemShut {NoStop}%
    \bibitem [{\citenamefont {Xue}\ \emph {et~al.}(2021)\citenamefont {Xue}, \citenamefont {Li}, \citenamefont {Hu}, \citenamefont {Song},\ and\ \citenamefont {Wang}}]{xue2021simple}%
      \BibitemOpen
      \bibfield  {author} {\bibinfo {author} {\bibfnamefont {W.-T.}\ \bibnamefont {Xue}}, \bibinfo {author} {\bibfnamefont {M.-R.}\ \bibnamefont {Li}}, \bibinfo {author} {\bibfnamefont {Y.-M.}\ \bibnamefont {Hu}}, \bibinfo {author} {\bibfnamefont {F.}~\bibnamefont {Song}}, \ and\ \bibinfo {author} {\bibfnamefont {Z.}~\bibnamefont {Wang}},\ }\href@noop {} {\bibfield  {journal} {\bibinfo  {journal} {Physical Review B}\ }\textbf {\bibinfo {volume} {103}},\ \bibinfo {pages} {L241408} (\bibinfo {year} {2021})}\BibitemShut {NoStop}%
    \bibitem [{\citenamefont {Longhi}(2022{\natexlab{a}})}]{longhi2022self}%
      \BibitemOpen
      \bibfield  {author} {\bibinfo {author} {\bibfnamefont {S.}~\bibnamefont {Longhi}},\ }\href {\doibase 10.1103/PhysRevLett.128.157601} {\bibfield  {journal} {\bibinfo  {journal} {Phys. Rev. Lett.}\ }\textbf {\bibinfo {volume} {128}},\ \bibinfo {pages} {157601} (\bibinfo {year} {2022}{\natexlab{a}})}\BibitemShut {NoStop}%
    \bibitem [{\citenamefont {Xue}\ \emph {et~al.}(2022)\citenamefont {Xue}, \citenamefont {Hu}, \citenamefont {Song},\ and\ \citenamefont {Wang}}]{xue2022non}%
      \BibitemOpen
      \bibfield  {author} {\bibinfo {author} {\bibfnamefont {W.-T.}\ \bibnamefont {Xue}}, \bibinfo {author} {\bibfnamefont {Y.-M.}\ \bibnamefont {Hu}}, \bibinfo {author} {\bibfnamefont {F.}~\bibnamefont {Song}}, \ and\ \bibinfo {author} {\bibfnamefont {Z.}~\bibnamefont {Wang}},\ }\href {\doibase 10.1103/PhysRevLett.128.120401} {\bibfield  {journal} {\bibinfo  {journal} {Phys. Rev. Lett.}\ }\textbf {\bibinfo {volume} {128}},\ \bibinfo {pages} {120401} (\bibinfo {year} {2022})}\BibitemShut {NoStop}%
    \bibitem [{\citenamefont {Budich}\ and\ \citenamefont {Bergholtz}(2020)}]{budich2020sensor}%
      \BibitemOpen
      \bibfield  {author} {\bibinfo {author} {\bibfnamefont {J.~C.}\ \bibnamefont {Budich}}\ and\ \bibinfo {author} {\bibfnamefont {E.~J.}\ \bibnamefont {Bergholtz}},\ }\href {\doibase 10.1103/PhysRevLett.125.180403} {\bibfield  {journal} {\bibinfo  {journal} {Phys. Rev. Lett.}\ }\textbf {\bibinfo {volume} {125}},\ \bibinfo {pages} {180403} (\bibinfo {year} {2020})}\BibitemShut {NoStop}%
    \bibitem [{\citenamefont {Guo}\ \emph {et~al.}(2021)\citenamefont {Guo}, \citenamefont {Liu}, \citenamefont {Zhao}, \citenamefont {Liu},\ and\ \citenamefont {Chen}}]{guo2021exact}%
      \BibitemOpen
      \bibfield  {author} {\bibinfo {author} {\bibfnamefont {C.-X.}\ \bibnamefont {Guo}}, \bibinfo {author} {\bibfnamefont {C.-H.}\ \bibnamefont {Liu}}, \bibinfo {author} {\bibfnamefont {X.-M.}\ \bibnamefont {Zhao}}, \bibinfo {author} {\bibfnamefont {Y.}~\bibnamefont {Liu}}, \ and\ \bibinfo {author} {\bibfnamefont {S.}~\bibnamefont {Chen}},\ }\href {\doibase 10.1103/PhysRevLett.127.116801} {\bibfield  {journal} {\bibinfo  {journal} {Phys. Rev. Lett.}\ }\textbf {\bibinfo {volume} {127}},\ \bibinfo {pages} {116801} (\bibinfo {year} {2021})}\BibitemShut {NoStop}%
    \bibitem [{\citenamefont {Li}\ \emph {et~al.}(2021{\natexlab{a}})\citenamefont {Li}, \citenamefont {Lee},\ and\ \citenamefont {Gong}}]{li2021impurity}%
      \BibitemOpen
      \bibfield  {author} {\bibinfo {author} {\bibfnamefont {L.}~\bibnamefont {Li}}, \bibinfo {author} {\bibfnamefont {C.~H.}\ \bibnamefont {Lee}}, \ and\ \bibinfo {author} {\bibfnamefont {J.}~\bibnamefont {Gong}},\ }\href@noop {} {\bibfield  {journal} {\bibinfo  {journal} {Communications Physics}\ }\textbf {\bibinfo {volume} {4}},\ \bibinfo {pages} {42} (\bibinfo {year} {2021}{\natexlab{a}})}\BibitemShut {NoStop}%
    \bibitem [{\citenamefont {Li}\ \emph {et~al.}(2020{\natexlab{a}})\citenamefont {Li}, \citenamefont {Lee}, \citenamefont {Mu},\ and\ \citenamefont {Gong}}]{li2020critical}%
      \BibitemOpen
      \bibfield  {author} {\bibinfo {author} {\bibfnamefont {L.}~\bibnamefont {Li}}, \bibinfo {author} {\bibfnamefont {C.~H.}\ \bibnamefont {Lee}}, \bibinfo {author} {\bibfnamefont {S.}~\bibnamefont {Mu}}, \ and\ \bibinfo {author} {\bibfnamefont {J.}~\bibnamefont {Gong}},\ }\href@noop {} {\bibfield  {journal} {\bibinfo  {journal} {Nature communications}\ }\textbf {\bibinfo {volume} {11}} (\bibinfo {year} {2020}{\natexlab{a}})}\BibitemShut {NoStop}%
    \bibitem [{\citenamefont {Yokomizo}\ and\ \citenamefont {Murakami}(2021)}]{yokomizo2021scaling}%
      \BibitemOpen
      \bibfield  {author} {\bibinfo {author} {\bibfnamefont {K.}~\bibnamefont {Yokomizo}}\ and\ \bibinfo {author} {\bibfnamefont {S.}~\bibnamefont {Murakami}},\ }\href {\doibase 10.1103/PhysRevB.104.165117} {\bibfield  {journal} {\bibinfo  {journal} {Phys. Rev. B}\ }\textbf {\bibinfo {volume} {104}},\ \bibinfo {pages} {165117} (\bibinfo {year} {2021})}\BibitemShut {NoStop}%
    \bibitem [{\citenamefont {Liu}\ \emph {et~al.}(2020{\natexlab{a}})\citenamefont {Liu}, \citenamefont {Zhang}, \citenamefont {Yang},\ and\ \citenamefont {Chen}}]{liu2020helical}%
      \BibitemOpen
      \bibfield  {author} {\bibinfo {author} {\bibfnamefont {C.-H.}\ \bibnamefont {Liu}}, \bibinfo {author} {\bibfnamefont {K.}~\bibnamefont {Zhang}}, \bibinfo {author} {\bibfnamefont {Z.}~\bibnamefont {Yang}}, \ and\ \bibinfo {author} {\bibfnamefont {S.}~\bibnamefont {Chen}},\ }\href {\doibase 10.1103/PhysRevResearch.2.043167} {\bibfield  {journal} {\bibinfo  {journal} {Phys. Rev. Research}\ }\textbf {\bibinfo {volume} {2}},\ \bibinfo {pages} {043167} (\bibinfo {year} {2020}{\natexlab{a}})}\BibitemShut {NoStop}%
    \bibitem [{\citenamefont {Sun}\ \emph {et~al.}(2021)\citenamefont {Sun}, \citenamefont {Zhu},\ and\ \citenamefont {Hughes}}]{sun2021geometric}%
      \BibitemOpen
      \bibfield  {author} {\bibinfo {author} {\bibfnamefont {X.-Q.}\ \bibnamefont {Sun}}, \bibinfo {author} {\bibfnamefont {P.}~\bibnamefont {Zhu}}, \ and\ \bibinfo {author} {\bibfnamefont {T.~L.}\ \bibnamefont {Hughes}},\ }\href {\doibase 10.1103/PhysRevLett.127.066401} {\bibfield  {journal} {\bibinfo  {journal} {Phys. Rev. Lett.}\ }\textbf {\bibinfo {volume} {127}},\ \bibinfo {pages} {066401} (\bibinfo {year} {2021})}\BibitemShut {NoStop}%
    \bibitem [{\citenamefont {Bhargava}\ \emph {et~al.}(2021)\citenamefont {Bhargava}, \citenamefont {Fulga}, \citenamefont {Van Den~Brink},\ and\ \citenamefont {Moghaddam}}]{bhargava2021non}%
      \BibitemOpen
      \bibfield  {author} {\bibinfo {author} {\bibfnamefont {B.~A.}\ \bibnamefont {Bhargava}}, \bibinfo {author} {\bibfnamefont {I.~C.}\ \bibnamefont {Fulga}}, \bibinfo {author} {\bibfnamefont {J.}~\bibnamefont {Van Den~Brink}}, \ and\ \bibinfo {author} {\bibfnamefont {A.~G.}\ \bibnamefont {Moghaddam}},\ }\href@noop {} {\bibfield  {journal} {\bibinfo  {journal} {Physical Review B}\ }\textbf {\bibinfo {volume} {104}},\ \bibinfo {pages} {L241402} (\bibinfo {year} {2021})}\BibitemShut {NoStop}%
    \bibitem [{\citenamefont {Schindler}\ and\ \citenamefont {Prem}(2021)}]{schindler2021dislocation}%
      \BibitemOpen
      \bibfield  {author} {\bibinfo {author} {\bibfnamefont {F.}~\bibnamefont {Schindler}}\ and\ \bibinfo {author} {\bibfnamefont {A.}~\bibnamefont {Prem}},\ }\href {\doibase 10.1103/PhysRevB.104.L161106} {\bibfield  {journal} {\bibinfo  {journal} {Phys. Rev. B}\ }\textbf {\bibinfo {volume} {104}},\ \bibinfo {pages} {L161106} (\bibinfo {year} {2021})}\BibitemShut {NoStop}%
    \bibitem [{\citenamefont {Zhang}\ \emph {et~al.}(2022{\natexlab{a}})\citenamefont {Zhang}, \citenamefont {Yang},\ and\ \citenamefont {Fang}}]{zhang2022universal}%
      \BibitemOpen
      \bibfield  {author} {\bibinfo {author} {\bibfnamefont {K.}~\bibnamefont {Zhang}}, \bibinfo {author} {\bibfnamefont {Z.}~\bibnamefont {Yang}}, \ and\ \bibinfo {author} {\bibfnamefont {C.}~\bibnamefont {Fang}},\ }\href@noop {} {\bibfield  {journal} {\bibinfo  {journal} {Nature communications}\ }\textbf {\bibinfo {volume} {13}},\ \bibinfo {pages} {1} (\bibinfo {year} {2022}{\natexlab{a}})}\BibitemShut {NoStop}%
    \bibitem [{\citenamefont {Panigrahi}\ \emph {et~al.}(2022)\citenamefont {Panigrahi}, \citenamefont {Moessner},\ and\ \citenamefont {Roy}}]{panigrahi2022non}%
      \BibitemOpen
      \bibfield  {author} {\bibinfo {author} {\bibfnamefont {A.}~\bibnamefont {Panigrahi}}, \bibinfo {author} {\bibfnamefont {R.}~\bibnamefont {Moessner}}, \ and\ \bibinfo {author} {\bibfnamefont {B.}~\bibnamefont {Roy}},\ }\href@noop {} {\bibfield  {journal} {\bibinfo  {journal} {Physical Review B}\ }\textbf {\bibinfo {volume} {106}},\ \bibinfo {pages} {L041302} (\bibinfo {year} {2022})}\BibitemShut {NoStop}%
    \bibitem [{\citenamefont {Manna}\ and\ \citenamefont {Roy}(2022)}]{manna2022inner}%
      \BibitemOpen
      \bibfield  {author} {\bibinfo {author} {\bibfnamefont {S.}~\bibnamefont {Manna}}\ and\ \bibinfo {author} {\bibfnamefont {B.}~\bibnamefont {Roy}},\ }\href@noop {} {\bibfield  {journal} {\bibinfo  {journal} {arXiv preprint arXiv:2202.07658}\ } (\bibinfo {year} {2022})}\BibitemShut {NoStop}%
    \bibitem [{\citenamefont {Helbig}\ \emph {et~al.}(2020)\citenamefont {Helbig}, \citenamefont {Hofmann}, \citenamefont {Imhof}, \citenamefont {Abdelghany}, \citenamefont {Kiessling}, \citenamefont {Molenkamp}, \citenamefont {Lee}, \citenamefont {Szameit}, \citenamefont {Greiter},\ and\ \citenamefont {Thomale}}]{helbig2020generalized}%
      \BibitemOpen
      \bibfield  {author} {\bibinfo {author} {\bibfnamefont {T.}~\bibnamefont {Helbig}}, \bibinfo {author} {\bibfnamefont {T.}~\bibnamefont {Hofmann}}, \bibinfo {author} {\bibfnamefont {S.}~\bibnamefont {Imhof}}, \bibinfo {author} {\bibfnamefont {M.}~\bibnamefont {Abdelghany}}, \bibinfo {author} {\bibfnamefont {T.}~\bibnamefont {Kiessling}}, \bibinfo {author} {\bibfnamefont {L.}~\bibnamefont {Molenkamp}}, \bibinfo {author} {\bibfnamefont {C.}~\bibnamefont {Lee}}, \bibinfo {author} {\bibfnamefont {A.}~\bibnamefont {Szameit}}, \bibinfo {author} {\bibfnamefont {M.}~\bibnamefont {Greiter}}, \ and\ \bibinfo {author} {\bibfnamefont {R.}~\bibnamefont {Thomale}},\ }\href@noop {} {\bibfield  {journal} {\bibinfo  {journal} {Nature Physics}\ }\textbf {\bibinfo {volume} {16}},\ \bibinfo {pages} {747–750} (\bibinfo {year} {2020})}\BibitemShut {NoStop}%
    \bibitem [{\citenamefont {Hofmann}\ \emph {et~al.}(2020)\citenamefont {Hofmann}, \citenamefont {Helbig}, \citenamefont {Schindler}, \citenamefont {Salgo}, \citenamefont {Brzezi{\'n}ska}, \citenamefont {Greiter}, \citenamefont {Kiessling}, \citenamefont {Wolf}, \citenamefont {Vollhardt}, \citenamefont {Kaba{\v{s}}i} \emph {et~al.}}]{hofmann2020reciprocal}%
      \BibitemOpen
      \bibfield  {author} {\bibinfo {author} {\bibfnamefont {T.}~\bibnamefont {Hofmann}}, \bibinfo {author} {\bibfnamefont {T.}~\bibnamefont {Helbig}}, \bibinfo {author} {\bibfnamefont {F.}~\bibnamefont {Schindler}}, \bibinfo {author} {\bibfnamefont {N.}~\bibnamefont {Salgo}}, \bibinfo {author} {\bibfnamefont {M.}~\bibnamefont {Brzezi{\'n}ska}}, \bibinfo {author} {\bibfnamefont {M.}~\bibnamefont {Greiter}}, \bibinfo {author} {\bibfnamefont {T.}~\bibnamefont {Kiessling}}, \bibinfo {author} {\bibfnamefont {D.}~\bibnamefont {Wolf}}, \bibinfo {author} {\bibfnamefont {A.}~\bibnamefont {Vollhardt}}, \bibinfo {author} {\bibfnamefont {A.}~\bibnamefont {Kaba{\v{s}}i}},  \emph {et~al.},\ }\href@noop {} {\bibfield  {journal} {\bibinfo  {journal} {Physical Review Research}\ }\textbf {\bibinfo {volume} {2}},\ \bibinfo {pages} {023265} (\bibinfo {year} {2020})}\BibitemShut {NoStop}%
    \bibitem [{\citenamefont {Liu}\ \emph {et~al.}(2021{\natexlab{a}})\citenamefont {Liu}, \citenamefont {Shao}, \citenamefont {Ma}, \citenamefont {Zhang}, \citenamefont {You}, \citenamefont {Wu}, \citenamefont {Xiang}, \citenamefont {Cui},\ and\ \citenamefont {Zhang}}]{liu2021non}%
      \BibitemOpen
      \bibfield  {author} {\bibinfo {author} {\bibfnamefont {S.}~\bibnamefont {Liu}}, \bibinfo {author} {\bibfnamefont {R.}~\bibnamefont {Shao}}, \bibinfo {author} {\bibfnamefont {S.}~\bibnamefont {Ma}}, \bibinfo {author} {\bibfnamefont {L.}~\bibnamefont {Zhang}}, \bibinfo {author} {\bibfnamefont {O.}~\bibnamefont {You}}, \bibinfo {author} {\bibfnamefont {H.}~\bibnamefont {Wu}}, \bibinfo {author} {\bibfnamefont {Y.~J.}\ \bibnamefont {Xiang}}, \bibinfo {author} {\bibfnamefont {T.~J.}\ \bibnamefont {Cui}}, \ and\ \bibinfo {author} {\bibfnamefont {S.}~\bibnamefont {Zhang}},\ }\href@noop {} {\bibfield  {journal} {\bibinfo  {journal} {Research}\ }\textbf {\bibinfo {volume} {2021}},\ \bibinfo {pages} {5608038} (\bibinfo {year} {2021}{\natexlab{a}})}\BibitemShut {NoStop}%
    \bibitem [{\citenamefont {Zou}\ \emph {et~al.}(2021)\citenamefont {Zou}, \citenamefont {Chen}, \citenamefont {He}, \citenamefont {Bao}, \citenamefont {Lee}, \citenamefont {Sun},\ and\ \citenamefont {Zhang}}]{zou2021observation}%
      \BibitemOpen
      \bibfield  {author} {\bibinfo {author} {\bibfnamefont {D.}~\bibnamefont {Zou}}, \bibinfo {author} {\bibfnamefont {T.}~\bibnamefont {Chen}}, \bibinfo {author} {\bibfnamefont {W.}~\bibnamefont {He}}, \bibinfo {author} {\bibfnamefont {J.}~\bibnamefont {Bao}}, \bibinfo {author} {\bibfnamefont {C.~H.}\ \bibnamefont {Lee}}, \bibinfo {author} {\bibfnamefont {H.}~\bibnamefont {Sun}}, \ and\ \bibinfo {author} {\bibfnamefont {X.}~\bibnamefont {Zhang}},\ }\href@noop {} {\bibfield  {journal} {\bibinfo  {journal} {Nature Communications}\ }\textbf {\bibinfo {volume} {12}},\ \bibinfo {pages} {7201} (\bibinfo {year} {2021})}\BibitemShut {NoStop}%
    \bibitem [{\citenamefont {Zhang}\ \emph {et~al.}(2021{\natexlab{a}})\citenamefont {Zhang}, \citenamefont {Tian}, \citenamefont {Jiang}, \citenamefont {Lu},\ and\ \citenamefont {Chen}}]{zhang2021observation}%
      \BibitemOpen
      \bibfield  {author} {\bibinfo {author} {\bibfnamefont {X.}~\bibnamefont {Zhang}}, \bibinfo {author} {\bibfnamefont {Y.}~\bibnamefont {Tian}}, \bibinfo {author} {\bibfnamefont {J.-H.}\ \bibnamefont {Jiang}}, \bibinfo {author} {\bibfnamefont {M.-H.}\ \bibnamefont {Lu}}, \ and\ \bibinfo {author} {\bibfnamefont {Y.-F.}\ \bibnamefont {Chen}},\ }\href@noop {} {\bibfield  {journal} {\bibinfo  {journal} {Nature communications}\ }\textbf {\bibinfo {volume} {12}},\ \bibinfo {pages} {5377} (\bibinfo {year} {2021}{\natexlab{a}})}\BibitemShut {NoStop}%
    \bibitem [{\citenamefont {Shang}\ \emph {et~al.}(2022)\citenamefont {Shang}, \citenamefont {Liu}, \citenamefont {Shao}, \citenamefont {Han}, \citenamefont {Zang}, \citenamefont {Zhang}, \citenamefont {Salama}, \citenamefont {Gao}, \citenamefont {Lee}, \citenamefont {Thomale}, \citenamefont {Manchon}, \citenamefont {Zhang}, \citenamefont {Cui},\ and\ \citenamefont {Schwingenschlögl}}]{shang2022experimental}%
      \BibitemOpen
      \bibfield  {author} {\bibinfo {author} {\bibfnamefont {C.}~\bibnamefont {Shang}}, \bibinfo {author} {\bibfnamefont {S.}~\bibnamefont {Liu}}, \bibinfo {author} {\bibfnamefont {R.}~\bibnamefont {Shao}}, \bibinfo {author} {\bibfnamefont {P.}~\bibnamefont {Han}}, \bibinfo {author} {\bibfnamefont {X.}~\bibnamefont {Zang}}, \bibinfo {author} {\bibfnamefont {X.}~\bibnamefont {Zhang}}, \bibinfo {author} {\bibfnamefont {K.~N.}\ \bibnamefont {Salama}}, \bibinfo {author} {\bibfnamefont {W.}~\bibnamefont {Gao}}, \bibinfo {author} {\bibfnamefont {C.~H.}\ \bibnamefont {Lee}}, \bibinfo {author} {\bibfnamefont {R.}~\bibnamefont {Thomale}}, \bibinfo {author} {\bibfnamefont {A.}~\bibnamefont {Manchon}}, \bibinfo {author} {\bibfnamefont {Z.}~\bibnamefont {Zhang}}, \bibinfo {author} {\bibfnamefont {T.~J.}\ \bibnamefont {Cui}}, \ and\ \bibinfo {author} {\bibfnamefont {U.}~\bibnamefont {Schwingenschlögl}},\ }\href {\doibase 10.1002/advs.202202922} {\bibfield  {journal} {\bibinfo  {journal} {Advanced Science}\ }\textbf {\bibinfo {volume} {9}},\ \bibinfo {pages} {2202922} (\bibinfo {year} {2022})}\BibitemShut {NoStop}%
    \bibitem [{\citenamefont {Zhang}\ \emph {et~al.}(2021{\natexlab{b}})\citenamefont {Zhang}, \citenamefont {Yang}, \citenamefont {Ge}, \citenamefont {Guan}, \citenamefont {Chen}, \citenamefont {Yan}, \citenamefont {Chen}, \citenamefont {Xi}, \citenamefont {Li}, \citenamefont {Jia} \emph {et~al.}}]{zhang2021acoustic}%
      \BibitemOpen
      \bibfield  {author} {\bibinfo {author} {\bibfnamefont {L.}~\bibnamefont {Zhang}}, \bibinfo {author} {\bibfnamefont {Y.}~\bibnamefont {Yang}}, \bibinfo {author} {\bibfnamefont {Y.}~\bibnamefont {Ge}}, \bibinfo {author} {\bibfnamefont {Y.-J.}\ \bibnamefont {Guan}}, \bibinfo {author} {\bibfnamefont {Q.}~\bibnamefont {Chen}}, \bibinfo {author} {\bibfnamefont {Q.}~\bibnamefont {Yan}}, \bibinfo {author} {\bibfnamefont {F.}~\bibnamefont {Chen}}, \bibinfo {author} {\bibfnamefont {R.}~\bibnamefont {Xi}}, \bibinfo {author} {\bibfnamefont {Y.}~\bibnamefont {Li}}, \bibinfo {author} {\bibfnamefont {D.}~\bibnamefont {Jia}},  \emph {et~al.},\ }\href@noop {} {\bibfield  {journal} {\bibinfo  {journal} {Nature communications}\ }\textbf {\bibinfo {volume} {12}},\ \bibinfo {pages} {6297} (\bibinfo {year} {2021}{\natexlab{b}})}\BibitemShut {NoStop}%
    \bibitem [{\citenamefont {Gao}\ \emph {et~al.}(2022)\citenamefont {Gao}, \citenamefont {Xue}, \citenamefont {Gu}, \citenamefont {Li}, \citenamefont {Zhu}, \citenamefont {Su}, \citenamefont {Zhu}, \citenamefont {Zhang},\ and\ \citenamefont {Chong}}]{gao2022non}%
      \BibitemOpen
      \bibfield  {author} {\bibinfo {author} {\bibfnamefont {H.}~\bibnamefont {Gao}}, \bibinfo {author} {\bibfnamefont {H.}~\bibnamefont {Xue}}, \bibinfo {author} {\bibfnamefont {Z.}~\bibnamefont {Gu}}, \bibinfo {author} {\bibfnamefont {L.}~\bibnamefont {Li}}, \bibinfo {author} {\bibfnamefont {W.}~\bibnamefont {Zhu}}, \bibinfo {author} {\bibfnamefont {Z.}~\bibnamefont {Su}}, \bibinfo {author} {\bibfnamefont {J.}~\bibnamefont {Zhu}}, \bibinfo {author} {\bibfnamefont {B.}~\bibnamefont {Zhang}}, \ and\ \bibinfo {author} {\bibfnamefont {Y.}~\bibnamefont {Chong}},\ }\href@noop {} {\bibfield  {journal} {\bibinfo  {journal} {arXiv preprint arXiv:2205.14824}\ } (\bibinfo {year} {2022})}\BibitemShut {NoStop}%
    \bibitem [{\citenamefont {Weidemann}\ \emph {et~al.}(2020)\citenamefont {Weidemann}, \citenamefont {Kremer}, \citenamefont {Helbig}, \citenamefont {Hofmann}, \citenamefont {Stegmaier}, \citenamefont {Greiter}, \citenamefont {Thomale},\ and\ \citenamefont {Szameit}}]{weidemann2020topological}%
      \BibitemOpen
      \bibfield  {author} {\bibinfo {author} {\bibfnamefont {S.}~\bibnamefont {Weidemann}}, \bibinfo {author} {\bibfnamefont {M.}~\bibnamefont {Kremer}}, \bibinfo {author} {\bibfnamefont {T.}~\bibnamefont {Helbig}}, \bibinfo {author} {\bibfnamefont {T.}~\bibnamefont {Hofmann}}, \bibinfo {author} {\bibfnamefont {A.}~\bibnamefont {Stegmaier}}, \bibinfo {author} {\bibfnamefont {M.}~\bibnamefont {Greiter}}, \bibinfo {author} {\bibfnamefont {R.}~\bibnamefont {Thomale}}, \ and\ \bibinfo {author} {\bibfnamefont {A.}~\bibnamefont {Szameit}},\ }\href@noop {} {\bibfield  {journal} {\bibinfo  {journal} {Science}\ }\textbf {\bibinfo {volume} {368}},\ \bibinfo {pages} {311} (\bibinfo {year} {2020})}\BibitemShut {NoStop}%
    \bibitem [{\citenamefont {Song}\ \emph {et~al.}(2020)\citenamefont {Song}, \citenamefont {Liu}, \citenamefont {Zheng}, \citenamefont {Zhang}, \citenamefont {Wang},\ and\ \citenamefont {Lu}}]{song2020two}%
      \BibitemOpen
      \bibfield  {author} {\bibinfo {author} {\bibfnamefont {Y.}~\bibnamefont {Song}}, \bibinfo {author} {\bibfnamefont {W.}~\bibnamefont {Liu}}, \bibinfo {author} {\bibfnamefont {L.}~\bibnamefont {Zheng}}, \bibinfo {author} {\bibfnamefont {Y.}~\bibnamefont {Zhang}}, \bibinfo {author} {\bibfnamefont {B.}~\bibnamefont {Wang}}, \ and\ \bibinfo {author} {\bibfnamefont {P.}~\bibnamefont {Lu}},\ }\href {\doibase 10.1103/PhysRevApplied.14.064076} {\bibfield  {journal} {\bibinfo  {journal} {Phys. Rev. Applied}\ }\textbf {\bibinfo {volume} {14}},\ \bibinfo {pages} {064076} (\bibinfo {year} {2020})}\BibitemShut {NoStop}%
    \bibitem [{\citenamefont {Xiao}\ \emph {et~al.}(2021)\citenamefont {Xiao}, \citenamefont {Deng}, \citenamefont {Wang}, \citenamefont {Wang}, \citenamefont {Yi},\ and\ \citenamefont {Xue}}]{xiao2021observation}%
      \BibitemOpen
      \bibfield  {author} {\bibinfo {author} {\bibfnamefont {L.}~\bibnamefont {Xiao}}, \bibinfo {author} {\bibfnamefont {T.}~\bibnamefont {Deng}}, \bibinfo {author} {\bibfnamefont {K.}~\bibnamefont {Wang}}, \bibinfo {author} {\bibfnamefont {Z.}~\bibnamefont {Wang}}, \bibinfo {author} {\bibfnamefont {W.}~\bibnamefont {Yi}}, \ and\ \bibinfo {author} {\bibfnamefont {P.}~\bibnamefont {Xue}},\ }\href@noop {} {\bibfield  {journal} {\bibinfo  {journal} {Physical Review Letters}\ }\textbf {\bibinfo {volume} {126}},\ \bibinfo {pages} {230402} (\bibinfo {year} {2021})}\BibitemShut {NoStop}%
    \bibitem [{\citenamefont {Wang}\ \emph {et~al.}(2021{\natexlab{a}})\citenamefont {Wang}, \citenamefont {Li}, \citenamefont {Xiao}, \citenamefont {Han}, \citenamefont {Yi},\ and\ \citenamefont {Xue}}]{wang2021detecting}%
      \BibitemOpen
      \bibfield  {author} {\bibinfo {author} {\bibfnamefont {K.}~\bibnamefont {Wang}}, \bibinfo {author} {\bibfnamefont {T.}~\bibnamefont {Li}}, \bibinfo {author} {\bibfnamefont {L.}~\bibnamefont {Xiao}}, \bibinfo {author} {\bibfnamefont {Y.}~\bibnamefont {Han}}, \bibinfo {author} {\bibfnamefont {W.}~\bibnamefont {Yi}}, \ and\ \bibinfo {author} {\bibfnamefont {P.}~\bibnamefont {Xue}},\ }\href@noop {} {\bibfield  {journal} {\bibinfo  {journal} {Physical Review Letters}\ }\textbf {\bibinfo {volume} {127}},\ \bibinfo {pages} {270602} (\bibinfo {year} {2021}{\natexlab{a}})}\BibitemShut {NoStop}%
    \bibitem [{\citenamefont {Brandenbourger}\ \emph {et~al.}(2019)\citenamefont {Brandenbourger}, \citenamefont {Locsin}, \citenamefont {Lerner},\ and\ \citenamefont {Coulais}}]{Brandenbourger2019}%
      \BibitemOpen
      \bibfield  {author} {\bibinfo {author} {\bibfnamefont {M.}~\bibnamefont {Brandenbourger}}, \bibinfo {author} {\bibfnamefont {X.}~\bibnamefont {Locsin}}, \bibinfo {author} {\bibfnamefont {E.}~\bibnamefont {Lerner}}, \ and\ \bibinfo {author} {\bibfnamefont {C.}~\bibnamefont {Coulais}},\ }\href@noop {} {\bibfield  {journal} {\bibinfo  {journal} {Nature communications}\ }\textbf {\bibinfo {volume} {10}},\ \bibinfo {pages} {4608} (\bibinfo {year} {2019})}\BibitemShut {NoStop}%
    \bibitem [{\citenamefont {Ghatak}\ \emph {et~al.}(2020)\citenamefont {Ghatak}, \citenamefont {Brandenbourger}, \citenamefont {van Wezel},\ and\ \citenamefont {Coulais}}]{ghatak2020observation}%
      \BibitemOpen
      \bibfield  {author} {\bibinfo {author} {\bibfnamefont {A.}~\bibnamefont {Ghatak}}, \bibinfo {author} {\bibfnamefont {M.}~\bibnamefont {Brandenbourger}}, \bibinfo {author} {\bibfnamefont {J.}~\bibnamefont {van Wezel}}, \ and\ \bibinfo {author} {\bibfnamefont {C.}~\bibnamefont {Coulais}},\ }\href@noop {} {\bibfield  {journal} {\bibinfo  {journal} {Proceedings of the National Academy of Sciences}\ }\textbf {\bibinfo {volume} {117}},\ \bibinfo {pages} {29561} (\bibinfo {year} {2020})}\BibitemShut {NoStop}%
    \bibitem [{\citenamefont {Gou}\ \emph {et~al.}(2020)\citenamefont {Gou}, \citenamefont {Chen}, \citenamefont {Xie}, \citenamefont {Xiao}, \citenamefont {Deng}, \citenamefont {Gadway}, \citenamefont {Yi},\ and\ \citenamefont {Yan}}]{gou2020tunable}%
      \BibitemOpen
      \bibfield  {author} {\bibinfo {author} {\bibfnamefont {W.}~\bibnamefont {Gou}}, \bibinfo {author} {\bibfnamefont {T.}~\bibnamefont {Chen}}, \bibinfo {author} {\bibfnamefont {D.}~\bibnamefont {Xie}}, \bibinfo {author} {\bibfnamefont {T.}~\bibnamefont {Xiao}}, \bibinfo {author} {\bibfnamefont {T.-S.}\ \bibnamefont {Deng}}, \bibinfo {author} {\bibfnamefont {B.}~\bibnamefont {Gadway}}, \bibinfo {author} {\bibfnamefont {W.}~\bibnamefont {Yi}}, \ and\ \bibinfo {author} {\bibfnamefont {B.}~\bibnamefont {Yan}},\ }\href {\doibase 10.1103/PhysRevLett.124.070402} {\bibfield  {journal} {\bibinfo  {journal} {Phys. Rev. Lett.}\ }\textbf {\bibinfo {volume} {124}},\ \bibinfo {pages} {070402} (\bibinfo {year} {2020})}\BibitemShut {NoStop}%
    \bibitem [{\citenamefont {Liang}\ \emph {et~al.}(2022{\natexlab{a}})\citenamefont {Liang}, \citenamefont {Xie}, \citenamefont {Dong}, \citenamefont {Li}, \citenamefont {Li}, \citenamefont {Gadway}, \citenamefont {Yi},\ and\ \citenamefont {Yan}}]{liang2022observation}%
      \BibitemOpen
      \bibfield  {author} {\bibinfo {author} {\bibfnamefont {Q.}~\bibnamefont {Liang}}, \bibinfo {author} {\bibfnamefont {D.}~\bibnamefont {Xie}}, \bibinfo {author} {\bibfnamefont {Z.}~\bibnamefont {Dong}}, \bibinfo {author} {\bibfnamefont {H.}~\bibnamefont {Li}}, \bibinfo {author} {\bibfnamefont {H.}~\bibnamefont {Li}}, \bibinfo {author} {\bibfnamefont {B.}~\bibnamefont {Gadway}}, \bibinfo {author} {\bibfnamefont {W.}~\bibnamefont {Yi}}, \ and\ \bibinfo {author} {\bibfnamefont {B.}~\bibnamefont {Yan}},\ }\href {\doibase 10.1103/PhysRevLett.129.070401} {\bibfield  {journal} {\bibinfo  {journal} {Phys. Rev. Lett.}\ }\textbf {\bibinfo {volume} {129}},\ \bibinfo {pages} {070401} (\bibinfo {year} {2022}{\natexlab{a}})}\BibitemShut {NoStop}%
    \bibitem [{\citenamefont {Scheurer}\ and\ \citenamefont {Slager}(2020)}]{scheurer2020unsupervised}%
      \BibitemOpen
      \bibfield  {author} {\bibinfo {author} {\bibfnamefont {M.~S.}\ \bibnamefont {Scheurer}}\ and\ \bibinfo {author} {\bibfnamefont {R.-J.}\ \bibnamefont {Slager}},\ }\href@noop {} {\bibfield  {journal} {\bibinfo  {journal} {Physical review letters}\ }\textbf {\bibinfo {volume} {124}},\ \bibinfo {pages} {226401} (\bibinfo {year} {2020})}\BibitemShut {NoStop}%
    \bibitem [{\citenamefont {Yu}\ and\ \citenamefont {Deng}(2021)}]{yu2021unsupervised}%
      \BibitemOpen
      \bibfield  {author} {\bibinfo {author} {\bibfnamefont {L.-W.}\ \bibnamefont {Yu}}\ and\ \bibinfo {author} {\bibfnamefont {D.-L.}\ \bibnamefont {Deng}},\ }\href@noop {} {\bibfield  {journal} {\bibinfo  {journal} {Physical Review Letters}\ }\textbf {\bibinfo {volume} {126}},\ \bibinfo {pages} {240402} (\bibinfo {year} {2021})}\BibitemShut {NoStop}%
    \bibitem [{\citenamefont {Yang}\ \emph {et~al.}(2022{\natexlab{a}})\citenamefont {Yang}, \citenamefont {Tan}, \citenamefont {Tai}, \citenamefont {Koh}, \citenamefont {Li}, \citenamefont {Longhi},\ and\ \citenamefont {Lee}}]{yang2022designing}%
      \BibitemOpen
      \bibfield  {author} {\bibinfo {author} {\bibfnamefont {R.}~\bibnamefont {Yang}}, \bibinfo {author} {\bibfnamefont {J.~W.}\ \bibnamefont {Tan}}, \bibinfo {author} {\bibfnamefont {T.}~\bibnamefont {Tai}}, \bibinfo {author} {\bibfnamefont {J.~M.}\ \bibnamefont {Koh}}, \bibinfo {author} {\bibfnamefont {L.}~\bibnamefont {Li}}, \bibinfo {author} {\bibfnamefont {S.}~\bibnamefont {Longhi}}, \ and\ \bibinfo {author} {\bibfnamefont {C.~H.}\ \bibnamefont {Lee}},\ }\href@noop {} {\bibfield  {journal} {\bibinfo  {journal} {Science Bulletin}\ } (\bibinfo {year} {2022}{\natexlab{a}})}\BibitemShut {NoStop}%
    \bibitem [{\citenamefont {Oztas}\ and\ \citenamefont {Candemir}(2019)}]{oztas2019schrieffer}%
      \BibitemOpen
      \bibfield  {author} {\bibinfo {author} {\bibfnamefont {Z.}~\bibnamefont {Oztas}}\ and\ \bibinfo {author} {\bibfnamefont {N.}~\bibnamefont {Candemir}},\ }\href@noop {} {\bibfield  {journal} {\bibinfo  {journal} {Physics Letters A}\ }\textbf {\bibinfo {volume} {383}},\ \bibinfo {pages} {1821} (\bibinfo {year} {2019})}\BibitemShut {NoStop}%
    \bibitem [{\citenamefont {Zhu}\ \emph {et~al.}(2021)\citenamefont {Zhu}, \citenamefont {Teo}, \citenamefont {Li},\ and\ \citenamefont {Gong}}]{zhu2021delocalization}%
      \BibitemOpen
      \bibfield  {author} {\bibinfo {author} {\bibfnamefont {W.}~\bibnamefont {Zhu}}, \bibinfo {author} {\bibfnamefont {W.~X.}\ \bibnamefont {Teo}}, \bibinfo {author} {\bibfnamefont {L.}~\bibnamefont {Li}}, \ and\ \bibinfo {author} {\bibfnamefont {J.}~\bibnamefont {Gong}},\ }\href {\doibase 10.1103/PhysRevB.103.195414} {\bibfield  {journal} {\bibinfo  {journal} {Phys. Rev. B}\ }\textbf {\bibinfo {volume} {103}},\ \bibinfo {pages} {195414} (\bibinfo {year} {2021})}\BibitemShut {NoStop}%
    \bibitem [{\citenamefont {Cheng}\ \emph {et~al.}(2022)\citenamefont {Cheng}, \citenamefont {Zhang}, \citenamefont {Lu},\ and\ \citenamefont {Chen}}]{cheng2022competition}%
      \BibitemOpen
      \bibfield  {author} {\bibinfo {author} {\bibfnamefont {J.}~\bibnamefont {Cheng}}, \bibinfo {author} {\bibfnamefont {X.}~\bibnamefont {Zhang}}, \bibinfo {author} {\bibfnamefont {M.-H.}\ \bibnamefont {Lu}}, \ and\ \bibinfo {author} {\bibfnamefont {Y.-F.}\ \bibnamefont {Chen}},\ }\href {\doibase 10.1103/PhysRevB.105.094103} {\bibfield  {journal} {\bibinfo  {journal} {Phys. Rev. B}\ }\textbf {\bibinfo {volume} {105}},\ \bibinfo {pages} {094103} (\bibinfo {year} {2022})}\BibitemShut {NoStop}%
    \bibitem [{\citenamefont {Okuma}\ and\ \citenamefont {Sato}(2022)}]{okuma2022non}%
      \BibitemOpen
      \bibfield  {author} {\bibinfo {author} {\bibfnamefont {N.}~\bibnamefont {Okuma}}\ and\ \bibinfo {author} {\bibfnamefont {M.}~\bibnamefont {Sato}},\ }\href@noop {} {\bibfield  {journal} {\bibinfo  {journal} {Annual Review of Condensed Matter Physics}\ }\textbf {\bibinfo {volume} {14}} (\bibinfo {year} {2022})}\BibitemShut {NoStop}%
    \bibitem [{\citenamefont {Wang}\ \emph {et~al.}(2020)\citenamefont {Wang}, \citenamefont {Guo},\ and\ \citenamefont {Kou}}]{wang2020defective}%
      \BibitemOpen
      \bibfield  {author} {\bibinfo {author} {\bibfnamefont {X.-R.}\ \bibnamefont {Wang}}, \bibinfo {author} {\bibfnamefont {C.-X.}\ \bibnamefont {Guo}}, \ and\ \bibinfo {author} {\bibfnamefont {S.-P.}\ \bibnamefont {Kou}},\ }\href {\doibase 10.1103/PhysRevB.101.121116} {\bibfield  {journal} {\bibinfo  {journal} {Phys. Rev. B}\ }\textbf {\bibinfo {volume} {101}},\ \bibinfo {pages} {121116} (\bibinfo {year} {2020})}\BibitemShut {NoStop}%
    \bibitem [{\citenamefont {Lee}\ \emph {et~al.}(2019)\citenamefont {Lee}, \citenamefont {Li},\ and\ \citenamefont {Gong}}]{Lee2019hybrid}%
      \BibitemOpen
      \bibfield  {author} {\bibinfo {author} {\bibfnamefont {C.~H.}\ \bibnamefont {Lee}}, \bibinfo {author} {\bibfnamefont {L.}~\bibnamefont {Li}}, \ and\ \bibinfo {author} {\bibfnamefont {J.}~\bibnamefont {Gong}},\ }\href {\doibase 10.1103/PhysRevLett.123.016805} {\bibfield  {journal} {\bibinfo  {journal} {Phys. Rev. Lett.}\ }\textbf {\bibinfo {volume} {123}},\ \bibinfo {pages} {016805} (\bibinfo {year} {2019})}\BibitemShut {NoStop}%
    \bibitem [{\citenamefont {Li}\ \emph {et~al.}(2020{\natexlab{b}})\citenamefont {Li}, \citenamefont {Lee},\ and\ \citenamefont {Gong}}]{li2020topological}%
      \BibitemOpen
      \bibfield  {author} {\bibinfo {author} {\bibfnamefont {L.}~\bibnamefont {Li}}, \bibinfo {author} {\bibfnamefont {C.~H.}\ \bibnamefont {Lee}}, \ and\ \bibinfo {author} {\bibfnamefont {J.}~\bibnamefont {Gong}},\ }\href@noop {} {\bibfield  {journal} {\bibinfo  {journal} {Physical Review Letters}\ }\textbf {\bibinfo {volume} {124}},\ \bibinfo {pages} {250402} (\bibinfo {year} {2020}{\natexlab{b}})}\BibitemShut {NoStop}%
    \bibitem [{\citenamefont {Li}\ \emph {et~al.}(2022{\natexlab{a}})\citenamefont {Li}, \citenamefont {Liang}, \citenamefont {Wang}, \citenamefont {Lu},\ and\ \citenamefont {Liu}}]{li2022gain}%
      \BibitemOpen
      \bibfield  {author} {\bibinfo {author} {\bibfnamefont {Y.}~\bibnamefont {Li}}, \bibinfo {author} {\bibfnamefont {C.}~\bibnamefont {Liang}}, \bibinfo {author} {\bibfnamefont {C.}~\bibnamefont {Wang}}, \bibinfo {author} {\bibfnamefont {C.}~\bibnamefont {Lu}}, \ and\ \bibinfo {author} {\bibfnamefont {Y.-C.}\ \bibnamefont {Liu}},\ }\href {\doibase 10.1103/PhysRevLett.128.223903} {\bibfield  {journal} {\bibinfo  {journal} {Phys. Rev. Lett.}\ }\textbf {\bibinfo {volume} {128}},\ \bibinfo {pages} {223903} (\bibinfo {year} {2022}{\natexlab{a}})}\BibitemShut {NoStop}%
    \bibitem [{\citenamefont {Zhu}\ and\ \citenamefont {Gong}(2022)}]{zhu2022hybrid}%
      \BibitemOpen
      \bibfield  {author} {\bibinfo {author} {\bibfnamefont {W.}~\bibnamefont {Zhu}}\ and\ \bibinfo {author} {\bibfnamefont {J.}~\bibnamefont {Gong}},\ }\href {\doibase 10.1103/PhysRevB.106.035425} {\bibfield  {journal} {\bibinfo  {journal} {Phys. Rev. B}\ }\textbf {\bibinfo {volume} {106}},\ \bibinfo {pages} {035425} (\bibinfo {year} {2022})}\BibitemShut {NoStop}%
    \bibitem [{\citenamefont {Kawabata}\ \emph {et~al.}(2020)\citenamefont {Kawabata}, \citenamefont {Sato},\ and\ \citenamefont {Shiozaki}}]{kawabata2020higher}%
      \BibitemOpen
      \bibfield  {author} {\bibinfo {author} {\bibfnamefont {K.}~\bibnamefont {Kawabata}}, \bibinfo {author} {\bibfnamefont {M.}~\bibnamefont {Sato}}, \ and\ \bibinfo {author} {\bibfnamefont {K.}~\bibnamefont {Shiozaki}},\ }\href {\doibase 10.1103/PhysRevB.102.205118} {\bibfield  {journal} {\bibinfo  {journal} {Phys. Rev. B}\ }\textbf {\bibinfo {volume} {102}},\ \bibinfo {pages} {205118} (\bibinfo {year} {2020})}\BibitemShut {NoStop}%
    \bibitem [{\citenamefont {Okugawa}\ \emph {et~al.}(2020)\citenamefont {Okugawa}, \citenamefont {Takahashi},\ and\ \citenamefont {Yokomizo}}]{okugawa2020second}%
      \BibitemOpen
      \bibfield  {author} {\bibinfo {author} {\bibfnamefont {R.}~\bibnamefont {Okugawa}}, \bibinfo {author} {\bibfnamefont {R.}~\bibnamefont {Takahashi}}, \ and\ \bibinfo {author} {\bibfnamefont {K.}~\bibnamefont {Yokomizo}},\ }\href {\doibase 10.1103/PhysRevB.102.241202} {\bibfield  {journal} {\bibinfo  {journal} {Phys. Rev. B}\ }\textbf {\bibinfo {volume} {102}},\ \bibinfo {pages} {241202} (\bibinfo {year} {2020})}\BibitemShut {NoStop}%
    \bibitem [{\citenamefont {Fu}\ \emph {et~al.}(2021)\citenamefont {Fu}, \citenamefont {Hu},\ and\ \citenamefont {Wan}}]{fu2021nonH}%
      \BibitemOpen
      \bibfield  {author} {\bibinfo {author} {\bibfnamefont {Y.}~\bibnamefont {Fu}}, \bibinfo {author} {\bibfnamefont {J.}~\bibnamefont {Hu}}, \ and\ \bibinfo {author} {\bibfnamefont {S.}~\bibnamefont {Wan}},\ }\href {\doibase 10.1103/PhysRevB.103.045420} {\bibfield  {journal} {\bibinfo  {journal} {Phys. Rev. B}\ }\textbf {\bibinfo {volume} {103}},\ \bibinfo {pages} {045420} (\bibinfo {year} {2021})}\BibitemShut {NoStop}%
    \bibitem [{\citenamefont {Zhang}\ \emph {et~al.}(2020)\citenamefont {Zhang}, \citenamefont {Yang},\ and\ \citenamefont {Fang}}]{zhang2020correspondence}%
      \BibitemOpen
      \bibfield  {author} {\bibinfo {author} {\bibfnamefont {K.}~\bibnamefont {Zhang}}, \bibinfo {author} {\bibfnamefont {Z.}~\bibnamefont {Yang}}, \ and\ \bibinfo {author} {\bibfnamefont {C.}~\bibnamefont {Fang}},\ }\href {\doibase 10.1103/PhysRevLett.125.126402} {\bibfield  {journal} {\bibinfo  {journal} {Phys. Rev. Lett.}\ }\textbf {\bibinfo {volume} {125}},\ \bibinfo {pages} {126402} (\bibinfo {year} {2020})}\BibitemShut {NoStop}%
    \bibitem [{\citenamefont {Borgnia}\ \emph {et~al.}(2020)\citenamefont {Borgnia}, \citenamefont {Kruchkov},\ and\ \citenamefont {Slager}}]{borgnia2020nonH}%
      \BibitemOpen
      \bibfield  {author} {\bibinfo {author} {\bibfnamefont {D.~S.}\ \bibnamefont {Borgnia}}, \bibinfo {author} {\bibfnamefont {A.~J.}\ \bibnamefont {Kruchkov}}, \ and\ \bibinfo {author} {\bibfnamefont {R.-J.}\ \bibnamefont {Slager}},\ }\href {\doibase 10.1103/PhysRevLett.124.056802} {\bibfield  {journal} {\bibinfo  {journal} {Phys. Rev. Lett.}\ }\textbf {\bibinfo {volume} {124}},\ \bibinfo {pages} {056802} (\bibinfo {year} {2020})}\BibitemShut {NoStop}%
    \bibitem [{\citenamefont {Okuma}\ \emph {et~al.}(2020)\citenamefont {Okuma}, \citenamefont {Kawabata}, \citenamefont {Shiozaki},\ and\ \citenamefont {Sato}}]{okuma2020topological}%
      \BibitemOpen
      \bibfield  {author} {\bibinfo {author} {\bibfnamefont {N.}~\bibnamefont {Okuma}}, \bibinfo {author} {\bibfnamefont {K.}~\bibnamefont {Kawabata}}, \bibinfo {author} {\bibfnamefont {K.}~\bibnamefont {Shiozaki}}, \ and\ \bibinfo {author} {\bibfnamefont {M.}~\bibnamefont {Sato}},\ }\href {\doibase 10.1103/PhysRevLett.124.086801} {\bibfield  {journal} {\bibinfo  {journal} {Phys. Rev. Lett.}\ }\textbf {\bibinfo {volume} {124}},\ \bibinfo {pages} {086801} (\bibinfo {year} {2020})}\BibitemShut {NoStop}%
    \bibitem [{\citenamefont {Li}\ \emph {et~al.}(2021{\natexlab{b}})\citenamefont {Li}, \citenamefont {Mu}, \citenamefont {Lee},\ and\ \citenamefont {Gong}}]{Li2021}%
      \BibitemOpen
      \bibfield  {author} {\bibinfo {author} {\bibfnamefont {L.}~\bibnamefont {Li}}, \bibinfo {author} {\bibfnamefont {S.}~\bibnamefont {Mu}}, \bibinfo {author} {\bibfnamefont {C.~H.}\ \bibnamefont {Lee}}, \ and\ \bibinfo {author} {\bibfnamefont {J.}~\bibnamefont {Gong}},\ }\href@noop {} {\bibfield  {journal} {\bibinfo  {journal} {Nature communications}\ }\textbf {\bibinfo {volume} {12}},\ \bibinfo {pages} {5294} (\bibinfo {year} {2021}{\natexlab{b}})}\BibitemShut {NoStop}%
    \bibitem [{\citenamefont {Liang}\ \emph {et~al.}(2022{\natexlab{b}})\citenamefont {Liang}, \citenamefont {Mu}, \citenamefont {Gong},\ and\ \citenamefont {Li}}]{liang2022anomalous}%
      \BibitemOpen
      \bibfield  {author} {\bibinfo {author} {\bibfnamefont {H.-Q.}\ \bibnamefont {Liang}}, \bibinfo {author} {\bibfnamefont {S.}~\bibnamefont {Mu}}, \bibinfo {author} {\bibfnamefont {J.}~\bibnamefont {Gong}}, \ and\ \bibinfo {author} {\bibfnamefont {L.}~\bibnamefont {Li}},\ }\href@noop {} {\bibfield  {journal} {\bibinfo  {journal} {Physical Review B}\ }\textbf {\bibinfo {volume} {105}},\ \bibinfo {pages} {L241402} (\bibinfo {year} {2022}{\natexlab{b}})}\BibitemShut {NoStop}%
    \bibitem [{\citenamefont {Longhi}(2019{\natexlab{a}})}]{longhi2019topological}%
      \BibitemOpen
      \bibfield  {author} {\bibinfo {author} {\bibfnamefont {S.}~\bibnamefont {Longhi}},\ }\href@noop {} {\bibfield  {journal} {\bibinfo  {journal} {Physical Review Letters}\ }\textbf {\bibinfo {volume} {122}},\ \bibinfo {pages} {237601} (\bibinfo {year} {2019}{\natexlab{a}})}\BibitemShut {NoStop}%
    \bibitem [{\citenamefont {Jiang}\ \emph {et~al.}(2019)\citenamefont {Jiang}, \citenamefont {Lang}, \citenamefont {Yang}, \citenamefont {Zhu},\ and\ \citenamefont {Chen}}]{jiang2019interplay}%
      \BibitemOpen
      \bibfield  {author} {\bibinfo {author} {\bibfnamefont {H.}~\bibnamefont {Jiang}}, \bibinfo {author} {\bibfnamefont {L.-J.}\ \bibnamefont {Lang}}, \bibinfo {author} {\bibfnamefont {C.}~\bibnamefont {Yang}}, \bibinfo {author} {\bibfnamefont {S.-L.}\ \bibnamefont {Zhu}}, \ and\ \bibinfo {author} {\bibfnamefont {S.}~\bibnamefont {Chen}},\ }\href@noop {} {\bibfield  {journal} {\bibinfo  {journal} {Physical Review B}\ }\textbf {\bibinfo {volume} {100}},\ \bibinfo {pages} {054301} (\bibinfo {year} {2019})}\BibitemShut {NoStop}%
    \bibitem [{\citenamefont {Longhi}(2019{\natexlab{b}})}]{longhi2019metal}%
      \BibitemOpen
      \bibfield  {author} {\bibinfo {author} {\bibfnamefont {S.}~\bibnamefont {Longhi}},\ }\href {\doibase 10.1103/PhysRevB.100.125157} {\bibfield  {journal} {\bibinfo  {journal} {Phys. Rev. B}\ }\textbf {\bibinfo {volume} {100}},\ \bibinfo {pages} {125157} (\bibinfo {year} {2019}{\natexlab{b}})}\BibitemShut {NoStop}%
    \bibitem [{\citenamefont {Zeng}\ and\ \citenamefont {Xu}(2020)}]{zeng2020winding}%
      \BibitemOpen
      \bibfield  {author} {\bibinfo {author} {\bibfnamefont {Q.-B.}\ \bibnamefont {Zeng}}\ and\ \bibinfo {author} {\bibfnamefont {Y.}~\bibnamefont {Xu}},\ }\href {\doibase 10.1103/PhysRevResearch.2.033052} {\bibfield  {journal} {\bibinfo  {journal} {Phys. Rev. Research}\ }\textbf {\bibinfo {volume} {2}},\ \bibinfo {pages} {033052} (\bibinfo {year} {2020})}\BibitemShut {NoStop}%
    \bibitem [{\citenamefont {Zeng}\ \emph {et~al.}(2020)\citenamefont {Zeng}, \citenamefont {Yang},\ and\ \citenamefont {Xu}}]{zeng2020topological}%
      \BibitemOpen
      \bibfield  {author} {\bibinfo {author} {\bibfnamefont {Q.-B.}\ \bibnamefont {Zeng}}, \bibinfo {author} {\bibfnamefont {Y.-B.}\ \bibnamefont {Yang}}, \ and\ \bibinfo {author} {\bibfnamefont {Y.}~\bibnamefont {Xu}},\ }\href@noop {} {\bibfield  {journal} {\bibinfo  {journal} {Physical Review B}\ }\textbf {\bibinfo {volume} {101}},\ \bibinfo {pages} {020201} (\bibinfo {year} {2020})}\BibitemShut {NoStop}%
    \bibitem [{\citenamefont {Liu}\ \emph {et~al.}(2020{\natexlab{b}})\citenamefont {Liu}, \citenamefont {Jiang}, \citenamefont {Cao},\ and\ \citenamefont {Chen}}]{liu2020nonH}%
      \BibitemOpen
      \bibfield  {author} {\bibinfo {author} {\bibfnamefont {Y.}~\bibnamefont {Liu}}, \bibinfo {author} {\bibfnamefont {X.-P.}\ \bibnamefont {Jiang}}, \bibinfo {author} {\bibfnamefont {J.}~\bibnamefont {Cao}}, \ and\ \bibinfo {author} {\bibfnamefont {S.}~\bibnamefont {Chen}},\ }\href {\doibase 10.1103/PhysRevB.101.174205} {\bibfield  {journal} {\bibinfo  {journal} {Phys. Rev. B}\ }\textbf {\bibinfo {volume} {101}},\ \bibinfo {pages} {174205} (\bibinfo {year} {2020}{\natexlab{b}})}\BibitemShut {NoStop}%
    \bibitem [{\citenamefont {Zhai}\ \emph {et~al.}(2020)\citenamefont {Zhai}, \citenamefont {Yin},\ and\ \citenamefont {Huang}}]{zhai2020many}%
      \BibitemOpen
      \bibfield  {author} {\bibinfo {author} {\bibfnamefont {L.-J.}\ \bibnamefont {Zhai}}, \bibinfo {author} {\bibfnamefont {S.}~\bibnamefont {Yin}}, \ and\ \bibinfo {author} {\bibfnamefont {G.-Y.}\ \bibnamefont {Huang}},\ }\href {\doibase 10.1103/PhysRevB.102.064206} {\bibfield  {journal} {\bibinfo  {journal} {Phys. Rev. B}\ }\textbf {\bibinfo {volume} {102}},\ \bibinfo {pages} {064206} (\bibinfo {year} {2020})}\BibitemShut {NoStop}%
    \bibitem [{\citenamefont {Cai}(2021)}]{cai2021boundary}%
      \BibitemOpen
      \bibfield  {author} {\bibinfo {author} {\bibfnamefont {X.}~\bibnamefont {Cai}},\ }\href {\doibase 10.1103/PhysRevB.103.014201} {\bibfield  {journal} {\bibinfo  {journal} {Phys. Rev. B}\ }\textbf {\bibinfo {volume} {103}},\ \bibinfo {pages} {014201} (\bibinfo {year} {2021})}\BibitemShut {NoStop}%
    \bibitem [{\citenamefont {Liu}\ \emph {et~al.}(2021{\natexlab{b}})\citenamefont {Liu}, \citenamefont {Wang}, \citenamefont {Liu}, \citenamefont {Zhou},\ and\ \citenamefont {Chen}}]{liu2021exact1}%
      \BibitemOpen
      \bibfield  {author} {\bibinfo {author} {\bibfnamefont {Y.}~\bibnamefont {Liu}}, \bibinfo {author} {\bibfnamefont {Y.}~\bibnamefont {Wang}}, \bibinfo {author} {\bibfnamefont {X.-J.}\ \bibnamefont {Liu}}, \bibinfo {author} {\bibfnamefont {Q.}~\bibnamefont {Zhou}}, \ and\ \bibinfo {author} {\bibfnamefont {S.}~\bibnamefont {Chen}},\ }\href {\doibase 10.1103/PhysRevB.103.014203} {\bibfield  {journal} {\bibinfo  {journal} {Phys. Rev. B}\ }\textbf {\bibinfo {volume} {103}},\ \bibinfo {pages} {014203} (\bibinfo {year} {2021}{\natexlab{b}})}\BibitemShut {NoStop}%
    \bibitem [{\citenamefont {Liu}\ \emph {et~al.}(2021{\natexlab{c}})\citenamefont {Liu}, \citenamefont {Zhou},\ and\ \citenamefont {Chen}}]{liu2021localization}%
      \BibitemOpen
      \bibfield  {author} {\bibinfo {author} {\bibfnamefont {Y.}~\bibnamefont {Liu}}, \bibinfo {author} {\bibfnamefont {Q.}~\bibnamefont {Zhou}}, \ and\ \bibinfo {author} {\bibfnamefont {S.}~\bibnamefont {Chen}},\ }\href {\doibase 10.1103/PhysRevB.104.024201} {\bibfield  {journal} {\bibinfo  {journal} {Phys. Rev. B}\ }\textbf {\bibinfo {volume} {104}},\ \bibinfo {pages} {024201} (\bibinfo {year} {2021}{\natexlab{c}})}\BibitemShut {NoStop}%
    \bibitem [{\citenamefont {Claes}\ and\ \citenamefont {Hughes}(2021)}]{claes2021skin}%
      \BibitemOpen
      \bibfield  {author} {\bibinfo {author} {\bibfnamefont {J.}~\bibnamefont {Claes}}\ and\ \bibinfo {author} {\bibfnamefont {T.~L.}\ \bibnamefont {Hughes}},\ }\href {\doibase 10.1103/PhysRevB.103.L140201} {\bibfield  {journal} {\bibinfo  {journal} {Phys. Rev. B}\ }\textbf {\bibinfo {volume} {103}},\ \bibinfo {pages} {L140201} (\bibinfo {year} {2021})}\BibitemShut {NoStop}%
    \bibitem [{\citenamefont {Longhi}(2022{\natexlab{b}})}]{longhi2022non4}%
      \BibitemOpen
      \bibfield  {author} {\bibinfo {author} {\bibfnamefont {S.}~\bibnamefont {Longhi}},\ }\href@noop {} {\bibfield  {journal} {\bibinfo  {journal} {Optics Letters}\ }\textbf {\bibinfo {volume} {47}},\ \bibinfo {pages} {2951} (\bibinfo {year} {2022}{\natexlab{b}})}\BibitemShut {NoStop}%
    \bibitem [{\citenamefont {Zhang}\ \emph {et~al.}(2021{\natexlab{c}})\citenamefont {Zhang}, \citenamefont {Li}, \citenamefont {Liu}, \citenamefont {Tai}, \citenamefont {Thomale},\ and\ \citenamefont {Lee}}]{zhang2021tidal}%
      \BibitemOpen
      \bibfield  {author} {\bibinfo {author} {\bibfnamefont {X.}~\bibnamefont {Zhang}}, \bibinfo {author} {\bibfnamefont {G.}~\bibnamefont {Li}}, \bibinfo {author} {\bibfnamefont {Y.}~\bibnamefont {Liu}}, \bibinfo {author} {\bibfnamefont {T.}~\bibnamefont {Tai}}, \bibinfo {author} {\bibfnamefont {R.}~\bibnamefont {Thomale}}, \ and\ \bibinfo {author} {\bibfnamefont {C.~H.}\ \bibnamefont {Lee}},\ }\href@noop {} {\bibfield  {journal} {\bibinfo  {journal} {Communications Physics}\ }\textbf {\bibinfo {volume} {4}},\ \bibinfo {pages} {1} (\bibinfo {year} {2021}{\natexlab{c}})}\BibitemShut {NoStop}%
    \bibitem [{\citenamefont {Herviou}\ \emph {et~al.}(2019{\natexlab{a}})\citenamefont {Herviou}, \citenamefont {Bardarson},\ and\ \citenamefont {Regnault}}]{herviou2019defining}%
      \BibitemOpen
      \bibfield  {author} {\bibinfo {author} {\bibfnamefont {L.}~\bibnamefont {Herviou}}, \bibinfo {author} {\bibfnamefont {J.~H.}\ \bibnamefont {Bardarson}}, \ and\ \bibinfo {author} {\bibfnamefont {N.}~\bibnamefont {Regnault}},\ }\href@noop {} {\bibfield  {journal} {\bibinfo  {journal} {Physical Review A}\ }\textbf {\bibinfo {volume} {99}},\ \bibinfo {pages} {052118} (\bibinfo {year} {2019}{\natexlab{a}})}\BibitemShut {NoStop}%
    \bibitem [{\citenamefont {Zirnstein}\ \emph {et~al.}(2021)\citenamefont {Zirnstein}, \citenamefont {Refael},\ and\ \citenamefont {Rosenow}}]{zirnstein2021bulk}%
      \BibitemOpen
      \bibfield  {author} {\bibinfo {author} {\bibfnamefont {H.-G.}\ \bibnamefont {Zirnstein}}, \bibinfo {author} {\bibfnamefont {G.}~\bibnamefont {Refael}}, \ and\ \bibinfo {author} {\bibfnamefont {B.}~\bibnamefont {Rosenow}},\ }\href@noop {} {\bibfield  {journal} {\bibinfo  {journal} {Physical review letters}\ }\textbf {\bibinfo {volume} {126}},\ \bibinfo {pages} {216407} (\bibinfo {year} {2021})}\BibitemShut {NoStop}%
    \bibitem [{\citenamefont {Su}\ \emph {et~al.}(1979)\citenamefont {Su}, \citenamefont {Schrieffer},\ and\ \citenamefont {Heeger}}]{su1979}%
      \BibitemOpen
      \bibfield  {author} {\bibinfo {author} {\bibfnamefont {W.~P.}\ \bibnamefont {Su}}, \bibinfo {author} {\bibfnamefont {J.~R.}\ \bibnamefont {Schrieffer}}, \ and\ \bibinfo {author} {\bibfnamefont {A.~J.}\ \bibnamefont {Heeger}},\ }\href {\doibase 10.1103/PhysRevLett.42.1698} {\bibfield  {journal} {\bibinfo  {journal} {Phys. Rev. Lett.}\ }\textbf {\bibinfo {volume} {42}},\ \bibinfo {pages} {1698} (\bibinfo {year} {1979})}\BibitemShut {NoStop}%
    \bibitem [{\citenamefont {Creutz}(1999)}]{Creutz1999}%
      \BibitemOpen
      \bibfield  {author} {\bibinfo {author} {\bibfnamefont {M.}~\bibnamefont {Creutz}},\ }\href {\doibase 10.1103/PhysRevLett.83.2636} {\bibfield  {journal} {\bibinfo  {journal} {Phys. Rev. Lett.}\ }\textbf {\bibinfo {volume} {83}},\ \bibinfo {pages} {2636} (\bibinfo {year} {1999})}\BibitemShut {NoStop}%
    \bibitem [{\citenamefont {Liang}\ and\ \citenamefont {Li}(2022)}]{liang2022topological}%
      \BibitemOpen
      \bibfield  {author} {\bibinfo {author} {\bibfnamefont {H.-Q.}\ \bibnamefont {Liang}}\ and\ \bibinfo {author} {\bibfnamefont {L.}~\bibnamefont {Li}},\ }\href@noop {} {\bibfield  {journal} {\bibinfo  {journal} {Chinese Physics B}\ }\textbf {\bibinfo {volume} {31}},\ \bibinfo {pages} {010310} (\bibinfo {year} {2022})}\BibitemShut {NoStop}%
    \bibitem [{\citenamefont {Edvardsson}\ \emph {et~al.}(2020)\citenamefont {Edvardsson}, \citenamefont {Kunst}, \citenamefont {Yoshida},\ and\ \citenamefont {Bergholtz}}]{PhysRevResearch.2.043046}%
      \BibitemOpen
      \bibfield  {author} {\bibinfo {author} {\bibfnamefont {E.}~\bibnamefont {Edvardsson}}, \bibinfo {author} {\bibfnamefont {F.~K.}\ \bibnamefont {Kunst}}, \bibinfo {author} {\bibfnamefont {T.}~\bibnamefont {Yoshida}}, \ and\ \bibinfo {author} {\bibfnamefont {E.~J.}\ \bibnamefont {Bergholtz}},\ }\href {\doibase 10.1103/PhysRevResearch.2.043046} {\bibfield  {journal} {\bibinfo  {journal} {Phys. Rev. Res.}\ }\textbf {\bibinfo {volume} {2}},\ \bibinfo {pages} {043046} (\bibinfo {year} {2020})}\BibitemShut {NoStop}%
    \bibitem [{\citenamefont {Masuda}\ and\ \citenamefont {Nakamura}(2022{\natexlab{a}})}]{masuda2022relationship}%
      \BibitemOpen
      \bibfield  {author} {\bibinfo {author} {\bibfnamefont {S.}~\bibnamefont {Masuda}}\ and\ \bibinfo {author} {\bibfnamefont {M.}~\bibnamefont {Nakamura}},\ }\href@noop {} {\bibfield  {journal} {\bibinfo  {journal} {Journal of the Physical Society of Japan}\ }\textbf {\bibinfo {volume} {91}},\ \bibinfo {pages} {043701} (\bibinfo {year} {2022}{\natexlab{a}})}\BibitemShut {NoStop}%
    \bibitem [{\citenamefont {Masuda}\ and\ \citenamefont {Nakamura}(2022{\natexlab{b}})}]{masuda2022electronic}%
      \BibitemOpen
      \bibfield  {author} {\bibinfo {author} {\bibfnamefont {S.}~\bibnamefont {Masuda}}\ and\ \bibinfo {author} {\bibfnamefont {M.}~\bibnamefont {Nakamura}},\ }\href@noop {} {\bibfield  {journal} {\bibinfo  {journal} {Journal of the Physical Society of Japan}\ }\textbf {\bibinfo {volume} {91}},\ \bibinfo {pages} {114705} (\bibinfo {year} {2022}{\natexlab{b}})}\BibitemShut {NoStop}%
    \bibitem [{\citenamefont {Deng}\ and\ \citenamefont {Yi}(2019)}]{deng2019non}%
      \BibitemOpen
      \bibfield  {author} {\bibinfo {author} {\bibfnamefont {T.-S.}\ \bibnamefont {Deng}}\ and\ \bibinfo {author} {\bibfnamefont {W.}~\bibnamefont {Yi}},\ }\href@noop {} {\bibfield  {journal} {\bibinfo  {journal} {Physical Review B}\ }\textbf {\bibinfo {volume} {100}},\ \bibinfo {pages} {035102} (\bibinfo {year} {2019})}\BibitemShut {NoStop}%
    \bibitem [{\citenamefont {Liu}\ \emph {et~al.}(2022{\natexlab{a}})\citenamefont {Liu}, \citenamefont {Lu}, \citenamefont {Zhang},\ and\ \citenamefont {Jiang}}]{liu2022modified}%
      \BibitemOpen
      \bibfield  {author} {\bibinfo {author} {\bibfnamefont {H.}~\bibnamefont {Liu}}, \bibinfo {author} {\bibfnamefont {M.}~\bibnamefont {Lu}}, \bibinfo {author} {\bibfnamefont {Z.-Q.}\ \bibnamefont {Zhang}}, \ and\ \bibinfo {author} {\bibfnamefont {H.}~\bibnamefont {Jiang}},\ }\href@noop {} {\bibfield  {journal} {\bibinfo  {journal} {arXiv preprint arXiv:2208.03013}\ } (\bibinfo {year} {2022}{\natexlab{a}})}\BibitemShut {NoStop}%
    \bibitem [{\citenamefont {Yang}\ \emph {et~al.}(2020{\natexlab{a}})\citenamefont {Yang}, \citenamefont {Zhang}, \citenamefont {Fang},\ and\ \citenamefont {Hu}}]{yang2020non}%
      \BibitemOpen
      \bibfield  {author} {\bibinfo {author} {\bibfnamefont {Z.}~\bibnamefont {Yang}}, \bibinfo {author} {\bibfnamefont {K.}~\bibnamefont {Zhang}}, \bibinfo {author} {\bibfnamefont {C.}~\bibnamefont {Fang}}, \ and\ \bibinfo {author} {\bibfnamefont {J.}~\bibnamefont {Hu}},\ }\href@noop {} {\bibfield  {journal} {\bibinfo  {journal} {Physical Review Letters}\ }\textbf {\bibinfo {volume} {125}},\ \bibinfo {pages} {226402} (\bibinfo {year} {2020}{\natexlab{a}})}\BibitemShut {NoStop}%
    \bibitem [{\citenamefont {Tai}\ and\ \citenamefont {Lee}(2023)}]{tai2022zoology}%
      \BibitemOpen
      \bibfield  {author} {\bibinfo {author} {\bibfnamefont {T.}~\bibnamefont {Tai}}\ and\ \bibinfo {author} {\bibfnamefont {C.~H.}\ \bibnamefont {Lee}},\ }\href@noop {} {\bibfield  {journal} {\bibinfo  {journal} {Physical Review B}\ }\textbf {\bibinfo {volume} {107}},\ \bibinfo {pages} {L220301} (\bibinfo {year} {2023})}\BibitemShut {NoStop}%
    \bibitem [{\citenamefont {Zhang}\ \emph {et~al.}(2023{\natexlab{a}})\citenamefont {Zhang}, \citenamefont {Liu}, \citenamefont {Liu}, \citenamefont {Jiang},\ and\ \citenamefont {Xie}}]{ZHANG2023}%
      \BibitemOpen
      \bibfield  {author} {\bibinfo {author} {\bibfnamefont {Z.-Q.}\ \bibnamefont {Zhang}}, \bibinfo {author} {\bibfnamefont {H.}~\bibnamefont {Liu}}, \bibinfo {author} {\bibfnamefont {H.}~\bibnamefont {Liu}}, \bibinfo {author} {\bibfnamefont {H.}~\bibnamefont {Jiang}}, \ and\ \bibinfo {author} {\bibfnamefont {X.}~\bibnamefont {Xie}},\ }\href {\doibase https://doi.org/10.1016/j.scib.2023.01.002} {\bibfield  {journal} {\bibinfo  {journal} {Science Bulletin}\ } (\bibinfo {year} {2023}{\natexlab{a}}),\ https://doi.org/10.1016/j.scib.2023.01.002}\BibitemShut {NoStop}%
    \bibitem [{\citenamefont {Chen}\ \emph {et~al.}(2019)\citenamefont {Chen}, \citenamefont {Chen}, \citenamefont {Zhou},\ and\ \citenamefont {Xu}}]{chen2019finite}%
      \BibitemOpen
      \bibfield  {author} {\bibinfo {author} {\bibfnamefont {R.}~\bibnamefont {Chen}}, \bibinfo {author} {\bibfnamefont {C.-Z.}\ \bibnamefont {Chen}}, \bibinfo {author} {\bibfnamefont {B.}~\bibnamefont {Zhou}}, \ and\ \bibinfo {author} {\bibfnamefont {D.-H.}\ \bibnamefont {Xu}},\ }\href {\doibase 10.1103/PhysRevB.99.155431} {\bibfield  {journal} {\bibinfo  {journal} {Phys. Rev. B}\ }\textbf {\bibinfo {volume} {99}},\ \bibinfo {pages} {155431} (\bibinfo {year} {2019})}\BibitemShut {NoStop}%
    \bibitem [{\citenamefont {Rafi-Ul-Islam}\ \emph {et~al.}(2022{\natexlab{a}})\citenamefont {Rafi-Ul-Islam}, \citenamefont {Siu}, \citenamefont {Sahin}, \citenamefont {Lee},\ and\ \citenamefont {Jalil}}]{rafi2022unconventional}%
      \BibitemOpen
      \bibfield  {author} {\bibinfo {author} {\bibfnamefont {S.}~\bibnamefont {Rafi-Ul-Islam}}, \bibinfo {author} {\bibfnamefont {Z.~B.}\ \bibnamefont {Siu}}, \bibinfo {author} {\bibfnamefont {H.}~\bibnamefont {Sahin}}, \bibinfo {author} {\bibfnamefont {C.~H.}\ \bibnamefont {Lee}}, \ and\ \bibinfo {author} {\bibfnamefont {M.~B.}\ \bibnamefont {Jalil}},\ }\href@noop {} {\bibfield  {journal} {\bibinfo  {journal} {Physical Review Research}\ }\textbf {\bibinfo {volume} {4}},\ \bibinfo {pages} {043108} (\bibinfo {year} {2022}{\natexlab{a}})}\BibitemShut {NoStop}%
    \bibitem [{\citenamefont {Wang}\ \emph {et~al.}(2022{\natexlab{a}})\citenamefont {Wang}, \citenamefont {Wang},\ and\ \citenamefont {Ma}}]{wang2022non1}%
      \BibitemOpen
      \bibfield  {author} {\bibinfo {author} {\bibfnamefont {W.}~\bibnamefont {Wang}}, \bibinfo {author} {\bibfnamefont {X.}~\bibnamefont {Wang}}, \ and\ \bibinfo {author} {\bibfnamefont {G.}~\bibnamefont {Ma}},\ }\href {\doibase 10.1038/s41586-022-04929-1} {\bibfield  {journal} {\bibinfo  {journal} {Nature}\ }\textbf {\bibinfo {volume} {608}},\ \bibinfo {pages} {50} (\bibinfo {year} {2022}{\natexlab{a}})}\BibitemShut {NoStop}%
    \bibitem [{\citenamefont {Wang}\ \emph {et~al.}(2022{\natexlab{b}})\citenamefont {Wang}, \citenamefont {Zhang}, \citenamefont {Yu}, \citenamefont {Ge}, \citenamefont {Liu}, \citenamefont {Wu}, \citenamefont {Sun}, \citenamefont {Liu}, \citenamefont {Chen}, \citenamefont {He} \emph {et~al.}}]{wang2022extended}%
      \BibitemOpen
      \bibfield  {author} {\bibinfo {author} {\bibfnamefont {J.-Q.}\ \bibnamefont {Wang}}, \bibinfo {author} {\bibfnamefont {Z.-D.}\ \bibnamefont {Zhang}}, \bibinfo {author} {\bibfnamefont {S.-Y.}\ \bibnamefont {Yu}}, \bibinfo {author} {\bibfnamefont {H.}~\bibnamefont {Ge}}, \bibinfo {author} {\bibfnamefont {K.-F.}\ \bibnamefont {Liu}}, \bibinfo {author} {\bibfnamefont {T.}~\bibnamefont {Wu}}, \bibinfo {author} {\bibfnamefont {X.-C.}\ \bibnamefont {Sun}}, \bibinfo {author} {\bibfnamefont {L.}~\bibnamefont {Liu}}, \bibinfo {author} {\bibfnamefont {H.-Y.}\ \bibnamefont {Chen}}, \bibinfo {author} {\bibfnamefont {C.}~\bibnamefont {He}},  \emph {et~al.},\ }\href@noop {} {\bibfield  {journal} {\bibinfo  {journal} {Nature communications}\ }\textbf {\bibinfo {volume} {13}},\ \bibinfo {pages} {1324} (\bibinfo {year} {2022}{\natexlab{b}})}\BibitemShut {NoStop}%
    \bibitem [{\citenamefont {Longhi}(2018{\natexlab{b}})}]{longhi2018non}%
      \BibitemOpen
      \bibfield  {author} {\bibinfo {author} {\bibfnamefont {S.}~\bibnamefont {Longhi}},\ }\href@noop {} {\bibfield  {journal} {\bibinfo  {journal} {Annalen der Physik}\ ,\ \bibinfo {pages} {1800023}} (\bibinfo {year} {2018}{\natexlab{b}})}\BibitemShut {NoStop}%
    \bibitem [{\citenamefont {Tang}\ \emph {et~al.}(2022{\natexlab{a}})\citenamefont {Tang}, \citenamefont {Wang}, \citenamefont {Valligatla}, \citenamefont {Saggau}, \citenamefont {Dong}, \citenamefont {Naz}, \citenamefont {Klembt}, \citenamefont {Lee}, \citenamefont {Thomale}, \citenamefont {Brink} \emph {et~al.}}]{tang2022symmetry}%
      \BibitemOpen
      \bibfield  {author} {\bibinfo {author} {\bibfnamefont {M.}~\bibnamefont {Tang}}, \bibinfo {author} {\bibfnamefont {J.}~\bibnamefont {Wang}}, \bibinfo {author} {\bibfnamefont {S.}~\bibnamefont {Valligatla}}, \bibinfo {author} {\bibfnamefont {C.~N.}\ \bibnamefont {Saggau}}, \bibinfo {author} {\bibfnamefont {H.}~\bibnamefont {Dong}}, \bibinfo {author} {\bibfnamefont {E.~S.~G.}\ \bibnamefont {Naz}}, \bibinfo {author} {\bibfnamefont {S.}~\bibnamefont {Klembt}}, \bibinfo {author} {\bibfnamefont {C.~H.}\ \bibnamefont {Lee}}, \bibinfo {author} {\bibfnamefont {R.}~\bibnamefont {Thomale}}, \bibinfo {author} {\bibfnamefont {J.~v.~d.}\ \bibnamefont {Brink}},  \emph {et~al.},\ }\href@noop {} {\bibfield  {journal} {\bibinfo  {journal} {arXiv preprint arXiv:2211.06228}\ } (\bibinfo {year} {2022}{\natexlab{a}})}\BibitemShut {NoStop}%
    \bibitem [{\citenamefont {Haldane}(1988)}]{haldane1988model}%
      \BibitemOpen
      \bibfield  {author} {\bibinfo {author} {\bibfnamefont {F.~D.~M.}\ \bibnamefont {Haldane}},\ }\href@noop {} {\bibfield  {journal} {\bibinfo  {journal} {Phys. Rev. Lett.}\ }\textbf {\bibinfo {volume} {61}},\ \bibinfo {pages} {2015} (\bibinfo {year} {1988})}\BibitemShut {NoStop}%
    \bibitem [{\citenamefont {Qi}\ \emph {et~al.}(2008)\citenamefont {Qi}, \citenamefont {Hughes},\ and\ \citenamefont {Zhang}}]{qi2008}%
      \BibitemOpen
      \bibfield  {author} {\bibinfo {author} {\bibfnamefont {X.-L.}\ \bibnamefont {Qi}}, \bibinfo {author} {\bibfnamefont {T.~L.}\ \bibnamefont {Hughes}}, \ and\ \bibinfo {author} {\bibfnamefont {S.-C.}\ \bibnamefont {Zhang}},\ }\href@noop {} {\bibfield  {journal} {\bibinfo  {journal} {Phys. Rev. B}\ }\textbf {\bibinfo {volume} {78}},\ \bibinfo {pages} {195424} (\bibinfo {year} {2008})}\BibitemShut {NoStop}%
    \bibitem [{\citenamefont {Chang}\ \emph {et~al.}(2013)\citenamefont {Chang}, \citenamefont {Zhang}, \citenamefont {Feng}, \citenamefont {Shen}, \citenamefont {Zhang}, \citenamefont {Guo}, \citenamefont {Li}, \citenamefont {Ou}, \citenamefont {Wei}, \citenamefont {Wang} \emph {et~al.}}]{chang2013experimental}%
      \BibitemOpen
      \bibfield  {author} {\bibinfo {author} {\bibfnamefont {C.-Z.}\ \bibnamefont {Chang}}, \bibinfo {author} {\bibfnamefont {J.}~\bibnamefont {Zhang}}, \bibinfo {author} {\bibfnamefont {X.}~\bibnamefont {Feng}}, \bibinfo {author} {\bibfnamefont {J.}~\bibnamefont {Shen}}, \bibinfo {author} {\bibfnamefont {Z.}~\bibnamefont {Zhang}}, \bibinfo {author} {\bibfnamefont {M.}~\bibnamefont {Guo}}, \bibinfo {author} {\bibfnamefont {K.}~\bibnamefont {Li}}, \bibinfo {author} {\bibfnamefont {Y.}~\bibnamefont {Ou}}, \bibinfo {author} {\bibfnamefont {P.}~\bibnamefont {Wei}}, \bibinfo {author} {\bibfnamefont {L.-L.}\ \bibnamefont {Wang}},  \emph {et~al.},\ }\href@noop {} {\bibfield  {journal} {\bibinfo  {journal} {Science}\ }\textbf {\bibinfo {volume} {340}},\ \bibinfo {pages} {167} (\bibinfo {year} {2013})}\BibitemShut {NoStop}%
    \bibitem [{\citenamefont {Lee}\ and\ \citenamefont {Qi}(2014)}]{lee2014lattice}%
      \BibitemOpen
      \bibfield  {author} {\bibinfo {author} {\bibfnamefont {C.~H.}\ \bibnamefont {Lee}}\ and\ \bibinfo {author} {\bibfnamefont {X.-L.}\ \bibnamefont {Qi}},\ }\href@noop {} {\bibfield  {journal} {\bibinfo  {journal} {Phys. Rev. B}\ }\textbf {\bibinfo {volume} {90}},\ \bibinfo {pages} {085103} (\bibinfo {year} {2014})}\BibitemShut {NoStop}%
    \bibitem [{\citenamefont {Neupert}\ \emph {et~al.}(2015)\citenamefont {Neupert}, \citenamefont {Chamon}, \citenamefont {Iadecola}, \citenamefont {Santos},\ and\ \citenamefont {Mudry}}]{neupert2015fractional}%
      \BibitemOpen
      \bibfield  {author} {\bibinfo {author} {\bibfnamefont {T.}~\bibnamefont {Neupert}}, \bibinfo {author} {\bibfnamefont {C.}~\bibnamefont {Chamon}}, \bibinfo {author} {\bibfnamefont {T.}~\bibnamefont {Iadecola}}, \bibinfo {author} {\bibfnamefont {L.~H.}\ \bibnamefont {Santos}}, \ and\ \bibinfo {author} {\bibfnamefont {C.}~\bibnamefont {Mudry}},\ }\href@noop {} {\bibfield  {journal} {\bibinfo  {journal} {Physica Scripta}\ }\textbf {\bibinfo {volume} {2015}},\ \bibinfo {pages} {014005} (\bibinfo {year} {2015})}\BibitemShut {NoStop}%
    \bibitem [{\citenamefont {Yao}\ \emph {et~al.}(2018)\citenamefont {Yao}, \citenamefont {Song},\ and\ \citenamefont {Wang}}]{Yao2018nonH2D}%
      \BibitemOpen
      \bibfield  {author} {\bibinfo {author} {\bibfnamefont {S.}~\bibnamefont {Yao}}, \bibinfo {author} {\bibfnamefont {F.}~\bibnamefont {Song}}, \ and\ \bibinfo {author} {\bibfnamefont {Z.}~\bibnamefont {Wang}},\ }\href {\doibase 10.1103/PhysRevLett.121.136802} {\bibfield  {journal} {\bibinfo  {journal} {Phys. Rev. Lett.}\ }\textbf {\bibinfo {volume} {121}},\ \bibinfo {pages} {136802} (\bibinfo {year} {2018})}\BibitemShut {NoStop}%
    \bibitem [{\citenamefont {Kawabata}\ \emph {et~al.}(2018)\citenamefont {Kawabata}, \citenamefont {Shiozaki},\ and\ \citenamefont {Ueda}}]{kawabata2018anomalous}%
      \BibitemOpen
      \bibfield  {author} {\bibinfo {author} {\bibfnamefont {K.}~\bibnamefont {Kawabata}}, \bibinfo {author} {\bibfnamefont {K.}~\bibnamefont {Shiozaki}}, \ and\ \bibinfo {author} {\bibfnamefont {M.}~\bibnamefont {Ueda}},\ }\href@noop {} {\bibfield  {journal} {\bibinfo  {journal} {Physical Review B}\ }\textbf {\bibinfo {volume} {98}},\ \bibinfo {pages} {165148} (\bibinfo {year} {2018})}\BibitemShut {NoStop}%
    \bibitem [{\citenamefont {Xiao}\ and\ \citenamefont {Chan}(2022)}]{xiao2022topology}%
      \BibitemOpen
      \bibfield  {author} {\bibinfo {author} {\bibfnamefont {Y.-X.}\ \bibnamefont {Xiao}}\ and\ \bibinfo {author} {\bibfnamefont {C.~T.}\ \bibnamefont {Chan}},\ }\href@noop {} {\bibfield  {journal} {\bibinfo  {journal} {Physical Review B}\ }\textbf {\bibinfo {volume} {105}},\ \bibinfo {pages} {075128} (\bibinfo {year} {2022})}\BibitemShut {NoStop}%
    \bibitem [{\citenamefont {Liu}\ \emph {et~al.}(2021{\natexlab{d}})\citenamefont {Liu}, \citenamefont {You}, \citenamefont {Ryu},\ and\ \citenamefont {Fulga}}]{liu2021supermetal}%
      \BibitemOpen
      \bibfield  {author} {\bibinfo {author} {\bibfnamefont {H.}~\bibnamefont {Liu}}, \bibinfo {author} {\bibfnamefont {J.-S.}\ \bibnamefont {You}}, \bibinfo {author} {\bibfnamefont {S.}~\bibnamefont {Ryu}}, \ and\ \bibinfo {author} {\bibfnamefont {I.~C.}\ \bibnamefont {Fulga}},\ }\href@noop {} {\bibfield  {journal} {\bibinfo  {journal} {Physical Review B}\ }\textbf {\bibinfo {volume} {104}},\ \bibinfo {pages} {155412} (\bibinfo {year} {2021}{\natexlab{d}})}\BibitemShut {NoStop}%
    \bibitem [{\citenamefont {Philip}\ \emph {et~al.}(2018)\citenamefont {Philip}, \citenamefont {Hirsbrunner},\ and\ \citenamefont {Gilbert}}]{philip2018loss}%
      \BibitemOpen
      \bibfield  {author} {\bibinfo {author} {\bibfnamefont {T.~M.}\ \bibnamefont {Philip}}, \bibinfo {author} {\bibfnamefont {M.~R.}\ \bibnamefont {Hirsbrunner}}, \ and\ \bibinfo {author} {\bibfnamefont {M.~J.}\ \bibnamefont {Gilbert}},\ }\href@noop {} {\bibfield  {journal} {\bibinfo  {journal} {Physical Review B}\ }\textbf {\bibinfo {volume} {98}},\ \bibinfo {pages} {155430} (\bibinfo {year} {2018})}\BibitemShut {NoStop}%
    \bibitem [{\citenamefont {Sayyad}\ \emph {et~al.}(2022)\citenamefont {Sayyad}, \citenamefont {Hannukainen},\ and\ \citenamefont {Grushin}}]{sayyad2022non}%
      \BibitemOpen
      \bibfield  {author} {\bibinfo {author} {\bibfnamefont {S.}~\bibnamefont {Sayyad}}, \bibinfo {author} {\bibfnamefont {J.~D.}\ \bibnamefont {Hannukainen}}, \ and\ \bibinfo {author} {\bibfnamefont {A.~G.}\ \bibnamefont {Grushin}},\ }\href@noop {} {\bibfield  {journal} {\bibinfo  {journal} {Physical Review Research}\ }\textbf {\bibinfo {volume} {4}},\ \bibinfo {pages} {L042004} (\bibinfo {year} {2022})}\BibitemShut {NoStop}%
    \bibitem [{\citenamefont {Bessho}\ and\ \citenamefont {Sato}(2021)}]{bessho2021nielsen}%
      \BibitemOpen
      \bibfield  {author} {\bibinfo {author} {\bibfnamefont {T.}~\bibnamefont {Bessho}}\ and\ \bibinfo {author} {\bibfnamefont {M.}~\bibnamefont {Sato}},\ }\href@noop {} {\bibfield  {journal} {\bibinfo  {journal} {Physical Review Letters}\ }\textbf {\bibinfo {volume} {127}},\ \bibinfo {pages} {196404} (\bibinfo {year} {2021})}\BibitemShut {NoStop}%
    \bibitem [{\citenamefont {Wang}\ and\ \citenamefont {Wang}(2022{\natexlab{a}})}]{wang2022hermitian}%
      \BibitemOpen
      \bibfield  {author} {\bibinfo {author} {\bibfnamefont {C.}~\bibnamefont {Wang}}\ and\ \bibinfo {author} {\bibfnamefont {X.}~\bibnamefont {Wang}},\ }\href@noop {} {\bibfield  {journal} {\bibinfo  {journal} {Physical Review B}\ }\textbf {\bibinfo {volume} {105}},\ \bibinfo {pages} {125103} (\bibinfo {year} {2022}{\natexlab{a}})}\BibitemShut {NoStop}%
    \bibitem [{\citenamefont {Lee}\ \emph {et~al.}(2018{\natexlab{a}})\citenamefont {Lee}, \citenamefont {Wang}, \citenamefont {Chen},\ and\ \citenamefont {Zhang}}]{lee2018electromagnetic}%
      \BibitemOpen
      \bibfield  {author} {\bibinfo {author} {\bibfnamefont {C.~H.}\ \bibnamefont {Lee}}, \bibinfo {author} {\bibfnamefont {Y.}~\bibnamefont {Wang}}, \bibinfo {author} {\bibfnamefont {Y.}~\bibnamefont {Chen}}, \ and\ \bibinfo {author} {\bibfnamefont {X.}~\bibnamefont {Zhang}},\ }\href@noop {} {\bibfield  {journal} {\bibinfo  {journal} {Physical Review B}\ }\textbf {\bibinfo {volume} {98}},\ \bibinfo {pages} {094434} (\bibinfo {year} {2018}{\natexlab{a}})}\BibitemShut {NoStop}%
    \bibitem [{\citenamefont {Petrides}\ \emph {et~al.}(2018)\citenamefont {Petrides}, \citenamefont {Price},\ and\ \citenamefont {Zilberberg}}]{petrides2018six}%
      \BibitemOpen
      \bibfield  {author} {\bibinfo {author} {\bibfnamefont {I.}~\bibnamefont {Petrides}}, \bibinfo {author} {\bibfnamefont {H.~M.}\ \bibnamefont {Price}}, \ and\ \bibinfo {author} {\bibfnamefont {O.}~\bibnamefont {Zilberberg}},\ }\href@noop {} {\bibfield  {journal} {\bibinfo  {journal} {Physical Review B}\ }\textbf {\bibinfo {volume} {98}},\ \bibinfo {pages} {125431} (\bibinfo {year} {2018})}\BibitemShut {NoStop}%
    \bibitem [{\citenamefont {Shao}\ \emph {et~al.}(2022)\citenamefont {Shao}, \citenamefont {Cai}, \citenamefont {Geng}, \citenamefont {Chen},\ and\ \citenamefont {Xing}}]{shao2022Cyclotron}%
      \BibitemOpen
      \bibfield  {author} {\bibinfo {author} {\bibfnamefont {K.}~\bibnamefont {Shao}}, \bibinfo {author} {\bibfnamefont {Z.-T.}\ \bibnamefont {Cai}}, \bibinfo {author} {\bibfnamefont {H.}~\bibnamefont {Geng}}, \bibinfo {author} {\bibfnamefont {W.}~\bibnamefont {Chen}}, \ and\ \bibinfo {author} {\bibfnamefont {D.~Y.}\ \bibnamefont {Xing}},\ }\href {\doibase 10.1103/PhysRevB.106.L081402} {\bibfield  {journal} {\bibinfo  {journal} {Phys. Rev. B}\ }\textbf {\bibinfo {volume} {106}},\ \bibinfo {pages} {L081402} (\bibinfo {year} {2022})}\BibitemShut {NoStop}%
    \bibitem [{\citenamefont {Deng}\ and\ \citenamefont {Flebus}(2022)}]{deng2022non}%
      \BibitemOpen
      \bibfield  {author} {\bibinfo {author} {\bibfnamefont {K.}~\bibnamefont {Deng}}\ and\ \bibinfo {author} {\bibfnamefont {B.}~\bibnamefont {Flebus}},\ }\href@noop {} {\bibfield  {journal} {\bibinfo  {journal} {Physical Review B}\ }\textbf {\bibinfo {volume} {105}},\ \bibinfo {pages} {L180406} (\bibinfo {year} {2022})}\BibitemShut {NoStop}%
    \bibitem [{\citenamefont {Denner}\ and\ \citenamefont {Schindler}(2022)}]{denner2022magnetic}%
      \BibitemOpen
      \bibfield  {author} {\bibinfo {author} {\bibfnamefont {M.~M.}\ \bibnamefont {Denner}}\ and\ \bibinfo {author} {\bibfnamefont {F.}~\bibnamefont {Schindler}},\ }\href@noop {} {\bibfield  {journal} {\bibinfo  {journal} {arXiv preprint arXiv:2208.11712}\ } (\bibinfo {year} {2022})}\BibitemShut {NoStop}%
    \bibitem [{\citenamefont {Lu}\ \emph {et~al.}(2021)\citenamefont {Lu}, \citenamefont {Zhang},\ and\ \citenamefont {Franz}}]{lu2021magnetic}%
      \BibitemOpen
      \bibfield  {author} {\bibinfo {author} {\bibfnamefont {M.}~\bibnamefont {Lu}}, \bibinfo {author} {\bibfnamefont {X.-X.}\ \bibnamefont {Zhang}}, \ and\ \bibinfo {author} {\bibfnamefont {M.}~\bibnamefont {Franz}},\ }\href@noop {} {\bibfield  {journal} {\bibinfo  {journal} {Physical review letters}\ }\textbf {\bibinfo {volume} {127}},\ \bibinfo {pages} {256402} (\bibinfo {year} {2021})}\BibitemShut {NoStop}%
    \bibitem [{\citenamefont {Yang}\ and\ \citenamefont {Lee}(2023)}]{yang2023percolation}%
      \BibitemOpen
      \bibfield  {author} {\bibinfo {author} {\bibfnamefont {M.}~\bibnamefont {Yang}}\ and\ \bibinfo {author} {\bibfnamefont {C.~H.}\ \bibnamefont {Lee}},\ }\href@noop {} {\bibfield  {journal} {\bibinfo  {journal} {arXiv preprint arXiv:2309.15008}\ } (\bibinfo {year} {2023})}\BibitemShut {NoStop}%
    \bibitem [{\citenamefont {Zhen}\ \emph {et~al.}(2015)\citenamefont {Zhen}, \citenamefont {Hsu}, \citenamefont {Igarashi}, \citenamefont {Lu}, \citenamefont {Kaminer}, \citenamefont {Pick}, \citenamefont {Chua}, \citenamefont {Joannopoulos},\ and\ \citenamefont {Solja{\v{c}}i{\'c}}}]{zhen2015spawning}%
      \BibitemOpen
      \bibfield  {author} {\bibinfo {author} {\bibfnamefont {B.}~\bibnamefont {Zhen}}, \bibinfo {author} {\bibfnamefont {C.~W.}\ \bibnamefont {Hsu}}, \bibinfo {author} {\bibfnamefont {Y.}~\bibnamefont {Igarashi}}, \bibinfo {author} {\bibfnamefont {L.}~\bibnamefont {Lu}}, \bibinfo {author} {\bibfnamefont {I.}~\bibnamefont {Kaminer}}, \bibinfo {author} {\bibfnamefont {A.}~\bibnamefont {Pick}}, \bibinfo {author} {\bibfnamefont {S.-L.}\ \bibnamefont {Chua}}, \bibinfo {author} {\bibfnamefont {J.~D.}\ \bibnamefont {Joannopoulos}}, \ and\ \bibinfo {author} {\bibfnamefont {M.}~\bibnamefont {Solja{\v{c}}i{\'c}}},\ }\href@noop {} {\bibfield  {journal} {\bibinfo  {journal} {Nature}\ }\textbf {\bibinfo {volume} {525}},\ \bibinfo {pages} {354} (\bibinfo {year} {2015})}\BibitemShut {NoStop}%
    \bibitem [{\citenamefont {Xu}\ \emph {et~al.}(2017)\citenamefont {Xu}, \citenamefont {Wang},\ and\ \citenamefont {Duan}}]{xu2017weyl}%
      \BibitemOpen
      \bibfield  {author} {\bibinfo {author} {\bibfnamefont {Y.}~\bibnamefont {Xu}}, \bibinfo {author} {\bibfnamefont {S.-T.}\ \bibnamefont {Wang}}, \ and\ \bibinfo {author} {\bibfnamefont {L.-M.}\ \bibnamefont {Duan}},\ }\href {\doibase 10.1103/PhysRevLett.118.045701} {\bibfield  {journal} {\bibinfo  {journal} {Phys. Rev. Lett.}\ }\textbf {\bibinfo {volume} {118}},\ \bibinfo {pages} {045701} (\bibinfo {year} {2017})}\BibitemShut {NoStop}%
    \bibitem [{\citenamefont {Liu}\ \emph {et~al.}(2022{\natexlab{b}})\citenamefont {Liu}, \citenamefont {Li}, \citenamefont {Chen}, \citenamefont {Tang}, \citenamefont {Chen}, \citenamefont {Liang}, \citenamefont {Ma},\ and\ \citenamefont {Cheng}}]{liu2022experimental}%
      \BibitemOpen
      \bibfield  {author} {\bibinfo {author} {\bibfnamefont {J.-j.}\ \bibnamefont {Liu}}, \bibinfo {author} {\bibfnamefont {Z.-w.}\ \bibnamefont {Li}}, \bibinfo {author} {\bibfnamefont {Z.-G.}\ \bibnamefont {Chen}}, \bibinfo {author} {\bibfnamefont {W.}~\bibnamefont {Tang}}, \bibinfo {author} {\bibfnamefont {A.}~\bibnamefont {Chen}}, \bibinfo {author} {\bibfnamefont {B.}~\bibnamefont {Liang}}, \bibinfo {author} {\bibfnamefont {G.}~\bibnamefont {Ma}}, \ and\ \bibinfo {author} {\bibfnamefont {J.-C.}\ \bibnamefont {Cheng}},\ }\href {\doibase 10.1103/PhysRevLett.129.084301} {\bibfield  {journal} {\bibinfo  {journal} {Phys. Rev. Lett.}\ }\textbf {\bibinfo {volume} {129}},\ \bibinfo {pages} {084301} (\bibinfo {year} {2022}{\natexlab{b}})}\BibitemShut {NoStop}%
    \bibitem [{\citenamefont {Cerjan}\ \emph {et~al.}(2019)\citenamefont {Cerjan}, \citenamefont {Huang}, \citenamefont {Wang}, \citenamefont {Chen}, \citenamefont {Chong},\ and\ \citenamefont {Rechtsman}}]{cerjan2019experimental}%
      \BibitemOpen
      \bibfield  {author} {\bibinfo {author} {\bibfnamefont {A.}~\bibnamefont {Cerjan}}, \bibinfo {author} {\bibfnamefont {S.}~\bibnamefont {Huang}}, \bibinfo {author} {\bibfnamefont {M.}~\bibnamefont {Wang}}, \bibinfo {author} {\bibfnamefont {K.~P.}\ \bibnamefont {Chen}}, \bibinfo {author} {\bibfnamefont {Y.}~\bibnamefont {Chong}}, \ and\ \bibinfo {author} {\bibfnamefont {M.~C.}\ \bibnamefont {Rechtsman}},\ }\href@noop {} {\bibfield  {journal} {\bibinfo  {journal} {Nature Photonics}\ }\textbf {\bibinfo {volume} {13}},\ \bibinfo {pages} {623} (\bibinfo {year} {2019})}\BibitemShut {NoStop}%
    \bibitem [{\citenamefont {Tang}\ \emph {et~al.}(2022{\natexlab{b}})\citenamefont {Tang}, \citenamefont {Ding},\ and\ \citenamefont {Ma}}]{tang2022realization}%
      \BibitemOpen
      \bibfield  {author} {\bibinfo {author} {\bibfnamefont {W.}~\bibnamefont {Tang}}, \bibinfo {author} {\bibfnamefont {K.}~\bibnamefont {Ding}}, \ and\ \bibinfo {author} {\bibfnamefont {G.}~\bibnamefont {Ma}},\ }\href@noop {} {\bibfield  {journal} {\bibinfo  {journal} {National Science Review}\ }\textbf {\bibinfo {volume} {9}},\ \bibinfo {pages} {nwac010} (\bibinfo {year} {2022}{\natexlab{b}})}\BibitemShut {NoStop}%
    \bibitem [{\citenamefont {Bzdu{\v{s}}ek}\ \emph {et~al.}(2016)\citenamefont {Bzdu{\v{s}}ek}, \citenamefont {Wu}, \citenamefont {R{\"u}egg}, \citenamefont {Sigrist},\ and\ \citenamefont {Soluyanov}}]{bzduvsek2016nodal}%
      \BibitemOpen
      \bibfield  {author} {\bibinfo {author} {\bibfnamefont {T.}~\bibnamefont {Bzdu{\v{s}}ek}}, \bibinfo {author} {\bibfnamefont {Q.}~\bibnamefont {Wu}}, \bibinfo {author} {\bibfnamefont {A.}~\bibnamefont {R{\"u}egg}}, \bibinfo {author} {\bibfnamefont {M.}~\bibnamefont {Sigrist}}, \ and\ \bibinfo {author} {\bibfnamefont {A.~A.}\ \bibnamefont {Soluyanov}},\ }\href@noop {} {\bibfield  {journal} {\bibinfo  {journal} {Nature}\ }\textbf {\bibinfo {volume} {538}},\ \bibinfo {pages} {75} (\bibinfo {year} {2016})}\BibitemShut {NoStop}%
    \bibitem [{\citenamefont {Huh}\ \emph {et~al.}(2016)\citenamefont {Huh}, \citenamefont {Moon},\ and\ \citenamefont {Kim}}]{huh2016long}%
      \BibitemOpen
      \bibfield  {author} {\bibinfo {author} {\bibfnamefont {Y.}~\bibnamefont {Huh}}, \bibinfo {author} {\bibfnamefont {E.-G.}\ \bibnamefont {Moon}}, \ and\ \bibinfo {author} {\bibfnamefont {Y.~B.}\ \bibnamefont {Kim}},\ }\href@noop {} {\bibfield  {journal} {\bibinfo  {journal} {Physical Review B}\ }\textbf {\bibinfo {volume} {93}},\ \bibinfo {pages} {035138} (\bibinfo {year} {2016})}\BibitemShut {NoStop}%
    \bibitem [{\citenamefont {Li}\ and\ \citenamefont {Ara\'ujo}(2016)}]{li2016topo}%
      \BibitemOpen
      \bibfield  {author} {\bibinfo {author} {\bibfnamefont {L.}~\bibnamefont {Li}}\ and\ \bibinfo {author} {\bibfnamefont {M.~A.~N.}\ \bibnamefont {Ara\'ujo}},\ }\href {\doibase 10.1103/PhysRevB.94.165117} {\bibfield  {journal} {\bibinfo  {journal} {Phys. Rev. B}\ }\textbf {\bibinfo {volume} {94}},\ \bibinfo {pages} {165117} (\bibinfo {year} {2016})}\BibitemShut {NoStop}%
    \bibitem [{\citenamefont {Li}\ \emph {et~al.}(2017{\natexlab{a}})\citenamefont {Li}, \citenamefont {Yin}, \citenamefont {Chen},\ and\ \citenamefont {Ara\'ujo}}]{li2017chiral}%
      \BibitemOpen
      \bibfield  {author} {\bibinfo {author} {\bibfnamefont {L.}~\bibnamefont {Li}}, \bibinfo {author} {\bibfnamefont {C.}~\bibnamefont {Yin}}, \bibinfo {author} {\bibfnamefont {S.}~\bibnamefont {Chen}}, \ and\ \bibinfo {author} {\bibfnamefont {M.~A.~N.}\ \bibnamefont {Ara\'ujo}},\ }\href {\doibase 10.1103/PhysRevB.95.121107} {\bibfield  {journal} {\bibinfo  {journal} {Phys. Rev. B}\ }\textbf {\bibinfo {volume} {95}},\ \bibinfo {pages} {121107} (\bibinfo {year} {2017}{\natexlab{a}})}\BibitemShut {NoStop}%
    \bibitem [{\citenamefont {Li}\ \emph {et~al.}(2017{\natexlab{b}})\citenamefont {Li}, \citenamefont {Chesi}, \citenamefont {Yin},\ and\ \citenamefont {Chen}}]{li20172pi}%
      \BibitemOpen
      \bibfield  {author} {\bibinfo {author} {\bibfnamefont {L.}~\bibnamefont {Li}}, \bibinfo {author} {\bibfnamefont {S.}~\bibnamefont {Chesi}}, \bibinfo {author} {\bibfnamefont {C.}~\bibnamefont {Yin}}, \ and\ \bibinfo {author} {\bibfnamefont {S.}~\bibnamefont {Chen}},\ }\href {\doibase 10.1103/PhysRevB.96.081116} {\bibfield  {journal} {\bibinfo  {journal} {Phys. Rev. B}\ }\textbf {\bibinfo {volume} {96}},\ \bibinfo {pages} {081116} (\bibinfo {year} {2017}{\natexlab{b}})}\BibitemShut {NoStop}%
    \bibitem [{\citenamefont {Yan}\ \emph {et~al.}(2017)\citenamefont {Yan}, \citenamefont {Bi}, \citenamefont {Shen}, \citenamefont {Lu}, \citenamefont {Zhang},\ and\ \citenamefont {Wang}}]{yan2017nodal}%
      \BibitemOpen
      \bibfield  {author} {\bibinfo {author} {\bibfnamefont {Z.}~\bibnamefont {Yan}}, \bibinfo {author} {\bibfnamefont {R.}~\bibnamefont {Bi}}, \bibinfo {author} {\bibfnamefont {H.}~\bibnamefont {Shen}}, \bibinfo {author} {\bibfnamefont {L.}~\bibnamefont {Lu}}, \bibinfo {author} {\bibfnamefont {S.-C.}\ \bibnamefont {Zhang}}, \ and\ \bibinfo {author} {\bibfnamefont {Z.}~\bibnamefont {Wang}},\ }\href@noop {} {\bibfield  {journal} {\bibinfo  {journal} {Physical Review B}\ }\textbf {\bibinfo {volume} {96}},\ \bibinfo {pages} {041103} (\bibinfo {year} {2017})}\BibitemShut {NoStop}%
    \bibitem [{\citenamefont {Li}\ \emph {et~al.}(2017{\natexlab{c}})\citenamefont {Li}, \citenamefont {Yu}, \citenamefont {Liu}, \citenamefont {Guan}, \citenamefont {Wang}, \citenamefont {Zhang}, \citenamefont {Yao},\ and\ \citenamefont {Yang}}]{li2017type}%
      \BibitemOpen
      \bibfield  {author} {\bibinfo {author} {\bibfnamefont {S.}~\bibnamefont {Li}}, \bibinfo {author} {\bibfnamefont {Z.-M.}\ \bibnamefont {Yu}}, \bibinfo {author} {\bibfnamefont {Y.}~\bibnamefont {Liu}}, \bibinfo {author} {\bibfnamefont {S.}~\bibnamefont {Guan}}, \bibinfo {author} {\bibfnamefont {S.-S.}\ \bibnamefont {Wang}}, \bibinfo {author} {\bibfnamefont {X.}~\bibnamefont {Zhang}}, \bibinfo {author} {\bibfnamefont {Y.}~\bibnamefont {Yao}}, \ and\ \bibinfo {author} {\bibfnamefont {S.~A.}\ \bibnamefont {Yang}},\ }\href@noop {} {\bibfield  {journal} {\bibinfo  {journal} {Physical Review B}\ }\textbf {\bibinfo {volume} {96}},\ \bibinfo {pages} {081106} (\bibinfo {year} {2017}{\natexlab{c}})}\BibitemShut {NoStop}%
    \bibitem [{\citenamefont {Li}\ \emph {et~al.}(2017{\natexlab{d}})\citenamefont {Li}, \citenamefont {Yap}, \citenamefont {Ara{\'u}jo},\ and\ \citenamefont {Gong}}]{li2017engineering}%
      \BibitemOpen
      \bibfield  {author} {\bibinfo {author} {\bibfnamefont {L.}~\bibnamefont {Li}}, \bibinfo {author} {\bibfnamefont {H.~H.}\ \bibnamefont {Yap}}, \bibinfo {author} {\bibfnamefont {M.~A.}\ \bibnamefont {Ara{\'u}jo}}, \ and\ \bibinfo {author} {\bibfnamefont {J.}~\bibnamefont {Gong}},\ }\href@noop {} {\bibfield  {journal} {\bibinfo  {journal} {Physical Review B}\ }\textbf {\bibinfo {volume} {96}},\ \bibinfo {pages} {235424} (\bibinfo {year} {2017}{\natexlab{d}})}\BibitemShut {NoStop}%
    \bibitem [{\citenamefont {Zhou}\ \emph {et~al.}(2018)\citenamefont {Zhou}, \citenamefont {Xiong}, \citenamefont {Wan},\ and\ \citenamefont {An}}]{zhou2018hopf}%
      \BibitemOpen
      \bibfield  {author} {\bibinfo {author} {\bibfnamefont {Y.}~\bibnamefont {Zhou}}, \bibinfo {author} {\bibfnamefont {F.}~\bibnamefont {Xiong}}, \bibinfo {author} {\bibfnamefont {X.}~\bibnamefont {Wan}}, \ and\ \bibinfo {author} {\bibfnamefont {J.}~\bibnamefont {An}},\ }\href@noop {} {\bibfield  {journal} {\bibinfo  {journal} {Physical Review B}\ }\textbf {\bibinfo {volume} {97}},\ \bibinfo {pages} {155140} (\bibinfo {year} {2018})}\BibitemShut {NoStop}%
    \bibitem [{\citenamefont {Pezzini}\ \emph {et~al.}(2018)\citenamefont {Pezzini}, \citenamefont {Van~Delft}, \citenamefont {Schoop}, \citenamefont {Lotsch}, \citenamefont {Carrington}, \citenamefont {Katsnelson}, \citenamefont {Hussey},\ and\ \citenamefont {Wiedmann}}]{pezzini2018unconventional}%
      \BibitemOpen
      \bibfield  {author} {\bibinfo {author} {\bibfnamefont {S.}~\bibnamefont {Pezzini}}, \bibinfo {author} {\bibfnamefont {M.}~\bibnamefont {Van~Delft}}, \bibinfo {author} {\bibfnamefont {L.}~\bibnamefont {Schoop}}, \bibinfo {author} {\bibfnamefont {B.}~\bibnamefont {Lotsch}}, \bibinfo {author} {\bibfnamefont {A.}~\bibnamefont {Carrington}}, \bibinfo {author} {\bibfnamefont {M.}~\bibnamefont {Katsnelson}}, \bibinfo {author} {\bibfnamefont {N.~E.}\ \bibnamefont {Hussey}}, \ and\ \bibinfo {author} {\bibfnamefont {S.}~\bibnamefont {Wiedmann}},\ }\href@noop {} {\bibfield  {journal} {\bibinfo  {journal} {Nature Physics}\ }\textbf {\bibinfo {volume} {14}},\ \bibinfo {pages} {178} (\bibinfo {year} {2018})}\BibitemShut {NoStop}%
    \bibitem [{\citenamefont {Zhang}\ \emph {et~al.}(2018)\citenamefont {Zhang}, \citenamefont {Yu}, \citenamefont {Zhu}, \citenamefont {Wu}, \citenamefont {Wang}, \citenamefont {Sheng},\ and\ \citenamefont {Yang}}]{zhang2018nodal}%
      \BibitemOpen
      \bibfield  {author} {\bibinfo {author} {\bibfnamefont {X.}~\bibnamefont {Zhang}}, \bibinfo {author} {\bibfnamefont {Z.-M.}\ \bibnamefont {Yu}}, \bibinfo {author} {\bibfnamefont {Z.}~\bibnamefont {Zhu}}, \bibinfo {author} {\bibfnamefont {W.}~\bibnamefont {Wu}}, \bibinfo {author} {\bibfnamefont {S.-S.}\ \bibnamefont {Wang}}, \bibinfo {author} {\bibfnamefont {X.-L.}\ \bibnamefont {Sheng}}, \ and\ \bibinfo {author} {\bibfnamefont {S.~A.}\ \bibnamefont {Yang}},\ }\href@noop {} {\bibfield  {journal} {\bibinfo  {journal} {Physical Review B}\ }\textbf {\bibinfo {volume} {97}},\ \bibinfo {pages} {235150} (\bibinfo {year} {2018})}\BibitemShut {NoStop}%
    \bibitem [{\citenamefont {Lee}\ \emph {et~al.}(2018{\natexlab{b}})\citenamefont {Lee}, \citenamefont {Ho}, \citenamefont {Yang}, \citenamefont {Gong},\ and\ \citenamefont {Papi{\'c}}}]{lee2018floquet}%
      \BibitemOpen
      \bibfield  {author} {\bibinfo {author} {\bibfnamefont {C.~H.}\ \bibnamefont {Lee}}, \bibinfo {author} {\bibfnamefont {W.~W.}\ \bibnamefont {Ho}}, \bibinfo {author} {\bibfnamefont {B.}~\bibnamefont {Yang}}, \bibinfo {author} {\bibfnamefont {J.}~\bibnamefont {Gong}}, \ and\ \bibinfo {author} {\bibfnamefont {Z.}~\bibnamefont {Papi{\'c}}},\ }\href@noop {} {\bibfield  {journal} {\bibinfo  {journal} {Physical review letters}\ }\textbf {\bibinfo {volume} {121}},\ \bibinfo {pages} {237401} (\bibinfo {year} {2018}{\natexlab{b}})}\BibitemShut {NoStop}%
    \bibitem [{\citenamefont {{\"U}nal}\ \emph {et~al.}(2020)\citenamefont {{\"U}nal}, \citenamefont {Bouhon},\ and\ \citenamefont {Slager}}]{unal2020topological}%
      \BibitemOpen
      \bibfield  {author} {\bibinfo {author} {\bibfnamefont {F.~N.}\ \bibnamefont {{\"U}nal}}, \bibinfo {author} {\bibfnamefont {A.}~\bibnamefont {Bouhon}}, \ and\ \bibinfo {author} {\bibfnamefont {R.-J.}\ \bibnamefont {Slager}},\ }\href@noop {} {\bibfield  {journal} {\bibinfo  {journal} {Physical review letters}\ }\textbf {\bibinfo {volume} {125}},\ \bibinfo {pages} {053601} (\bibinfo {year} {2020})}\BibitemShut {NoStop}%
    \bibitem [{\citenamefont {Bouhon}\ \emph {et~al.}(2020)\citenamefont {Bouhon}, \citenamefont {Wu}, \citenamefont {Slager}, \citenamefont {Weng}, \citenamefont {Yazyev},\ and\ \citenamefont {Bzdu{\v{s}}ek}}]{bouhon2020non}%
      \BibitemOpen
      \bibfield  {author} {\bibinfo {author} {\bibfnamefont {A.}~\bibnamefont {Bouhon}}, \bibinfo {author} {\bibfnamefont {Q.}~\bibnamefont {Wu}}, \bibinfo {author} {\bibfnamefont {R.-J.}\ \bibnamefont {Slager}}, \bibinfo {author} {\bibfnamefont {H.}~\bibnamefont {Weng}}, \bibinfo {author} {\bibfnamefont {O.~V.}\ \bibnamefont {Yazyev}}, \ and\ \bibinfo {author} {\bibfnamefont {T.}~\bibnamefont {Bzdu{\v{s}}ek}},\ }\href@noop {} {\bibfield  {journal} {\bibinfo  {journal} {Nature Physics}\ }\textbf {\bibinfo {volume} {16}},\ \bibinfo {pages} {1137} (\bibinfo {year} {2020})}\BibitemShut {NoStop}%
    \bibitem [{\citenamefont {Lee}\ \emph {et~al.}(2020{\natexlab{b}})\citenamefont {Lee}, \citenamefont {Yap}, \citenamefont {Tai}, \citenamefont {Xu}, \citenamefont {Zhang},\ and\ \citenamefont {Gong}}]{lee2020enhanced}%
      \BibitemOpen
      \bibfield  {author} {\bibinfo {author} {\bibfnamefont {C.~H.}\ \bibnamefont {Lee}}, \bibinfo {author} {\bibfnamefont {H.~H.}\ \bibnamefont {Yap}}, \bibinfo {author} {\bibfnamefont {T.}~\bibnamefont {Tai}}, \bibinfo {author} {\bibfnamefont {G.}~\bibnamefont {Xu}}, \bibinfo {author} {\bibfnamefont {X.}~\bibnamefont {Zhang}}, \ and\ \bibinfo {author} {\bibfnamefont {J.}~\bibnamefont {Gong}},\ }\href@noop {} {\bibfield  {journal} {\bibinfo  {journal} {Physical Review B}\ }\textbf {\bibinfo {volume} {102}},\ \bibinfo {pages} {035138} (\bibinfo {year} {2020}{\natexlab{b}})}\BibitemShut {NoStop}%
    \bibitem [{\citenamefont {Ang}\ \emph {et~al.}(2020)\citenamefont {Ang}, \citenamefont {Lee},\ and\ \citenamefont {Ang}}]{ang2020universal}%
      \BibitemOpen
      \bibfield  {author} {\bibinfo {author} {\bibfnamefont {Y.~S.}\ \bibnamefont {Ang}}, \bibinfo {author} {\bibfnamefont {C.~H.}\ \bibnamefont {Lee}}, \ and\ \bibinfo {author} {\bibfnamefont {L.}~\bibnamefont {Ang}},\ }\href@noop {} {\bibfield  {journal} {\bibinfo  {journal} {arXiv preprint arXiv:2003.14004}\ } (\bibinfo {year} {2020})}\BibitemShut {NoStop}%
    \bibitem [{\citenamefont {Yang}\ \emph {et~al.}(2020{\natexlab{b}})\citenamefont {Yang}, \citenamefont {Yang}, \citenamefont {You}, \citenamefont {Chan}, \citenamefont {Mao}, \citenamefont {Guo}, \citenamefont {Ma}, \citenamefont {Xia}, \citenamefont {Fan}, \citenamefont {Xiang} \emph {et~al.}}]{yang2020observation}%
      \BibitemOpen
      \bibfield  {author} {\bibinfo {author} {\bibfnamefont {E.}~\bibnamefont {Yang}}, \bibinfo {author} {\bibfnamefont {B.}~\bibnamefont {Yang}}, \bibinfo {author} {\bibfnamefont {O.}~\bibnamefont {You}}, \bibinfo {author} {\bibfnamefont {H.-C.}\ \bibnamefont {Chan}}, \bibinfo {author} {\bibfnamefont {P.}~\bibnamefont {Mao}}, \bibinfo {author} {\bibfnamefont {Q.}~\bibnamefont {Guo}}, \bibinfo {author} {\bibfnamefont {S.}~\bibnamefont {Ma}}, \bibinfo {author} {\bibfnamefont {L.}~\bibnamefont {Xia}}, \bibinfo {author} {\bibfnamefont {D.}~\bibnamefont {Fan}}, \bibinfo {author} {\bibfnamefont {Y.}~\bibnamefont {Xiang}},  \emph {et~al.},\ }\href@noop {} {\bibfield  {journal} {\bibinfo  {journal} {Physical Review Letters}\ }\textbf {\bibinfo {volume} {125}},\ \bibinfo {pages} {033901} (\bibinfo {year} {2020}{\natexlab{b}})}\BibitemShut {NoStop}%
    \bibitem [{\citenamefont {Lee}\ \emph {et~al.}(2020{\natexlab{c}})\citenamefont {Lee}, \citenamefont {Sutrisno}, \citenamefont {Hofmann}, \citenamefont {Helbig}, \citenamefont {Liu}, \citenamefont {Ang}, \citenamefont {Ang}, \citenamefont {Zhang}, \citenamefont {Greiter},\ and\ \citenamefont {Thomale}}]{lee2020imaging}%
      \BibitemOpen
      \bibfield  {author} {\bibinfo {author} {\bibfnamefont {C.~H.}\ \bibnamefont {Lee}}, \bibinfo {author} {\bibfnamefont {A.}~\bibnamefont {Sutrisno}}, \bibinfo {author} {\bibfnamefont {T.}~\bibnamefont {Hofmann}}, \bibinfo {author} {\bibfnamefont {T.}~\bibnamefont {Helbig}}, \bibinfo {author} {\bibfnamefont {Y.}~\bibnamefont {Liu}}, \bibinfo {author} {\bibfnamefont {Y.~S.}\ \bibnamefont {Ang}}, \bibinfo {author} {\bibfnamefont {L.~K.}\ \bibnamefont {Ang}}, \bibinfo {author} {\bibfnamefont {X.}~\bibnamefont {Zhang}}, \bibinfo {author} {\bibfnamefont {M.}~\bibnamefont {Greiter}}, \ and\ \bibinfo {author} {\bibfnamefont {R.}~\bibnamefont {Thomale}},\ }\href@noop {} {\bibfield  {journal} {\bibinfo  {journal} {Nature communications}\ }\textbf {\bibinfo {volume} {11}},\ \bibinfo {pages} {4385} (\bibinfo {year} {2020}{\natexlab{c}})}\BibitemShut {NoStop}%
    \bibitem [{\citenamefont {Tai}\ and\ \citenamefont {Lee}(2021)}]{tai2021anisotropic}%
      \BibitemOpen
      \bibfield  {author} {\bibinfo {author} {\bibfnamefont {T.}~\bibnamefont {Tai}}\ and\ \bibinfo {author} {\bibfnamefont {C.~H.}\ \bibnamefont {Lee}},\ }\href@noop {} {\bibfield  {journal} {\bibinfo  {journal} {Physical Review B}\ }\textbf {\bibinfo {volume} {103}},\ \bibinfo {pages} {195125} (\bibinfo {year} {2021})}\BibitemShut {NoStop}%
    \bibitem [{\citenamefont {Lenggenhager}\ \emph {et~al.}(2021)\citenamefont {Lenggenhager}, \citenamefont {Liu}, \citenamefont {Tsirkin}, \citenamefont {Neupert},\ and\ \citenamefont {Bzdu{\v{s}}ek}}]{lenggenhager2021triple}%
      \BibitemOpen
      \bibfield  {author} {\bibinfo {author} {\bibfnamefont {P.~M.}\ \bibnamefont {Lenggenhager}}, \bibinfo {author} {\bibfnamefont {X.}~\bibnamefont {Liu}}, \bibinfo {author} {\bibfnamefont {S.~S.}\ \bibnamefont {Tsirkin}}, \bibinfo {author} {\bibfnamefont {T.}~\bibnamefont {Neupert}}, \ and\ \bibinfo {author} {\bibfnamefont {T.}~\bibnamefont {Bzdu{\v{s}}ek}},\ }\href@noop {} {\bibfield  {journal} {\bibinfo  {journal} {Physical Review B}\ }\textbf {\bibinfo {volume} {103}},\ \bibinfo {pages} {L121101} (\bibinfo {year} {2021})}\BibitemShut {NoStop}%
    \bibitem [{\citenamefont {Wang}\ \emph {et~al.}(2022{\natexlab{c}})\citenamefont {Wang}, \citenamefont {Liu}, \citenamefont {Ma}, \citenamefont {Zhang}, \citenamefont {Wang}, \citenamefont {Guo}, \citenamefont {Yang}, \citenamefont {Ke}, \citenamefont {Liu},\ and\ \citenamefont {Chan}}]{wang2022experimentala}%
      \BibitemOpen
      \bibfield  {author} {\bibinfo {author} {\bibfnamefont {M.}~\bibnamefont {Wang}}, \bibinfo {author} {\bibfnamefont {S.}~\bibnamefont {Liu}}, \bibinfo {author} {\bibfnamefont {Q.}~\bibnamefont {Ma}}, \bibinfo {author} {\bibfnamefont {R.-Y.}\ \bibnamefont {Zhang}}, \bibinfo {author} {\bibfnamefont {D.}~\bibnamefont {Wang}}, \bibinfo {author} {\bibfnamefont {Q.}~\bibnamefont {Guo}}, \bibinfo {author} {\bibfnamefont {B.}~\bibnamefont {Yang}}, \bibinfo {author} {\bibfnamefont {M.}~\bibnamefont {Ke}}, \bibinfo {author} {\bibfnamefont {Z.}~\bibnamefont {Liu}}, \ and\ \bibinfo {author} {\bibfnamefont {C.}~\bibnamefont {Chan}},\ }\href@noop {} {\bibfield  {journal} {\bibinfo  {journal} {Physical Review Letters}\ }\textbf {\bibinfo {volume} {128}},\ \bibinfo {pages} {246601} (\bibinfo {year} {2022}{\natexlab{c}})}\BibitemShut {NoStop}%
    \bibitem [{\citenamefont {Slager}\ \emph {et~al.}(2022)\citenamefont {Slager}, \citenamefont {Bouhon},\ and\ \citenamefont {{\"U}nal}}]{slager2022floquet}%
      \BibitemOpen
      \bibfield  {author} {\bibinfo {author} {\bibfnamefont {R.-J.}\ \bibnamefont {Slager}}, \bibinfo {author} {\bibfnamefont {A.}~\bibnamefont {Bouhon}}, \ and\ \bibinfo {author} {\bibfnamefont {F.~N.}\ \bibnamefont {{\"U}nal}},\ }\href@noop {} {\bibfield  {journal} {\bibinfo  {journal} {arXiv preprint arXiv:2208.12824}\ } (\bibinfo {year} {2022})}\BibitemShut {NoStop}%
    \bibitem [{\citenamefont {Peng}\ \emph {et~al.}(2022{\natexlab{a}})\citenamefont {Peng}, \citenamefont {Bouhon}, \citenamefont {Monserrat},\ and\ \citenamefont {Slager}}]{peng2022phonons}%
      \BibitemOpen
      \bibfield  {author} {\bibinfo {author} {\bibfnamefont {B.}~\bibnamefont {Peng}}, \bibinfo {author} {\bibfnamefont {A.}~\bibnamefont {Bouhon}}, \bibinfo {author} {\bibfnamefont {B.}~\bibnamefont {Monserrat}}, \ and\ \bibinfo {author} {\bibfnamefont {R.-J.}\ \bibnamefont {Slager}},\ }\href@noop {} {\bibfield  {journal} {\bibinfo  {journal} {Nature Communications}\ }\textbf {\bibinfo {volume} {13}},\ \bibinfo {pages} {423} (\bibinfo {year} {2022}{\natexlab{a}})}\BibitemShut {NoStop}%
    \bibitem [{\citenamefont {Bouhon}\ and\ \citenamefont {Slager}(2022)}]{bouhon2022multi}%
      \BibitemOpen
      \bibfield  {author} {\bibinfo {author} {\bibfnamefont {A.}~\bibnamefont {Bouhon}}\ and\ \bibinfo {author} {\bibfnamefont {R.-J.}\ \bibnamefont {Slager}},\ }\href@noop {} {\bibfield  {journal} {\bibinfo  {journal} {arXiv preprint arXiv:2203.16741}\ } (\bibinfo {year} {2022})}\BibitemShut {NoStop}%
    \bibitem [{\citenamefont {Park}\ \emph {et~al.}(2022)\citenamefont {Park}, \citenamefont {Wong}, \citenamefont {Bouhon}, \citenamefont {Slager},\ and\ \citenamefont {Oh}}]{park2022topological}%
      \BibitemOpen
      \bibfield  {author} {\bibinfo {author} {\bibfnamefont {H.}~\bibnamefont {Park}}, \bibinfo {author} {\bibfnamefont {S.}~\bibnamefont {Wong}}, \bibinfo {author} {\bibfnamefont {A.}~\bibnamefont {Bouhon}}, \bibinfo {author} {\bibfnamefont {R.-J.}\ \bibnamefont {Slager}}, \ and\ \bibinfo {author} {\bibfnamefont {S.~S.}\ \bibnamefont {Oh}},\ }\href@noop {} {\bibfield  {journal} {\bibinfo  {journal} {Physical Review B}\ }\textbf {\bibinfo {volume} {105}},\ \bibinfo {pages} {214108} (\bibinfo {year} {2022})}\BibitemShut {NoStop}%
    \bibitem [{\citenamefont {Yang}\ \emph {et~al.}(2022{\natexlab{b}})\citenamefont {Yang}, \citenamefont {Cao},\ and\ \citenamefont {Zhai}}]{yang2022non}%
      \BibitemOpen
      \bibfield  {author} {\bibinfo {author} {\bibfnamefont {X.}~\bibnamefont {Yang}}, \bibinfo {author} {\bibfnamefont {Y.}~\bibnamefont {Cao}}, \ and\ \bibinfo {author} {\bibfnamefont {Y.}~\bibnamefont {Zhai}},\ }\href@noop {} {\bibfield  {journal} {\bibinfo  {journal} {Chinese Physics B}\ }\textbf {\bibinfo {volume} {31}},\ \bibinfo {pages} {010308} (\bibinfo {year} {2022}{\natexlab{b}})}\BibitemShut {NoStop}%
    \bibitem [{\citenamefont {Li}\ \emph {et~al.}(2018)\citenamefont {Li}, \citenamefont {Lee},\ and\ \citenamefont {Gong}}]{li2018realistic}%
      \BibitemOpen
      \bibfield  {author} {\bibinfo {author} {\bibfnamefont {L.}~\bibnamefont {Li}}, \bibinfo {author} {\bibfnamefont {C.~H.}\ \bibnamefont {Lee}}, \ and\ \bibinfo {author} {\bibfnamefont {J.}~\bibnamefont {Gong}},\ }\href@noop {} {\bibfield  {journal} {\bibinfo  {journal} {Physical review letters}\ }\textbf {\bibinfo {volume} {121}},\ \bibinfo {pages} {036401} (\bibinfo {year} {2018})}\BibitemShut {NoStop}%
    \bibitem [{\citenamefont {Yan}\ and\ \citenamefont {Wang}(2017)}]{yan2017floquet}%
      \BibitemOpen
      \bibfield  {author} {\bibinfo {author} {\bibfnamefont {Z.}~\bibnamefont {Yan}}\ and\ \bibinfo {author} {\bibfnamefont {Z.}~\bibnamefont {Wang}},\ }\href@noop {} {\bibfield  {journal} {\bibinfo  {journal} {Physical Review B}\ }\textbf {\bibinfo {volume} {96}},\ \bibinfo {pages} {041206} (\bibinfo {year} {2017})}\BibitemShut {NoStop}%
    \bibitem [{\citenamefont {Carlstr\"om}\ and\ \citenamefont {Bergholtz}(2018)}]{carlstrom2018EL}%
      \BibitemOpen
      \bibfield  {author} {\bibinfo {author} {\bibfnamefont {J.}~\bibnamefont {Carlstr\"om}}\ and\ \bibinfo {author} {\bibfnamefont {E.~J.}\ \bibnamefont {Bergholtz}},\ }\href {\doibase 10.1103/PhysRevA.98.042114} {\bibfield  {journal} {\bibinfo  {journal} {Phys. Rev. A}\ }\textbf {\bibinfo {volume} {98}},\ \bibinfo {pages} {042114} (\bibinfo {year} {2018})}\BibitemShut {NoStop}%
    \bibitem [{\citenamefont {Chen}\ \emph {et~al.}(2018)\citenamefont {Chen}, \citenamefont {Zhou},\ and\ \citenamefont {Xu}}]{chen2018floquet}%
      \BibitemOpen
      \bibfield  {author} {\bibinfo {author} {\bibfnamefont {R.}~\bibnamefont {Chen}}, \bibinfo {author} {\bibfnamefont {B.}~\bibnamefont {Zhou}}, \ and\ \bibinfo {author} {\bibfnamefont {D.-H.}\ \bibnamefont {Xu}},\ }\href@noop {} {\bibfield  {journal} {\bibinfo  {journal} {Physical Review B}\ }\textbf {\bibinfo {volume} {97}},\ \bibinfo {pages} {155152} (\bibinfo {year} {2018})}\BibitemShut {NoStop}%
    \bibitem [{\citenamefont {Carlstr\"om}\ \emph {et~al.}(2019)\citenamefont {Carlstr\"om}, \citenamefont {St\aa{}lhammar}, \citenamefont {Budich},\ and\ \citenamefont {Bergholtz}}]{carlstrom2019EL}%
      \BibitemOpen
      \bibfield  {author} {\bibinfo {author} {\bibfnamefont {J.}~\bibnamefont {Carlstr\"om}}, \bibinfo {author} {\bibfnamefont {M.}~\bibnamefont {St\aa{}lhammar}}, \bibinfo {author} {\bibfnamefont {J.~C.}\ \bibnamefont {Budich}}, \ and\ \bibinfo {author} {\bibfnamefont {E.~J.}\ \bibnamefont {Bergholtz}},\ }\href {\doibase 10.1103/PhysRevB.99.161115} {\bibfield  {journal} {\bibinfo  {journal} {Phys. Rev. B}\ }\textbf {\bibinfo {volume} {99}},\ \bibinfo {pages} {161115} (\bibinfo {year} {2019})}\BibitemShut {NoStop}%
    \bibitem [{\citenamefont {Kim}\ \emph {et~al.}(2019)\citenamefont {Kim}, \citenamefont {Kwon},\ and\ \citenamefont {Park}}]{kim2019floquet}%
      \BibitemOpen
      \bibfield  {author} {\bibinfo {author} {\bibfnamefont {K.~W.}\ \bibnamefont {Kim}}, \bibinfo {author} {\bibfnamefont {H.}~\bibnamefont {Kwon}}, \ and\ \bibinfo {author} {\bibfnamefont {K.}~\bibnamefont {Park}},\ }\href@noop {} {\bibfield  {journal} {\bibinfo  {journal} {Physical Review B}\ }\textbf {\bibinfo {volume} {99}},\ \bibinfo {pages} {115136} (\bibinfo {year} {2019})}\BibitemShut {NoStop}%
    \bibitem [{\citenamefont {St{\aa}lhammar}\ \emph {et~al.}(2019)\citenamefont {St{\aa}lhammar}, \citenamefont {R{\o}dland}, \citenamefont {Arone}, \citenamefont {Budich},\ and\ \citenamefont {Bergholtz}}]{staalhammar2019hyperbolic}%
      \BibitemOpen
      \bibfield  {author} {\bibinfo {author} {\bibfnamefont {M.}~\bibnamefont {St{\aa}lhammar}}, \bibinfo {author} {\bibfnamefont {L.}~\bibnamefont {R{\o}dland}}, \bibinfo {author} {\bibfnamefont {G.}~\bibnamefont {Arone}}, \bibinfo {author} {\bibfnamefont {J.~C.}\ \bibnamefont {Budich}}, \ and\ \bibinfo {author} {\bibfnamefont {E.}~\bibnamefont {Bergholtz}},\ }\href@noop {} {\bibfield  {journal} {\bibinfo  {journal} {SciPost Physics}\ }\textbf {\bibinfo {volume} {7}},\ \bibinfo {pages} {019} (\bibinfo {year} {2019})}\BibitemShut {NoStop}%
    \bibitem [{\citenamefont {Salerno}\ \emph {et~al.}(2020)\citenamefont {Salerno}, \citenamefont {Goldman},\ and\ \citenamefont {Palumbo}}]{salerno2020floquet}%
      \BibitemOpen
      \bibfield  {author} {\bibinfo {author} {\bibfnamefont {G.}~\bibnamefont {Salerno}}, \bibinfo {author} {\bibfnamefont {N.}~\bibnamefont {Goldman}}, \ and\ \bibinfo {author} {\bibfnamefont {G.}~\bibnamefont {Palumbo}},\ }\href@noop {} {\bibfield  {journal} {\bibinfo  {journal} {Physical Review Research}\ }\textbf {\bibinfo {volume} {2}},\ \bibinfo {pages} {013224} (\bibinfo {year} {2020})}\BibitemShut {NoStop}%
    \bibitem [{\citenamefont {Meng}\ \emph {et~al.}(2022)\citenamefont {Meng}, \citenamefont {Wang}, \citenamefont {Lee},\ and\ \citenamefont {Ang}}]{meng2022terahertz}%
      \BibitemOpen
      \bibfield  {author} {\bibinfo {author} {\bibfnamefont {H.}~\bibnamefont {Meng}}, \bibinfo {author} {\bibfnamefont {L.}~\bibnamefont {Wang}}, \bibinfo {author} {\bibfnamefont {C.~H.}\ \bibnamefont {Lee}}, \ and\ \bibinfo {author} {\bibfnamefont {Y.~S.}\ \bibnamefont {Ang}},\ }\href@noop {} {\bibfield  {journal} {\bibinfo  {journal} {Applied Physics Letters}\ }\textbf {\bibinfo {volume} {121}},\ \bibinfo {pages} {193102} (\bibinfo {year} {2022})}\BibitemShut {NoStop}%
    \bibitem [{\citenamefont {Qin}\ \emph {et~al.}(2022)\citenamefont {Qin}, \citenamefont {Lee}, \citenamefont {Chen} \emph {et~al.}}]{qin2022light}%
      \BibitemOpen
      \bibfield  {author} {\bibinfo {author} {\bibfnamefont {F.}~\bibnamefont {Qin}}, \bibinfo {author} {\bibfnamefont {C.~H.}\ \bibnamefont {Lee}}, \bibinfo {author} {\bibfnamefont {R.}~\bibnamefont {Chen}},  \emph {et~al.},\ }\href@noop {} {\bibfield  {journal} {\bibinfo  {journal} {Physical Review B}\ }\textbf {\bibinfo {volume} {106}},\ \bibinfo {pages} {235405} (\bibinfo {year} {2022})}\BibitemShut {NoStop}%
    \bibitem [{\citenamefont {Wu}\ and\ \citenamefont {An}(2022)}]{wu2022non}%
      \BibitemOpen
      \bibfield  {author} {\bibinfo {author} {\bibfnamefont {H.}~\bibnamefont {Wu}}\ and\ \bibinfo {author} {\bibfnamefont {J.-H.}\ \bibnamefont {An}},\ }\href@noop {} {\bibfield  {journal} {\bibinfo  {journal} {Physical Review B}\ }\textbf {\bibinfo {volume} {105}},\ \bibinfo {pages} {L121113} (\bibinfo {year} {2022})}\BibitemShut {NoStop}%
    \bibitem [{\citenamefont {Wang}\ \emph {et~al.}(2021{\natexlab{b}})\citenamefont {Wang}, \citenamefont {Xiao}, \citenamefont {Budich}, \citenamefont {Yi},\ and\ \citenamefont {Xue}}]{PhysRevLett.127.026404}%
      \BibitemOpen
      \bibfield  {author} {\bibinfo {author} {\bibfnamefont {K.}~\bibnamefont {Wang}}, \bibinfo {author} {\bibfnamefont {L.}~\bibnamefont {Xiao}}, \bibinfo {author} {\bibfnamefont {J.~C.}\ \bibnamefont {Budich}}, \bibinfo {author} {\bibfnamefont {W.}~\bibnamefont {Yi}}, \ and\ \bibinfo {author} {\bibfnamefont {P.}~\bibnamefont {Xue}},\ }\href {\doibase 10.1103/PhysRevLett.127.026404} {\bibfield  {journal} {\bibinfo  {journal} {Phys. Rev. Lett.}\ }\textbf {\bibinfo {volume} {127}},\ \bibinfo {pages} {026404} (\bibinfo {year} {2021}{\natexlab{b}})}\BibitemShut {NoStop}%
    \bibitem [{\citenamefont {Benalcazar}\ \emph {et~al.}(2017)\citenamefont {Benalcazar}, \citenamefont {Bernevig},\ and\ \citenamefont {Hughes}}]{benalcazar2017HOTI}%
      \BibitemOpen
      \bibfield  {author} {\bibinfo {author} {\bibfnamefont {W.~A.}\ \bibnamefont {Benalcazar}}, \bibinfo {author} {\bibfnamefont {B.~A.}\ \bibnamefont {Bernevig}}, \ and\ \bibinfo {author} {\bibfnamefont {T.~L.}\ \bibnamefont {Hughes}},\ }\href@noop {} {\bibfield  {journal} {\bibinfo  {journal} {Science}\ }\textbf {\bibinfo {volume} {357}},\ \bibinfo {pages} {61} (\bibinfo {year} {2017})}\BibitemShut {NoStop}%
    \bibitem [{\citenamefont {Edvardsson}\ \emph {et~al.}(2019)\citenamefont {Edvardsson}, \citenamefont {Kunst},\ and\ \citenamefont {Bergholtz}}]{PhysRevB.99.081302}%
      \BibitemOpen
      \bibfield  {author} {\bibinfo {author} {\bibfnamefont {E.}~\bibnamefont {Edvardsson}}, \bibinfo {author} {\bibfnamefont {F.~K.}\ \bibnamefont {Kunst}}, \ and\ \bibinfo {author} {\bibfnamefont {E.~J.}\ \bibnamefont {Bergholtz}},\ }\href {\doibase 10.1103/PhysRevB.99.081302} {\bibfield  {journal} {\bibinfo  {journal} {Phys. Rev. B}\ }\textbf {\bibinfo {volume} {99}},\ \bibinfo {pages} {081302} (\bibinfo {year} {2019})}\BibitemShut {NoStop}%
    \bibitem [{\citenamefont {Ghosh}\ and\ \citenamefont {Nag}(2022)}]{ghosh2022non}%
      \BibitemOpen
      \bibfield  {author} {\bibinfo {author} {\bibfnamefont {A.~K.}\ \bibnamefont {Ghosh}}\ and\ \bibinfo {author} {\bibfnamefont {T.}~\bibnamefont {Nag}},\ }\href {\doibase 10.1103/PhysRevB.106.L140303} {\bibfield  {journal} {\bibinfo  {journal} {Phys. Rev. B}\ }\textbf {\bibinfo {volume} {106}},\ \bibinfo {pages} {L140303} (\bibinfo {year} {2022})}\BibitemShut {NoStop}%
    \bibitem [{\citenamefont {Kim}\ and\ \citenamefont {Park}(2021)}]{kim2021disorder}%
      \BibitemOpen
      \bibfield  {author} {\bibinfo {author} {\bibfnamefont {K.-M.}\ \bibnamefont {Kim}}\ and\ \bibinfo {author} {\bibfnamefont {M.~J.}\ \bibnamefont {Park}},\ }\href@noop {} {\bibfield  {journal} {\bibinfo  {journal} {Physical Review B}\ }\textbf {\bibinfo {volume} {104}},\ \bibinfo {pages} {L121101} (\bibinfo {year} {2021})}\BibitemShut {NoStop}%
    \bibitem [{Note1()}]{Note1}%
      \BibitemOpen
      \bibinfo {note} {Note that asymmetric couplings are not always need to break reciprocity~\cite {zhu2022hybrid}.}\BibitemShut {Stop}%
    \bibitem [{\citenamefont {Lei}\ \emph {et~al.}(2024)\citenamefont {Lei}, \citenamefont {Lee},\ and\ \citenamefont {Li}}]{lei2024activating}%
      \BibitemOpen
      \bibfield  {author} {\bibinfo {author} {\bibfnamefont {Z.}~\bibnamefont {Lei}}, \bibinfo {author} {\bibfnamefont {C.~H.}\ \bibnamefont {Lee}}, \ and\ \bibinfo {author} {\bibfnamefont {L.}~\bibnamefont {Li}},\ }\href@noop {} {\bibfield  {journal} {\bibinfo  {journal} {Communications Physics}\ }\textbf {\bibinfo {volume} {7}},\ \bibinfo {pages} {100} (\bibinfo {year} {2024})}\BibitemShut {NoStop}%
    \bibitem [{\citenamefont {Li}\ \emph {et~al.}(2019{\natexlab{b}})\citenamefont {Li}, \citenamefont {Harter}, \citenamefont {Liu}, \citenamefont {de~Melo}, \citenamefont {Joglekar},\ and\ \citenamefont {Luo}}]{Jiaming2019gainloss}%
      \BibitemOpen
      \bibfield  {author} {\bibinfo {author} {\bibfnamefont {J.}~\bibnamefont {Li}}, \bibinfo {author} {\bibfnamefont {A.~K.}\ \bibnamefont {Harter}}, \bibinfo {author} {\bibfnamefont {J.}~\bibnamefont {Liu}}, \bibinfo {author} {\bibfnamefont {L.}~\bibnamefont {de~Melo}}, \bibinfo {author} {\bibfnamefont {Y.~N.}\ \bibnamefont {Joglekar}}, \ and\ \bibinfo {author} {\bibfnamefont {L.}~\bibnamefont {Luo}},\ }\href {\doibase 10.1038/s41467-019-08596-1} {\bibfield  {journal} {\bibinfo  {journal} {Nat. Comm.}\ }\textbf {\bibinfo {volume} {10}},\ \bibinfo {pages} {855} (\bibinfo {year} {2019}{\natexlab{b}})}\BibitemShut {NoStop}%
    \bibitem [{\citenamefont {Wu}\ \emph {et~al.}(2021)\citenamefont {Wu}, \citenamefont {Wang},\ and\ \citenamefont {An}}]{wu2021floquet}%
      \BibitemOpen
      \bibfield  {author} {\bibinfo {author} {\bibfnamefont {H.}~\bibnamefont {Wu}}, \bibinfo {author} {\bibfnamefont {B.-Q.}\ \bibnamefont {Wang}}, \ and\ \bibinfo {author} {\bibfnamefont {J.-H.}\ \bibnamefont {An}},\ }\href@noop {} {\bibfield  {journal} {\bibinfo  {journal} {Physical Review B}\ }\textbf {\bibinfo {volume} {103}},\ \bibinfo {pages} {L041115} (\bibinfo {year} {2021})}\BibitemShut {NoStop}%
    \bibitem [{\citenamefont {Zhou}\ \emph {et~al.}(2022{\natexlab{a}})\citenamefont {Zhou}, \citenamefont {Bomantara},\ and\ \citenamefont {Wu}}]{zhou2022q}%
      \BibitemOpen
      \bibfield  {author} {\bibinfo {author} {\bibfnamefont {L.}~\bibnamefont {Zhou}}, \bibinfo {author} {\bibfnamefont {R.~W.}\ \bibnamefont {Bomantara}}, \ and\ \bibinfo {author} {\bibfnamefont {S.}~\bibnamefont {Wu}},\ }\href {\doibase 10.21468/SciPostPhys.13.2.015} {\bibfield  {journal} {\bibinfo  {journal} {SciPost Phys.}\ }\textbf {\bibinfo {volume} {13}},\ \bibinfo {pages} {015} (\bibinfo {year} {2022}{\natexlab{a}})}\BibitemShut {NoStop}%
    \bibitem [{\citenamefont {Cao}\ \emph {et~al.}(2021{\natexlab{a}})\citenamefont {Cao}, \citenamefont {Li},\ and\ \citenamefont {Yang}}]{cao2021non}%
      \BibitemOpen
      \bibfield  {author} {\bibinfo {author} {\bibfnamefont {Y.}~\bibnamefont {Cao}}, \bibinfo {author} {\bibfnamefont {Y.}~\bibnamefont {Li}}, \ and\ \bibinfo {author} {\bibfnamefont {X.}~\bibnamefont {Yang}},\ }\href@noop {} {\bibfield  {journal} {\bibinfo  {journal} {Physical Review B}\ }\textbf {\bibinfo {volume} {103}},\ \bibinfo {pages} {075126} (\bibinfo {year} {2021}{\natexlab{a}})}\BibitemShut {NoStop}%
    \bibitem [{\citenamefont {Ezawa}(2019{\natexlab{a}})}]{PhysRevB.99.201411}%
      \BibitemOpen
      \bibfield  {author} {\bibinfo {author} {\bibfnamefont {M.}~\bibnamefont {Ezawa}},\ }\href {\doibase 10.1103/PhysRevB.99.201411} {\bibfield  {journal} {\bibinfo  {journal} {Phys. Rev. B}\ }\textbf {\bibinfo {volume} {99}},\ \bibinfo {pages} {201411} (\bibinfo {year} {2019}{\natexlab{a}})}\BibitemShut {NoStop}%
    \bibitem [{\citenamefont {Ezawa}(2019{\natexlab{b}})}]{PhysRevB.99.121411}%
      \BibitemOpen
      \bibfield  {author} {\bibinfo {author} {\bibfnamefont {M.}~\bibnamefont {Ezawa}},\ }\href {\doibase 10.1103/PhysRevB.99.121411} {\bibfield  {journal} {\bibinfo  {journal} {Phys. Rev. B}\ }\textbf {\bibinfo {volume} {99}},\ \bibinfo {pages} {121411} (\bibinfo {year} {2019}{\natexlab{b}})}\BibitemShut {NoStop}%
    \bibitem [{\citenamefont {Song}\ \emph {et~al.}(2017)\citenamefont {Song}, \citenamefont {Fang},\ and\ \citenamefont {Fang}}]{song2017HOTI}%
      \BibitemOpen
      \bibfield  {author} {\bibinfo {author} {\bibfnamefont {Z.}~\bibnamefont {Song}}, \bibinfo {author} {\bibfnamefont {Z.}~\bibnamefont {Fang}}, \ and\ \bibinfo {author} {\bibfnamefont {C.}~\bibnamefont {Fang}},\ }\href {\doibase 10.1103/PhysRevLett.119.246402} {\bibfield  {journal} {\bibinfo  {journal} {Phys. Rev. Lett.}\ }\textbf {\bibinfo {volume} {119}},\ \bibinfo {pages} {246402} (\bibinfo {year} {2017})}\BibitemShut {NoStop}%
    \bibitem [{\citenamefont {Khalaf}(2018)}]{khalaf2018HOTI}%
      \BibitemOpen
      \bibfield  {author} {\bibinfo {author} {\bibfnamefont {E.}~\bibnamefont {Khalaf}},\ }\href {\doibase 10.1103/PhysRevB.97.205136} {\bibfield  {journal} {\bibinfo  {journal} {Phys. Rev. B}\ }\textbf {\bibinfo {volume} {97}},\ \bibinfo {pages} {205136} (\bibinfo {year} {2018})}\BibitemShut {NoStop}%
    \bibitem [{\citenamefont {Trifunovic}\ and\ \citenamefont {Brouwer}(2021)}]{trifunovic2021higher}%
      \BibitemOpen
      \bibfield  {author} {\bibinfo {author} {\bibfnamefont {L.}~\bibnamefont {Trifunovic}}\ and\ \bibinfo {author} {\bibfnamefont {P.~W.}\ \bibnamefont {Brouwer}},\ }\href@noop {} {\bibfield  {journal} {\bibinfo  {journal} {physica status solidi (b)}\ }\textbf {\bibinfo {volume} {258}},\ \bibinfo {pages} {2000090} (\bibinfo {year} {2021})}\BibitemShut {NoStop}%
    \bibitem [{\citenamefont {Li}\ \emph {et~al.}()\citenamefont {Li}, \citenamefont {Trauzettel}, \citenamefont {Neupert},\ and\ \citenamefont {Zhang}}]{li2022enahncement}%
      \BibitemOpen
      \bibfield  {author} {\bibinfo {author} {\bibfnamefont {C.-A.}\ \bibnamefont {Li}}, \bibinfo {author} {\bibfnamefont {B.}~\bibnamefont {Trauzettel}}, \bibinfo {author} {\bibfnamefont {T.}~\bibnamefont {Neupert}}, \ and\ \bibinfo {author} {\bibfnamefont {S.-B.}\ \bibnamefont {Zhang}},\ }\href@noop {} {\ }\Eprint {http://arxiv.org/abs/2212.14691v1} {2212.14691v1} \BibitemShut {NoStop}%
    \bibitem [{\citenamefont {Jiang}\ and\ \citenamefont {Lee}(2023)}]{jiang2022dimensional}%
      \BibitemOpen
      \bibfield  {author} {\bibinfo {author} {\bibfnamefont {H.}~\bibnamefont {Jiang}}\ and\ \bibinfo {author} {\bibfnamefont {C.~H.}\ \bibnamefont {Lee}},\ }\href@noop {} {\bibfield  {journal} {\bibinfo  {journal} {Physical Review Letters}\ }\textbf {\bibinfo {volume} {131}},\ \bibinfo {pages} {076401} (\bibinfo {year} {2023})}\BibitemShut {NoStop}%
    \bibitem [{\citenamefont {Zhu}\ \emph {et~al.}(2022{\natexlab{a}})\citenamefont {Zhu}, \citenamefont {Sun}, \citenamefont {Hughes},\ and\ \citenamefont {Bahl}}]{zhu2022higher}%
      \BibitemOpen
      \bibfield  {author} {\bibinfo {author} {\bibfnamefont {P.}~\bibnamefont {Zhu}}, \bibinfo {author} {\bibfnamefont {X.-Q.}\ \bibnamefont {Sun}}, \bibinfo {author} {\bibfnamefont {T.~L.}\ \bibnamefont {Hughes}}, \ and\ \bibinfo {author} {\bibfnamefont {G.}~\bibnamefont {Bahl}},\ }\href@noop {} {\bibfield  {journal} {\bibinfo  {journal} {arXiv preprint arXiv:2207.02228}\ } (\bibinfo {year} {2022}{\natexlab{a}})}\BibitemShut {NoStop}%
    \bibitem [{\citenamefont {Song}\ \emph {et~al.}(2022)\citenamefont {Song}, \citenamefont {Wang},\ and\ \citenamefont {Wang}}]{song2021non}%
      \BibitemOpen
      \bibfield  {author} {\bibinfo {author} {\bibfnamefont {F.}~\bibnamefont {Song}}, \bibinfo {author} {\bibfnamefont {H.-Y.}\ \bibnamefont {Wang}}, \ and\ \bibinfo {author} {\bibfnamefont {Z.}~\bibnamefont {Wang}},\ }\href@noop {} {\bibfield  {journal} {\bibinfo  {journal} {A Festschrift in Honor of the C N Yang Centenary (arXiv:2102.02230)}\ } (\bibinfo {year} {2022})}\BibitemShut {NoStop}%
    \bibitem [{\citenamefont {Jiang}\ and\ \citenamefont {Lee}(2022)}]{jiang2022filling}%
      \BibitemOpen
      \bibfield  {author} {\bibinfo {author} {\bibfnamefont {H.}~\bibnamefont {Jiang}}\ and\ \bibinfo {author} {\bibfnamefont {C.~H.}\ \bibnamefont {Lee}},\ }\href@noop {} {\bibfield  {journal} {\bibinfo  {journal} {Chinese Physics B}\ }\textbf {\bibinfo {volume} {31}},\ \bibinfo {pages} {050307} (\bibinfo {year} {2022})}\BibitemShut {NoStop}%
    \bibitem [{\citenamefont {Kawabata}\ \emph {et~al.}(2019)\citenamefont {Kawabata}, \citenamefont {Shiozaki}, \citenamefont {Ueda},\ and\ \citenamefont {Sato}}]{kawabata2019symmetry}%
      \BibitemOpen
      \bibfield  {author} {\bibinfo {author} {\bibfnamefont {K.}~\bibnamefont {Kawabata}}, \bibinfo {author} {\bibfnamefont {K.}~\bibnamefont {Shiozaki}}, \bibinfo {author} {\bibfnamefont {M.}~\bibnamefont {Ueda}}, \ and\ \bibinfo {author} {\bibfnamefont {M.}~\bibnamefont {Sato}},\ }\href@noop {} {\bibfield  {journal} {\bibinfo  {journal} {Physical Review X}\ }\textbf {\bibinfo {volume} {9}},\ \bibinfo {pages} {041015} (\bibinfo {year} {2019})}\BibitemShut {NoStop}%
    \bibitem [{\citenamefont {Theiler}(1990)}]{theiler1990estimating}%
      \BibitemOpen
      \bibfield  {author} {\bibinfo {author} {\bibfnamefont {J.}~\bibnamefont {Theiler}},\ }\href@noop {} {\bibfield  {journal} {\bibinfo  {journal} {JOSA A}\ }\textbf {\bibinfo {volume} {7}},\ \bibinfo {pages} {1055} (\bibinfo {year} {1990})}\BibitemShut {NoStop}%
    \bibitem [{\citenamefont {Qi}(2013)}]{qi2013exact}%
      \BibitemOpen
      \bibfield  {author} {\bibinfo {author} {\bibfnamefont {X.-L.}\ \bibnamefont {Qi}},\ }\href@noop {} {\bibfield  {journal} {\bibinfo  {journal} {arXiv preprint arXiv:1309.6282}\ } (\bibinfo {year} {2013})}\BibitemShut {NoStop}%
    \bibitem [{\citenamefont {Lee}\ and\ \citenamefont {Qi}(2016)}]{lee2016exact}%
      \BibitemOpen
      \bibfield  {author} {\bibinfo {author} {\bibfnamefont {C.~H.}\ \bibnamefont {Lee}}\ and\ \bibinfo {author} {\bibfnamefont {X.-L.}\ \bibnamefont {Qi}},\ }\href@noop {} {\bibfield  {journal} {\bibinfo  {journal} {Physical Review B}\ }\textbf {\bibinfo {volume} {93}},\ \bibinfo {pages} {035112} (\bibinfo {year} {2016})}\BibitemShut {NoStop}%
    \bibitem [{\citenamefont {Gu}\ \emph {et~al.}(2016)\citenamefont {Gu}, \citenamefont {Lee}, \citenamefont {Wen}, \citenamefont {Cho}, \citenamefont {Ryu},\ and\ \citenamefont {Qi}}]{gu2016holographic}%
      \BibitemOpen
      \bibfield  {author} {\bibinfo {author} {\bibfnamefont {Y.}~\bibnamefont {Gu}}, \bibinfo {author} {\bibfnamefont {C.~H.}\ \bibnamefont {Lee}}, \bibinfo {author} {\bibfnamefont {X.}~\bibnamefont {Wen}}, \bibinfo {author} {\bibfnamefont {G.~Y.}\ \bibnamefont {Cho}}, \bibinfo {author} {\bibfnamefont {S.}~\bibnamefont {Ryu}}, \ and\ \bibinfo {author} {\bibfnamefont {X.-L.}\ \bibnamefont {Qi}},\ }\href@noop {} {\bibfield  {journal} {\bibinfo  {journal} {Phys. Rev. B}\ }\textbf {\bibinfo {volume} {94}},\ \bibinfo {pages} {125107} (\bibinfo {year} {2016})}\BibitemShut {NoStop}%
    \bibitem [{\citenamefont {Yoshida}\ \emph {et~al.}(2020)\citenamefont {Yoshida}, \citenamefont {Mizoguchi},\ and\ \citenamefont {Hatsugai}}]{PhysRevResearch.2.022062}%
      \BibitemOpen
      \bibfield  {author} {\bibinfo {author} {\bibfnamefont {T.}~\bibnamefont {Yoshida}}, \bibinfo {author} {\bibfnamefont {T.}~\bibnamefont {Mizoguchi}}, \ and\ \bibinfo {author} {\bibfnamefont {Y.}~\bibnamefont {Hatsugai}},\ }\href {\doibase 10.1103/PhysRevResearch.2.022062} {\bibfield  {journal} {\bibinfo  {journal} {Phys. Rev. Res.}\ }\textbf {\bibinfo {volume} {2}},\ \bibinfo {pages} {022062} (\bibinfo {year} {2020})}\BibitemShut {NoStop}%
    \bibitem [{\citenamefont {Liu}\ and\ \citenamefont {Chen}(2019)}]{liu2019topological}%
      \BibitemOpen
      \bibfield  {author} {\bibinfo {author} {\bibfnamefont {C.-H.}\ \bibnamefont {Liu}}\ and\ \bibinfo {author} {\bibfnamefont {S.}~\bibnamefont {Chen}},\ }\href {\doibase 10.1103/PhysRevB.100.144106} {\bibfield  {journal} {\bibinfo  {journal} {Phys. Rev. B}\ }\textbf {\bibinfo {volume} {100}},\ \bibinfo {pages} {144106} (\bibinfo {year} {2019})}\BibitemShut {NoStop}%
    \bibitem [{\citenamefont {Liu}\ \emph {et~al.}(2022{\natexlab{c}})\citenamefont {Liu}, \citenamefont {Hu},\ and\ \citenamefont {Chen}}]{liu2022symmetry}%
      \BibitemOpen
      \bibfield  {author} {\bibinfo {author} {\bibfnamefont {C.-H.}\ \bibnamefont {Liu}}, \bibinfo {author} {\bibfnamefont {H.}~\bibnamefont {Hu}}, \ and\ \bibinfo {author} {\bibfnamefont {S.}~\bibnamefont {Chen}},\ }\href {\doibase 10.1103/PhysRevB.105.214305} {\bibfield  {journal} {\bibinfo  {journal} {Phys. Rev. B}\ }\textbf {\bibinfo {volume} {105}},\ \bibinfo {pages} {214305} (\bibinfo {year} {2022}{\natexlab{c}})}\BibitemShut {NoStop}%
    \bibitem [{\citenamefont {Okugawa}\ \emph {et~al.}(2021)\citenamefont {Okugawa}, \citenamefont {Takahashi},\ and\ \citenamefont {Yokomizo}}]{okugawa2021non}%
      \BibitemOpen
      \bibfield  {author} {\bibinfo {author} {\bibfnamefont {R.}~\bibnamefont {Okugawa}}, \bibinfo {author} {\bibfnamefont {R.}~\bibnamefont {Takahashi}}, \ and\ \bibinfo {author} {\bibfnamefont {K.}~\bibnamefont {Yokomizo}},\ }\href@noop {} {\bibfield  {journal} {\bibinfo  {journal} {Physical Review B}\ }\textbf {\bibinfo {volume} {103}},\ \bibinfo {pages} {205205} (\bibinfo {year} {2021})}\BibitemShut {NoStop}%
    \bibitem [{\citenamefont {Vecsei}\ \emph {et~al.}(2021)\citenamefont {Vecsei}, \citenamefont {Denner}, \citenamefont {Neupert},\ and\ \citenamefont {Schindler}}]{PhysRevB.103.L201114}%
      \BibitemOpen
      \bibfield  {author} {\bibinfo {author} {\bibfnamefont {P.~M.}\ \bibnamefont {Vecsei}}, \bibinfo {author} {\bibfnamefont {M.~M.}\ \bibnamefont {Denner}}, \bibinfo {author} {\bibfnamefont {T.}~\bibnamefont {Neupert}}, \ and\ \bibinfo {author} {\bibfnamefont {F.}~\bibnamefont {Schindler}},\ }\href {\doibase 10.1103/PhysRevB.103.L201114} {\bibfield  {journal} {\bibinfo  {journal} {Phys. Rev. B}\ }\textbf {\bibinfo {volume} {103}},\ \bibinfo {pages} {L201114} (\bibinfo {year} {2021})}\BibitemShut {NoStop}%
    \bibitem [{\citenamefont {Gong}\ \emph {et~al.}(2018)\citenamefont {Gong}, \citenamefont {Ashida}, \citenamefont {Kawabata}, \citenamefont {Takasan}, \citenamefont {Higashikawa},\ and\ \citenamefont {Ueda}}]{gong2018topological}%
      \BibitemOpen
      \bibfield  {author} {\bibinfo {author} {\bibfnamefont {Z.}~\bibnamefont {Gong}}, \bibinfo {author} {\bibfnamefont {Y.}~\bibnamefont {Ashida}}, \bibinfo {author} {\bibfnamefont {K.}~\bibnamefont {Kawabata}}, \bibinfo {author} {\bibfnamefont {K.}~\bibnamefont {Takasan}}, \bibinfo {author} {\bibfnamefont {S.}~\bibnamefont {Higashikawa}}, \ and\ \bibinfo {author} {\bibfnamefont {M.}~\bibnamefont {Ueda}},\ }\href@noop {} {\bibfield  {journal} {\bibinfo  {journal} {Physical Review X}\ }\textbf {\bibinfo {volume} {8}},\ \bibinfo {pages} {031079} (\bibinfo {year} {2018})}\BibitemShut {NoStop}%
    \bibitem [{\citenamefont {Denner}\ \emph {et~al.}(2021)\citenamefont {Denner}, \citenamefont {Skurativska}, \citenamefont {Schindler}, \citenamefont {Fischer}, \citenamefont {Thomale}, \citenamefont {Bzdu{\v{s}}ek},\ and\ \citenamefont {Neupert}}]{denner2021exceptional}%
      \BibitemOpen
      \bibfield  {author} {\bibinfo {author} {\bibfnamefont {M.~M.}\ \bibnamefont {Denner}}, \bibinfo {author} {\bibfnamefont {A.}~\bibnamefont {Skurativska}}, \bibinfo {author} {\bibfnamefont {F.}~\bibnamefont {Schindler}}, \bibinfo {author} {\bibfnamefont {M.~H.}\ \bibnamefont {Fischer}}, \bibinfo {author} {\bibfnamefont {R.}~\bibnamefont {Thomale}}, \bibinfo {author} {\bibfnamefont {T.}~\bibnamefont {Bzdu{\v{s}}ek}}, \ and\ \bibinfo {author} {\bibfnamefont {T.}~\bibnamefont {Neupert}},\ }\href@noop {} {\bibfield  {journal} {\bibinfo  {journal} {Nature communications}\ }\textbf {\bibinfo {volume} {12}},\ \bibinfo {pages} {5681} (\bibinfo {year} {2021})}\BibitemShut {NoStop}%
    \bibitem [{\citenamefont {Nakamura}\ \emph {et~al.}(2022)\citenamefont {Nakamura}, \citenamefont {Bessho},\ and\ \citenamefont {Sato}}]{nakamura2022bulk}%
      \BibitemOpen
      \bibfield  {author} {\bibinfo {author} {\bibfnamefont {D.}~\bibnamefont {Nakamura}}, \bibinfo {author} {\bibfnamefont {T.}~\bibnamefont {Bessho}}, \ and\ \bibinfo {author} {\bibfnamefont {M.}~\bibnamefont {Sato}},\ }\href@noop {} {\bibfield  {journal} {\bibinfo  {journal} {arXiv preprint arXiv:2205.15635}\ } (\bibinfo {year} {2022})}\BibitemShut {NoStop}%
    \bibitem [{\citenamefont {Song}\ \emph {et~al.}(2019{\natexlab{b}})\citenamefont {Song}, \citenamefont {Yao},\ and\ \citenamefont {Wang}}]{song2019realspace}%
      \BibitemOpen
      \bibfield  {author} {\bibinfo {author} {\bibfnamefont {F.}~\bibnamefont {Song}}, \bibinfo {author} {\bibfnamefont {S.}~\bibnamefont {Yao}}, \ and\ \bibinfo {author} {\bibfnamefont {Z.}~\bibnamefont {Wang}},\ }\href@noop {} {\bibfield  {journal} {\bibinfo  {journal} {Physical Review Letters}\ }\textbf {\bibinfo {volume} {123}},\ \bibinfo {pages} {246801} (\bibinfo {year} {2019}{\natexlab{b}})}\BibitemShut {NoStop}%
    \bibitem [{\citenamefont {Tang}\ \emph {et~al.}(2020{\natexlab{a}})\citenamefont {Tang}, \citenamefont {Zhang}, \citenamefont {Zhang},\ and\ \citenamefont {Zhang}}]{Tang2020}%
      \BibitemOpen
      \bibfield  {author} {\bibinfo {author} {\bibfnamefont {L.-Z.}\ \bibnamefont {Tang}}, \bibinfo {author} {\bibfnamefont {L.-F.}\ \bibnamefont {Zhang}}, \bibinfo {author} {\bibfnamefont {G.-Q.}\ \bibnamefont {Zhang}}, \ and\ \bibinfo {author} {\bibfnamefont {D.-W.}\ \bibnamefont {Zhang}},\ }\href {\doibase 10.1103/PhysRevA.101.063612} {\bibfield  {journal} {\bibinfo  {journal} {Phys. Rev. A}\ }\textbf {\bibinfo {volume} {101}},\ \bibinfo {pages} {063612} (\bibinfo {year} {2020}{\natexlab{a}})}\BibitemShut {NoStop}%
    \bibitem [{\citenamefont {Ghorashi}\ \emph {et~al.}(2021{\natexlab{a}})\citenamefont {Ghorashi}, \citenamefont {Li}, \citenamefont {Sato},\ and\ \citenamefont {Hughes}}]{PhysRevB.104.L161116}%
      \BibitemOpen
      \bibfield  {author} {\bibinfo {author} {\bibfnamefont {S.~A.~A.}\ \bibnamefont {Ghorashi}}, \bibinfo {author} {\bibfnamefont {T.}~\bibnamefont {Li}}, \bibinfo {author} {\bibfnamefont {M.}~\bibnamefont {Sato}}, \ and\ \bibinfo {author} {\bibfnamefont {T.~L.}\ \bibnamefont {Hughes}},\ }\href {\doibase 10.1103/PhysRevB.104.L161116} {\bibfield  {journal} {\bibinfo  {journal} {Phys. Rev. B}\ }\textbf {\bibinfo {volume} {104}},\ \bibinfo {pages} {L161116} (\bibinfo {year} {2021}{\natexlab{a}})}\BibitemShut {NoStop}%
    \bibitem [{\citenamefont {Ghorashi}\ \emph {et~al.}(2021{\natexlab{b}})\citenamefont {Ghorashi}, \citenamefont {Li},\ and\ \citenamefont {Sato}}]{PhysRevB.104.L161117}%
      \BibitemOpen
      \bibfield  {author} {\bibinfo {author} {\bibfnamefont {S.~A.~A.}\ \bibnamefont {Ghorashi}}, \bibinfo {author} {\bibfnamefont {T.}~\bibnamefont {Li}}, \ and\ \bibinfo {author} {\bibfnamefont {M.}~\bibnamefont {Sato}},\ }\href {\doibase 10.1103/PhysRevB.104.L161117} {\bibfield  {journal} {\bibinfo  {journal} {Phys. Rev. B}\ }\textbf {\bibinfo {volume} {104}},\ \bibinfo {pages} {L161117} (\bibinfo {year} {2021}{\natexlab{b}})}\BibitemShut {NoStop}%
    \bibitem [{\citenamefont {Edvardsson}\ and\ \citenamefont {Ardonne}(2022)}]{edvardsson2022stability}%
      \BibitemOpen
      \bibfield  {author} {\bibinfo {author} {\bibfnamefont {E.}~\bibnamefont {Edvardsson}}\ and\ \bibinfo {author} {\bibfnamefont {E.}~\bibnamefont {Ardonne}},\ }\href {\doibase 10.1103/PhysRevB.106.115107} {\bibfield  {journal} {\bibinfo  {journal} {Phys. Rev. B}\ }\textbf {\bibinfo {volume} {106}},\ \bibinfo {pages} {115107} (\bibinfo {year} {2022})}\BibitemShut {NoStop}%
    \bibitem [{\citenamefont {B{\"o}ttcher}\ and\ \citenamefont {Grudsky}(2005)}]{bottcher2005spectral}%
      \BibitemOpen
      \bibfield  {author} {\bibinfo {author} {\bibfnamefont {A.}~\bibnamefont {B{\"o}ttcher}}\ and\ \bibinfo {author} {\bibfnamefont {S.~M.}\ \bibnamefont {Grudsky}},\ }\href@noop {} {\emph {\bibinfo {title} {Spectral properties of banded Toeplitz matrices}}}\ (\bibinfo  {publisher} {SIAM},\ \bibinfo {year} {2005})\BibitemShut {NoStop}%
    \bibitem [{\citenamefont {Bartlett}\ \emph {et~al.}(2021)\citenamefont {Bartlett}, \citenamefont {Hu},\ and\ \citenamefont {Zhao}}]{bartlett2021illuminating}%
      \BibitemOpen
      \bibfield  {author} {\bibinfo {author} {\bibfnamefont {J.}~\bibnamefont {Bartlett}}, \bibinfo {author} {\bibfnamefont {H.}~\bibnamefont {Hu}}, \ and\ \bibinfo {author} {\bibfnamefont {E.}~\bibnamefont {Zhao}},\ }\href@noop {} {\bibfield  {journal} {\bibinfo  {journal} {Physical Review B}\ }\textbf {\bibinfo {volume} {104}},\ \bibinfo {pages} {195131} (\bibinfo {year} {2021})}\BibitemShut {NoStop}%
    \bibitem [{\citenamefont {Teo}\ \emph {et~al.}(2020)\citenamefont {Teo}, \citenamefont {Li}, \citenamefont {Zhang},\ and\ \citenamefont {Gong}}]{teo2020topological}%
      \BibitemOpen
      \bibfield  {author} {\bibinfo {author} {\bibfnamefont {W.~X.~T.}\ \bibnamefont {Teo}}, \bibinfo {author} {\bibfnamefont {L.}~\bibnamefont {Li}}, \bibinfo {author} {\bibfnamefont {X.}~\bibnamefont {Zhang}}, \ and\ \bibinfo {author} {\bibfnamefont {J.}~\bibnamefont {Gong}},\ }\href@noop {} {\bibfield  {journal} {\bibinfo  {journal} {Physical Review B}\ }\textbf {\bibinfo {volume} {101}},\ \bibinfo {pages} {205309} (\bibinfo {year} {2020})}\BibitemShut {NoStop}%
    \bibitem [{\citenamefont {Pan}\ \emph {et~al.}(2021)\citenamefont {Pan}, \citenamefont {Li},\ and\ \citenamefont {Gong}}]{pan2021point}%
      \BibitemOpen
      \bibfield  {author} {\bibinfo {author} {\bibfnamefont {J.-S.}\ \bibnamefont {Pan}}, \bibinfo {author} {\bibfnamefont {L.}~\bibnamefont {Li}}, \ and\ \bibinfo {author} {\bibfnamefont {J.}~\bibnamefont {Gong}},\ }\href@noop {} {\bibfield  {journal} {\bibinfo  {journal} {Physical Review B}\ }\textbf {\bibinfo {volume} {103}},\ \bibinfo {pages} {205425} (\bibinfo {year} {2021})}\BibitemShut {NoStop}%
    \bibitem [{\citenamefont {Wang}\ \emph {et~al.}(2021{\natexlab{c}})\citenamefont {Wang}, \citenamefont {Dutt}, \citenamefont {Yang}, \citenamefont {Wojcik}, \citenamefont {Vu{\v{c}}kovi{\'c}},\ and\ \citenamefont {Fan}}]{wang2021generating}%
      \BibitemOpen
      \bibfield  {author} {\bibinfo {author} {\bibfnamefont {K.}~\bibnamefont {Wang}}, \bibinfo {author} {\bibfnamefont {A.}~\bibnamefont {Dutt}}, \bibinfo {author} {\bibfnamefont {K.~Y.}\ \bibnamefont {Yang}}, \bibinfo {author} {\bibfnamefont {C.~C.}\ \bibnamefont {Wojcik}}, \bibinfo {author} {\bibfnamefont {J.}~\bibnamefont {Vu{\v{c}}kovi{\'c}}}, \ and\ \bibinfo {author} {\bibfnamefont {S.}~\bibnamefont {Fan}},\ }\href@noop {} {\bibfield  {journal} {\bibinfo  {journal} {Science}\ }\textbf {\bibinfo {volume} {371}},\ \bibinfo {pages} {1240} (\bibinfo {year} {2021}{\natexlab{c}})}\BibitemShut {NoStop}%
    \bibitem [{\citenamefont {Lee}\ \emph {et~al.}(2018{\natexlab{c}})\citenamefont {Lee}, \citenamefont {Imhof}, \citenamefont {Berger}, \citenamefont {Bayer}, \citenamefont {Brehm}, \citenamefont {Molenkamp}, \citenamefont {Kiessling},\ and\ \citenamefont {Thomale}}]{lee2018topolectrical}%
      \BibitemOpen
      \bibfield  {author} {\bibinfo {author} {\bibfnamefont {C.~H.}\ \bibnamefont {Lee}}, \bibinfo {author} {\bibfnamefont {S.}~\bibnamefont {Imhof}}, \bibinfo {author} {\bibfnamefont {C.}~\bibnamefont {Berger}}, \bibinfo {author} {\bibfnamefont {F.}~\bibnamefont {Bayer}}, \bibinfo {author} {\bibfnamefont {J.}~\bibnamefont {Brehm}}, \bibinfo {author} {\bibfnamefont {L.~W.}\ \bibnamefont {Molenkamp}}, \bibinfo {author} {\bibfnamefont {T.}~\bibnamefont {Kiessling}}, \ and\ \bibinfo {author} {\bibfnamefont {R.}~\bibnamefont {Thomale}},\ }\href@noop {} {\bibfield  {journal} {\bibinfo  {journal} {Communications Physics}\ }\textbf {\bibinfo {volume} {1}},\ \bibinfo {pages} {1} (\bibinfo {year} {2018}{\natexlab{c}})}\BibitemShut {NoStop}%
    \bibitem [{\citenamefont {Liu}\ \emph {et~al.}(2021{\natexlab{e}})\citenamefont {Liu}, \citenamefont {Zeng}, \citenamefont {Li},\ and\ \citenamefont {Chen}}]{liu2021exact2}%
      \BibitemOpen
      \bibfield  {author} {\bibinfo {author} {\bibfnamefont {Y.}~\bibnamefont {Liu}}, \bibinfo {author} {\bibfnamefont {Y.}~\bibnamefont {Zeng}}, \bibinfo {author} {\bibfnamefont {L.}~\bibnamefont {Li}}, \ and\ \bibinfo {author} {\bibfnamefont {S.}~\bibnamefont {Chen}},\ }\href {\doibase 10.1103/PhysRevB.104.085401} {\bibfield  {journal} {\bibinfo  {journal} {Phys. Rev. B}\ }\textbf {\bibinfo {volume} {104}},\ \bibinfo {pages} {085401} (\bibinfo {year} {2021}{\natexlab{e}})}\BibitemShut {NoStop}%
    \bibitem [{\citenamefont {Ou}\ \emph {et~al.}(2023)\citenamefont {Ou}, \citenamefont {Wang},\ and\ \citenamefont {Li}}]{ou2023nonH.107.L161404}%
      \BibitemOpen
      \bibfield  {author} {\bibinfo {author} {\bibfnamefont {Z.}~\bibnamefont {Ou}}, \bibinfo {author} {\bibfnamefont {Y.}~\bibnamefont {Wang}}, \ and\ \bibinfo {author} {\bibfnamefont {L.}~\bibnamefont {Li}},\ }\href {\doibase 10.1103/PhysRevB.107.L161404} {\bibfield  {journal} {\bibinfo  {journal} {Phys. Rev. B}\ }\textbf {\bibinfo {volume} {107}},\ \bibinfo {pages} {L161404} (\bibinfo {year} {2023})}\BibitemShut {NoStop}%
    \bibitem [{\citenamefont {Okuma}(2022)}]{okuma2022boundary}%
      \BibitemOpen
      \bibfield  {author} {\bibinfo {author} {\bibfnamefont {N.}~\bibnamefont {Okuma}},\ }\href@noop {} {\bibfield  {journal} {\bibinfo  {journal} {Physical Review B}\ }\textbf {\bibinfo {volume} {105}},\ \bibinfo {pages} {224301} (\bibinfo {year} {2022})}\BibitemShut {NoStop}%
    \bibitem [{\citenamefont {Mao}\ \emph {et~al.}(2021)\citenamefont {Mao}, \citenamefont {Deng},\ and\ \citenamefont {Zhang}}]{mao2021boundary}%
      \BibitemOpen
      \bibfield  {author} {\bibinfo {author} {\bibfnamefont {L.}~\bibnamefont {Mao}}, \bibinfo {author} {\bibfnamefont {T.}~\bibnamefont {Deng}}, \ and\ \bibinfo {author} {\bibfnamefont {P.}~\bibnamefont {Zhang}},\ }\href@noop {} {\bibfield  {journal} {\bibinfo  {journal} {Physical Review B}\ }\textbf {\bibinfo {volume} {104}},\ \bibinfo {pages} {125435} (\bibinfo {year} {2021})}\BibitemShut {NoStop}%
    \bibitem [{\citenamefont {Hu}\ and\ \citenamefont {Zhao}(2021)}]{hu2021knots}%
      \BibitemOpen
      \bibfield  {author} {\bibinfo {author} {\bibfnamefont {H.}~\bibnamefont {Hu}}\ and\ \bibinfo {author} {\bibfnamefont {E.}~\bibnamefont {Zhao}},\ }\href {\doibase 10.1103/PhysRevLett.126.010401} {\bibfield  {journal} {\bibinfo  {journal} {Phys. Rev. Lett.}\ }\textbf {\bibinfo {volume} {126}},\ \bibinfo {pages} {010401} (\bibinfo {year} {2021})}\BibitemShut {NoStop}%
    \bibitem [{\citenamefont {Li}\ \emph {et~al.}(2022{\natexlab{b}})\citenamefont {Li}, \citenamefont {Ji}, \citenamefont {Chen}, \citenamefont {Yan},\ and\ \citenamefont {Yang}}]{PhysRevB.106.195425}%
      \BibitemOpen
      \bibfield  {author} {\bibinfo {author} {\bibfnamefont {Y.}~\bibnamefont {Li}}, \bibinfo {author} {\bibfnamefont {X.}~\bibnamefont {Ji}}, \bibinfo {author} {\bibfnamefont {Y.}~\bibnamefont {Chen}}, \bibinfo {author} {\bibfnamefont {X.}~\bibnamefont {Yan}}, \ and\ \bibinfo {author} {\bibfnamefont {X.}~\bibnamefont {Yang}},\ }\href {\doibase 10.1103/PhysRevB.106.195425} {\bibfield  {journal} {\bibinfo  {journal} {Phys. Rev. B}\ }\textbf {\bibinfo {volume} {106}},\ \bibinfo {pages} {195425} (\bibinfo {year} {2022}{\natexlab{b}})}\BibitemShut {NoStop}%
    \bibitem [{\citenamefont {Kauffman}(1991)}]{kauffman1991knots}%
      \BibitemOpen
      \bibfield  {author} {\bibinfo {author} {\bibfnamefont {L.~H.}\ \bibnamefont {Kauffman}},\ }\href@noop {} {\emph {\bibinfo {title} {Knots and physics}}},\ Vol.~\bibinfo {volume} {1}\ (\bibinfo  {publisher} {World scientific},\ \bibinfo {year} {1991})\BibitemShut {NoStop}%
    \bibitem [{\citenamefont {Hu}\ \emph {et~al.}(2022)\citenamefont {Hu}, \citenamefont {Sun},\ and\ \citenamefont {Chen}}]{hu2022knot}%
      \BibitemOpen
      \bibfield  {author} {\bibinfo {author} {\bibfnamefont {H.}~\bibnamefont {Hu}}, \bibinfo {author} {\bibfnamefont {S.}~\bibnamefont {Sun}}, \ and\ \bibinfo {author} {\bibfnamefont {S.}~\bibnamefont {Chen}},\ }\href@noop {} {\bibfield  {journal} {\bibinfo  {journal} {Physical Review Research}\ }\textbf {\bibinfo {volume} {4}},\ \bibinfo {pages} {L022064} (\bibinfo {year} {2022})}\BibitemShut {NoStop}%
    \bibitem [{\citenamefont {Li}\ \emph {et~al.}(2022{\natexlab{c}})\citenamefont {Li}, \citenamefont {Ding},\ and\ \citenamefont {Ma}}]{li2022non_knots}%
      \BibitemOpen
      \bibfield  {author} {\bibinfo {author} {\bibfnamefont {Z.}~\bibnamefont {Li}}, \bibinfo {author} {\bibfnamefont {K.}~\bibnamefont {Ding}}, \ and\ \bibinfo {author} {\bibfnamefont {G.}~\bibnamefont {Ma}},\ }\href@noop {} {\bibfield  {journal} {\bibinfo  {journal} {arXiv preprint arXiv:2211.13906}\ } (\bibinfo {year} {2022}{\natexlab{c}})}\BibitemShut {NoStop}%
    \bibitem [{\citenamefont {Wang}\ \emph {et~al.}(2021{\natexlab{d}})\citenamefont {Wang}, \citenamefont {Dutt}, \citenamefont {Wojcik},\ and\ \citenamefont {Fan}}]{wang2021topological}%
      \BibitemOpen
      \bibfield  {author} {\bibinfo {author} {\bibfnamefont {K.}~\bibnamefont {Wang}}, \bibinfo {author} {\bibfnamefont {A.}~\bibnamefont {Dutt}}, \bibinfo {author} {\bibfnamefont {C.~C.}\ \bibnamefont {Wojcik}}, \ and\ \bibinfo {author} {\bibfnamefont {S.}~\bibnamefont {Fan}},\ }\href@noop {} {\bibfield  {journal} {\bibinfo  {journal} {Nature}\ }\textbf {\bibinfo {volume} {598}},\ \bibinfo {pages} {59} (\bibinfo {year} {2021}{\natexlab{d}})}\BibitemShut {NoStop}%
    \bibitem [{\citenamefont {Tang}\ \emph {et~al.}(2022{\natexlab{c}})\citenamefont {Tang}, \citenamefont {Ding},\ and\ \citenamefont {Ma}}]{tang2022experimental}%
      \BibitemOpen
      \bibfield  {author} {\bibinfo {author} {\bibfnamefont {W.}~\bibnamefont {Tang}}, \bibinfo {author} {\bibfnamefont {K.}~\bibnamefont {Ding}}, \ and\ \bibinfo {author} {\bibfnamefont {G.}~\bibnamefont {Ma}},\ }\href@noop {} {\bibfield  {journal} {\bibinfo  {journal} {National Science Review}\ }\textbf {\bibinfo {volume} {9}},\ \bibinfo {pages} {nwac010} (\bibinfo {year} {2022}{\natexlab{c}})}\BibitemShut {NoStop}%
    \bibitem [{\citenamefont {Tang}\ \emph {et~al.}(2020{\natexlab{b}})\citenamefont {Tang}, \citenamefont {Jiang}, \citenamefont {Ding}, \citenamefont {Xiao}, \citenamefont {Zhang}, \citenamefont {Chan},\ and\ \citenamefont {Ma}}]{tang2020exceptional}%
      \BibitemOpen
      \bibfield  {author} {\bibinfo {author} {\bibfnamefont {W.}~\bibnamefont {Tang}}, \bibinfo {author} {\bibfnamefont {X.}~\bibnamefont {Jiang}}, \bibinfo {author} {\bibfnamefont {K.}~\bibnamefont {Ding}}, \bibinfo {author} {\bibfnamefont {Y.-X.}\ \bibnamefont {Xiao}}, \bibinfo {author} {\bibfnamefont {Z.-Q.}\ \bibnamefont {Zhang}}, \bibinfo {author} {\bibfnamefont {C.~T.}\ \bibnamefont {Chan}}, \ and\ \bibinfo {author} {\bibfnamefont {G.}~\bibnamefont {Ma}},\ }\href@noop {} {\bibfield  {journal} {\bibinfo  {journal} {Science}\ }\textbf {\bibinfo {volume} {370}},\ \bibinfo {pages} {1077} (\bibinfo {year} {2020}{\natexlab{b}})}\BibitemShut {NoStop}%
    \bibitem [{\citenamefont {Fu}\ and\ \citenamefont {Zhang}(2022)}]{fu2022anatomy}%
      \BibitemOpen
      \bibfield  {author} {\bibinfo {author} {\bibfnamefont {Y.}~\bibnamefont {Fu}}\ and\ \bibinfo {author} {\bibfnamefont {Y.}~\bibnamefont {Zhang}},\ }\href@noop {} {\bibfield  {journal} {\bibinfo  {journal} {arXiv preprint arXiv:2212.13753}\ } (\bibinfo {year} {2022})}\BibitemShut {NoStop}%
    \bibitem [{\citenamefont {Kohn}(1959)}]{kohn1959analytic}%
      \BibitemOpen
      \bibfield  {author} {\bibinfo {author} {\bibfnamefont {W.}~\bibnamefont {Kohn}},\ }\href@noop {} {\bibfield  {journal} {\bibinfo  {journal} {Physical Review}\ }\textbf {\bibinfo {volume} {115}},\ \bibinfo {pages} {809} (\bibinfo {year} {1959})}\BibitemShut {NoStop}%
    \bibitem [{\citenamefont {He}\ and\ \citenamefont {Vanderbilt}(2001)}]{he2001exponential}%
      \BibitemOpen
      \bibfield  {author} {\bibinfo {author} {\bibfnamefont {L.}~\bibnamefont {He}}\ and\ \bibinfo {author} {\bibfnamefont {D.}~\bibnamefont {Vanderbilt}},\ }\href@noop {} {\bibfield  {journal} {\bibinfo  {journal} {Phys. Rev. Lett.}\ }\textbf {\bibinfo {volume} {86}},\ \bibinfo {pages} {5341} (\bibinfo {year} {2001})}\BibitemShut {NoStop}%
    \bibitem [{\citenamefont {Brouder}\ \emph {et~al.}(2007)\citenamefont {Brouder}, \citenamefont {Panati}, \citenamefont {Calandra}, \citenamefont {Mourougane},\ and\ \citenamefont {Marzari}}]{brouder2007exponential}%
      \BibitemOpen
      \bibfield  {author} {\bibinfo {author} {\bibfnamefont {C.}~\bibnamefont {Brouder}}, \bibinfo {author} {\bibfnamefont {G.}~\bibnamefont {Panati}}, \bibinfo {author} {\bibfnamefont {M.}~\bibnamefont {Calandra}}, \bibinfo {author} {\bibfnamefont {C.}~\bibnamefont {Mourougane}}, \ and\ \bibinfo {author} {\bibfnamefont {N.}~\bibnamefont {Marzari}},\ }\href@noop {} {\bibfield  {journal} {\bibinfo  {journal} {Physical review letters}\ }\textbf {\bibinfo {volume} {98}},\ \bibinfo {pages} {046402} (\bibinfo {year} {2007})}\BibitemShut {NoStop}%
    \bibitem [{\citenamefont {Lee}\ \emph {et~al.}(2016)\citenamefont {Lee}, \citenamefont {Arovas},\ and\ \citenamefont {Thomale}}]{lee2016band}%
      \BibitemOpen
      \bibfield  {author} {\bibinfo {author} {\bibfnamefont {C.~H.}\ \bibnamefont {Lee}}, \bibinfo {author} {\bibfnamefont {D.~P.}\ \bibnamefont {Arovas}}, \ and\ \bibinfo {author} {\bibfnamefont {R.}~\bibnamefont {Thomale}},\ }\href@noop {} {\bibfield  {journal} {\bibinfo  {journal} {Phys. Rev. B}\ }\textbf {\bibinfo {volume} {93}},\ \bibinfo {pages} {155155} (\bibinfo {year} {2016})}\BibitemShut {NoStop}%
    \bibitem [{\citenamefont {Monaco}\ \emph {et~al.}(2018)\citenamefont {Monaco}, \citenamefont {Panati}, \citenamefont {Pisante},\ and\ \citenamefont {Teufel}}]{monaco2018optimal}%
      \BibitemOpen
      \bibfield  {author} {\bibinfo {author} {\bibfnamefont {D.}~\bibnamefont {Monaco}}, \bibinfo {author} {\bibfnamefont {G.}~\bibnamefont {Panati}}, \bibinfo {author} {\bibfnamefont {A.}~\bibnamefont {Pisante}}, \ and\ \bibinfo {author} {\bibfnamefont {S.}~\bibnamefont {Teufel}},\ }\href@noop {} {\bibfield  {journal} {\bibinfo  {journal} {Communications in Mathematical Physics}\ }\textbf {\bibinfo {volume} {359}},\ \bibinfo {pages} {61} (\bibinfo {year} {2018})}\BibitemShut {NoStop}%
    \bibitem [{\citenamefont {Qin}\ \emph {et~al.}(2023{\natexlab{a}})\citenamefont {Qin}, \citenamefont {Shen}, \citenamefont {Li},\ and\ \citenamefont {Lee}}]{qin2023kinked}%
      \BibitemOpen
      \bibfield  {author} {\bibinfo {author} {\bibfnamefont {F.}~\bibnamefont {Qin}}, \bibinfo {author} {\bibfnamefont {R.}~\bibnamefont {Shen}}, \bibinfo {author} {\bibfnamefont {L.}~\bibnamefont {Li}}, \ and\ \bibinfo {author} {\bibfnamefont {C.~H.}\ \bibnamefont {Lee}},\ }\href@noop {} {\bibfield  {journal} {\bibinfo  {journal} {arXiv preprint arXiv:2306.13139}\ } (\bibinfo {year} {2023}{\natexlab{a}})}\BibitemShut {NoStop}%
    \bibitem [{\citenamefont {Longhi}(2020{\natexlab{a}})}]{PhysRevLett.124.066602}%
      \BibitemOpen
      \bibfield  {author} {\bibinfo {author} {\bibfnamefont {S.}~\bibnamefont {Longhi}},\ }\href {\doibase 10.1103/PhysRevLett.124.066602} {\bibfield  {journal} {\bibinfo  {journal} {Phys. Rev. Lett.}\ }\textbf {\bibinfo {volume} {124}},\ \bibinfo {pages} {066602} (\bibinfo {year} {2020}{\natexlab{a}})}\BibitemShut {NoStop}%
    \bibitem [{\citenamefont {Li}\ and\ \citenamefont {Lee}(2022)}]{li2022non}%
      \BibitemOpen
      \bibfield  {author} {\bibinfo {author} {\bibfnamefont {L.}~\bibnamefont {Li}}\ and\ \bibinfo {author} {\bibfnamefont {C.~H.}\ \bibnamefont {Lee}},\ }\href@noop {} {\bibfield  {journal} {\bibinfo  {journal} {Science Bulletin}\ } (\bibinfo {year} {2022})}\BibitemShut {NoStop}%
    \bibitem [{\citenamefont {Qin}\ \emph {et~al.}(2023{\natexlab{b}})\citenamefont {Qin}, \citenamefont {Ma}, \citenamefont {Shen}, \citenamefont {Lee} \emph {et~al.}}]{qin2022universal}%
      \BibitemOpen
      \bibfield  {author} {\bibinfo {author} {\bibfnamefont {F.}~\bibnamefont {Qin}}, \bibinfo {author} {\bibfnamefont {Y.}~\bibnamefont {Ma}}, \bibinfo {author} {\bibfnamefont {R.}~\bibnamefont {Shen}}, \bibinfo {author} {\bibfnamefont {C.~H.}\ \bibnamefont {Lee}},  \emph {et~al.},\ }\href@noop {} {\bibfield  {journal} {\bibinfo  {journal} {Physical Review B}\ }\textbf {\bibinfo {volume} {107}},\ \bibinfo {pages} {155430} (\bibinfo {year} {2023}{\natexlab{b}})}\BibitemShut {NoStop}%
    \bibitem [{\citenamefont {Rafi-Ul-Islam}\ \emph {et~al.}(2022{\natexlab{b}})\citenamefont {Rafi-Ul-Islam}, \citenamefont {Siu}, \citenamefont {Sahin}, \citenamefont {Lee},\ and\ \citenamefont {Jalil}}]{rafi2022critical}%
      \BibitemOpen
      \bibfield  {author} {\bibinfo {author} {\bibfnamefont {S.}~\bibnamefont {Rafi-Ul-Islam}}, \bibinfo {author} {\bibfnamefont {Z.~B.}\ \bibnamefont {Siu}}, \bibinfo {author} {\bibfnamefont {H.}~\bibnamefont {Sahin}}, \bibinfo {author} {\bibfnamefont {C.~H.}\ \bibnamefont {Lee}}, \ and\ \bibinfo {author} {\bibfnamefont {M.~B.}\ \bibnamefont {Jalil}},\ }\href@noop {} {\bibfield  {journal} {\bibinfo  {journal} {Physical Review Research}\ }\textbf {\bibinfo {volume} {4}},\ \bibinfo {pages} {013243} (\bibinfo {year} {2022}{\natexlab{b}})}\BibitemShut {NoStop}%
    \bibitem [{\citenamefont {Rafi-Ul-Islam}\ \emph {et~al.}(2022{\natexlab{c}})\citenamefont {Rafi-Ul-Islam}, \citenamefont {Siu}, \citenamefont {Sahin}, \citenamefont {Lee},\ and\ \citenamefont {Jalil}}]{rafi2022system}%
      \BibitemOpen
      \bibfield  {author} {\bibinfo {author} {\bibfnamefont {S.}~\bibnamefont {Rafi-Ul-Islam}}, \bibinfo {author} {\bibfnamefont {Z.~B.}\ \bibnamefont {Siu}}, \bibinfo {author} {\bibfnamefont {H.}~\bibnamefont {Sahin}}, \bibinfo {author} {\bibfnamefont {C.~H.}\ \bibnamefont {Lee}}, \ and\ \bibinfo {author} {\bibfnamefont {M.~B.}\ \bibnamefont {Jalil}},\ }\href@noop {} {\bibfield  {journal} {\bibinfo  {journal} {Physical Review B}\ }\textbf {\bibinfo {volume} {106}},\ \bibinfo {pages} {075158} (\bibinfo {year} {2022}{\natexlab{c}})}\BibitemShut {NoStop}%
    \bibitem [{\citenamefont {Sun}\ \emph {et~al.}(2022)\citenamefont {Sun}, \citenamefont {Tang},\ and\ \citenamefont {Kou}}]{sun2022biorthogonal}%
      \BibitemOpen
      \bibfield  {author} {\bibinfo {author} {\bibfnamefont {G.}~\bibnamefont {Sun}}, \bibinfo {author} {\bibfnamefont {J.-C.}\ \bibnamefont {Tang}}, \ and\ \bibinfo {author} {\bibfnamefont {S.-P.}\ \bibnamefont {Kou}},\ }\href@noop {} {\bibfield  {journal} {\bibinfo  {journal} {Frontiers of Physics}\ }\textbf {\bibinfo {volume} {17}},\ \bibinfo {pages} {1} (\bibinfo {year} {2022})}\BibitemShut {NoStop}%
    \bibitem [{\citenamefont {Bao}\ \emph {et~al.}(2021)\citenamefont {Bao}, \citenamefont {Guo}, \citenamefont {Du}, \citenamefont {Gu},\ and\ \citenamefont {Tan}}]{bao2021topological}%
      \BibitemOpen
      \bibfield  {author} {\bibinfo {author} {\bibfnamefont {X.-X.}\ \bibnamefont {Bao}}, \bibinfo {author} {\bibfnamefont {G.-F.}\ \bibnamefont {Guo}}, \bibinfo {author} {\bibfnamefont {X.-P.}\ \bibnamefont {Du}}, \bibinfo {author} {\bibfnamefont {H.-Q.}\ \bibnamefont {Gu}}, \ and\ \bibinfo {author} {\bibfnamefont {L.}~\bibnamefont {Tan}},\ }\href@noop {} {\bibfield  {journal} {\bibinfo  {journal} {Journal of Physics: Condensed Matter}\ }\textbf {\bibinfo {volume} {33}},\ \bibinfo {pages} {185401} (\bibinfo {year} {2021})}\BibitemShut {NoStop}%
    \bibitem [{\citenamefont {Arouca}\ \emph {et~al.}(2020)\citenamefont {Arouca}, \citenamefont {Lee},\ and\ \citenamefont {Smith}}]{arouca2020unconventional}%
      \BibitemOpen
      \bibfield  {author} {\bibinfo {author} {\bibfnamefont {R.}~\bibnamefont {Arouca}}, \bibinfo {author} {\bibfnamefont {C.}~\bibnamefont {Lee}}, \ and\ \bibinfo {author} {\bibfnamefont {C.~M.}\ \bibnamefont {Smith}},\ }\href@noop {} {\bibfield  {journal} {\bibinfo  {journal} {Physical Review B}\ }\textbf {\bibinfo {volume} {102}},\ \bibinfo {pages} {245145} (\bibinfo {year} {2020})}\BibitemShut {NoStop}%
    \bibitem [{\citenamefont {Aquino}\ \emph {et~al.}(2022)\citenamefont {Aquino}, \citenamefont {Lopes},\ and\ \citenamefont {Barci}}]{aquino2022critical}%
      \BibitemOpen
      \bibfield  {author} {\bibinfo {author} {\bibfnamefont {R.}~\bibnamefont {Aquino}}, \bibinfo {author} {\bibfnamefont {N.}~\bibnamefont {Lopes}}, \ and\ \bibinfo {author} {\bibfnamefont {D.~G.}\ \bibnamefont {Barci}},\ }\href@noop {} {\bibfield  {journal} {\bibinfo  {journal} {arXiv preprint arXiv:2208.14400}\ } (\bibinfo {year} {2022})}\BibitemShut {NoStop}%
    \bibitem [{\citenamefont {Rahul}\ and\ \citenamefont {Sarkar}(2022)}]{rahul2022topological}%
      \BibitemOpen
      \bibfield  {author} {\bibinfo {author} {\bibfnamefont {S.}~\bibnamefont {Rahul}}\ and\ \bibinfo {author} {\bibfnamefont {S.}~\bibnamefont {Sarkar}},\ }\href@noop {} {\bibfield  {journal} {\bibinfo  {journal} {Scientific Reports}\ }\textbf {\bibinfo {volume} {12}},\ \bibinfo {pages} {1} (\bibinfo {year} {2022})}\BibitemShut {NoStop}%
    \bibitem [{\citenamefont {Jian}\ \emph {et~al.}(2021)\citenamefont {Jian}, \citenamefont {Yang}, \citenamefont {Bi},\ and\ \citenamefont {Chen}}]{jian2021yang}%
      \BibitemOpen
      \bibfield  {author} {\bibinfo {author} {\bibfnamefont {S.-K.}\ \bibnamefont {Jian}}, \bibinfo {author} {\bibfnamefont {Z.-C.}\ \bibnamefont {Yang}}, \bibinfo {author} {\bibfnamefont {Z.}~\bibnamefont {Bi}}, \ and\ \bibinfo {author} {\bibfnamefont {X.}~\bibnamefont {Chen}},\ }\href@noop {} {\bibfield  {journal} {\bibinfo  {journal} {Physical Review B}\ }\textbf {\bibinfo {volume} {104}},\ \bibinfo {pages} {L161107} (\bibinfo {year} {2021})}\BibitemShut {NoStop}%
    \bibitem [{\citenamefont {Shen}\ \emph {et~al.}(2023{\natexlab{a}})\citenamefont {Shen}, \citenamefont {Chen}, \citenamefont {Aliyu}, \citenamefont {Qin}, \citenamefont {Zhong}, \citenamefont {Loh}, \citenamefont {Lee} \emph {et~al.}}]{shen2023proposal}%
      \BibitemOpen
      \bibfield  {author} {\bibinfo {author} {\bibfnamefont {R.}~\bibnamefont {Shen}}, \bibinfo {author} {\bibfnamefont {T.}~\bibnamefont {Chen}}, \bibinfo {author} {\bibfnamefont {M.~M.}\ \bibnamefont {Aliyu}}, \bibinfo {author} {\bibfnamefont {F.}~\bibnamefont {Qin}}, \bibinfo {author} {\bibfnamefont {Y.}~\bibnamefont {Zhong}}, \bibinfo {author} {\bibfnamefont {H.}~\bibnamefont {Loh}}, \bibinfo {author} {\bibfnamefont {C.~H.}\ \bibnamefont {Lee}},  \emph {et~al.},\ }\href@noop {} {\bibfield  {journal} {\bibinfo  {journal} {Physical Review Letters}\ }\textbf {\bibinfo {volume} {131}},\ \bibinfo {pages} {080403} (\bibinfo {year} {2023}{\natexlab{a}})}\BibitemShut {NoStop}%
    \bibitem [{\citenamefont {Zhou}\ \emph {et~al.}(2020)\citenamefont {Zhou}, \citenamefont {Wang},\ and\ \citenamefont {Wang}}]{PhysRevB.102.205116}%
      \BibitemOpen
      \bibfield  {author} {\bibinfo {author} {\bibfnamefont {B.}~\bibnamefont {Zhou}}, \bibinfo {author} {\bibfnamefont {R.}~\bibnamefont {Wang}}, \ and\ \bibinfo {author} {\bibfnamefont {B.}~\bibnamefont {Wang}},\ }\href {\doibase 10.1103/PhysRevB.102.205116} {\bibfield  {journal} {\bibinfo  {journal} {Phys. Rev. B}\ }\textbf {\bibinfo {volume} {102}},\ \bibinfo {pages} {205116} (\bibinfo {year} {2020})}\BibitemShut {NoStop}%
    \bibitem [{\citenamefont {Yin}\ \emph {et~al.}(2017)\citenamefont {Yin}, \citenamefont {Huang}, \citenamefont {Lo},\ and\ \citenamefont {Chen}}]{yin2017KZ}%
      \BibitemOpen
      \bibfield  {author} {\bibinfo {author} {\bibfnamefont {S.}~\bibnamefont {Yin}}, \bibinfo {author} {\bibfnamefont {G.-Y.}\ \bibnamefont {Huang}}, \bibinfo {author} {\bibfnamefont {C.-Y.}\ \bibnamefont {Lo}}, \ and\ \bibinfo {author} {\bibfnamefont {P.}~\bibnamefont {Chen}},\ }\href {\doibase 10.1103/PhysRevLett.118.065701} {\bibfield  {journal} {\bibinfo  {journal} {Phys. Rev. Lett.}\ }\textbf {\bibinfo {volume} {118}},\ \bibinfo {pages} {065701} (\bibinfo {year} {2017})}\BibitemShut {NoStop}%
    \bibitem [{\citenamefont {Okuma}\ and\ \citenamefont {Sato}(2021)}]{okuma2021quantum}%
      \BibitemOpen
      \bibfield  {author} {\bibinfo {author} {\bibfnamefont {N.}~\bibnamefont {Okuma}}\ and\ \bibinfo {author} {\bibfnamefont {M.}~\bibnamefont {Sato}},\ }\href@noop {} {\bibfield  {journal} {\bibinfo  {journal} {Physical Review B}\ }\textbf {\bibinfo {volume} {103}},\ \bibinfo {pages} {085428} (\bibinfo {year} {2021})}\BibitemShut {NoStop}%
    \bibitem [{\citenamefont {Kawabata}\ \emph {et~al.}(2022{\natexlab{a}})\citenamefont {Kawabata}, \citenamefont {Numasawa},\ and\ \citenamefont {Ryu}}]{kawabata2022entanglement}%
      \BibitemOpen
      \bibfield  {author} {\bibinfo {author} {\bibfnamefont {K.}~\bibnamefont {Kawabata}}, \bibinfo {author} {\bibfnamefont {T.}~\bibnamefont {Numasawa}}, \ and\ \bibinfo {author} {\bibfnamefont {S.}~\bibnamefont {Ryu}},\ }\href@noop {} {\bibfield  {journal} {\bibinfo  {journal} {arXiv preprint arXiv:2206.05384}\ } (\bibinfo {year} {2022}{\natexlab{a}})}\BibitemShut {NoStop}%
    \bibitem [{\citenamefont {Lee}(2022)}]{lee2022exceptional}%
      \BibitemOpen
      \bibfield  {author} {\bibinfo {author} {\bibfnamefont {C.~H.}\ \bibnamefont {Lee}},\ }\href@noop {} {\bibfield  {journal} {\bibinfo  {journal} {Physical Review Letters}\ }\textbf {\bibinfo {volume} {128}},\ \bibinfo {pages} {010402} (\bibinfo {year} {2022})}\BibitemShut {NoStop}%
    \bibitem [{\citenamefont {Peschel}\ and\ \citenamefont {Eisler}(2009)}]{peschel2009reduced}%
      \BibitemOpen
      \bibfield  {author} {\bibinfo {author} {\bibfnamefont {I.}~\bibnamefont {Peschel}}\ and\ \bibinfo {author} {\bibfnamefont {V.}~\bibnamefont {Eisler}},\ }\href@noop {} {\bibfield  {journal} {\bibinfo  {journal} {Journal of physics a: mathematical and theoretical}\ }\textbf {\bibinfo {volume} {42}},\ \bibinfo {pages} {504003} (\bibinfo {year} {2009})}\BibitemShut {NoStop}%
    \bibitem [{\citenamefont {Qi}(2011)}]{qi2011generic}%
      \BibitemOpen
      \bibfield  {author} {\bibinfo {author} {\bibfnamefont {X.-L.}\ \bibnamefont {Qi}},\ }\href@noop {} {\bibfield  {journal} {\bibinfo  {journal} {Phys. Rev. Lett.}\ }\textbf {\bibinfo {volume} {107}},\ \bibinfo {pages} {126803} (\bibinfo {year} {2011})}\BibitemShut {NoStop}%
    \bibitem [{\citenamefont {Hughes}\ \emph {et~al.}(2011)\citenamefont {Hughes}, \citenamefont {Prodan},\ and\ \citenamefont {Bernevig}}]{hughes2011inversion}%
      \BibitemOpen
      \bibfield  {author} {\bibinfo {author} {\bibfnamefont {T.~L.}\ \bibnamefont {Hughes}}, \bibinfo {author} {\bibfnamefont {E.}~\bibnamefont {Prodan}}, \ and\ \bibinfo {author} {\bibfnamefont {B.~A.}\ \bibnamefont {Bernevig}},\ }\href@noop {} {\bibfield  {journal} {\bibinfo  {journal} {Physical Review B}\ }\textbf {\bibinfo {volume} {83}},\ \bibinfo {pages} {245132} (\bibinfo {year} {2011})}\BibitemShut {NoStop}%
    \bibitem [{\citenamefont {Alexandradinata}\ \emph {et~al.}(2011)\citenamefont {Alexandradinata}, \citenamefont {Hughes},\ and\ \citenamefont {Bernevig}}]{alexandradinata2011trace}%
      \BibitemOpen
      \bibfield  {author} {\bibinfo {author} {\bibfnamefont {A.}~\bibnamefont {Alexandradinata}}, \bibinfo {author} {\bibfnamefont {T.~L.}\ \bibnamefont {Hughes}}, \ and\ \bibinfo {author} {\bibfnamefont {B.~A.}\ \bibnamefont {Bernevig}},\ }\href@noop {} {\bibfield  {journal} {\bibinfo  {journal} {Phys. Rev. B}\ }\textbf {\bibinfo {volume} {84}},\ \bibinfo {pages} {195103} (\bibinfo {year} {2011})}\BibitemShut {NoStop}%
    \bibitem [{\citenamefont {Lee}\ and\ \citenamefont {Ye}(2015)}]{lee2015free}%
      \BibitemOpen
      \bibfield  {author} {\bibinfo {author} {\bibfnamefont {C.~H.}\ \bibnamefont {Lee}}\ and\ \bibinfo {author} {\bibfnamefont {P.}~\bibnamefont {Ye}},\ }\href@noop {} {\bibfield  {journal} {\bibinfo  {journal} {Physical Review B}\ }\textbf {\bibinfo {volume} {91}},\ \bibinfo {pages} {085119} (\bibinfo {year} {2015})}\BibitemShut {NoStop}%
    \bibitem [{\citenamefont {Herviou}\ \emph {et~al.}(2019{\natexlab{b}})\citenamefont {Herviou}, \citenamefont {Regnault},\ and\ \citenamefont {Bardarson}}]{herviou2019entanglement}%
      \BibitemOpen
      \bibfield  {author} {\bibinfo {author} {\bibfnamefont {L.}~\bibnamefont {Herviou}}, \bibinfo {author} {\bibfnamefont {N.}~\bibnamefont {Regnault}}, \ and\ \bibinfo {author} {\bibfnamefont {J.~H.}\ \bibnamefont {Bardarson}},\ }\href@noop {} {\bibfield  {journal} {\bibinfo  {journal} {SciPost Physics}\ }\textbf {\bibinfo {volume} {7}},\ \bibinfo {pages} {069} (\bibinfo {year} {2019}{\natexlab{b}})}\BibitemShut {NoStop}%
    \bibitem [{\citenamefont {Chen}\ \emph {et~al.}(2021)\citenamefont {Chen}, \citenamefont {Chen},\ and\ \citenamefont {Ye}}]{chen2021entanglement}%
      \BibitemOpen
      \bibfield  {author} {\bibinfo {author} {\bibfnamefont {L.-M.}\ \bibnamefont {Chen}}, \bibinfo {author} {\bibfnamefont {S.~A.}\ \bibnamefont {Chen}}, \ and\ \bibinfo {author} {\bibfnamefont {P.}~\bibnamefont {Ye}},\ }\href@noop {} {\bibfield  {journal} {\bibinfo  {journal} {SciPost Physics}\ }\textbf {\bibinfo {volume} {11}},\ \bibinfo {pages} {003} (\bibinfo {year} {2021})}\BibitemShut {NoStop}%
    \bibitem [{\citenamefont {Chang}\ \emph {et~al.}(2020)\citenamefont {Chang}, \citenamefont {You}, \citenamefont {Wen},\ and\ \citenamefont {Ryu}}]{chang2020entanglement}%
      \BibitemOpen
      \bibfield  {author} {\bibinfo {author} {\bibfnamefont {P.-Y.}\ \bibnamefont {Chang}}, \bibinfo {author} {\bibfnamefont {J.-S.}\ \bibnamefont {You}}, \bibinfo {author} {\bibfnamefont {X.}~\bibnamefont {Wen}}, \ and\ \bibinfo {author} {\bibfnamefont {S.}~\bibnamefont {Ryu}},\ }\href@noop {} {\bibfield  {journal} {\bibinfo  {journal} {Physical Review Research}\ }\textbf {\bibinfo {volume} {2}},\ \bibinfo {pages} {033069} (\bibinfo {year} {2020})}\BibitemShut {NoStop}%
    \bibitem [{\citenamefont {Xue}\ and\ \citenamefont {Lee}(2024)}]{xue2024topologically}%
      \BibitemOpen
      \bibfield  {author} {\bibinfo {author} {\bibfnamefont {W.-T.}\ \bibnamefont {Xue}}\ and\ \bibinfo {author} {\bibfnamefont {C.~H.}\ \bibnamefont {Lee}},\ }\href@noop {} {\bibfield  {journal} {\bibinfo  {journal} {arXiv preprint arXiv:2403.03259}\ } (\bibinfo {year} {2024})}\BibitemShut {NoStop}%
    \bibitem [{\citenamefont {Zou}\ \emph {et~al.}(2023)\citenamefont {Zou}, \citenamefont {Chen}, \citenamefont {Meng}, \citenamefont {Ang}, \citenamefont {Zhang},\ and\ \citenamefont {Lee}}]{zou2023experimental}%
      \BibitemOpen
      \bibfield  {author} {\bibinfo {author} {\bibfnamefont {D.}~\bibnamefont {Zou}}, \bibinfo {author} {\bibfnamefont {T.}~\bibnamefont {Chen}}, \bibinfo {author} {\bibfnamefont {H.}~\bibnamefont {Meng}}, \bibinfo {author} {\bibfnamefont {Y.~S.}\ \bibnamefont {Ang}}, \bibinfo {author} {\bibfnamefont {X.}~\bibnamefont {Zhang}}, \ and\ \bibinfo {author} {\bibfnamefont {C.~H.}\ \bibnamefont {Lee}},\ }\href@noop {} {\bibfield  {journal} {\bibinfo  {journal} {arXiv preprint arXiv:2308.01970}\ } (\bibinfo {year} {2023})}\BibitemShut {NoStop}%
    \bibitem [{\citenamefont {Li}\ and\ \citenamefont {Wan}(2022)}]{li2022wave}%
      \BibitemOpen
      \bibfield  {author} {\bibinfo {author} {\bibfnamefont {H.}~\bibnamefont {Li}}\ and\ \bibinfo {author} {\bibfnamefont {S.}~\bibnamefont {Wan}},\ }\href {\doibase 10.1103/PhysRevB.106.L241112} {\bibfield  {journal} {\bibinfo  {journal} {Phys. Rev. B}\ }\textbf {\bibinfo {volume} {106}},\ \bibinfo {pages} {L241112} (\bibinfo {year} {2022})}\BibitemShut {NoStop}%
    \bibitem [{\citenamefont {Longhi}(2022{\natexlab{c}})}]{PhysRevB.105.245143}%
      \BibitemOpen
      \bibfield  {author} {\bibinfo {author} {\bibfnamefont {S.}~\bibnamefont {Longhi}},\ }\href {\doibase 10.1103/PhysRevB.105.245143} {\bibfield  {journal} {\bibinfo  {journal} {Phys. Rev. B}\ }\textbf {\bibinfo {volume} {105}},\ \bibinfo {pages} {245143} (\bibinfo {year} {2022}{\natexlab{c}})}\BibitemShut {NoStop}%
    \bibitem [{\citenamefont {Li}\ \emph {et~al.}(2021{\natexlab{c}})\citenamefont {Li}, \citenamefont {Sun}, \citenamefont {Zhang},\ and\ \citenamefont {Yi}}]{PhysRevResearch3023022}%
      \BibitemOpen
      \bibfield  {author} {\bibinfo {author} {\bibfnamefont {T.}~\bibnamefont {Li}}, \bibinfo {author} {\bibfnamefont {J.-Z.}\ \bibnamefont {Sun}}, \bibinfo {author} {\bibfnamefont {Y.-S.}\ \bibnamefont {Zhang}}, \ and\ \bibinfo {author} {\bibfnamefont {W.}~\bibnamefont {Yi}},\ }\href {\doibase 10.1103/PhysRevResearch.3.023022} {\bibfield  {journal} {\bibinfo  {journal} {Phys. Rev. Res.}\ }\textbf {\bibinfo {volume} {3}},\ \bibinfo {pages} {023022} (\bibinfo {year} {2021}{\natexlab{c}})}\BibitemShut {NoStop}%
    \bibitem [{\citenamefont {Lee}\ and\ \citenamefont {Longhi}(2020)}]{lee2020ultrafast}%
      \BibitemOpen
      \bibfield  {author} {\bibinfo {author} {\bibfnamefont {C.~H.}\ \bibnamefont {Lee}}\ and\ \bibinfo {author} {\bibfnamefont {S.}~\bibnamefont {Longhi}},\ }\href@noop {} {\bibfield  {journal} {\bibinfo  {journal} {Communications Physics}\ }\textbf {\bibinfo {volume} {3}},\ \bibinfo {pages} {1} (\bibinfo {year} {2020})}\BibitemShut {NoStop}%
    \bibitem [{\citenamefont {Li}\ \emph {et~al.}(2022{\natexlab{d}})\citenamefont {Li}, \citenamefont {Teo}, \citenamefont {Mu},\ and\ \citenamefont {Gong}}]{li2022direction}%
      \BibitemOpen
      \bibfield  {author} {\bibinfo {author} {\bibfnamefont {L.}~\bibnamefont {Li}}, \bibinfo {author} {\bibfnamefont {W.~X.}\ \bibnamefont {Teo}}, \bibinfo {author} {\bibfnamefont {S.}~\bibnamefont {Mu}}, \ and\ \bibinfo {author} {\bibfnamefont {J.}~\bibnamefont {Gong}},\ }\href {\doibase 10.1103/PhysRevB.106.085427} {\bibfield  {journal} {\bibinfo  {journal} {Phys. Rev. B}\ }\textbf {\bibinfo {volume} {106}},\ \bibinfo {pages} {085427} (\bibinfo {year} {2022}{\natexlab{d}})}\BibitemShut {NoStop}%
    \bibitem [{\citenamefont {Peng}\ \emph {et~al.}(2022{\natexlab{b}})\citenamefont {Peng}, \citenamefont {Jie}, \citenamefont {Yu},\ and\ \citenamefont {Wang}}]{peng2022manipulating}%
      \BibitemOpen
      \bibfield  {author} {\bibinfo {author} {\bibfnamefont {Y.}~\bibnamefont {Peng}}, \bibinfo {author} {\bibfnamefont {J.}~\bibnamefont {Jie}}, \bibinfo {author} {\bibfnamefont {D.}~\bibnamefont {Yu}}, \ and\ \bibinfo {author} {\bibfnamefont {Y.}~\bibnamefont {Wang}},\ }\href {\doibase 10.1103/PhysRevB.106.L161402} {\bibfield  {journal} {\bibinfo  {journal} {Phys. Rev. B}\ }\textbf {\bibinfo {volume} {106}},\ \bibinfo {pages} {L161402} (\bibinfo {year} {2022}{\natexlab{b}})}\BibitemShut {NoStop}%
    \bibitem [{\citenamefont {Zhang}\ and\ \citenamefont {Gong}(2020)}]{zhang2020non}%
      \BibitemOpen
      \bibfield  {author} {\bibinfo {author} {\bibfnamefont {X.}~\bibnamefont {Zhang}}\ and\ \bibinfo {author} {\bibfnamefont {J.}~\bibnamefont {Gong}},\ }\href@noop {} {\bibfield  {journal} {\bibinfo  {journal} {Physical Review B}\ }\textbf {\bibinfo {volume} {101}},\ \bibinfo {pages} {045415} (\bibinfo {year} {2020})}\BibitemShut {NoStop}%
    \bibitem [{\citenamefont {Longhi}(2019{\natexlab{c}})}]{longhi2019probing}%
      \BibitemOpen
      \bibfield  {author} {\bibinfo {author} {\bibfnamefont {S.}~\bibnamefont {Longhi}},\ }\href@noop {} {\bibfield  {journal} {\bibinfo  {journal} {Physical Review Research}\ }\textbf {\bibinfo {volume} {1}},\ \bibinfo {pages} {023013} (\bibinfo {year} {2019}{\natexlab{c}})}\BibitemShut {NoStop}%
    \bibitem [{\citenamefont {Yang}\ \emph {et~al.}(2022{\natexlab{c}})\citenamefont {Yang}, \citenamefont {Jiang},\ and\ \citenamefont {Bergholtz}}]{PhysRevResearch.4.023160}%
      \BibitemOpen
      \bibfield  {author} {\bibinfo {author} {\bibfnamefont {F.}~\bibnamefont {Yang}}, \bibinfo {author} {\bibfnamefont {Q.-D.}\ \bibnamefont {Jiang}}, \ and\ \bibinfo {author} {\bibfnamefont {E.~J.}\ \bibnamefont {Bergholtz}},\ }\href {\doibase 10.1103/PhysRevResearch.4.023160} {\bibfield  {journal} {\bibinfo  {journal} {Phys. Rev. Res.}\ }\textbf {\bibinfo {volume} {4}},\ \bibinfo {pages} {023160} (\bibinfo {year} {2022}{\natexlab{c}})}\BibitemShut {NoStop}%
    \bibitem [{\citenamefont {Longhi}(2020{\natexlab{b}})}]{PhysRevB.102.201103}%
      \BibitemOpen
      \bibfield  {author} {\bibinfo {author} {\bibfnamefont {S.}~\bibnamefont {Longhi}},\ }\href {\doibase 10.1103/PhysRevB.102.201103} {\bibfield  {journal} {\bibinfo  {journal} {Phys. Rev. B}\ }\textbf {\bibinfo {volume} {102}},\ \bibinfo {pages} {201103} (\bibinfo {year} {2020}{\natexlab{b}})}\BibitemShut {NoStop}%
    \bibitem [{\citenamefont {Longhi}(2020{\natexlab{c}})}]{longhi2020stochastic}%
      \BibitemOpen
      \bibfield  {author} {\bibinfo {author} {\bibfnamefont {S.}~\bibnamefont {Longhi}},\ }\href@noop {} {\bibfield  {journal} {\bibinfo  {journal} {Optics Letters}\ }\textbf {\bibinfo {volume} {45}},\ \bibinfo {pages} {5250} (\bibinfo {year} {2020}{\natexlab{c}})}\BibitemShut {NoStop}%
    \bibitem [{\citenamefont {Longhi}(2021{\natexlab{a}})}]{longhi2021bulk}%
      \BibitemOpen
      \bibfield  {author} {\bibinfo {author} {\bibfnamefont {S.}~\bibnamefont {Longhi}},\ }\href@noop {} {\bibfield  {journal} {\bibinfo  {journal} {Optics Letters}\ }\textbf {\bibinfo {volume} {46}},\ \bibinfo {pages} {6107} (\bibinfo {year} {2021}{\natexlab{a}})}\BibitemShut {NoStop}%
    \bibitem [{\citenamefont {Longhi}(2022{\natexlab{d}})}]{longhi2022hartman}%
      \BibitemOpen
      \bibfield  {author} {\bibinfo {author} {\bibfnamefont {S.}~\bibnamefont {Longhi}},\ }\href@noop {} {\bibfield  {journal} {\bibinfo  {journal} {Annalen der Physik}\ ,\ \bibinfo {pages} {2200250}} (\bibinfo {year} {2022}{\natexlab{d}})}\BibitemShut {NoStop}%
    \bibitem [{\citenamefont {Stegmaier}\ \emph {et~al.}(2021)\citenamefont {Stegmaier}, \citenamefont {Imhof}, \citenamefont {Helbig}, \citenamefont {Hofmann}, \citenamefont {Lee}, \citenamefont {Kremer}, \citenamefont {Fritzsche}, \citenamefont {Feichtner}, \citenamefont {Klembt}, \citenamefont {H{\"o}fling} \emph {et~al.}}]{stegmaier2021topological}%
      \BibitemOpen
      \bibfield  {author} {\bibinfo {author} {\bibfnamefont {A.}~\bibnamefont {Stegmaier}}, \bibinfo {author} {\bibfnamefont {S.}~\bibnamefont {Imhof}}, \bibinfo {author} {\bibfnamefont {T.}~\bibnamefont {Helbig}}, \bibinfo {author} {\bibfnamefont {T.}~\bibnamefont {Hofmann}}, \bibinfo {author} {\bibfnamefont {C.~H.}\ \bibnamefont {Lee}}, \bibinfo {author} {\bibfnamefont {M.}~\bibnamefont {Kremer}}, \bibinfo {author} {\bibfnamefont {A.}~\bibnamefont {Fritzsche}}, \bibinfo {author} {\bibfnamefont {T.}~\bibnamefont {Feichtner}}, \bibinfo {author} {\bibfnamefont {S.}~\bibnamefont {Klembt}}, \bibinfo {author} {\bibfnamefont {S.}~\bibnamefont {H{\"o}fling}},  \emph {et~al.},\ }\href@noop {} {\bibfield  {journal} {\bibinfo  {journal} {Physical Review Letters}\ }\textbf {\bibinfo {volume} {126}},\ \bibinfo {pages} {215302} (\bibinfo {year} {2021})}\BibitemShut {NoStop}%
    \bibitem [{\citenamefont {Longhi}(2021{\natexlab{b}})}]{PhysRevB.103.054203}%
      \BibitemOpen
      \bibfield  {author} {\bibinfo {author} {\bibfnamefont {S.}~\bibnamefont {Longhi}},\ }\href {\doibase 10.1103/PhysRevB.103.054203} {\bibfield  {journal} {\bibinfo  {journal} {Phys. Rev. B}\ }\textbf {\bibinfo {volume} {103}},\ \bibinfo {pages} {054203} (\bibinfo {year} {2021}{\natexlab{b}})}\BibitemShut {NoStop}%
    \bibitem [{\citenamefont {Suthar}\ \emph {et~al.}(2022)\citenamefont {Suthar}, \citenamefont {Wang}, \citenamefont {Huang}, \citenamefont {Jen},\ and\ \citenamefont {You}}]{suthar2022non}%
      \BibitemOpen
      \bibfield  {author} {\bibinfo {author} {\bibfnamefont {K.}~\bibnamefont {Suthar}}, \bibinfo {author} {\bibfnamefont {Y.-C.}\ \bibnamefont {Wang}}, \bibinfo {author} {\bibfnamefont {Y.-P.}\ \bibnamefont {Huang}}, \bibinfo {author} {\bibfnamefont {H.~H.}\ \bibnamefont {Jen}}, \ and\ \bibinfo {author} {\bibfnamefont {J.-S.}\ \bibnamefont {You}},\ }\href {\doibase 10.1103/PhysRevB.106.064208} {\bibfield  {journal} {\bibinfo  {journal} {Phys. Rev. B}\ }\textbf {\bibinfo {volume} {106}},\ \bibinfo {pages} {064208} (\bibinfo {year} {2022})}\BibitemShut {NoStop}%
    \bibitem [{\citenamefont {Zhang}\ \emph {et~al.}(2022{\natexlab{b}})\citenamefont {Zhang}, \citenamefont {Li}, \citenamefont {Zhang},\ and\ \citenamefont {Lee}}]{zhang2022real}%
      \BibitemOpen
      \bibfield  {author} {\bibinfo {author} {\bibfnamefont {B.}~\bibnamefont {Zhang}}, \bibinfo {author} {\bibfnamefont {Q.}~\bibnamefont {Li}}, \bibinfo {author} {\bibfnamefont {X.}~\bibnamefont {Zhang}}, \ and\ \bibinfo {author} {\bibfnamefont {C.~H.}\ \bibnamefont {Lee}},\ }\href@noop {} {\bibfield  {journal} {\bibinfo  {journal} {Chinese Physics B}\ } (\bibinfo {year} {2022}{\natexlab{b}})}\BibitemShut {NoStop}%
    \bibitem [{\citenamefont {Rechtsman}\ \emph {et~al.}(2013)\citenamefont {Rechtsman}, \citenamefont {Zeuner}, \citenamefont {Plotnik}, \citenamefont {Lumer}, \citenamefont {Podolsky}, \citenamefont {Dreisow}, \citenamefont {Nolte}, \citenamefont {Segev},\ and\ \citenamefont {Szameit}}]{rechtsman2013photonic}%
      \BibitemOpen
      \bibfield  {author} {\bibinfo {author} {\bibfnamefont {M.~C.}\ \bibnamefont {Rechtsman}}, \bibinfo {author} {\bibfnamefont {J.~M.}\ \bibnamefont {Zeuner}}, \bibinfo {author} {\bibfnamefont {Y.}~\bibnamefont {Plotnik}}, \bibinfo {author} {\bibfnamefont {Y.}~\bibnamefont {Lumer}}, \bibinfo {author} {\bibfnamefont {D.}~\bibnamefont {Podolsky}}, \bibinfo {author} {\bibfnamefont {F.}~\bibnamefont {Dreisow}}, \bibinfo {author} {\bibfnamefont {S.}~\bibnamefont {Nolte}}, \bibinfo {author} {\bibfnamefont {M.}~\bibnamefont {Segev}}, \ and\ \bibinfo {author} {\bibfnamefont {A.}~\bibnamefont {Szameit}},\ }\href@noop {} {\bibfield  {journal} {\bibinfo  {journal} {Nature}\ }\textbf {\bibinfo {volume} {496}},\ \bibinfo {pages} {196} (\bibinfo {year} {2013})}\BibitemShut {NoStop}%
    \bibitem [{\citenamefont {Fl{\"a}schner}\ \emph {et~al.}(2016)\citenamefont {Fl{\"a}schner}, \citenamefont {Rem}, \citenamefont {Tarnowski}, \citenamefont {Vogel}, \citenamefont {L{\"u}hmann}, \citenamefont {Sengstock},\ and\ \citenamefont {Weitenberg}}]{flaschner2016experimental}%
      \BibitemOpen
      \bibfield  {author} {\bibinfo {author} {\bibfnamefont {N.}~\bibnamefont {Fl{\"a}schner}}, \bibinfo {author} {\bibfnamefont {B.}~\bibnamefont {Rem}}, \bibinfo {author} {\bibfnamefont {M.}~\bibnamefont {Tarnowski}}, \bibinfo {author} {\bibfnamefont {D.}~\bibnamefont {Vogel}}, \bibinfo {author} {\bibfnamefont {D.-S.}\ \bibnamefont {L{\"u}hmann}}, \bibinfo {author} {\bibfnamefont {K.}~\bibnamefont {Sengstock}}, \ and\ \bibinfo {author} {\bibfnamefont {C.}~\bibnamefont {Weitenberg}},\ }\href@noop {} {\bibfield  {journal} {\bibinfo  {journal} {Science}\ }\textbf {\bibinfo {volume} {352}},\ \bibinfo {pages} {1091} (\bibinfo {year} {2016})}\BibitemShut {NoStop}%
    \bibitem [{\citenamefont {Zhou}\ and\ \citenamefont {Gong}(2018)}]{Longwen2018nonHFloquet}%
      \BibitemOpen
      \bibfield  {author} {\bibinfo {author} {\bibfnamefont {L.}~\bibnamefont {Zhou}}\ and\ \bibinfo {author} {\bibfnamefont {J.}~\bibnamefont {Gong}},\ }\href {\doibase 10.1103/PhysRevB.98.205417} {\bibfield  {journal} {\bibinfo  {journal} {Phys. Rev. B}\ }\textbf {\bibinfo {volume} {98}},\ \bibinfo {pages} {205417} (\bibinfo {year} {2018})}\BibitemShut {NoStop}%
    \bibitem [{\citenamefont {Ma}\ and\ \citenamefont {Gong}(2022)}]{PhysRevResearch.4.013234}%
      \BibitemOpen
      \bibfield  {author} {\bibinfo {author} {\bibfnamefont {N.}~\bibnamefont {Ma}}\ and\ \bibinfo {author} {\bibfnamefont {J.}~\bibnamefont {Gong}},\ }\href {\doibase 10.1103/PhysRevResearch.4.013234} {\bibfield  {journal} {\bibinfo  {journal} {Phys. Rev. Res.}\ }\textbf {\bibinfo {volume} {4}},\ \bibinfo {pages} {013234} (\bibinfo {year} {2022})}\BibitemShut {NoStop}%
    \bibitem [{\citenamefont {Wu}\ and\ \citenamefont {An}(2020)}]{PhysRevB.102.041119}%
      \BibitemOpen
      \bibfield  {author} {\bibinfo {author} {\bibfnamefont {H.}~\bibnamefont {Wu}}\ and\ \bibinfo {author} {\bibfnamefont {J.-H.}\ \bibnamefont {An}},\ }\href {\doibase 10.1103/PhysRevB.102.041119} {\bibfield  {journal} {\bibinfo  {journal} {Phys. Rev. B}\ }\textbf {\bibinfo {volume} {102}},\ \bibinfo {pages} {041119} (\bibinfo {year} {2020})}\BibitemShut {NoStop}%
    \bibitem [{\citenamefont {Zhang}\ \emph {et~al.}(2021{\natexlab{d}})\citenamefont {Zhang}, \citenamefont {Xu}, \citenamefont {Liu},\ and\ \citenamefont {Yi}}]{zhang2021floquet}%
      \BibitemOpen
      \bibfield  {author} {\bibinfo {author} {\bibfnamefont {Y.-N.}\ \bibnamefont {Zhang}}, \bibinfo {author} {\bibfnamefont {S.}~\bibnamefont {Xu}}, \bibinfo {author} {\bibfnamefont {H.-D.}\ \bibnamefont {Liu}}, \ and\ \bibinfo {author} {\bibfnamefont {X.-X.}\ \bibnamefont {Yi}},\ }\href@noop {} {\bibfield  {journal} {\bibinfo  {journal} {International Journal of Theoretical Physics}\ }\textbf {\bibinfo {volume} {60}},\ \bibinfo {pages} {355} (\bibinfo {year} {2021}{\natexlab{d}})}\BibitemShut {NoStop}%
    \bibitem [{\citenamefont {Zhou}\ and\ \citenamefont {Pan}(2019)}]{PhysRevA.100.053608}%
      \BibitemOpen
      \bibfield  {author} {\bibinfo {author} {\bibfnamefont {L.}~\bibnamefont {Zhou}}\ and\ \bibinfo {author} {\bibfnamefont {J.}~\bibnamefont {Pan}},\ }\href {\doibase 10.1103/PhysRevA.100.053608} {\bibfield  {journal} {\bibinfo  {journal} {Phys. Rev. A}\ }\textbf {\bibinfo {volume} {100}},\ \bibinfo {pages} {053608} (\bibinfo {year} {2019})}\BibitemShut {NoStop}%
    \bibitem [{\citenamefont {Zhou}(2019)}]{PhysRevB.100.184314}%
      \BibitemOpen
      \bibfield  {author} {\bibinfo {author} {\bibfnamefont {L.}~\bibnamefont {Zhou}},\ }\href {\doibase 10.1103/PhysRevB.100.184314} {\bibfield  {journal} {\bibinfo  {journal} {Phys. Rev. B}\ }\textbf {\bibinfo {volume} {100}},\ \bibinfo {pages} {184314} (\bibinfo {year} {2019})}\BibitemShut {NoStop}%
    \bibitem [{\citenamefont {Rudner}\ \emph {et~al.}(2013)\citenamefont {Rudner}, \citenamefont {Lindner}, \citenamefont {Berg},\ and\ \citenamefont {Levin}}]{rudner2013anomalous}%
      \BibitemOpen
      \bibfield  {author} {\bibinfo {author} {\bibfnamefont {M.~S.}\ \bibnamefont {Rudner}}, \bibinfo {author} {\bibfnamefont {N.~H.}\ \bibnamefont {Lindner}}, \bibinfo {author} {\bibfnamefont {E.}~\bibnamefont {Berg}}, \ and\ \bibinfo {author} {\bibfnamefont {M.}~\bibnamefont {Levin}},\ }\href@noop {} {\bibfield  {journal} {\bibinfo  {journal} {Physical Review X}\ }\textbf {\bibinfo {volume} {3}},\ \bibinfo {pages} {031005} (\bibinfo {year} {2013})}\BibitemShut {NoStop}%
    \bibitem [{\citenamefont {Wang}\ \emph {et~al.}(2022{\natexlab{d}})\citenamefont {Wang}, \citenamefont {Zhao}, \citenamefont {Zhuang},\ and\ \citenamefont {Liu}}]{wang2022nonf}%
      \BibitemOpen
      \bibfield  {author} {\bibinfo {author} {\bibfnamefont {H.-Y.}\ \bibnamefont {Wang}}, \bibinfo {author} {\bibfnamefont {X.-M.}\ \bibnamefont {Zhao}}, \bibinfo {author} {\bibfnamefont {L.}~\bibnamefont {Zhuang}}, \ and\ \bibinfo {author} {\bibfnamefont {W.-M.}\ \bibnamefont {Liu}},\ }\href@noop {} {\bibfield  {journal} {\bibinfo  {journal} {Journal of Physics: Condensed Matter}\ }\textbf {\bibinfo {volume} {34}},\ \bibinfo {pages} {365402} (\bibinfo {year} {2022}{\natexlab{d}})}\BibitemShut {NoStop}%
    \bibitem [{\citenamefont {Zhou}\ \emph {et~al.}(2021)\citenamefont {Zhou}, \citenamefont {Gu},\ and\ \citenamefont {Gong}}]{PhysRevB.103.L041404}%
      \BibitemOpen
      \bibfield  {author} {\bibinfo {author} {\bibfnamefont {L.}~\bibnamefont {Zhou}}, \bibinfo {author} {\bibfnamefont {Y.}~\bibnamefont {Gu}}, \ and\ \bibinfo {author} {\bibfnamefont {J.}~\bibnamefont {Gong}},\ }\href {\doibase 10.1103/PhysRevB.103.L041404} {\bibfield  {journal} {\bibinfo  {journal} {Phys. Rev. B}\ }\textbf {\bibinfo {volume} {103}},\ \bibinfo {pages} {L041404} (\bibinfo {year} {2021})}\BibitemShut {NoStop}%
    \bibitem [{\citenamefont {Zhang}\ \emph {et~al.}(2021{\natexlab{e}})\citenamefont {Zhang}, \citenamefont {Delplace},\ and\ \citenamefont {Fleury}}]{zhang2021superior}%
      \BibitemOpen
      \bibfield  {author} {\bibinfo {author} {\bibfnamefont {Z.}~\bibnamefont {Zhang}}, \bibinfo {author} {\bibfnamefont {P.}~\bibnamefont {Delplace}}, \ and\ \bibinfo {author} {\bibfnamefont {R.}~\bibnamefont {Fleury}},\ }\href@noop {} {\bibfield  {journal} {\bibinfo  {journal} {Nature}\ }\textbf {\bibinfo {volume} {598}},\ \bibinfo {pages} {293} (\bibinfo {year} {2021}{\natexlab{e}})}\BibitemShut {NoStop}%
    \bibitem [{\citenamefont {Zhou}(2020)}]{PhysRevB.101.014306}%
      \BibitemOpen
      \bibfield  {author} {\bibinfo {author} {\bibfnamefont {L.}~\bibnamefont {Zhou}},\ }\href {\doibase 10.1103/PhysRevB.101.014306} {\bibfield  {journal} {\bibinfo  {journal} {Phys. Rev. B}\ }\textbf {\bibinfo {volume} {101}},\ \bibinfo {pages} {014306} (\bibinfo {year} {2020})}\BibitemShut {NoStop}%
    \bibitem [{\citenamefont {Pan}\ and\ \citenamefont {Zhou}(2020)}]{PhysRevB.102.094305}%
      \BibitemOpen
      \bibfield  {author} {\bibinfo {author} {\bibfnamefont {J.}~\bibnamefont {Pan}}\ and\ \bibinfo {author} {\bibfnamefont {L.}~\bibnamefont {Zhou}},\ }\href {\doibase 10.1103/PhysRevB.102.094305} {\bibfield  {journal} {\bibinfo  {journal} {Phys. Rev. B}\ }\textbf {\bibinfo {volume} {102}},\ \bibinfo {pages} {094305} (\bibinfo {year} {2020})}\BibitemShut {NoStop}%
    \bibitem [{\citenamefont {Liu}\ and\ \citenamefont {Fulga}(2022)}]{liu2022mixed}%
      \BibitemOpen
      \bibfield  {author} {\bibinfo {author} {\bibfnamefont {H.}~\bibnamefont {Liu}}\ and\ \bibinfo {author} {\bibfnamefont {I.~C.}\ \bibnamefont {Fulga}},\ }\href@noop {} {\bibfield  {journal} {\bibinfo  {journal} {arXiv preprint arXiv:2210.03097}\ } (\bibinfo {year} {2022})}\BibitemShut {NoStop}%
    \bibitem [{\citenamefont {Longhi}(2021{\natexlab{c}})}]{longhi2021non}%
      \BibitemOpen
      \bibfield  {author} {\bibinfo {author} {\bibfnamefont {S.}~\bibnamefont {Longhi}},\ }\href@noop {} {\bibfield  {journal} {\bibinfo  {journal} {Physical Review B}\ }\textbf {\bibinfo {volume} {104}},\ \bibinfo {pages} {125109} (\bibinfo {year} {2021}{\natexlab{c}})}\BibitemShut {NoStop}%
    \bibitem [{\citenamefont {Lang}\ \emph {et~al.}(2021)\citenamefont {Lang}, \citenamefont {Zhu},\ and\ \citenamefont {Chong}}]{lang2021non}%
      \BibitemOpen
      \bibfield  {author} {\bibinfo {author} {\bibfnamefont {L.-J.}\ \bibnamefont {Lang}}, \bibinfo {author} {\bibfnamefont {S.-L.}\ \bibnamefont {Zhu}}, \ and\ \bibinfo {author} {\bibfnamefont {Y.}~\bibnamefont {Chong}},\ }\href@noop {} {\bibfield  {journal} {\bibinfo  {journal} {Physical Review B}\ }\textbf {\bibinfo {volume} {104}},\ \bibinfo {pages} {L020303} (\bibinfo {year} {2021})}\BibitemShut {NoStop}%
    \bibitem [{\citenamefont {Lee}(2021)}]{lee2021many}%
      \BibitemOpen
      \bibfield  {author} {\bibinfo {author} {\bibfnamefont {C.~H.}\ \bibnamefont {Lee}},\ }\href@noop {} {\bibfield  {journal} {\bibinfo  {journal} {Physical Review B}\ }\textbf {\bibinfo {volume} {104}},\ \bibinfo {pages} {195102} (\bibinfo {year} {2021})}\BibitemShut {NoStop}%
    \bibitem [{\citenamefont {Shen}\ and\ \citenamefont {Lee}(2022)}]{shen2022non}%
      \BibitemOpen
      \bibfield  {author} {\bibinfo {author} {\bibfnamefont {R.}~\bibnamefont {Shen}}\ and\ \bibinfo {author} {\bibfnamefont {C.~H.}\ \bibnamefont {Lee}},\ }\href@noop {} {\bibfield  {journal} {\bibinfo  {journal} {Communications Physics}\ }\textbf {\bibinfo {volume} {5}},\ \bibinfo {pages} {1} (\bibinfo {year} {2022})}\BibitemShut {NoStop}%
    \bibitem [{\citenamefont {Sarkar}\ \emph {et~al.}(2022)\citenamefont {Sarkar}, \citenamefont {Hegde},\ and\ \citenamefont {Narayan}}]{sarkar2022interplay}%
      \BibitemOpen
      \bibfield  {author} {\bibinfo {author} {\bibfnamefont {R.}~\bibnamefont {Sarkar}}, \bibinfo {author} {\bibfnamefont {S.~S.}\ \bibnamefont {Hegde}}, \ and\ \bibinfo {author} {\bibfnamefont {A.}~\bibnamefont {Narayan}},\ }\href {\doibase 10.1103/PhysRevB.106.014207} {\bibfield  {journal} {\bibinfo  {journal} {Phys. Rev. B}\ }\textbf {\bibinfo {volume} {106}},\ \bibinfo {pages} {014207} (\bibinfo {year} {2022})}\BibitemShut {NoStop}%
    \bibitem [{\citenamefont {Yuce}\ and\ \citenamefont {Ramezani}(2022)}]{yuce2022coexistence}%
      \BibitemOpen
      \bibfield  {author} {\bibinfo {author} {\bibfnamefont {C.}~\bibnamefont {Yuce}}\ and\ \bibinfo {author} {\bibfnamefont {H.}~\bibnamefont {Ramezani}},\ }\href {\doibase 10.1103/PhysRevB.106.024202} {\bibfield  {journal} {\bibinfo  {journal} {Phys. Rev. B}\ }\textbf {\bibinfo {volume} {106}},\ \bibinfo {pages} {024202} (\bibinfo {year} {2022})}\BibitemShut {NoStop}%
    \bibitem [{\citenamefont {Roccati}(2021)}]{roccati2021non}%
      \BibitemOpen
      \bibfield  {author} {\bibinfo {author} {\bibfnamefont {F.}~\bibnamefont {Roccati}},\ }\href@noop {} {\bibfield  {journal} {\bibinfo  {journal} {Physical Review A}\ }\textbf {\bibinfo {volume} {104}},\ \bibinfo {pages} {022215} (\bibinfo {year} {2021})}\BibitemShut {NoStop}%
    \bibitem [{\citenamefont {Wang}\ and\ \citenamefont {Wang}(2022{\natexlab{b}})}]{wang2022chiral}%
      \BibitemOpen
      \bibfield  {author} {\bibinfo {author} {\bibfnamefont {C.}~\bibnamefont {Wang}}\ and\ \bibinfo {author} {\bibfnamefont {X.~R.}\ \bibnamefont {Wang}},\ }\href {\doibase 10.1103/PhysRevB.106.045142} {\bibfield  {journal} {\bibinfo  {journal} {Phys. Rev. B}\ }\textbf {\bibinfo {volume} {106}},\ \bibinfo {pages} {045142} (\bibinfo {year} {2022}{\natexlab{b}})}\BibitemShut {NoStop}%
    \bibitem [{\citenamefont {Longhi}(2021{\natexlab{d}})}]{longhi2021spectral}%
      \BibitemOpen
      \bibfield  {author} {\bibinfo {author} {\bibfnamefont {S.}~\bibnamefont {Longhi}},\ }\href@noop {} {\bibfield  {journal} {\bibinfo  {journal} {Physical Review B}\ }\textbf {\bibinfo {volume} {103}},\ \bibinfo {pages} {144202} (\bibinfo {year} {2021}{\natexlab{d}})}\BibitemShut {NoStop}%
    \bibitem [{\citenamefont {Chen}\ \emph {et~al.}(2022{\natexlab{a}})\citenamefont {Chen}, \citenamefont {Zhou}, \citenamefont {Chen},\ and\ \citenamefont {Ye}}]{PhysRevB.105.L121115}%
      \BibitemOpen
      \bibfield  {author} {\bibinfo {author} {\bibfnamefont {L.-M.}\ \bibnamefont {Chen}}, \bibinfo {author} {\bibfnamefont {Y.}~\bibnamefont {Zhou}}, \bibinfo {author} {\bibfnamefont {S.~A.}\ \bibnamefont {Chen}}, \ and\ \bibinfo {author} {\bibfnamefont {P.}~\bibnamefont {Ye}},\ }\href {\doibase 10.1103/PhysRevB.105.L121115} {\bibfield  {journal} {\bibinfo  {journal} {Phys. Rev. B}\ }\textbf {\bibinfo {volume} {105}},\ \bibinfo {pages} {L121115} (\bibinfo {year} {2022}{\natexlab{a}})}\BibitemShut {NoStop}%
    \bibitem [{\citenamefont {Chakrabarty}\ and\ \citenamefont {Datta}(2022)}]{chakrabarty2022skin}%
      \BibitemOpen
      \bibfield  {author} {\bibinfo {author} {\bibfnamefont {A.}~\bibnamefont {Chakrabarty}}\ and\ \bibinfo {author} {\bibfnamefont {S.}~\bibnamefont {Datta}},\ }\href@noop {} {\bibfield  {journal} {\bibinfo  {journal} {arXiv preprint arXiv:2208.10359}\ } (\bibinfo {year} {2022})}\BibitemShut {NoStop}%
    \bibitem [{\citenamefont {Wang}\ \emph {et~al.}(2023)\citenamefont {Wang}, \citenamefont {Hu}, \citenamefont {Wang}, \citenamefont {Ma},\ and\ \citenamefont {Ding}}]{wang2023experimental}%
      \BibitemOpen
      \bibfield  {author} {\bibinfo {author} {\bibfnamefont {W.}~\bibnamefont {Wang}}, \bibinfo {author} {\bibfnamefont {M.}~\bibnamefont {Hu}}, \bibinfo {author} {\bibfnamefont {X.}~\bibnamefont {Wang}}, \bibinfo {author} {\bibfnamefont {G.}~\bibnamefont {Ma}}, \ and\ \bibinfo {author} {\bibfnamefont {K.}~\bibnamefont {Ding}},\ }\href@noop {} {\bibfield  {journal} {\bibinfo  {journal} {arXiv preprint arXiv:2302.06314}\ } (\bibinfo {year} {2023})}\BibitemShut {NoStop}%
    \bibitem [{\citenamefont {Lv}\ \emph {et~al.}(2022)\citenamefont {Lv}, \citenamefont {Zhang}, \citenamefont {Zhai},\ and\ \citenamefont {Zhou}}]{lv2022curving}%
      \BibitemOpen
      \bibfield  {author} {\bibinfo {author} {\bibfnamefont {C.}~\bibnamefont {Lv}}, \bibinfo {author} {\bibfnamefont {R.}~\bibnamefont {Zhang}}, \bibinfo {author} {\bibfnamefont {Z.}~\bibnamefont {Zhai}}, \ and\ \bibinfo {author} {\bibfnamefont {Q.}~\bibnamefont {Zhou}},\ }\href@noop {} {\bibfield  {journal} {\bibinfo  {journal} {Nature communications}\ }\textbf {\bibinfo {volume} {13}},\ \bibinfo {pages} {2184} (\bibinfo {year} {2022})}\BibitemShut {NoStop}%
    \bibitem [{\citenamefont {Wang}\ and\ \citenamefont {Wan}(2022)}]{wang2022duality}%
      \BibitemOpen
      \bibfield  {author} {\bibinfo {author} {\bibfnamefont {S.-X.}\ \bibnamefont {Wang}}\ and\ \bibinfo {author} {\bibfnamefont {S.}~\bibnamefont {Wan}},\ }\href@noop {} {\bibfield  {journal} {\bibinfo  {journal} {Physical Review B}\ }\textbf {\bibinfo {volume} {106}},\ \bibinfo {pages} {075112} (\bibinfo {year} {2022})}\BibitemShut {NoStop}%
    \bibitem [{\citenamefont {Lv}\ and\ \citenamefont {Zhou}(2022)}]{lv2022emergent}%
      \BibitemOpen
      \bibfield  {author} {\bibinfo {author} {\bibfnamefont {C.}~\bibnamefont {Lv}}\ and\ \bibinfo {author} {\bibfnamefont {Q.}~\bibnamefont {Zhou}},\ }\href@noop {} {\bibfield  {journal} {\bibinfo  {journal} {arXiv preprint arXiv:2205.07429}\ } (\bibinfo {year} {2022})}\BibitemShut {NoStop}%
    \bibitem [{\citenamefont {Gao}\ \emph {et~al.}(2015{\natexlab{b}})\citenamefont {Gao}, \citenamefont {Lawrence}, \citenamefont {Yang}, \citenamefont {Liu}, \citenamefont {Fang}, \citenamefont {B{\'e}ri}, \citenamefont {Li},\ and\ \citenamefont {Zhang}}]{gao2015topological}%
      \BibitemOpen
      \bibfield  {author} {\bibinfo {author} {\bibfnamefont {W.}~\bibnamefont {Gao}}, \bibinfo {author} {\bibfnamefont {M.}~\bibnamefont {Lawrence}}, \bibinfo {author} {\bibfnamefont {B.}~\bibnamefont {Yang}}, \bibinfo {author} {\bibfnamefont {F.}~\bibnamefont {Liu}}, \bibinfo {author} {\bibfnamefont {F.}~\bibnamefont {Fang}}, \bibinfo {author} {\bibfnamefont {B.}~\bibnamefont {B{\'e}ri}}, \bibinfo {author} {\bibfnamefont {J.}~\bibnamefont {Li}}, \ and\ \bibinfo {author} {\bibfnamefont {S.}~\bibnamefont {Zhang}},\ }\href@noop {} {\bibfield  {journal} {\bibinfo  {journal} {Phys. Rev. Lett.}\ }\textbf {\bibinfo {volume} {114}},\ \bibinfo {pages} {037402} (\bibinfo {year} {2015}{\natexlab{b}})}\BibitemShut {NoStop}%
    \bibitem [{\citenamefont {Koll{\'a}r}\ \emph {et~al.}(2019)\citenamefont {Koll{\'a}r}, \citenamefont {Fitzpatrick},\ and\ \citenamefont {Houck}}]{kollar2019hyperbolic}%
      \BibitemOpen
      \bibfield  {author} {\bibinfo {author} {\bibfnamefont {A.~J.}\ \bibnamefont {Koll{\'a}r}}, \bibinfo {author} {\bibfnamefont {M.}~\bibnamefont {Fitzpatrick}}, \ and\ \bibinfo {author} {\bibfnamefont {A.~A.}\ \bibnamefont {Houck}},\ }\href@noop {} {\bibfield  {journal} {\bibinfo  {journal} {Nature}\ }\textbf {\bibinfo {volume} {571}},\ \bibinfo {pages} {45} (\bibinfo {year} {2019})}\BibitemShut {NoStop}%
    \bibitem [{\citenamefont {Lenggenhager}\ \emph {et~al.}(2022)\citenamefont {Lenggenhager}, \citenamefont {Stegmaier}, \citenamefont {Upreti}, \citenamefont {Hofmann}, \citenamefont {Helbig}, \citenamefont {Vollhardt}, \citenamefont {Greiter}, \citenamefont {Lee}, \citenamefont {Imhof}, \citenamefont {Brand} \emph {et~al.}}]{lenggenhager2022simulating}%
      \BibitemOpen
      \bibfield  {author} {\bibinfo {author} {\bibfnamefont {P.~M.}\ \bibnamefont {Lenggenhager}}, \bibinfo {author} {\bibfnamefont {A.}~\bibnamefont {Stegmaier}}, \bibinfo {author} {\bibfnamefont {L.~K.}\ \bibnamefont {Upreti}}, \bibinfo {author} {\bibfnamefont {T.}~\bibnamefont {Hofmann}}, \bibinfo {author} {\bibfnamefont {T.}~\bibnamefont {Helbig}}, \bibinfo {author} {\bibfnamefont {A.}~\bibnamefont {Vollhardt}}, \bibinfo {author} {\bibfnamefont {M.}~\bibnamefont {Greiter}}, \bibinfo {author} {\bibfnamefont {C.~H.}\ \bibnamefont {Lee}}, \bibinfo {author} {\bibfnamefont {S.}~\bibnamefont {Imhof}}, \bibinfo {author} {\bibfnamefont {H.}~\bibnamefont {Brand}},  \emph {et~al.},\ }\href@noop {} {\bibfield  {journal} {\bibinfo  {journal} {Nature Communications}\ }\textbf {\bibinfo {volume} {13}},\ \bibinfo {pages} {4373} (\bibinfo {year} {2022})}\BibitemShut {NoStop}%
    \bibitem [{\citenamefont {Ezawa}(2022{\natexlab{a}})}]{ezawa2022dynamical}%
      \BibitemOpen
      \bibfield  {author} {\bibinfo {author} {\bibfnamefont {M.}~\bibnamefont {Ezawa}},\ }\href@noop {} {\bibfield  {journal} {\bibinfo  {journal} {Physical Review B}\ }\textbf {\bibinfo {volume} {105}},\ \bibinfo {pages} {125421} (\bibinfo {year} {2022}{\natexlab{a}})}\BibitemShut {NoStop}%
    \bibitem [{\citenamefont {Yuce}(2021)}]{yuce2021nonlinear}%
      \BibitemOpen
      \bibfield  {author} {\bibinfo {author} {\bibfnamefont {C.}~\bibnamefont {Yuce}},\ }\href@noop {} {\bibfield  {journal} {\bibinfo  {journal} {Physics Letters A}\ }\textbf {\bibinfo {volume} {408}},\ \bibinfo {pages} {127484} (\bibinfo {year} {2021})}\BibitemShut {NoStop}%
    \bibitem [{\citenamefont {Tuloup}\ \emph {et~al.}(2020)\citenamefont {Tuloup}, \citenamefont {Bomantara}, \citenamefont {Lee},\ and\ \citenamefont {Gong}}]{tuloup2020nonlinearity}%
      \BibitemOpen
      \bibfield  {author} {\bibinfo {author} {\bibfnamefont {T.}~\bibnamefont {Tuloup}}, \bibinfo {author} {\bibfnamefont {R.~W.}\ \bibnamefont {Bomantara}}, \bibinfo {author} {\bibfnamefont {C.~H.}\ \bibnamefont {Lee}}, \ and\ \bibinfo {author} {\bibfnamefont {J.}~\bibnamefont {Gong}},\ }\href@noop {} {\bibfield  {journal} {\bibinfo  {journal} {Physical Review B}\ }\textbf {\bibinfo {volume} {102}},\ \bibinfo {pages} {115411} (\bibinfo {year} {2020})}\BibitemShut {NoStop}%
    \bibitem [{\citenamefont {Hadad}\ \emph {et~al.}(2018)\citenamefont {Hadad}, \citenamefont {Soric}, \citenamefont {Khanikaev},\ and\ \citenamefont {Alu}}]{hadad2018self}%
      \BibitemOpen
      \bibfield  {author} {\bibinfo {author} {\bibfnamefont {Y.}~\bibnamefont {Hadad}}, \bibinfo {author} {\bibfnamefont {J.~C.}\ \bibnamefont {Soric}}, \bibinfo {author} {\bibfnamefont {A.~B.}\ \bibnamefont {Khanikaev}}, \ and\ \bibinfo {author} {\bibfnamefont {A.}~\bibnamefont {Alu}},\ }\href@noop {} {\bibfield  {journal} {\bibinfo  {journal} {Nature Electronics}\ }\textbf {\bibinfo {volume} {1}},\ \bibinfo {pages} {178} (\bibinfo {year} {2018})}\BibitemShut {NoStop}%
    \bibitem [{\citenamefont {Wang}\ \emph {et~al.}(2019)\citenamefont {Wang}, \citenamefont {Lang}, \citenamefont {Lee}, \citenamefont {Zhang},\ and\ \citenamefont {Chong}}]{wang2019topologically}%
      \BibitemOpen
      \bibfield  {author} {\bibinfo {author} {\bibfnamefont {Y.}~\bibnamefont {Wang}}, \bibinfo {author} {\bibfnamefont {L.-J.}\ \bibnamefont {Lang}}, \bibinfo {author} {\bibfnamefont {C.~H.}\ \bibnamefont {Lee}}, \bibinfo {author} {\bibfnamefont {B.}~\bibnamefont {Zhang}}, \ and\ \bibinfo {author} {\bibfnamefont {Y.}~\bibnamefont {Chong}},\ }\href@noop {} {\bibfield  {journal} {\bibinfo  {journal} {Nature communications}\ }\textbf {\bibinfo {volume} {10}},\ \bibinfo {pages} {1102} (\bibinfo {year} {2019})}\BibitemShut {NoStop}%
    \bibitem [{\citenamefont {Kotwal}\ \emph {et~al.}(2021)\citenamefont {Kotwal}, \citenamefont {Moseley}, \citenamefont {Stegmaier}, \citenamefont {Imhof}, \citenamefont {Brand}, \citenamefont {Kie{\ss}ling}, \citenamefont {Thomale}, \citenamefont {Ronellenfitsch},\ and\ \citenamefont {Dunkel}}]{kotwal2021active}%
      \BibitemOpen
      \bibfield  {author} {\bibinfo {author} {\bibfnamefont {T.}~\bibnamefont {Kotwal}}, \bibinfo {author} {\bibfnamefont {F.}~\bibnamefont {Moseley}}, \bibinfo {author} {\bibfnamefont {A.}~\bibnamefont {Stegmaier}}, \bibinfo {author} {\bibfnamefont {S.}~\bibnamefont {Imhof}}, \bibinfo {author} {\bibfnamefont {H.}~\bibnamefont {Brand}}, \bibinfo {author} {\bibfnamefont {T.}~\bibnamefont {Kie{\ss}ling}}, \bibinfo {author} {\bibfnamefont {R.}~\bibnamefont {Thomale}}, \bibinfo {author} {\bibfnamefont {H.}~\bibnamefont {Ronellenfitsch}}, \ and\ \bibinfo {author} {\bibfnamefont {J.}~\bibnamefont {Dunkel}},\ }\href@noop {} {\bibfield  {journal} {\bibinfo  {journal} {Proceedings of the National Academy of Sciences}\ }\textbf {\bibinfo {volume} {118}},\ \bibinfo {pages} {e2106411118} (\bibinfo {year} {2021})}\BibitemShut {NoStop}%
    \bibitem [{\citenamefont {Hohmann}\ \emph {et~al.}(2023)\citenamefont {Hohmann}, \citenamefont {Hofmann}, \citenamefont {Helbig}, \citenamefont {Imhof}, \citenamefont {Brand}, \citenamefont {Upreti}, \citenamefont {Stegmaier}, \citenamefont {Fritzsche}, \citenamefont {M{\"u}ller}, \citenamefont {Schwingenschl{\"o}gl} \emph {et~al.}}]{hohmann2022observation}%
      \BibitemOpen
      \bibfield  {author} {\bibinfo {author} {\bibfnamefont {H.}~\bibnamefont {Hohmann}}, \bibinfo {author} {\bibfnamefont {T.}~\bibnamefont {Hofmann}}, \bibinfo {author} {\bibfnamefont {T.}~\bibnamefont {Helbig}}, \bibinfo {author} {\bibfnamefont {S.}~\bibnamefont {Imhof}}, \bibinfo {author} {\bibfnamefont {H.}~\bibnamefont {Brand}}, \bibinfo {author} {\bibfnamefont {L.~K.}\ \bibnamefont {Upreti}}, \bibinfo {author} {\bibfnamefont {A.}~\bibnamefont {Stegmaier}}, \bibinfo {author} {\bibfnamefont {A.}~\bibnamefont {Fritzsche}}, \bibinfo {author} {\bibfnamefont {T.}~\bibnamefont {M{\"u}ller}}, \bibinfo {author} {\bibfnamefont {U.}~\bibnamefont {Schwingenschl{\"o}gl}},  \emph {et~al.},\ }\href@noop {} {\bibfield  {journal} {\bibinfo  {journal} {Physical Review Research}\ }\textbf {\bibinfo {volume} {5}},\ \bibinfo {pages} {L012041} (\bibinfo {year} {2023})}\BibitemShut {NoStop}%
    \bibitem [{\citenamefont {Dobrykh}\ \emph {et~al.}(2018)\citenamefont {Dobrykh}, \citenamefont {Yulin}, \citenamefont {Slobozhanyuk}, \citenamefont {Poddubny},\ and\ \citenamefont {Kivshar}}]{dobrykh2018nonlinear}%
      \BibitemOpen
      \bibfield  {author} {\bibinfo {author} {\bibfnamefont {D.}~\bibnamefont {Dobrykh}}, \bibinfo {author} {\bibfnamefont {A.}~\bibnamefont {Yulin}}, \bibinfo {author} {\bibfnamefont {A.}~\bibnamefont {Slobozhanyuk}}, \bibinfo {author} {\bibfnamefont {A.}~\bibnamefont {Poddubny}}, \ and\ \bibinfo {author} {\bibfnamefont {Y.~S.}\ \bibnamefont {Kivshar}},\ }\href@noop {} {\bibfield  {journal} {\bibinfo  {journal} {Physical review letters}\ }\textbf {\bibinfo {volume} {121}},\ \bibinfo {pages} {163901} (\bibinfo {year} {2018})}\BibitemShut {NoStop}%
    \bibitem [{\citenamefont {Fu}\ \emph {et~al.}(2022)\citenamefont {Fu}, \citenamefont {Xu}, \citenamefont {Lee}, \citenamefont {Ang},\ and\ \citenamefont {Liu}}]{fu2022gate}%
      \BibitemOpen
      \bibfield  {author} {\bibinfo {author} {\bibfnamefont {P.-H.}\ \bibnamefont {Fu}}, \bibinfo {author} {\bibfnamefont {Y.}~\bibnamefont {Xu}}, \bibinfo {author} {\bibfnamefont {C.~H.}\ \bibnamefont {Lee}}, \bibinfo {author} {\bibfnamefont {Y.~S.}\ \bibnamefont {Ang}}, \ and\ \bibinfo {author} {\bibfnamefont {J.-F.}\ \bibnamefont {Liu}},\ }\href@noop {} {\bibfield  {journal} {\bibinfo  {journal} {arXiv preprint arXiv:2212.01980}\ } (\bibinfo {year} {2022})}\BibitemShut {NoStop}%
    \bibitem [{\citenamefont {Ezawa}(2022{\natexlab{b}})}]{ezawa2022nonlinear}%
      \BibitemOpen
      \bibfield  {author} {\bibinfo {author} {\bibfnamefont {M.}~\bibnamefont {Ezawa}},\ }\href {\doibase 10.7566/JPSJ.91.084703} {\bibfield  {journal} {\bibinfo  {journal} {J. Phys. Soc. Jpn.}\ }\textbf {\bibinfo {volume} {91}},\ \bibinfo {pages} {084703} (\bibinfo {year} {2022}{\natexlab{b}})}\BibitemShut {NoStop}%
    \bibitem [{\citenamefont {Hofmann}\ \emph {et~al.}(2019)\citenamefont {Hofmann}, \citenamefont {Helbig}, \citenamefont {Lee}, \citenamefont {Greiter},\ and\ \citenamefont {Thomale}}]{hofmann2019chiral}%
      \BibitemOpen
      \bibfield  {author} {\bibinfo {author} {\bibfnamefont {T.}~\bibnamefont {Hofmann}}, \bibinfo {author} {\bibfnamefont {T.}~\bibnamefont {Helbig}}, \bibinfo {author} {\bibfnamefont {C.~H.}\ \bibnamefont {Lee}}, \bibinfo {author} {\bibfnamefont {M.}~\bibnamefont {Greiter}}, \ and\ \bibinfo {author} {\bibfnamefont {R.}~\bibnamefont {Thomale}},\ }\href@noop {} {\bibfield  {journal} {\bibinfo  {journal} {Physical review letters}\ }\textbf {\bibinfo {volume} {122}},\ \bibinfo {pages} {247702} (\bibinfo {year} {2019})}\BibitemShut {NoStop}%
    \bibitem [{\citenamefont {Mu}\ \emph {et~al.}(2020)\citenamefont {Mu}, \citenamefont {Lee}, \citenamefont {Li},\ and\ \citenamefont {Gong}}]{mu2020emergent}%
      \BibitemOpen
      \bibfield  {author} {\bibinfo {author} {\bibfnamefont {S.}~\bibnamefont {Mu}}, \bibinfo {author} {\bibfnamefont {C.~H.}\ \bibnamefont {Lee}}, \bibinfo {author} {\bibfnamefont {L.}~\bibnamefont {Li}}, \ and\ \bibinfo {author} {\bibfnamefont {J.}~\bibnamefont {Gong}},\ }\href {\doibase 10.1103/PhysRevB.102.081115} {\bibfield  {journal} {\bibinfo  {journal} {Phys. Rev. B}\ }\textbf {\bibinfo {volume} {102}},\ \bibinfo {pages} {081115} (\bibinfo {year} {2020})}\BibitemShut {NoStop}%
    \bibitem [{\citenamefont {Zhang}\ \emph {et~al.}(2022{\natexlab{c}})\citenamefont {Zhang}, \citenamefont {Denner}, \citenamefont {Bzdu\ifmmode~\check{s}\else \v{s}\fi{}ek}, \citenamefont {Sentef},\ and\ \citenamefont {Neupert}}]{zhang2022symmetry}%
      \BibitemOpen
      \bibfield  {author} {\bibinfo {author} {\bibfnamefont {S.-B.}\ \bibnamefont {Zhang}}, \bibinfo {author} {\bibfnamefont {M.~M.}\ \bibnamefont {Denner}}, \bibinfo {author} {\bibfnamefont {T.~c.~v.}\ \bibnamefont {Bzdu\ifmmode~\check{s}\else \v{s}\fi{}ek}}, \bibinfo {author} {\bibfnamefont {M.~A.}\ \bibnamefont {Sentef}}, \ and\ \bibinfo {author} {\bibfnamefont {T.}~\bibnamefont {Neupert}},\ }\href {\doibase 10.1103/PhysRevB.106.L121102} {\bibfield  {journal} {\bibinfo  {journal} {Phys. Rev. B}\ }\textbf {\bibinfo {volume} {106}},\ \bibinfo {pages} {L121102} (\bibinfo {year} {2022}{\natexlab{c}})}\BibitemShut {NoStop}%
    \bibitem [{\citenamefont {Hamanaka}\ and\ \citenamefont {Kawabata}(2024)}]{hamanaka2024multifractality}%
      \BibitemOpen
      \bibfield  {author} {\bibinfo {author} {\bibfnamefont {S.}~\bibnamefont {Hamanaka}}\ and\ \bibinfo {author} {\bibfnamefont {K.}~\bibnamefont {Kawabata}},\ }\href@noop {} {\bibfield  {journal} {\bibinfo  {journal} {arXiv preprint arXiv:2401.08304}\ } (\bibinfo {year} {2024})}\BibitemShut {NoStop}%
    \bibitem [{\citenamefont {Alsallom}\ \emph {et~al.}(2022)\citenamefont {Alsallom}, \citenamefont {Herviou}, \citenamefont {Yazyev},\ and\ \citenamefont {Brzezi{\'n}ska}}]{alsallom2022fate}%
      \BibitemOpen
      \bibfield  {author} {\bibinfo {author} {\bibfnamefont {F.}~\bibnamefont {Alsallom}}, \bibinfo {author} {\bibfnamefont {L.}~\bibnamefont {Herviou}}, \bibinfo {author} {\bibfnamefont {O.~V.}\ \bibnamefont {Yazyev}}, \ and\ \bibinfo {author} {\bibfnamefont {M.}~\bibnamefont {Brzezi{\'n}ska}},\ }\href@noop {} {\bibfield  {journal} {\bibinfo  {journal} {Physical Review Research}\ }\textbf {\bibinfo {volume} {4}},\ \bibinfo {pages} {033122} (\bibinfo {year} {2022})}\BibitemShut {NoStop}%
    \bibitem [{\citenamefont {Shen}\ \emph {et~al.}(2023{\natexlab{b}})\citenamefont {Shen}, \citenamefont {Chen}, \citenamefont {Yang},\ and\ \citenamefont {Lee}}]{shen2023observation}%
      \BibitemOpen
      \bibfield  {author} {\bibinfo {author} {\bibfnamefont {R.}~\bibnamefont {Shen}}, \bibinfo {author} {\bibfnamefont {T.}~\bibnamefont {Chen}}, \bibinfo {author} {\bibfnamefont {B.}~\bibnamefont {Yang}}, \ and\ \bibinfo {author} {\bibfnamefont {C.~H.}\ \bibnamefont {Lee}},\ }\href@noop {} {\bibfield  {journal} {\bibinfo  {journal} {arXiv preprint arXiv:2311.10143}\ } (\bibinfo {year} {2023}{\natexlab{b}})}\BibitemShut {NoStop}%
    \bibitem [{\citenamefont {D\'ora}\ and\ \citenamefont {Moca}(2022)}]{dora2022full}%
      \BibitemOpen
      \bibfield  {author} {\bibinfo {author} {\bibfnamefont {B.}~\bibnamefont {D\'ora}}\ and\ \bibinfo {author} {\bibfnamefont {C.~u. u. u. u. P. m.~c.}\ \bibnamefont {Moca}},\ }\href {\doibase 10.1103/PhysRevB.106.235125} {\bibfield  {journal} {\bibinfo  {journal} {Phys. Rev. B}\ }\textbf {\bibinfo {volume} {106}},\ \bibinfo {pages} {235125} (\bibinfo {year} {2022})}\BibitemShut {NoStop}%
    \bibitem [{\citenamefont {Yoshida}\ \emph {et~al.}(2019)\citenamefont {Yoshida}, \citenamefont {Kudo},\ and\ \citenamefont {Hatsugai}}]{yoshida2019non}%
      \BibitemOpen
      \bibfield  {author} {\bibinfo {author} {\bibfnamefont {T.}~\bibnamefont {Yoshida}}, \bibinfo {author} {\bibfnamefont {K.}~\bibnamefont {Kudo}}, \ and\ \bibinfo {author} {\bibfnamefont {Y.}~\bibnamefont {Hatsugai}},\ }\href@noop {} {\bibfield  {journal} {\bibinfo  {journal} {Scientific reports}\ }\textbf {\bibinfo {volume} {9}},\ \bibinfo {pages} {1} (\bibinfo {year} {2019})}\BibitemShut {NoStop}%
    \bibitem [{\citenamefont {Wang}\ \emph {et~al.}(2022{\natexlab{e}})\citenamefont {Wang}, \citenamefont {You},\ and\ \citenamefont {Sun}}]{wang2022non2}%
      \BibitemOpen
      \bibfield  {author} {\bibinfo {author} {\bibfnamefont {Y.-N.}\ \bibnamefont {Wang}}, \bibinfo {author} {\bibfnamefont {W.-L.}\ \bibnamefont {You}}, \ and\ \bibinfo {author} {\bibfnamefont {G.}~\bibnamefont {Sun}},\ }\href {\doibase 10.1103/PhysRevA.106.053315} {\bibfield  {journal} {\bibinfo  {journal} {Phys. Rev. A}\ }\textbf {\bibinfo {volume} {106}},\ \bibinfo {pages} {053315} (\bibinfo {year} {2022}{\natexlab{e}})}\BibitemShut {NoStop}%
    \bibitem [{\citenamefont {Mao}\ \emph {et~al.}(2022)\citenamefont {Mao}, \citenamefont {Hao},\ and\ \citenamefont {Pan}}]{mao2022non}%
      \BibitemOpen
      \bibfield  {author} {\bibinfo {author} {\bibfnamefont {L.}~\bibnamefont {Mao}}, \bibinfo {author} {\bibfnamefont {Y.}~\bibnamefont {Hao}}, \ and\ \bibinfo {author} {\bibfnamefont {L.}~\bibnamefont {Pan}},\ }\href@noop {} {\bibfield  {journal} {\bibinfo  {journal} {arXiv preprint arXiv:2207.12637}\ } (\bibinfo {year} {2022})}\BibitemShut {NoStop}%
    \bibitem [{\citenamefont {Chen}\ \emph {et~al.}(2022{\natexlab{b}})\citenamefont {Chen}, \citenamefont {Song},\ and\ \citenamefont {Lado}}]{chen2022topological}%
      \BibitemOpen
      \bibfield  {author} {\bibinfo {author} {\bibfnamefont {G.}~\bibnamefont {Chen}}, \bibinfo {author} {\bibfnamefont {F.}~\bibnamefont {Song}}, \ and\ \bibinfo {author} {\bibfnamefont {J.~L.}\ \bibnamefont {Lado}},\ }\href@noop {} {\bibfield  {journal} {\bibinfo  {journal} {arXiv preprint arXiv:2208.06425}\ } (\bibinfo {year} {2022}{\natexlab{b}})}\BibitemShut {NoStop}%
    \bibitem [{\citenamefont {Yoshida}\ and\ \citenamefont {Hatsugai}(2022)}]{yoshida2022reduction}%
      \BibitemOpen
      \bibfield  {author} {\bibinfo {author} {\bibfnamefont {T.}~\bibnamefont {Yoshida}}\ and\ \bibinfo {author} {\bibfnamefont {Y.}~\bibnamefont {Hatsugai}},\ }\href {\doibase 10.1103/PhysRevB.106.205147} {\bibfield  {journal} {\bibinfo  {journal} {Phys. Rev. B}\ }\textbf {\bibinfo {volume} {106}},\ \bibinfo {pages} {205147} (\bibinfo {year} {2022})}\BibitemShut {NoStop}%
    \bibitem [{\citenamefont {Zhang}\ \emph {et~al.}(2022{\natexlab{d}})\citenamefont {Zhang}, \citenamefont {Di}, \citenamefont {Yuan}, \citenamefont {Wang}, \citenamefont {Zheng}, \citenamefont {He}, \citenamefont {Sun},\ and\ \citenamefont {Zhang}}]{zhang2022observation}%
      \BibitemOpen
      \bibfield  {author} {\bibinfo {author} {\bibfnamefont {W.}~\bibnamefont {Zhang}}, \bibinfo {author} {\bibfnamefont {F.}~\bibnamefont {Di}}, \bibinfo {author} {\bibfnamefont {H.}~\bibnamefont {Yuan}}, \bibinfo {author} {\bibfnamefont {H.}~\bibnamefont {Wang}}, \bibinfo {author} {\bibfnamefont {X.}~\bibnamefont {Zheng}}, \bibinfo {author} {\bibfnamefont {L.}~\bibnamefont {He}}, \bibinfo {author} {\bibfnamefont {H.}~\bibnamefont {Sun}}, \ and\ \bibinfo {author} {\bibfnamefont {X.}~\bibnamefont {Zhang}},\ }\href@noop {} {\bibfield  {journal} {\bibinfo  {journal} {Physical Review B}\ }\textbf {\bibinfo {volume} {105}},\ \bibinfo {pages} {195131} (\bibinfo {year} {2022}{\natexlab{d}})}\BibitemShut {NoStop}%
    \bibitem [{\citenamefont {Shen}\ \emph {et~al.}(2024)\citenamefont {Shen}, \citenamefont {Qin}, \citenamefont {Desaules}, \citenamefont {Papi{\'c}},\ and\ \citenamefont {Lee}}]{shen2024enhanced}%
      \BibitemOpen
      \bibfield  {author} {\bibinfo {author} {\bibfnamefont {R.}~\bibnamefont {Shen}}, \bibinfo {author} {\bibfnamefont {F.}~\bibnamefont {Qin}}, \bibinfo {author} {\bibfnamefont {J.-Y.}\ \bibnamefont {Desaules}}, \bibinfo {author} {\bibfnamefont {Z.}~\bibnamefont {Papi{\'c}}}, \ and\ \bibinfo {author} {\bibfnamefont {C.~H.}\ \bibnamefont {Lee}},\ }\href@noop {} {\bibfield  {journal} {\bibinfo  {journal} {arXiv preprint arXiv:2403.02395}\ } (\bibinfo {year} {2024})}\BibitemShut {NoStop}%
    \bibitem [{\citenamefont {Kawabata}\ \emph {et~al.}(2022{\natexlab{b}})\citenamefont {Kawabata}, \citenamefont {Shiozaki},\ and\ \citenamefont {Ryu}}]{kawabata2022many}%
      \BibitemOpen
      \bibfield  {author} {\bibinfo {author} {\bibfnamefont {K.}~\bibnamefont {Kawabata}}, \bibinfo {author} {\bibfnamefont {K.}~\bibnamefont {Shiozaki}}, \ and\ \bibinfo {author} {\bibfnamefont {S.}~\bibnamefont {Ryu}},\ }\href@noop {} {\bibfield  {journal} {\bibinfo  {journal} {Physical Review B}\ }\textbf {\bibinfo {volume} {105}},\ \bibinfo {pages} {165137} (\bibinfo {year} {2022}{\natexlab{b}})}\BibitemShut {NoStop}%
    \bibitem [{\citenamefont {Yoshida}(2021)}]{yoshida2021real}%
      \BibitemOpen
      \bibfield  {author} {\bibinfo {author} {\bibfnamefont {T.}~\bibnamefont {Yoshida}},\ }\href@noop {} {\bibfield  {journal} {\bibinfo  {journal} {Physical Review B}\ }\textbf {\bibinfo {volume} {103}},\ \bibinfo {pages} {125145} (\bibinfo {year} {2021})}\BibitemShut {NoStop}%
    \bibitem [{\citenamefont {Arkhipov}\ and\ \citenamefont {Minganti}(2023)}]{arkhipov2023emergent}%
      \BibitemOpen
      \bibfield  {author} {\bibinfo {author} {\bibfnamefont {I.~I.}\ \bibnamefont {Arkhipov}}\ and\ \bibinfo {author} {\bibfnamefont {F.}~\bibnamefont {Minganti}},\ }\href@noop {} {\bibfield  {journal} {\bibinfo  {journal} {Physical Review A}\ }\textbf {\bibinfo {volume} {107}},\ \bibinfo {pages} {012202} (\bibinfo {year} {2023})}\BibitemShut {NoStop}%
    \bibitem [{\citenamefont {Qin}\ \emph {et~al.}(2023{\natexlab{c}})\citenamefont {Qin}, \citenamefont {Shen},\ and\ \citenamefont {Lee}}]{qin2022non}%
      \BibitemOpen
      \bibfield  {author} {\bibinfo {author} {\bibfnamefont {F.}~\bibnamefont {Qin}}, \bibinfo {author} {\bibfnamefont {R.}~\bibnamefont {Shen}}, \ and\ \bibinfo {author} {\bibfnamefont {C.~H.}\ \bibnamefont {Lee}},\ }\href {\doibase 10.1103/PhysRevA.107.L010202} {\bibfield  {journal} {\bibinfo  {journal} {Phys. Rev. A}\ }\textbf {\bibinfo {volume} {107}},\ \bibinfo {pages} {L010202} (\bibinfo {year} {2023}{\natexlab{c}})}\BibitemShut {NoStop}%
    \bibitem [{\citenamefont {Gliozzi}\ \emph {et~al.}(2024)\citenamefont {Gliozzi}, \citenamefont {De~Tomasi},\ and\ \citenamefont {Hughes}}]{gliozzi2024many}%
      \BibitemOpen
      \bibfield  {author} {\bibinfo {author} {\bibfnamefont {J.}~\bibnamefont {Gliozzi}}, \bibinfo {author} {\bibfnamefont {G.}~\bibnamefont {De~Tomasi}}, \ and\ \bibinfo {author} {\bibfnamefont {T.~L.}\ \bibnamefont {Hughes}},\ }\href@noop {} {\bibfield  {journal} {\bibinfo  {journal} {arXiv preprint arXiv:2401.04162}\ } (\bibinfo {year} {2024})}\BibitemShut {NoStop}%
    \bibitem [{\citenamefont {Lee}\ \emph {et~al.}(2015)\citenamefont {Lee}, \citenamefont {Papi{\'c}},\ and\ \citenamefont {Thomale}}]{lee2015geometric}%
      \BibitemOpen
      \bibfield  {author} {\bibinfo {author} {\bibfnamefont {C.~H.}\ \bibnamefont {Lee}}, \bibinfo {author} {\bibfnamefont {Z.}~\bibnamefont {Papi{\'c}}}, \ and\ \bibinfo {author} {\bibfnamefont {R.}~\bibnamefont {Thomale}},\ }\href@noop {} {\bibfield  {journal} {\bibinfo  {journal} {Physical Review X}\ }\textbf {\bibinfo {volume} {5}},\ \bibinfo {pages} {041003} (\bibinfo {year} {2015})}\BibitemShut {NoStop}%
    \bibitem [{\citenamefont {Micallo}\ \emph {et~al.}(2023)\citenamefont {Micallo}, \citenamefont {Lehmann},\ and\ \citenamefont {Budich}}]{micallo2023correlation}%
      \BibitemOpen
      \bibfield  {author} {\bibinfo {author} {\bibfnamefont {T.}~\bibnamefont {Micallo}}, \bibinfo {author} {\bibfnamefont {C.}~\bibnamefont {Lehmann}}, \ and\ \bibinfo {author} {\bibfnamefont {J.~C.}\ \bibnamefont {Budich}},\ }\href@noop {} {\bibfield  {journal} {\bibinfo  {journal} {arXiv preprint arXiv:2302.00019}\ } (\bibinfo {year} {2023})}\BibitemShut {NoStop}%
    \bibitem [{\citenamefont {Yang}\ \emph {et~al.}(2021)\citenamefont {Yang}, \citenamefont {Morampudi},\ and\ \citenamefont {Bergholtz}}]{PhysRevLett.126.077201}%
      \BibitemOpen
      \bibfield  {author} {\bibinfo {author} {\bibfnamefont {K.}~\bibnamefont {Yang}}, \bibinfo {author} {\bibfnamefont {S.~C.}\ \bibnamefont {Morampudi}}, \ and\ \bibinfo {author} {\bibfnamefont {E.~J.}\ \bibnamefont {Bergholtz}},\ }\href {\doibase 10.1103/PhysRevLett.126.077201} {\bibfield  {journal} {\bibinfo  {journal} {Phys. Rev. Lett.}\ }\textbf {\bibinfo {volume} {126}},\ \bibinfo {pages} {077201} (\bibinfo {year} {2021})}\BibitemShut {NoStop}%
    \bibitem [{\citenamefont {{\v{Z}}nidari{\v{c}}}(2022)}]{vznidarivc2022solvable}%
      \BibitemOpen
      \bibfield  {author} {\bibinfo {author} {\bibfnamefont {M.}~\bibnamefont {{\v{Z}}nidari{\v{c}}}},\ }\href@noop {} {\bibfield  {journal} {\bibinfo  {journal} {Physical Review Research}\ }\textbf {\bibinfo {volume} {4}},\ \bibinfo {pages} {033041} (\bibinfo {year} {2022})}\BibitemShut {NoStop}%
    \bibitem [{\citenamefont {Gong}\ \emph {et~al.}(2022{\natexlab{a}})\citenamefont {Gong}, \citenamefont {Bello}, \citenamefont {Malz},\ and\ \citenamefont {Kunst}}]{gong2022anomalous}%
      \BibitemOpen
      \bibfield  {author} {\bibinfo {author} {\bibfnamefont {Z.}~\bibnamefont {Gong}}, \bibinfo {author} {\bibfnamefont {M.}~\bibnamefont {Bello}}, \bibinfo {author} {\bibfnamefont {D.}~\bibnamefont {Malz}}, \ and\ \bibinfo {author} {\bibfnamefont {F.~K.}\ \bibnamefont {Kunst}},\ }\href {\doibase 10.1103/PhysRevLett.129.223601} {\bibfield  {journal} {\bibinfo  {journal} {Phys. Rev. Lett.}\ }\textbf {\bibinfo {volume} {129}},\ \bibinfo {pages} {223601} (\bibinfo {year} {2022}{\natexlab{a}})}\BibitemShut {NoStop}%
    \bibitem [{\citenamefont {Gong}\ \emph {et~al.}(2022{\natexlab{b}})\citenamefont {Gong}, \citenamefont {Bello}, \citenamefont {Malz},\ and\ \citenamefont {Kunst}}]{gong2022bound}%
      \BibitemOpen
      \bibfield  {author} {\bibinfo {author} {\bibfnamefont {Z.}~\bibnamefont {Gong}}, \bibinfo {author} {\bibfnamefont {M.}~\bibnamefont {Bello}}, \bibinfo {author} {\bibfnamefont {D.}~\bibnamefont {Malz}}, \ and\ \bibinfo {author} {\bibfnamefont {F.~K.}\ \bibnamefont {Kunst}},\ }\href@noop {} {\bibfield  {journal} {\bibinfo  {journal} {Physical Review A}\ }\textbf {\bibinfo {volume} {106}},\ \bibinfo {pages} {053517} (\bibinfo {year} {2022}{\natexlab{b}})}\BibitemShut {NoStop}%
    \bibitem [{\citenamefont {Roccati}\ \emph {et~al.}(2022)\citenamefont {Roccati}, \citenamefont {Lorenzo}, \citenamefont {Calaj{\`o}}, \citenamefont {Palma}, \citenamefont {Carollo},\ and\ \citenamefont {Ciccarello}}]{roccati2022exotic}%
      \BibitemOpen
      \bibfield  {author} {\bibinfo {author} {\bibfnamefont {F.}~\bibnamefont {Roccati}}, \bibinfo {author} {\bibfnamefont {S.}~\bibnamefont {Lorenzo}}, \bibinfo {author} {\bibfnamefont {G.}~\bibnamefont {Calaj{\`o}}}, \bibinfo {author} {\bibfnamefont {G.~M.}\ \bibnamefont {Palma}}, \bibinfo {author} {\bibfnamefont {A.}~\bibnamefont {Carollo}}, \ and\ \bibinfo {author} {\bibfnamefont {F.}~\bibnamefont {Ciccarello}},\ }\href@noop {} {\bibfield  {journal} {\bibinfo  {journal} {Optica}\ }\textbf {\bibinfo {volume} {9}},\ \bibinfo {pages} {565} (\bibinfo {year} {2022})}\BibitemShut {NoStop}%
    \bibitem [{\citenamefont {Cao}\ \emph {et~al.}(2021{\natexlab{b}})\citenamefont {Cao}, \citenamefont {Du}, \citenamefont {Wang},\ and\ \citenamefont {Kou}}]{cao2021physics}%
      \BibitemOpen
      \bibfield  {author} {\bibinfo {author} {\bibfnamefont {K.}~\bibnamefont {Cao}}, \bibinfo {author} {\bibfnamefont {Q.}~\bibnamefont {Du}}, \bibinfo {author} {\bibfnamefont {X.-R.}\ \bibnamefont {Wang}}, \ and\ \bibinfo {author} {\bibfnamefont {S.-P.}\ \bibnamefont {Kou}},\ }\href@noop {} {\bibfield  {journal} {\bibinfo  {journal} {arXiv preprint arXiv:2109.03690}\ } (\bibinfo {year} {2021}{\natexlab{b}})}\BibitemShut {NoStop}%
    \bibitem [{\citenamefont {Zhou}\ \emph {et~al.}(2022{\natexlab{b}})\citenamefont {Zhou}, \citenamefont {Zhou}, \citenamefont {Zhang},\ and\ \citenamefont {Zhai}}]{zhou2022space}%
      \BibitemOpen
      \bibfield  {author} {\bibinfo {author} {\bibfnamefont {T.-G.}\ \bibnamefont {Zhou}}, \bibinfo {author} {\bibfnamefont {Y.-N.}\ \bibnamefont {Zhou}}, \bibinfo {author} {\bibfnamefont {P.}~\bibnamefont {Zhang}}, \ and\ \bibinfo {author} {\bibfnamefont {H.}~\bibnamefont {Zhai}},\ }\href@noop {} {\bibfield  {journal} {\bibinfo  {journal} {Physical Review Research}\ }\textbf {\bibinfo {volume} {4}},\ \bibinfo {pages} {L022039} (\bibinfo {year} {2022}{\natexlab{b}})}\BibitemShut {NoStop}%
    \bibitem [{\citenamefont {Xu}\ \emph {et~al.}(2021)\citenamefont {Xu}, \citenamefont {Zhang}, \citenamefont {Luo}, \citenamefont {Yu}, \citenamefont {Li},\ and\ \citenamefont {Zhang}}]{xu2021coexistence}%
      \BibitemOpen
      \bibfield  {author} {\bibinfo {author} {\bibfnamefont {K.}~\bibnamefont {Xu}}, \bibinfo {author} {\bibfnamefont {X.}~\bibnamefont {Zhang}}, \bibinfo {author} {\bibfnamefont {K.}~\bibnamefont {Luo}}, \bibinfo {author} {\bibfnamefont {R.}~\bibnamefont {Yu}}, \bibinfo {author} {\bibfnamefont {D.}~\bibnamefont {Li}}, \ and\ \bibinfo {author} {\bibfnamefont {H.}~\bibnamefont {Zhang}},\ }\href@noop {} {\bibfield  {journal} {\bibinfo  {journal} {Physical Review B}\ }\textbf {\bibinfo {volume} {103}},\ \bibinfo {pages} {125411} (\bibinfo {year} {2021})}\BibitemShut {NoStop}%
    \bibitem [{\citenamefont {Deng}\ \emph {et~al.}(2022)\citenamefont {Deng}, \citenamefont {Chen},\ and\ \citenamefont {Zhang}}]{deng2022nth}%
      \BibitemOpen
      \bibfield  {author} {\bibinfo {author} {\bibfnamefont {W.}~\bibnamefont {Deng}}, \bibinfo {author} {\bibfnamefont {T.}~\bibnamefont {Chen}}, \ and\ \bibinfo {author} {\bibfnamefont {X.}~\bibnamefont {Zhang}},\ }\href {\doibase 10.1103/PhysRevResearch.4.033109} {\bibfield  {journal} {\bibinfo  {journal} {Phys. Rev. Res.}\ }\textbf {\bibinfo {volume} {4}},\ \bibinfo {pages} {033109} (\bibinfo {year} {2022})}\BibitemShut {NoStop}%
    \bibitem [{\citenamefont {Zhang}\ \emph {et~al.}(2023{\natexlab{b}})\citenamefont {Zhang}, \citenamefont {Chen}, \citenamefont {Li}, \citenamefont {Lee},\ and\ \citenamefont {Zhang}}]{zhang2023electrical}%
      \BibitemOpen
      \bibfield  {author} {\bibinfo {author} {\bibfnamefont {H.}~\bibnamefont {Zhang}}, \bibinfo {author} {\bibfnamefont {T.}~\bibnamefont {Chen}}, \bibinfo {author} {\bibfnamefont {L.}~\bibnamefont {Li}}, \bibinfo {author} {\bibfnamefont {C.~H.}\ \bibnamefont {Lee}}, \ and\ \bibinfo {author} {\bibfnamefont {X.}~\bibnamefont {Zhang}},\ }\href@noop {} {\bibfield  {journal} {\bibinfo  {journal} {Physical Review B}\ }\textbf {\bibinfo {volume} {107}},\ \bibinfo {pages} {085426} (\bibinfo {year} {2023}{\natexlab{b}})}\BibitemShut {NoStop}%
    \bibitem [{\citenamefont {Stegmaier}\ \emph {et~al.}(2024)\citenamefont {Stegmaier}, \citenamefont {Brand}, \citenamefont {Imhof}, \citenamefont {Fritzsche}, \citenamefont {Helbig}, \citenamefont {Hofmann}, \citenamefont {Boettcher}, \citenamefont {Greiter}, \citenamefont {Lee}, \citenamefont {Bahl} \emph {et~al.}}]{stegmaier2024realizing}%
      \BibitemOpen
      \bibfield  {author} {\bibinfo {author} {\bibfnamefont {A.}~\bibnamefont {Stegmaier}}, \bibinfo {author} {\bibfnamefont {H.}~\bibnamefont {Brand}}, \bibinfo {author} {\bibfnamefont {S.}~\bibnamefont {Imhof}}, \bibinfo {author} {\bibfnamefont {A.}~\bibnamefont {Fritzsche}}, \bibinfo {author} {\bibfnamefont {T.}~\bibnamefont {Helbig}}, \bibinfo {author} {\bibfnamefont {T.}~\bibnamefont {Hofmann}}, \bibinfo {author} {\bibfnamefont {I.}~\bibnamefont {Boettcher}}, \bibinfo {author} {\bibfnamefont {M.}~\bibnamefont {Greiter}}, \bibinfo {author} {\bibfnamefont {C.~H.}\ \bibnamefont {Lee}}, \bibinfo {author} {\bibfnamefont {G.}~\bibnamefont {Bahl}},  \emph {et~al.},\ }\href@noop {} {\bibfield  {journal} {\bibinfo  {journal} {Physical Review Research}\ }\textbf {\bibinfo {volume} {6}},\ \bibinfo {pages} {023010} (\bibinfo {year} {2024})}\BibitemShut {NoStop}%
    \bibitem [{\citenamefont {Li}\ \emph {et~al.}(2019{\natexlab{c}})\citenamefont {Li}, \citenamefont {Lee},\ and\ \citenamefont {Gong}}]{li2019emergence}%
      \BibitemOpen
      \bibfield  {author} {\bibinfo {author} {\bibfnamefont {L.}~\bibnamefont {Li}}, \bibinfo {author} {\bibfnamefont {C.~H.}\ \bibnamefont {Lee}}, \ and\ \bibinfo {author} {\bibfnamefont {J.}~\bibnamefont {Gong}},\ }\href@noop {} {\bibfield  {journal} {\bibinfo  {journal} {Communications physics}\ }\textbf {\bibinfo {volume} {2}},\ \bibinfo {pages} {1} (\bibinfo {year} {2019}{\natexlab{c}})}\BibitemShut {NoStop}%
    \bibitem [{\citenamefont {Shang}\ \emph {et~al.}(2024)\citenamefont {Shang}, \citenamefont {Liu}, \citenamefont {Jiang}, \citenamefont {Shao}, \citenamefont {Zang}, \citenamefont {Lee}, \citenamefont {Thomale}, \citenamefont {Manchon}, \citenamefont {Cui},\ and\ \citenamefont {Schwingenschl{\"o}gl}}]{shang2024observation}%
      \BibitemOpen
      \bibfield  {author} {\bibinfo {author} {\bibfnamefont {C.}~\bibnamefont {Shang}}, \bibinfo {author} {\bibfnamefont {S.}~\bibnamefont {Liu}}, \bibinfo {author} {\bibfnamefont {C.}~\bibnamefont {Jiang}}, \bibinfo {author} {\bibfnamefont {R.}~\bibnamefont {Shao}}, \bibinfo {author} {\bibfnamefont {X.}~\bibnamefont {Zang}}, \bibinfo {author} {\bibfnamefont {C.~H.}\ \bibnamefont {Lee}}, \bibinfo {author} {\bibfnamefont {R.}~\bibnamefont {Thomale}}, \bibinfo {author} {\bibfnamefont {A.}~\bibnamefont {Manchon}}, \bibinfo {author} {\bibfnamefont {T.~J.}\ \bibnamefont {Cui}}, \ and\ \bibinfo {author} {\bibfnamefont {U.}~\bibnamefont {Schwingenschl{\"o}gl}},\ }\href@noop {} {\bibfield  {journal} {\bibinfo  {journal} {Advanced Science}\ ,\ \bibinfo {pages} {2303222}} (\bibinfo {year} {2024})}\BibitemShut {NoStop}%
    \bibitem [{\citenamefont {Lin}\ \emph {et~al.}(2022{\natexlab{a}})\citenamefont {Lin}, \citenamefont {Li}, \citenamefont {Xiao}, \citenamefont {Wang}, \citenamefont {Yi},\ and\ \citenamefont {Xue}}]{lin2022simulating}%
      \BibitemOpen
      \bibfield  {author} {\bibinfo {author} {\bibfnamefont {Q.}~\bibnamefont {Lin}}, \bibinfo {author} {\bibfnamefont {T.}~\bibnamefont {Li}}, \bibinfo {author} {\bibfnamefont {L.}~\bibnamefont {Xiao}}, \bibinfo {author} {\bibfnamefont {K.}~\bibnamefont {Wang}}, \bibinfo {author} {\bibfnamefont {W.}~\bibnamefont {Yi}}, \ and\ \bibinfo {author} {\bibfnamefont {P.}~\bibnamefont {Xue}},\ }\href {\doibase 10.1103/PhysRevLett.129.113601} {\bibfield  {journal} {\bibinfo  {journal} {Phys. Rev. Lett.}\ }\textbf {\bibinfo {volume} {129}},\ \bibinfo {pages} {113601} (\bibinfo {year} {2022}{\natexlab{a}})}\BibitemShut {NoStop}%
    \bibitem [{\citenamefont {Lin}\ \emph {et~al.}(2022{\natexlab{b}})\citenamefont {Lin}, \citenamefont {Li}, \citenamefont {Xiao}, \citenamefont {Wang}, \citenamefont {Yi},\ and\ \citenamefont {Xue}}]{lin2022observation}%
      \BibitemOpen
      \bibfield  {author} {\bibinfo {author} {\bibfnamefont {Q.}~\bibnamefont {Lin}}, \bibinfo {author} {\bibfnamefont {T.}~\bibnamefont {Li}}, \bibinfo {author} {\bibfnamefont {L.}~\bibnamefont {Xiao}}, \bibinfo {author} {\bibfnamefont {K.}~\bibnamefont {Wang}}, \bibinfo {author} {\bibfnamefont {W.}~\bibnamefont {Yi}}, \ and\ \bibinfo {author} {\bibfnamefont {P.}~\bibnamefont {Xue}},\ }\href@noop {} {\bibfield  {journal} {\bibinfo  {journal} {Nature Communications}\ }\textbf {\bibinfo {volume} {13}},\ \bibinfo {pages} {3229} (\bibinfo {year} {2022}{\natexlab{b}})}\BibitemShut {NoStop}%
    \bibitem [{\citenamefont {Lucas S.~Palacios}(2021)}]{palacios2021guided}%
      \BibitemOpen
      \bibfield  {author} {\bibinfo {author} {\bibfnamefont {M.~G. I. P. S. S. A. G.~G.}\ \bibnamefont {Lucas S.~Palacios}, \bibfnamefont {Serguei~Tchoumakov}},\ }\href@noop {} {\bibfield  {journal} {\bibinfo  {journal} {Nature communications}\ }\textbf {\bibinfo {volume} {12}},\ \bibinfo {pages} {4691} (\bibinfo {year} {2021})}\BibitemShut {NoStop}%
    \bibitem [{\citenamefont {Yu}\ \emph {et~al.}(2022)\citenamefont {Yu}, \citenamefont {Yu}, \citenamefont {Zhang}, \citenamefont {Zhang}, \citenamefont {Ouyang}, \citenamefont {Liu}, \citenamefont {Deng},\ and\ \citenamefont {Duan}}]{yu2021experimental}%
      \BibitemOpen
      \bibfield  {author} {\bibinfo {author} {\bibfnamefont {Y.}~\bibnamefont {Yu}}, \bibinfo {author} {\bibfnamefont {L.-W.}\ \bibnamefont {Yu}}, \bibinfo {author} {\bibfnamefont {W.}~\bibnamefont {Zhang}}, \bibinfo {author} {\bibfnamefont {H.}~\bibnamefont {Zhang}}, \bibinfo {author} {\bibfnamefont {X.}~\bibnamefont {Ouyang}}, \bibinfo {author} {\bibfnamefont {Y.}~\bibnamefont {Liu}}, \bibinfo {author} {\bibfnamefont {D.-L.}\ \bibnamefont {Deng}}, \ and\ \bibinfo {author} {\bibfnamefont {L.-M.}\ \bibnamefont {Duan}},\ }\href@noop {} {\bibfield  {journal} {\bibinfo  {journal} {npj Quantum Information}\ }\textbf {\bibinfo {volume} {8}},\ \bibinfo {pages} {1} (\bibinfo {year} {2022})}\BibitemShut {NoStop}%
    \bibitem [{\citenamefont {Bergholtz}\ \emph {et~al.}(2021)\citenamefont {Bergholtz}, \citenamefont {Budich},\ and\ \citenamefont {Kunst}}]{bergholtz2021exceptional}%
      \BibitemOpen
      \bibfield  {author} {\bibinfo {author} {\bibfnamefont {E.~J.}\ \bibnamefont {Bergholtz}}, \bibinfo {author} {\bibfnamefont {J.~C.}\ \bibnamefont {Budich}}, \ and\ \bibinfo {author} {\bibfnamefont {F.~K.}\ \bibnamefont {Kunst}},\ }\href@noop {} {\bibfield  {journal} {\bibinfo  {journal} {Reviews of Modern Physics}\ }\textbf {\bibinfo {volume} {93}},\ \bibinfo {pages} {015005} (\bibinfo {year} {2021})}\BibitemShut {NoStop}%
    \bibitem [{\citenamefont {Zhang}\ \emph {et~al.}(2022{\natexlab{e}})\citenamefont {Zhang}, \citenamefont {Zhang}, \citenamefont {Lu},\ and\ \citenamefont {Chen}}]{zhang2022review}%
      \BibitemOpen
      \bibfield  {author} {\bibinfo {author} {\bibfnamefont {X.}~\bibnamefont {Zhang}}, \bibinfo {author} {\bibfnamefont {T.}~\bibnamefont {Zhang}}, \bibinfo {author} {\bibfnamefont {M.-H.}\ \bibnamefont {Lu}}, \ and\ \bibinfo {author} {\bibfnamefont {Y.-F.}\ \bibnamefont {Chen}},\ }\href {\doibase 10.1080/23746149.2022.2109431} {\bibfield  {journal} {\bibinfo  {journal} {Advances in Physics: X}\ }\textbf {\bibinfo {volume} {7}},\ \bibinfo {pages} {2109431} (\bibinfo {year} {2022}{\natexlab{e}})},\ \Eprint {http://arxiv.org/abs/https://doi.org/10.1080/23746149.2022.2109431} {https://doi.org/10.1080/23746149.2022.2109431} \BibitemShut {NoStop}%
    \bibitem [{\citenamefont {Zhu}\ \emph {et~al.}(2022{\natexlab{b}})\citenamefont {Zhu}, \citenamefont {Wang}, \citenamefont {Leykam}, \citenamefont {Xue}, \citenamefont {Wang},\ and\ \citenamefont {Chong}}]{zhu2022anomalous}%
      \BibitemOpen
      \bibfield  {author} {\bibinfo {author} {\bibfnamefont {B.}~\bibnamefont {Zhu}}, \bibinfo {author} {\bibfnamefont {Q.}~\bibnamefont {Wang}}, \bibinfo {author} {\bibfnamefont {D.}~\bibnamefont {Leykam}}, \bibinfo {author} {\bibfnamefont {H.}~\bibnamefont {Xue}}, \bibinfo {author} {\bibfnamefont {Q.~J.}\ \bibnamefont {Wang}}, \ and\ \bibinfo {author} {\bibfnamefont {Y.}~\bibnamefont {Chong}},\ }\href@noop {} {\bibfield  {journal} {\bibinfo  {journal} {Physical Review Letters}\ }\textbf {\bibinfo {volume} {129}},\ \bibinfo {pages} {013903} (\bibinfo {year} {2022}{\natexlab{b}})}\BibitemShut {NoStop}%
    \bibitem [{\citenamefont {Mandal}\ \emph {et~al.}(2022)\citenamefont {Mandal}, \citenamefont {Banerjee},\ and\ \citenamefont {Liew}}]{mandal2022topological}%
      \BibitemOpen
      \bibfield  {author} {\bibinfo {author} {\bibfnamefont {S.}~\bibnamefont {Mandal}}, \bibinfo {author} {\bibfnamefont {R.}~\bibnamefont {Banerjee}}, \ and\ \bibinfo {author} {\bibfnamefont {T.~C.}\ \bibnamefont {Liew}},\ }\href@noop {} {\bibfield  {journal} {\bibinfo  {journal} {ACS Photonics}\ }\textbf {\bibinfo {volume} {9}},\ \bibinfo {pages} {527} (\bibinfo {year} {2022})}\BibitemShut {NoStop}%
    \bibitem [{\citenamefont {Xu}\ \emph {et~al.}(2022)\citenamefont {Xu}, \citenamefont {Bao},\ and\ \citenamefont {Liew}}]{xu2022non}%
      \BibitemOpen
      \bibfield  {author} {\bibinfo {author} {\bibfnamefont {X.}~\bibnamefont {Xu}}, \bibinfo {author} {\bibfnamefont {R.}~\bibnamefont {Bao}}, \ and\ \bibinfo {author} {\bibfnamefont {T.~C.~H.}\ \bibnamefont {Liew}},\ }\href {\doibase 10.1103/PhysRevB.106.L201302} {\bibfield  {journal} {\bibinfo  {journal} {Phys. Rev. B}\ }\textbf {\bibinfo {volume} {106}},\ \bibinfo {pages} {L201302} (\bibinfo {year} {2022})}\BibitemShut {NoStop}%
    \bibitem [{\citenamefont {Yu}\ \emph {et~al.}(2021)\citenamefont {Yu}, \citenamefont {Xue}, \citenamefont {Zhuang}, \citenamefont {Zhao},\ and\ \citenamefont {Liu}}]{yu2021non}%
      \BibitemOpen
      \bibfield  {author} {\bibinfo {author} {\bibfnamefont {Z.-F.}\ \bibnamefont {Yu}}, \bibinfo {author} {\bibfnamefont {J.-K.}\ \bibnamefont {Xue}}, \bibinfo {author} {\bibfnamefont {L.}~\bibnamefont {Zhuang}}, \bibinfo {author} {\bibfnamefont {J.}~\bibnamefont {Zhao}}, \ and\ \bibinfo {author} {\bibfnamefont {W.-M.}\ \bibnamefont {Liu}},\ }\href@noop {} {\bibfield  {journal} {\bibinfo  {journal} {Physical Review B}\ }\textbf {\bibinfo {volume} {104}},\ \bibinfo {pages} {235408} (\bibinfo {year} {2021})}\BibitemShut {NoStop}%
    \bibitem [{\citenamefont {Yang}\ \emph {et~al.}(2022{\natexlab{d}})\citenamefont {Yang}, \citenamefont {Wang}, \citenamefont {Wu}, \citenamefont {Xiao}, \citenamefont {Yu}, \citenamefont {Yuan},\ and\ \citenamefont {Chen}}]{yang2022concentrated}%
      \BibitemOpen
      \bibfield  {author} {\bibinfo {author} {\bibfnamefont {M.}~\bibnamefont {Yang}}, \bibinfo {author} {\bibfnamefont {L.}~\bibnamefont {Wang}}, \bibinfo {author} {\bibfnamefont {X.}~\bibnamefont {Wu}}, \bibinfo {author} {\bibfnamefont {H.}~\bibnamefont {Xiao}}, \bibinfo {author} {\bibfnamefont {D.}~\bibnamefont {Yu}}, \bibinfo {author} {\bibfnamefont {L.}~\bibnamefont {Yuan}}, \ and\ \bibinfo {author} {\bibfnamefont {X.}~\bibnamefont {Chen}},\ }\href {\doibase 10.1103/PhysRevA.106.043717} {\bibfield  {journal} {\bibinfo  {journal} {Phys. Rev. A}\ }\textbf {\bibinfo {volume} {106}},\ \bibinfo {pages} {043717} (\bibinfo {year} {2022}{\natexlab{d}})}\BibitemShut {NoStop}%
    \bibitem [{\citenamefont {Song}\ \emph {et~al.}(2023)\citenamefont {Song}, \citenamefont {Wu}, \citenamefont {Chen}, \citenamefont {Chen}, \citenamefont {Huang}, \citenamefont {Yuan}, \citenamefont {Zhu},\ and\ \citenamefont {Li}}]{song2023observation}%
      \BibitemOpen
      \bibfield  {author} {\bibinfo {author} {\bibfnamefont {W.}~\bibnamefont {Song}}, \bibinfo {author} {\bibfnamefont {S.}~\bibnamefont {Wu}}, \bibinfo {author} {\bibfnamefont {C.}~\bibnamefont {Chen}}, \bibinfo {author} {\bibfnamefont {Y.}~\bibnamefont {Chen}}, \bibinfo {author} {\bibfnamefont {C.}~\bibnamefont {Huang}}, \bibinfo {author} {\bibfnamefont {L.}~\bibnamefont {Yuan}}, \bibinfo {author} {\bibfnamefont {S.}~\bibnamefont {Zhu}}, \ and\ \bibinfo {author} {\bibfnamefont {T.}~\bibnamefont {Li}},\ }\href@noop {} {\bibfield  {journal} {\bibinfo  {journal} {Physical Review Letters}\ }\textbf {\bibinfo {volume} {130}},\ \bibinfo {pages} {043803} (\bibinfo {year} {2023})}\BibitemShut {NoStop}%
    \bibitem [{\citenamefont {Roccati}\ \emph {et~al.}(2024)\citenamefont {Roccati}, \citenamefont {Bello}, \citenamefont {Gong}, \citenamefont {Ueda}, \citenamefont {Ciccarello}, \citenamefont {Chenu},\ and\ \citenamefont {Carollo}}]{roccati2024hermitian}%
      \BibitemOpen
      \bibfield  {author} {\bibinfo {author} {\bibfnamefont {F.}~\bibnamefont {Roccati}}, \bibinfo {author} {\bibfnamefont {M.}~\bibnamefont {Bello}}, \bibinfo {author} {\bibfnamefont {Z.}~\bibnamefont {Gong}}, \bibinfo {author} {\bibfnamefont {M.}~\bibnamefont {Ueda}}, \bibinfo {author} {\bibfnamefont {F.}~\bibnamefont {Ciccarello}}, \bibinfo {author} {\bibfnamefont {A.}~\bibnamefont {Chenu}}, \ and\ \bibinfo {author} {\bibfnamefont {A.}~\bibnamefont {Carollo}},\ }\href@noop {} {\bibfield  {journal} {\bibinfo  {journal} {Nature Communications}\ }\textbf {\bibinfo {volume} {15}},\ \bibinfo {pages} {2400} (\bibinfo {year} {2024})}\BibitemShut {NoStop}%
    \bibitem [{\citenamefont {Lin}\ \emph {et~al.}(2021{\natexlab{a}})\citenamefont {Lin}, \citenamefont {Ding}, \citenamefont {Ke},\ and\ \citenamefont {Li}}]{lin2021steering}%
      \BibitemOpen
      \bibfield  {author} {\bibinfo {author} {\bibfnamefont {Z.}~\bibnamefont {Lin}}, \bibinfo {author} {\bibfnamefont {L.}~\bibnamefont {Ding}}, \bibinfo {author} {\bibfnamefont {S.}~\bibnamefont {Ke}}, \ and\ \bibinfo {author} {\bibfnamefont {X.}~\bibnamefont {Li}},\ }\href@noop {} {\bibfield  {journal} {\bibinfo  {journal} {Optics Letters}\ }\textbf {\bibinfo {volume} {46}},\ \bibinfo {pages} {3512} (\bibinfo {year} {2021}{\natexlab{a}})}\BibitemShut {NoStop}%
    \bibitem [{\citenamefont {Zhong}\ \emph {et~al.}(2021)\citenamefont {Zhong}, \citenamefont {Wang}, \citenamefont {Park}, \citenamefont {Asadchy}, \citenamefont {Wojcik}, \citenamefont {Dutt},\ and\ \citenamefont {Fan}}]{zhong2021nontrivial}%
      \BibitemOpen
      \bibfield  {author} {\bibinfo {author} {\bibfnamefont {J.}~\bibnamefont {Zhong}}, \bibinfo {author} {\bibfnamefont {K.}~\bibnamefont {Wang}}, \bibinfo {author} {\bibfnamefont {Y.}~\bibnamefont {Park}}, \bibinfo {author} {\bibfnamefont {V.}~\bibnamefont {Asadchy}}, \bibinfo {author} {\bibfnamefont {C.~C.}\ \bibnamefont {Wojcik}}, \bibinfo {author} {\bibfnamefont {A.}~\bibnamefont {Dutt}}, \ and\ \bibinfo {author} {\bibfnamefont {S.}~\bibnamefont {Fan}},\ }\href@noop {} {\bibfield  {journal} {\bibinfo  {journal} {Physical Review B}\ }\textbf {\bibinfo {volume} {104}},\ \bibinfo {pages} {125416} (\bibinfo {year} {2021})}\BibitemShut {NoStop}%
    \bibitem [{\citenamefont {Price}\ \emph {et~al.}(2022)\citenamefont {Price}, \citenamefont {Chong}, \citenamefont {Khanikaev}, \citenamefont {Schomerus}, \citenamefont {Maczewsky}, \citenamefont {Kremer}, \citenamefont {Heinrich}, \citenamefont {Szameit}, \citenamefont {Zilberberg}, \citenamefont {Yang} \emph {et~al.}}]{price2022roadmap}%
      \BibitemOpen
      \bibfield  {author} {\bibinfo {author} {\bibfnamefont {H.}~\bibnamefont {Price}}, \bibinfo {author} {\bibfnamefont {Y.}~\bibnamefont {Chong}}, \bibinfo {author} {\bibfnamefont {A.}~\bibnamefont {Khanikaev}}, \bibinfo {author} {\bibfnamefont {H.}~\bibnamefont {Schomerus}}, \bibinfo {author} {\bibfnamefont {L.~J.}\ \bibnamefont {Maczewsky}}, \bibinfo {author} {\bibfnamefont {M.}~\bibnamefont {Kremer}}, \bibinfo {author} {\bibfnamefont {M.}~\bibnamefont {Heinrich}}, \bibinfo {author} {\bibfnamefont {A.}~\bibnamefont {Szameit}}, \bibinfo {author} {\bibfnamefont {O.}~\bibnamefont {Zilberberg}}, \bibinfo {author} {\bibfnamefont {Y.}~\bibnamefont {Yang}},  \emph {et~al.},\ }\href@noop {} {\bibfield  {journal} {\bibinfo  {journal} {Journal of Physics: Photonics}\ }\textbf {\bibinfo {volume} {4}},\ \bibinfo {pages} {032501} (\bibinfo {year} {2022})}\BibitemShut {NoStop}%
    \bibitem [{\citenamefont {Yan}\ \emph {et~al.}(2021)\citenamefont {Yan}, \citenamefont {Chen},\ and\ \citenamefont {Yang}}]{yan2021non}%
      \BibitemOpen
      \bibfield  {author} {\bibinfo {author} {\bibfnamefont {Q.}~\bibnamefont {Yan}}, \bibinfo {author} {\bibfnamefont {H.}~\bibnamefont {Chen}}, \ and\ \bibinfo {author} {\bibfnamefont {Y.}~\bibnamefont {Yang}},\ }\href@noop {} {\bibfield  {journal} {\bibinfo  {journal} {arXiv preprint arXiv:2111.08213}\ } (\bibinfo {year} {2021})}\BibitemShut {NoStop}%
    \bibitem [{\citenamefont {Xie}\ \emph {et~al.}(2023)\citenamefont {Xie}, \citenamefont {Jin},\ and\ \citenamefont {Song}}]{xie2023antihelical}%
      \BibitemOpen
      \bibfield  {author} {\bibinfo {author} {\bibfnamefont {L.}~\bibnamefont {Xie}}, \bibinfo {author} {\bibfnamefont {L.}~\bibnamefont {Jin}}, \ and\ \bibinfo {author} {\bibfnamefont {Z.}~\bibnamefont {Song}},\ }\href@noop {} {\bibfield  {journal} {\bibinfo  {journal} {arXiv preprint arXiv:2302.05842}\ } (\bibinfo {year} {2023})}\BibitemShut {NoStop}%
    \bibitem [{\citenamefont {Lin}\ \emph {et~al.}(2022{\natexlab{c}})\citenamefont {Lin}, \citenamefont {Yi},\ and\ \citenamefont {Xue}}]{lin2022manipulating}%
      \BibitemOpen
      \bibfield  {author} {\bibinfo {author} {\bibfnamefont {Q.}~\bibnamefont {Lin}}, \bibinfo {author} {\bibfnamefont {W.}~\bibnamefont {Yi}}, \ and\ \bibinfo {author} {\bibfnamefont {P.}~\bibnamefont {Xue}},\ }\href@noop {} {\bibfield  {journal} {\bibinfo  {journal} {arXiv preprint arXiv:2212.00387}\ } (\bibinfo {year} {2022}{\natexlab{c}})}\BibitemShut {NoStop}%
    \bibitem [{\citenamefont {Li}\ \emph {et~al.}(2021{\natexlab{d}})\citenamefont {Li}, \citenamefont {Zhang},\ and\ \citenamefont {Yi}}]{li2021two}%
      \BibitemOpen
      \bibfield  {author} {\bibinfo {author} {\bibfnamefont {T.}~\bibnamefont {Li}}, \bibinfo {author} {\bibfnamefont {Y.-S.}\ \bibnamefont {Zhang}}, \ and\ \bibinfo {author} {\bibfnamefont {W.}~\bibnamefont {Yi}},\ }\href@noop {} {\bibfield  {journal} {\bibinfo  {journal} {Chinese Physics Letters}\ }\textbf {\bibinfo {volume} {38}},\ \bibinfo {pages} {030301} (\bibinfo {year} {2021}{\natexlab{d}})}\BibitemShut {NoStop}%
    \bibitem [{\citenamefont {Pyrialakos}\ \emph {et~al.}(2022)\citenamefont {Pyrialakos}, \citenamefont {Ren}, \citenamefont {Jung}, \citenamefont {Khajavikhan},\ and\ \citenamefont {Christodoulides}}]{pyrialakos2022thermalization}%
      \BibitemOpen
      \bibfield  {author} {\bibinfo {author} {\bibfnamefont {G.~G.}\ \bibnamefont {Pyrialakos}}, \bibinfo {author} {\bibfnamefont {H.}~\bibnamefont {Ren}}, \bibinfo {author} {\bibfnamefont {P.~S.}\ \bibnamefont {Jung}}, \bibinfo {author} {\bibfnamefont {M.}~\bibnamefont {Khajavikhan}}, \ and\ \bibinfo {author} {\bibfnamefont {D.~N.}\ \bibnamefont {Christodoulides}},\ }\href@noop {} {\bibfield  {journal} {\bibinfo  {journal} {Physical Review Letters}\ }\textbf {\bibinfo {volume} {128}},\ \bibinfo {pages} {213901} (\bibinfo {year} {2022})}\BibitemShut {NoStop}%
    \bibitem [{\citenamefont {Fleckenstein}\ \emph {et~al.}(2022)\citenamefont {Fleckenstein}, \citenamefont {Zorzato}, \citenamefont {Varjas}, \citenamefont {Bergholtz}, \citenamefont {Bardarson},\ and\ \citenamefont {Tiwari}}]{fleckenstein2022non}%
      \BibitemOpen
      \bibfield  {author} {\bibinfo {author} {\bibfnamefont {C.}~\bibnamefont {Fleckenstein}}, \bibinfo {author} {\bibfnamefont {A.}~\bibnamefont {Zorzato}}, \bibinfo {author} {\bibfnamefont {D.}~\bibnamefont {Varjas}}, \bibinfo {author} {\bibfnamefont {E.~J.}\ \bibnamefont {Bergholtz}}, \bibinfo {author} {\bibfnamefont {J.~H.}\ \bibnamefont {Bardarson}}, \ and\ \bibinfo {author} {\bibfnamefont {A.}~\bibnamefont {Tiwari}},\ }\href {\doibase 10.1103/PhysRevResearch.4.L032026} {\bibfield  {journal} {\bibinfo  {journal} {Phys. Rev. Res.}\ }\textbf {\bibinfo {volume} {4}},\ \bibinfo {pages} {L032026} (\bibinfo {year} {2022})}\BibitemShut {NoStop}%
    \bibitem [{\citenamefont {Lin}\ \emph {et~al.}(2021{\natexlab{b}})\citenamefont {Lin}, \citenamefont {Dilip}, \citenamefont {Green}, \citenamefont {Smith},\ and\ \citenamefont {Pollmann}}]{lin2021real}%
      \BibitemOpen
      \bibfield  {author} {\bibinfo {author} {\bibfnamefont {S.-H.}\ \bibnamefont {Lin}}, \bibinfo {author} {\bibfnamefont {R.}~\bibnamefont {Dilip}}, \bibinfo {author} {\bibfnamefont {A.~G.}\ \bibnamefont {Green}}, \bibinfo {author} {\bibfnamefont {A.}~\bibnamefont {Smith}}, \ and\ \bibinfo {author} {\bibfnamefont {F.}~\bibnamefont {Pollmann}},\ }\href@noop {} {\bibfield  {journal} {\bibinfo  {journal} {PRX Quantum}\ }\textbf {\bibinfo {volume} {2}},\ \bibinfo {pages} {010342} (\bibinfo {year} {2021}{\natexlab{b}})}\BibitemShut {NoStop}%
    \bibitem [{\citenamefont {Liu}\ \emph {et~al.}(2021{\natexlab{f}})\citenamefont {Liu}, \citenamefont {Liu},\ and\ \citenamefont {Fan}}]{liu2021probabilistic}%
      \BibitemOpen
      \bibfield  {author} {\bibinfo {author} {\bibfnamefont {T.}~\bibnamefont {Liu}}, \bibinfo {author} {\bibfnamefont {J.-G.}\ \bibnamefont {Liu}}, \ and\ \bibinfo {author} {\bibfnamefont {H.}~\bibnamefont {Fan}},\ }\href@noop {} {\bibfield  {journal} {\bibinfo  {journal} {Quantum Information Processing}\ }\textbf {\bibinfo {volume} {20}},\ \bibinfo {pages} {1} (\bibinfo {year} {2021}{\natexlab{f}})}\BibitemShut {NoStop}%
    \bibitem [{\citenamefont {Kamakari}\ \emph {et~al.}(2022)\citenamefont {Kamakari}, \citenamefont {Sun}, \citenamefont {Motta},\ and\ \citenamefont {Minnich}}]{kamakari2022digital}%
      \BibitemOpen
      \bibfield  {author} {\bibinfo {author} {\bibfnamefont {H.}~\bibnamefont {Kamakari}}, \bibinfo {author} {\bibfnamefont {S.-N.}\ \bibnamefont {Sun}}, \bibinfo {author} {\bibfnamefont {M.}~\bibnamefont {Motta}}, \ and\ \bibinfo {author} {\bibfnamefont {A.~J.}\ \bibnamefont {Minnich}},\ }\href@noop {} {\bibfield  {journal} {\bibinfo  {journal} {PRX Quantum}\ }\textbf {\bibinfo {volume} {3}},\ \bibinfo {pages} {010320} (\bibinfo {year} {2022})}\BibitemShut {NoStop}%
    \bibitem [{\citenamefont {Smith}\ \emph {et~al.}(2019)\citenamefont {Smith}, \citenamefont {Kim}, \citenamefont {Pollmann},\ and\ \citenamefont {Knolle}}]{smith2019simulating}%
      \BibitemOpen
      \bibfield  {author} {\bibinfo {author} {\bibfnamefont {A.}~\bibnamefont {Smith}}, \bibinfo {author} {\bibfnamefont {M.}~\bibnamefont {Kim}}, \bibinfo {author} {\bibfnamefont {F.}~\bibnamefont {Pollmann}}, \ and\ \bibinfo {author} {\bibfnamefont {J.}~\bibnamefont {Knolle}},\ }\href@noop {} {\bibfield  {journal} {\bibinfo  {journal} {npj Quantum Information}\ }\textbf {\bibinfo {volume} {5}},\ \bibinfo {pages} {1} (\bibinfo {year} {2019})}\BibitemShut {NoStop}%
    \bibitem [{\citenamefont {Koh}\ \emph {et~al.}(2022{\natexlab{a}})\citenamefont {Koh}, \citenamefont {Tai}, \citenamefont {Phee}, \citenamefont {Ng},\ and\ \citenamefont {Lee}}]{koh2022stabilizing}%
      \BibitemOpen
      \bibfield  {author} {\bibinfo {author} {\bibfnamefont {J.~M.}\ \bibnamefont {Koh}}, \bibinfo {author} {\bibfnamefont {T.}~\bibnamefont {Tai}}, \bibinfo {author} {\bibfnamefont {Y.~H.}\ \bibnamefont {Phee}}, \bibinfo {author} {\bibfnamefont {W.~E.}\ \bibnamefont {Ng}}, \ and\ \bibinfo {author} {\bibfnamefont {C.~H.}\ \bibnamefont {Lee}},\ }\href@noop {} {\bibfield  {journal} {\bibinfo  {journal} {npj Quantum Information}\ }\textbf {\bibinfo {volume} {8}},\ \bibinfo {pages} {1} (\bibinfo {year} {2022}{\natexlab{a}})}\BibitemShut {NoStop}%
    \bibitem [{\citenamefont {Frey}\ and\ \citenamefont {Rachel}(2022)}]{frey2022realization}%
      \BibitemOpen
      \bibfield  {author} {\bibinfo {author} {\bibfnamefont {P.}~\bibnamefont {Frey}}\ and\ \bibinfo {author} {\bibfnamefont {S.}~\bibnamefont {Rachel}},\ }\href@noop {} {\bibfield  {journal} {\bibinfo  {journal} {Science advances}\ }\textbf {\bibinfo {volume} {8}},\ \bibinfo {pages} {eabm7652} (\bibinfo {year} {2022})}\BibitemShut {NoStop}%
    \bibitem [{\citenamefont {Chen}\ \emph {et~al.}(2023{\natexlab{a}})\citenamefont {Chen}, \citenamefont {Shen}, \citenamefont {Lee}, \citenamefont {Yang},\ and\ \citenamefont {Bomantara}}]{chen2023robust}%
      \BibitemOpen
      \bibfield  {author} {\bibinfo {author} {\bibfnamefont {T.}~\bibnamefont {Chen}}, \bibinfo {author} {\bibfnamefont {R.}~\bibnamefont {Shen}}, \bibinfo {author} {\bibfnamefont {C.~H.}\ \bibnamefont {Lee}}, \bibinfo {author} {\bibfnamefont {B.}~\bibnamefont {Yang}}, \ and\ \bibinfo {author} {\bibfnamefont {R.~W.}\ \bibnamefont {Bomantara}},\ }\href@noop {} {\bibfield  {journal} {\bibinfo  {journal} {arXiv preprint arXiv:2309.11560}\ } (\bibinfo {year} {2023}{\natexlab{a}})}\BibitemShut {NoStop}%
    \bibitem [{\citenamefont {Koh}\ \emph {et~al.}(2022{\natexlab{b}})\citenamefont {Koh}, \citenamefont {Tai},\ and\ \citenamefont {Lee}}]{koh2022simulation}%
      \BibitemOpen
      \bibfield  {author} {\bibinfo {author} {\bibfnamefont {J.~M.}\ \bibnamefont {Koh}}, \bibinfo {author} {\bibfnamefont {T.}~\bibnamefont {Tai}}, \ and\ \bibinfo {author} {\bibfnamefont {C.~H.}\ \bibnamefont {Lee}},\ }\href@noop {} {\bibfield  {journal} {\bibinfo  {journal} {Physical Review Letters}\ }\textbf {\bibinfo {volume} {129}},\ \bibinfo {pages} {140502} (\bibinfo {year} {2022}{\natexlab{b}})}\BibitemShut {NoStop}%
    \bibitem [{\citenamefont {Koh}\ \emph {et~al.}(2023)\citenamefont {Koh}, \citenamefont {Tai},\ and\ \citenamefont {Lee}}]{koh2023observation}%
      \BibitemOpen
      \bibfield  {author} {\bibinfo {author} {\bibfnamefont {J.~M.}\ \bibnamefont {Koh}}, \bibinfo {author} {\bibfnamefont {T.}~\bibnamefont {Tai}}, \ and\ \bibinfo {author} {\bibfnamefont {C.~H.}\ \bibnamefont {Lee}},\ }\href@noop {} {\bibfield  {journal} {\bibinfo  {journal} {arXiv preprint arXiv:2303.02179}\ } (\bibinfo {year} {2023})}\BibitemShut {NoStop}%
    \bibitem [{\citenamefont {Chen}\ \emph {et~al.}(2023{\natexlab{b}})\citenamefont {Chen}, \citenamefont {Shen}, \citenamefont {Lee},\ and\ \citenamefont {Yang}}]{chen2022high}%
      \BibitemOpen
      \bibfield  {author} {\bibinfo {author} {\bibfnamefont {T.}~\bibnamefont {Chen}}, \bibinfo {author} {\bibfnamefont {R.}~\bibnamefont {Shen}}, \bibinfo {author} {\bibfnamefont {C.~H.}\ \bibnamefont {Lee}}, \ and\ \bibinfo {author} {\bibfnamefont {B.}~\bibnamefont {Yang}},\ }\href@noop {} {\bibfield  {journal} {\bibinfo  {journal} {SciPost Physics}\ }\textbf {\bibinfo {volume} {15}},\ \bibinfo {pages} {170} (\bibinfo {year} {2023}{\natexlab{b}})}\BibitemShut {NoStop}%
    \bibitem [{\citenamefont {Schomerus}(2020)}]{schomerus2020nonreciprocal}%
      \BibitemOpen
      \bibfield  {author} {\bibinfo {author} {\bibfnamefont {H.}~\bibnamefont {Schomerus}},\ }\href@noop {} {\bibfield  {journal} {\bibinfo  {journal} {Physical Review Research}\ }\textbf {\bibinfo {volume} {2}},\ \bibinfo {pages} {013058} (\bibinfo {year} {2020})}\BibitemShut {NoStop}%
    \bibitem [{\citenamefont {Braghini}\ \emph {et~al.}(2021)\citenamefont {Braghini}, \citenamefont {Villani}, \citenamefont {Rosa},\ and\ \citenamefont {de~F~Arruda}}]{braghini2021non}%
      \BibitemOpen
      \bibfield  {author} {\bibinfo {author} {\bibfnamefont {D.}~\bibnamefont {Braghini}}, \bibinfo {author} {\bibfnamefont {L.~G.}\ \bibnamefont {Villani}}, \bibinfo {author} {\bibfnamefont {M.~I.}\ \bibnamefont {Rosa}}, \ and\ \bibinfo {author} {\bibfnamefont {J.~R.}\ \bibnamefont {de~F~Arruda}},\ }\href@noop {} {\bibfield  {journal} {\bibinfo  {journal} {Journal of Physics D: Applied Physics}\ }\textbf {\bibinfo {volume} {54}},\ \bibinfo {pages} {285302} (\bibinfo {year} {2021})}\BibitemShut {NoStop}%
    \bibitem [{\citenamefont {Jin}\ \emph {et~al.}(2022)\citenamefont {Jin}, \citenamefont {Zhong}, \citenamefont {Cai}, \citenamefont {Zhuang}, \citenamefont {Pennec},\ and\ \citenamefont {Djafari-Rouhani}}]{jin2022non}%
      \BibitemOpen
      \bibfield  {author} {\bibinfo {author} {\bibfnamefont {Y.}~\bibnamefont {Jin}}, \bibinfo {author} {\bibfnamefont {W.}~\bibnamefont {Zhong}}, \bibinfo {author} {\bibfnamefont {R.}~\bibnamefont {Cai}}, \bibinfo {author} {\bibfnamefont {X.}~\bibnamefont {Zhuang}}, \bibinfo {author} {\bibfnamefont {Y.}~\bibnamefont {Pennec}}, \ and\ \bibinfo {author} {\bibfnamefont {B.}~\bibnamefont {Djafari-Rouhani}},\ }\href@noop {} {\bibfield  {journal} {\bibinfo  {journal} {Applied Physics Letters}\ }\textbf {\bibinfo {volume} {121}},\ \bibinfo {pages} {022202} (\bibinfo {year} {2022})}\BibitemShut {NoStop}%
    \bibitem [{\citenamefont {Wen}\ \emph {et~al.}(2022)\citenamefont {Wen}, \citenamefont {Wang},\ and\ \citenamefont {Long}}]{wen2022optomechanically}%
      \BibitemOpen
      \bibfield  {author} {\bibinfo {author} {\bibfnamefont {P.}~\bibnamefont {Wen}}, \bibinfo {author} {\bibfnamefont {M.}~\bibnamefont {Wang}}, \ and\ \bibinfo {author} {\bibfnamefont {G.-L.}\ \bibnamefont {Long}},\ }\href {\doibase 10.1364/OE.473652} {\bibfield  {journal} {\bibinfo  {journal} {Optics Express}\ }\textbf {\bibinfo {volume} {30}},\ \bibinfo {pages} {41012} (\bibinfo {year} {2022})}\BibitemShut {NoStop}%
    \bibitem [{\citenamefont {Ren}\ \emph {et~al.}(2022)\citenamefont {Ren}, \citenamefont {Liu}, \citenamefont {Zhao}, \citenamefont {He}, \citenamefont {Pak}, \citenamefont {Li},\ and\ \citenamefont {Jo}}]{ren2022chiral}%
      \BibitemOpen
      \bibfield  {author} {\bibinfo {author} {\bibfnamefont {Z.}~\bibnamefont {Ren}}, \bibinfo {author} {\bibfnamefont {D.}~\bibnamefont {Liu}}, \bibinfo {author} {\bibfnamefont {E.}~\bibnamefont {Zhao}}, \bibinfo {author} {\bibfnamefont {C.}~\bibnamefont {He}}, \bibinfo {author} {\bibfnamefont {K.~K.}\ \bibnamefont {Pak}}, \bibinfo {author} {\bibfnamefont {J.}~\bibnamefont {Li}}, \ and\ \bibinfo {author} {\bibfnamefont {G.-B.}\ \bibnamefont {Jo}},\ }\href@noop {} {\bibfield  {journal} {\bibinfo  {journal} {Nature Physics}\ }\textbf {\bibinfo {volume} {18}},\ \bibinfo {pages} {385} (\bibinfo {year} {2022})}\BibitemShut {NoStop}%
    \bibitem [{\citenamefont {Guo}\ \emph {et~al.}(2022)\citenamefont {Guo}, \citenamefont {Dong}, \citenamefont {Zhang}, \citenamefont {Hu},\ and\ \citenamefont {Yang}}]{guo2022theoretical}%
      \BibitemOpen
      \bibfield  {author} {\bibinfo {author} {\bibfnamefont {S.}~\bibnamefont {Guo}}, \bibinfo {author} {\bibfnamefont {C.}~\bibnamefont {Dong}}, \bibinfo {author} {\bibfnamefont {F.}~\bibnamefont {Zhang}}, \bibinfo {author} {\bibfnamefont {J.}~\bibnamefont {Hu}}, \ and\ \bibinfo {author} {\bibfnamefont {Z.}~\bibnamefont {Yang}},\ }\href {\doibase 10.1103/PhysRevA.106.L061302} {\bibfield  {journal} {\bibinfo  {journal} {Phys. Rev. A}\ }\textbf {\bibinfo {volume} {106}},\ \bibinfo {pages} {L061302} (\bibinfo {year} {2022})}\BibitemShut {NoStop}%
    \bibitem [{\citenamefont {Zhou}\ \emph {et~al.}(2022{\natexlab{c}})\citenamefont {Zhou}, \citenamefont {Li}, \citenamefont {Yi},\ and\ \citenamefont {Cui}}]{zhou2022engineering}%
      \BibitemOpen
      \bibfield  {author} {\bibinfo {author} {\bibfnamefont {L.}~\bibnamefont {Zhou}}, \bibinfo {author} {\bibfnamefont {H.}~\bibnamefont {Li}}, \bibinfo {author} {\bibfnamefont {W.}~\bibnamefont {Yi}}, \ and\ \bibinfo {author} {\bibfnamefont {X.}~\bibnamefont {Cui}},\ }\href@noop {} {\bibfield  {journal} {\bibinfo  {journal} {Communications Physics}\ }\textbf {\bibinfo {volume} {5}},\ \bibinfo {pages} {252} (\bibinfo {year} {2022}{\natexlab{c}})}\BibitemShut {NoStop}%
    \bibitem [{\citenamefont {Li}\ \emph {et~al.}(2022{\natexlab{e}})\citenamefont {Li}, \citenamefont {Cui},\ and\ \citenamefont {Yi}}]{li2022BEC}%
      \BibitemOpen
      \bibfield  {author} {\bibinfo {author} {\bibfnamefont {H.}~\bibnamefont {Li}}, \bibinfo {author} {\bibfnamefont {X.}~\bibnamefont {Cui}}, \ and\ \bibinfo {author} {\bibfnamefont {W.}~\bibnamefont {Yi}},\ }\href {\doibase 10.52396/JUSTC-2022-0003} {\bibfield  {journal} {\bibinfo  {journal} {JUSTC}\ }\textbf {\bibinfo {volume} {52}},\ \bibinfo {pages} {2} (\bibinfo {year} {2022}{\natexlab{e}})}\BibitemShut {NoStop}%
    \bibitem [{\citenamefont {Yoshida}\ \emph {et~al.}(2022)\citenamefont {Yoshida}, \citenamefont {Mizoguchi},\ and\ \citenamefont {Hatsugai}}]{yoshida2022non}%
      \BibitemOpen
      \bibfield  {author} {\bibinfo {author} {\bibfnamefont {T.}~\bibnamefont {Yoshida}}, \bibinfo {author} {\bibfnamefont {T.}~\bibnamefont {Mizoguchi}}, \ and\ \bibinfo {author} {\bibfnamefont {Y.}~\bibnamefont {Hatsugai}},\ }\href@noop {} {\bibfield  {journal} {\bibinfo  {journal} {Scientific reports}\ }\textbf {\bibinfo {volume} {12}},\ \bibinfo {pages} {1} (\bibinfo {year} {2022})}\BibitemShut {NoStop}%
    \bibitem [{\citenamefont {Dobrinevski}\ and\ \citenamefont {Frey}(2012)}]{dobrinevski2012extinction}%
      \BibitemOpen
      \bibfield  {author} {\bibinfo {author} {\bibfnamefont {A.}~\bibnamefont {Dobrinevski}}\ and\ \bibinfo {author} {\bibfnamefont {E.}~\bibnamefont {Frey}},\ }\href@noop {} {\bibfield  {journal} {\bibinfo  {journal} {Physical Review E}\ }\textbf {\bibinfo {volume} {85}},\ \bibinfo {pages} {051903} (\bibinfo {year} {2012})}\BibitemShut {NoStop}%
    \bibitem [{\citenamefont {Knebel}\ \emph {et~al.}(2013)\citenamefont {Knebel}, \citenamefont {Kr{\"u}ger}, \citenamefont {Weber},\ and\ \citenamefont {Frey}}]{knebel2013coexistence}%
      \BibitemOpen
      \bibfield  {author} {\bibinfo {author} {\bibfnamefont {J.}~\bibnamefont {Knebel}}, \bibinfo {author} {\bibfnamefont {T.}~\bibnamefont {Kr{\"u}ger}}, \bibinfo {author} {\bibfnamefont {M.~F.}\ \bibnamefont {Weber}}, \ and\ \bibinfo {author} {\bibfnamefont {E.}~\bibnamefont {Frey}},\ }\href@noop {} {\bibfield  {journal} {\bibinfo  {journal} {Physical review letters}\ }\textbf {\bibinfo {volume} {110}},\ \bibinfo {pages} {168106} (\bibinfo {year} {2013})}\BibitemShut {NoStop}%
    \bibitem [{\citenamefont {Knebel}\ \emph {et~al.}(2020)\citenamefont {Knebel}, \citenamefont {Geiger},\ and\ \citenamefont {Frey}}]{knebel2020topological}%
      \BibitemOpen
      \bibfield  {author} {\bibinfo {author} {\bibfnamefont {J.}~\bibnamefont {Knebel}}, \bibinfo {author} {\bibfnamefont {P.~M.}\ \bibnamefont {Geiger}}, \ and\ \bibinfo {author} {\bibfnamefont {E.}~\bibnamefont {Frey}},\ }\href@noop {} {\bibfield  {journal} {\bibinfo  {journal} {Physical Review Letters}\ }\textbf {\bibinfo {volume} {125}},\ \bibinfo {pages} {258301} (\bibinfo {year} {2020})}\BibitemShut {NoStop}%
    \bibitem [{\citenamefont {Yoshida}\ \emph {et~al.}(2021)\citenamefont {Yoshida}, \citenamefont {Mizoguchi},\ and\ \citenamefont {Hatsugai}}]{yoshida2021chiral}%
      \BibitemOpen
      \bibfield  {author} {\bibinfo {author} {\bibfnamefont {T.}~\bibnamefont {Yoshida}}, \bibinfo {author} {\bibfnamefont {T.}~\bibnamefont {Mizoguchi}}, \ and\ \bibinfo {author} {\bibfnamefont {Y.}~\bibnamefont {Hatsugai}},\ }\href@noop {} {\bibfield  {journal} {\bibinfo  {journal} {Physical Review E}\ }\textbf {\bibinfo {volume} {104}},\ \bibinfo {pages} {025003} (\bibinfo {year} {2021})}\BibitemShut {NoStop}%
    \bibitem [{\citenamefont {Umer}\ and\ \citenamefont {Gong}(2022)}]{umer2022topologically}%
      \BibitemOpen
      \bibfield  {author} {\bibinfo {author} {\bibfnamefont {M.}~\bibnamefont {Umer}}\ and\ \bibinfo {author} {\bibfnamefont {J.}~\bibnamefont {Gong}},\ }\href {\doibase 10.1103/PhysRevB.106.L241403} {\bibfield  {journal} {\bibinfo  {journal} {Phys. Rev. B}\ }\textbf {\bibinfo {volume} {106}},\ \bibinfo {pages} {L241403} (\bibinfo {year} {2022})}\BibitemShut {NoStop}%
    \bibitem [{\citenamefont {Scheibner}\ \emph {et~al.}(2020)\citenamefont {Scheibner}, \citenamefont {Irvine},\ and\ \citenamefont {Vitelli}}]{scheibner2020non}%
      \BibitemOpen
      \bibfield  {author} {\bibinfo {author} {\bibfnamefont {C.}~\bibnamefont {Scheibner}}, \bibinfo {author} {\bibfnamefont {W.~T.}\ \bibnamefont {Irvine}}, \ and\ \bibinfo {author} {\bibfnamefont {V.}~\bibnamefont {Vitelli}},\ }\href@noop {} {\bibfield  {journal} {\bibinfo  {journal} {Physical Review Letters}\ }\textbf {\bibinfo {volume} {125}},\ \bibinfo {pages} {118001} (\bibinfo {year} {2020})}\BibitemShut {NoStop}%
    \bibitem [{\citenamefont {Fruchart}\ \emph {et~al.}(2021)\citenamefont {Fruchart}, \citenamefont {Hanai}, \citenamefont {Littlewood},\ and\ \citenamefont {Vitelli}}]{fruchart2021non}%
      \BibitemOpen
      \bibfield  {author} {\bibinfo {author} {\bibfnamefont {M.}~\bibnamefont {Fruchart}}, \bibinfo {author} {\bibfnamefont {R.}~\bibnamefont {Hanai}}, \bibinfo {author} {\bibfnamefont {P.~B.}\ \bibnamefont {Littlewood}}, \ and\ \bibinfo {author} {\bibfnamefont {V.}~\bibnamefont {Vitelli}},\ }\href@noop {} {\bibfield  {journal} {\bibinfo  {journal} {Nature}\ }\textbf {\bibinfo {volume} {592}},\ \bibinfo {pages} {363} (\bibinfo {year} {2021})}\BibitemShut {NoStop}%
    \bibitem [{\citenamefont {Yu}\ and\ \citenamefont {Zeng}(2022)}]{yu2022giant}%
      \BibitemOpen
      \bibfield  {author} {\bibinfo {author} {\bibfnamefont {T.}~\bibnamefont {Yu}}\ and\ \bibinfo {author} {\bibfnamefont {B.}~\bibnamefont {Zeng}},\ }\href@noop {} {\bibfield  {journal} {\bibinfo  {journal} {Physical Review B}\ }\textbf {\bibinfo {volume} {105}},\ \bibinfo {pages} {L180401} (\bibinfo {year} {2022})}\BibitemShut {NoStop}%
    \bibitem [{\citenamefont {Zeng}\ and\ \citenamefont {Yu}(2023)}]{zeng2023radiation}%
      \BibitemOpen
      \bibfield  {author} {\bibinfo {author} {\bibfnamefont {B.}~\bibnamefont {Zeng}}\ and\ \bibinfo {author} {\bibfnamefont {T.}~\bibnamefont {Yu}},\ }\href@noop {} {\bibfield  {journal} {\bibinfo  {journal} {Physical Review Research}\ }\textbf {\bibinfo {volume} {5}},\ \bibinfo {pages} {013003} (\bibinfo {year} {2023})}\BibitemShut {NoStop}%
    \bibitem [{\citenamefont {Franca}\ \emph {et~al.}(2022)\citenamefont {Franca}, \citenamefont {K{\"o}nye}, \citenamefont {Hassler}, \citenamefont {van~den Brink},\ and\ \citenamefont {Fulga}}]{franca2022non}%
      \BibitemOpen
      \bibfield  {author} {\bibinfo {author} {\bibfnamefont {S.}~\bibnamefont {Franca}}, \bibinfo {author} {\bibfnamefont {V.}~\bibnamefont {K{\"o}nye}}, \bibinfo {author} {\bibfnamefont {F.}~\bibnamefont {Hassler}}, \bibinfo {author} {\bibfnamefont {J.}~\bibnamefont {van~den Brink}}, \ and\ \bibinfo {author} {\bibfnamefont {C.}~\bibnamefont {Fulga}},\ }\href@noop {} {\bibfield  {journal} {\bibinfo  {journal} {Physical Review Letters}\ }\textbf {\bibinfo {volume} {129}},\ \bibinfo {pages} {086601} (\bibinfo {year} {2022})}\BibitemShut {NoStop}%
    \bibitem [{\citenamefont {Zhang}\ \emph {et~al.}(2024)\citenamefont {Zhang}, \citenamefont {Zhang}, \citenamefont {Zhao},\ and\ \citenamefont {Lee}}]{zhang2022observationb}%
      \BibitemOpen
      \bibfield  {author} {\bibinfo {author} {\bibfnamefont {X.}~\bibnamefont {Zhang}}, \bibinfo {author} {\bibfnamefont {B.}~\bibnamefont {Zhang}}, \bibinfo {author} {\bibfnamefont {W.}~\bibnamefont {Zhao}}, \ and\ \bibinfo {author} {\bibfnamefont {C.~H.}\ \bibnamefont {Lee}},\ }\href@noop {} {\bibfield  {journal} {\bibinfo  {journal} {SciPost Physics}\ }\textbf {\bibinfo {volume} {16}},\ \bibinfo {pages} {002} (\bibinfo {year} {2024})}\BibitemShut {NoStop}%
    \bibitem [{\citenamefont {Geng}\ \emph {et~al.}(2023)\citenamefont {Geng}, \citenamefont {Wei}, \citenamefont {Zou}, \citenamefont {Sheng}, \citenamefont {Chen},\ and\ \citenamefont {Xing}}]{geng2023nonreciprocal}%
      \BibitemOpen
      \bibfield  {author} {\bibinfo {author} {\bibfnamefont {H.}~\bibnamefont {Geng}}, \bibinfo {author} {\bibfnamefont {J.~Y.}\ \bibnamefont {Wei}}, \bibinfo {author} {\bibfnamefont {M.~H.}\ \bibnamefont {Zou}}, \bibinfo {author} {\bibfnamefont {L.}~\bibnamefont {Sheng}}, \bibinfo {author} {\bibfnamefont {W.}~\bibnamefont {Chen}}, \ and\ \bibinfo {author} {\bibfnamefont {D.~Y.}\ \bibnamefont {Xing}},\ }\href {\doibase 10.1103/PhysRevB.107.035306} {\bibfield  {journal} {\bibinfo  {journal} {Phys. Rev. B}\ }\textbf {\bibinfo {volume} {107}},\ \bibinfo {pages} {035306} (\bibinfo {year} {2023})}\BibitemShut {NoStop}%
    \bibitem [{\citenamefont {Ghaemi-Dizicheh}\ and\ \citenamefont {Schomerus}(2021)}]{PhysRevA.104.023515}%
      \BibitemOpen
      \bibfield  {author} {\bibinfo {author} {\bibfnamefont {H.}~\bibnamefont {Ghaemi-Dizicheh}}\ and\ \bibinfo {author} {\bibfnamefont {H.}~\bibnamefont {Schomerus}},\ }\href {\doibase 10.1103/PhysRevA.104.023515} {\bibfield  {journal} {\bibinfo  {journal} {Phys. Rev. A}\ }\textbf {\bibinfo {volume} {104}},\ \bibinfo {pages} {023515} (\bibinfo {year} {2021})}\BibitemShut {NoStop}%
    \bibitem [{\citenamefont {Schomerus}(2022)}]{schomerus2022fundamental}%
      \BibitemOpen
      \bibfield  {author} {\bibinfo {author} {\bibfnamefont {H.}~\bibnamefont {Schomerus}},\ }\href {\doibase 10.1103/PhysRevA.106.063509} {\bibfield  {journal} {\bibinfo  {journal} {Phys. Rev. A}\ }\textbf {\bibinfo {volume} {106}},\ \bibinfo {pages} {063509} (\bibinfo {year} {2022})}\BibitemShut {NoStop}%
    \end{thebibliography}
%
    
\end{document}